\documentclass[12pt]{iopart}

\usepackage{iopams}  
\usepackage{color}
\usepackage[dvipsnames]{xcolor} 
\usepackage{tikz}
\usetikzlibrary{decorations.pathmorphing,patterns} 
\usepackage{psfrag}
\usepackage{graphicx}
\usepackage{stackrel}
\usepackage{amsfonts}
\usepackage{bm}          
\usepackage{amssymb}
\usepackage{perpage}
\MakePerPage{footnote}

\newcommand{\be}{\begin{equation}}
\newcommand{\ee}{\end{equation}}
\newcommand{\ba}{\begin{eqnarray}}
\newcommand{\ea}{\end{eqnarray}}
\newcommand{\nn}{\nonumber}

\begin{document}

\title[High-precision percolation thresholds and Potts-model critical manifolds]
{High-precision percolation thresholds and Potts-model critical manifolds from graph polynomials}

\author{Jesper Lykke Jacobsen$^{1,2}$}
\address{${}^1$LPTENS, \'Ecole Normale Sup\'erieure, 24 rue Lhomond, 75231
Paris, France}
\address{${}^2$Universit\'e Pierre et Marie Curie, 4 place Jussieu, 75252 Paris,
France}

\eads{\mailto{jesper.jacobsen@ens.fr}}

\begin{abstract}

The critical curves of the $q$-state Potts model can be determined exactly
for regular two-dimensional lattices $G$ that are of the three-terminal type.
This comprises the square, triangular, hexagonal and bowtie lattices.
Jacobsen and Scullard have defined a graph polynomial $P_B(q,v)$ that gives
access to the critical manifold for general lattices. It
depends on a finite repeating part of the lattice, called the basis $B$, and
its real roots in the temperature variable $v = {\rm e}^K - 1$ provide
increasingly accurate approximations to the
critical manifolds upon increasing the size of $B$.
Using transfer matrix techniques, these authors computed $P_B(q,v)$ for large bases
(up to 243 edges),
obtaining determinations of the ferromagnetic critical point $v_{\rm c} > 0$ for the
$(4,8^2)$, kagome, and $(3,12^2)$ lattices to a precision (of the order $10^{-8}$)
slightly superior to that of the best available Monte Carlo simulations.

In this paper we describe a more efficient transfer matrix approach to the
computation of $P_B(q,v)$ that relies on a formulation within the periodic Temperley-Lieb
algebra. This makes possible computations for substantially larger
bases (up to 882 edges), and the precision on $v_{\rm c}$ is hence taken to the
range $10^{-13}$. We further show that a large variety of regular
lattices can be cast in a form suitable for this approach. This
includes all Archimedean lattices, their duals and their medials. For all these
lattices we tabulate high-precision estimates of the bond percolation thresholds
$p_{\rm c}$ and Potts critical points $v_{\rm c}$. We also trace and discuss the
full Potts critical manifold in the $(q,v)$ plane, paying special attention to the
antiferromagnetic region $v < 0$. Finally, we adapt the technique to site percolation as well,
and compute the polynomials $P_B(p)$ for certain Archimedean and dual
lattices (those having only cubic and quartic vertices), using very large bases (up to 243 vertices).
This produces the site percolation thresholds $p_{\rm c}$ to a precision of the order $10^{-9}$.

\end{abstract}

\noindent

\section{Introduction}

The notion of exact solvability plays a prominent role within the field of two-dimensional
statistical physics. The exact solutions of a certain number of lattice models---such as the
Ising model \cite{Onsager44}, dimer coverings \cite{Kasteleyn61}, the six-vertex \cite{Lieb67}
and eight-vertex models \cite{Baxter72}, and the Potts model \cite{Baxter73}---have
served as milestones by which the advance of the field
can be judged, and as benchmarks for analytical and numerical methods.

In this context, the question of what makes a non-trivial model exactly solvable is
obviously of high importance. One may ask what role does the choice of lattice play
in the solvability. All models cited were initially solved on the simplest possible,
square lattice. It quickly turned out that the solutions of the Ising
and dimer models could be extended to essentially any regular two-dimensional
lattice, whereas vertex and Potts models have only been solved on a few other
simple lattices. It is of interest to solve these models, or find accurate approximate
solutions, on more general lattices, such as the remaining Archimedean lattices.

We here examine this issue in the context of the $q$-state Potts model \cite{Potts52}.
Given a connected graph $G=(V,E)$ with vertex set $V$ and edge
set $E$, its partition function $Z$ is can be defined as \cite{FK1972}
\be
 Z = \sum_{A \subseteq E} v^{|A|} q^{k(A)} \,,
 \label{FK_repr}
\ee
where $|A|$ denotes the number of edges in the subset $A$, and $k(A)$
is the number of connected components (including isolated vertices)
in the subgraph $G_A = (V,A)$.
The temperature variable is denoted $v = {\rm e}^K - 1$, where $K$ is the reduced interaction
energy between adjacent $q$-component spins.
In the representation (\ref{FK_repr}) we shall formally allow both $q$ and $v$
to take arbitrary real values.

A first aspect to be addressed when solving the Potts model defined on some
lattice $G$ is the determination of the values, for any given $q$, of the
temperature $v$ where a phase transition takes place. We shall not be
concerned here with the nature (order) of phase transitions, and simply refer to
the set of transition temperatures in the real $(q,v)$ plane as the {\em critical
manifold}. The critical manifold has only been determined exactly 
when $G$ is the square \cite{Baxter73}, triangular \cite{Baxter78}, hexagonal
(the dual of the former), and bowtie \cite{SJbowtie} lattices, as well as certain
decorations of these lattices \cite{Wu06}. More precisely, the solvable lattices
are all of the three-terminal type, that is, they are regular arrangements of triangles, each
consisting of three boundary spins (or terminals) and an arbitrary number of internal spins.
The interactions inside each triangle can take any form, but distinct triangles
only interact through the terminals. The triangles can be disposed as the
up-pointing triangles in a triangular lattice \cite{WuLin80}, or in a bowtie
pattern \cite{SJbowtie}.

By contrast, lattices of the four-terminal type do not appear to be exactly solvable. Wu has
however shown that in a number of cases their critical manifolds can be well
approximated by a homogeneity assumption \cite{Wu79}. Very recently, Jacobsen and
Scullard have defined a graph polynomial $P_B(q,v)$ that depends on a
finite repeating part of the lattice, called the basis $B$, and reduces to Wu's expressions
for the smallest possible choices of $B$ \cite{Jacobsen12}. The real roots of $P_B(q,v)$ in the
temperature variable $v = {\rm e}^K - 1$ provide increasingly accurate
approximations to the critical manifolds upon increasing the size of $B$.
Using transfer matrix techniques, these authors computed $P_B(q,v)$ for large bases
(up to 243 edges),
obtaining determinations of the ferromagnetic critical point $v_{\rm c} > 0$ for the
four-eight, kagome, and three-twelve lattices to a precision of the order $10^{-8}$
\cite{Jacobsen13}, slightly superior to that of the best available Monte Carlo simulations.

For $q=1$ the polynomial $P_B(q,v)$ reduces to the bond percolation polynomial introduced by Scullard
and Ziff \cite{ScullardZiff06,ScullardZiff08,ScullardZiff10} and studied further in
\cite{Scullard11,Scullard11-2,Scullard12,SJ12}.

The goal of this paper is twofold. First, we extend the set of lattices that can be studied to all
Archimedean lattices, their dual (Laves) lattices, as well
as their medial (or surrounding) lattices.%
\footnote{The Archimedean lattices were previously considered in the special case $q=1$ by Scullard \cite{Scullard11-2,Scullard12},
using bases of size up to $36$ edges. These bases include our $n=2$ square bases for the kagome \cite{Scullard11-2},
four-eight, three-twelve, snub square, snub hexagonal and ruby lattices \cite{Scullard12}.}
Second, we describe a transfer matrix approach
to the computation of $P_B(q,v)$ that is more efficient than the one given in Ref.~\cite{Jacobsen13}.
On a technical level, this is done by representing all these lattices in a particular four-terminal
form, and writing the $\check{R}$-matrix of the fundamental building blocks in terms of operators acting
within the periodic Temperley-Lieb algebra. From a practical point of view, this makes possible
computations for substantially larger bases (up to 882 edges) than those used in Ref.~\cite{Jacobsen13}.
The precision on $v_{\rm c}$ is hence taken to the range $10^{-13}$, far ahead of any
competing perturbative or numerical technique.

We further compute graph polynomials $P_B(p)$ for site percolation problems on several
different lattices.%
\footnote{This generalises the earlier work \cite{SJ12} to several new lattices and to considerably larger bases.}
For practical reasons, we limit ourselves in this case to Archimedean and dual lattices
having only cubic and quartic vertices (i.e., no vertex has degree $\ge 5$)---but it will become clear that
this is not an essential limitation of the method. It turns out that the estimates for the percolation threshold $p_{\rm c}$
do not converge as fast as in the case of bond percolation, or for $v_{\rm c}$ in the Potts model.
Accordingly we obtain $p_{\rm c}$ to a precision which is typically of the order $10^{-8}$, and sometimes
even $10^{-9}$. This precision is however still superior to that of the best simulation results.

For the exactly solvable lattices, $P_B(q,v)$ was found to factorise in a number of cases \cite{Jacobsen12,Jacobsen13},
shedding a small factor that corresponds to the exactly known critical curve(s). But it was
also observed \cite{Jacobsen12,Jacobsen13} that the remaining, large factor contains pertinent 
information about the phase diagram in the region $v < 0$. We continue these investigations here,
by tracing the full Potts critical manifold in the $(q,v)$ plane, using the larger basis and the
substantially larger selection of lattices now at hand.

Throughout the paper the estimates for $v_{\rm c}$ coming from finite bases $B$ and their extrapolation
to the thermodynamic limit will be presented, for each lattice, in table form for easy perusal.
However, to give the reader a very concrete idea about the precision attained by the present method,
we now briefly present a few sample results. 
\begin{itemize}
\item For the bond percolation threshold on the three-twelve
lattice our graph polynomial result and the currently best available numerical calculation
(diagonalisation of the transfer matrix \cite{Ding10}) read respectively
\begin{equation}
 p_{\rm c} = \left \lbrace
 \begin{array}{ll}
 0.740\, 420\, 798\, 847\, 4(7) & \mbox{(Graph polynomial, this work)} \\
 0.740\, 420\, 800(2) & \mbox{(Graph polynomial, Ref.~\cite{SJ12})} \\
 0.740 \,420\, 77 (2) & \mbox{(Transfer matrices)}
 \end{array} \right.
\end{equation}
where the number in parentheses is the error bar on the last given digit.
We have also shown for comparison the best graph polynomial result \cite{SJ12}
prior to the improvements presented in this paper.
\item For the Ising model, $q=2$, it was previously observed \cite{Jacobsen12,Jacobsen13} that
$P_B(q,v)$ invariable factorises. Our results are thus exact in that case. For the Archimedean
lattices and their duals, all our results coincide with those obtained in Ref.~\cite{Codello10}
from the Feynman-Vdovichenko combinatorial approach. The Ising case therefore strongly
supports the correctness of the method for those lattices. However, for some of the medial
lattices our Ising results are new (although we believe they could easily be derived, e.g., using
the methods of \cite{Codello10}).
\item For the $q=3$ state Potts model on the kagome lattice we can
compare with the best available series estimate (67-term low-temperature series \cite{Jensen97}):
\begin{equation}
 v_{\rm c} = \left \lbrace
 \begin{array}{ll}
 1.876\,459\,574\,2(1) & \mbox{(Graph polynomial)} \\
 1.876 \,46(5) & \mbox{(Series expansion)}
 \end{array} \right.
\end{equation}
Our precision for $q=4$ is similar and notably does not suffer from the logarithmic corrections
usually associated with the presence of a marginally irrelevant operator.
\item For the site percolation threshold on the square lattice, the graph polynomial has a more
modest performance, but the precision is still better than that of the best available numerical calculation
(Monte Carlo simulation \cite{FengDengBlote08}):
\begin{equation}
 p_{\rm c} = \left \lbrace
 \begin{array}{ll}
 0.592\, 746\, 01(2) & \mbox{(Graph polynomial)} \\
 0.592 \,746\, 05(3) & \mbox{(Monte Carlo)}
 \end{array} \right.
\end{equation}
\end{itemize}

The organisation of the paper is as follows. In section~\ref{sec:lattices} we present the lattices to
be studied and introduce some useful terminology. The technical centrepiece  of this work is
section~\ref{sec:TL}, where we discuss how the graph polynomial $P_B(q,v)$ can be expressed
in terms of the periodic Temperley-Lieb algebra. Some readers might want to skip that section at
a first reading and go straight to the results. Those are presented in section~\ref{sec:res-archi}
for the Potts model and bond percolation on the Archimedean lattices, and in
sections~\ref{sec:res-dual}--\ref{sec:res-medial} for the same models on the dual and medial
lattices. In section~\ref{sec:site} we explain how to compute the graph polynomial $P_B(p)$ for
site percolation and we give results for the Archimedean and dual lattices with only cubic and quartic
vertices. Finally, section~\ref{sec:disc} contains the discussion and some concluding remarks.

\section{Archimedean lattices, their duals, and their medials}
\label{sec:lattices}

\begin{figure}
\begin{center}
 \includegraphics[width=12cm]{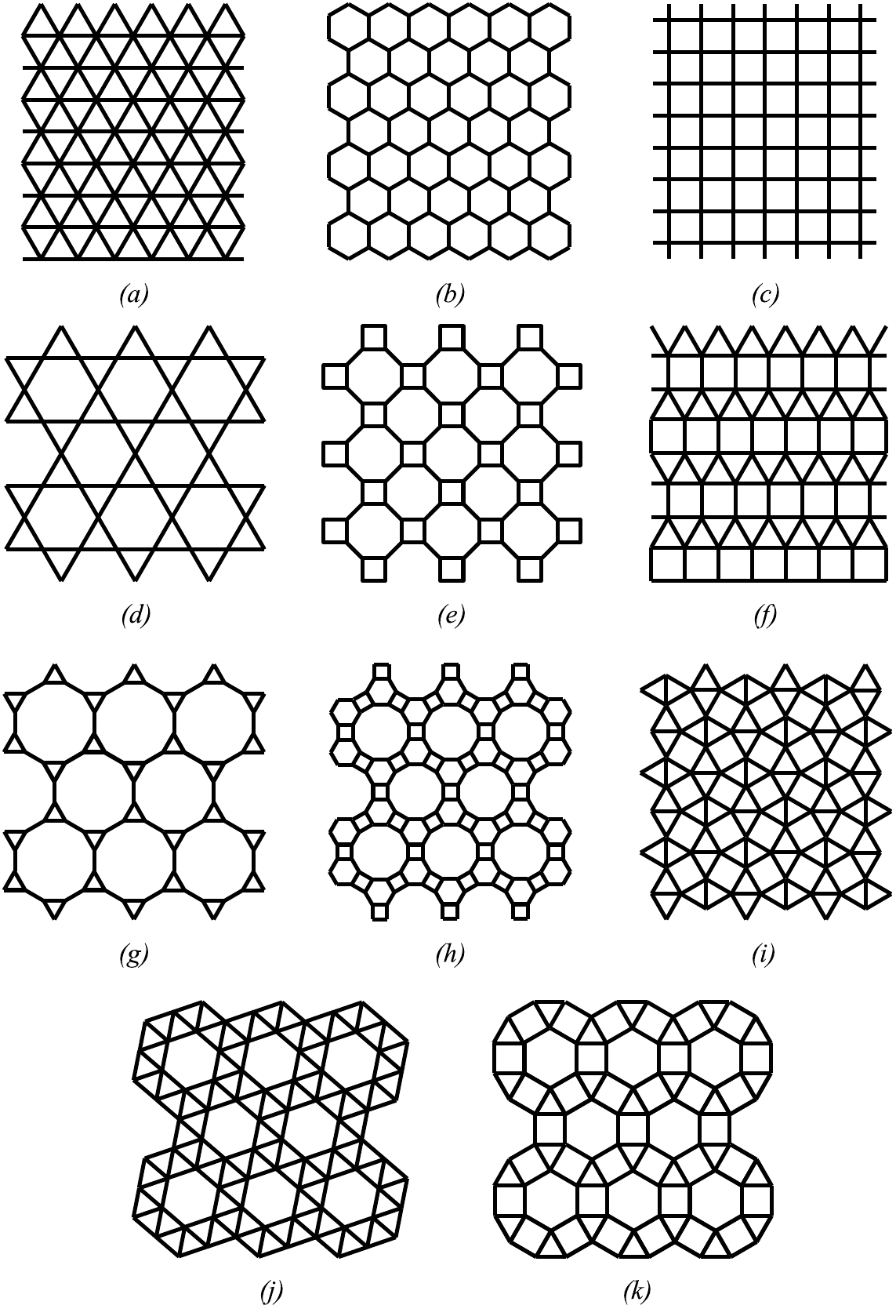}
 \caption{The eleven Archimedean lattices. Their names are given in Table~\ref{tab:archi}.}
 \label{fig:archi}
\end{center}
\end{figure}

The eleven Archimedean lattices are shown in Figure~\ref{fig:archi}. By definition, an Archimedean
lattice is such that each vertex is surrounded by the same types of faces, appearing in the
same cyclic order. For instance, each vertex of the lattice shown in Figure~\ref{fig:archi}(k) is surrounded by a triangle, a square,
a hexagon, and another square, so this lattice is called $(3,4,6,4)$ in the notation of Gr\"unbaum and
Shephard \cite{GrunbaumShephard87}. The corresponding dual lattice is denoted $D(3,4,6,4)$.
By definition, the dual lattice is obtained from the primal one by replacing vertices by faces,
and vice versa, and by replacing edges by intersecting dual edges. It follows in particular that an Archimedean
dual consists of identical faces (or tiles) and that the vertices bordering each tile have the degrees specified
by the labels.  So $D(3,4,6,4)$ is a quadrangulation (since there are four labels), and each quadrangle is bordered by vertices of degrees
3, 4, 6, and 4.

The Archimedean lattices
and their duals have convenient nicknames, shown in Table~\ref{tab:archi}, that we shall often
use throughout this work. For instance, $(3,4,6,4)$ is known as the ruby lattice.

Note that the square lattice is selfdual, while the triangular and hexagonal
lattices are mutually dual. In other words, $D(4^4) = (4^4)$ and $D(3^6) = (6^3)$.
These three are the only three-terminal (hence exactly solvable) lattices.
We show in this paper that the remaining eight lattices are of the four-terminal type, as required
by our specific transfer matrix setup.

The graph polynomial for the Potts model on the dual lattice is obtained from its primal
counterpart $P_B(q,v)$ by replacing $v$ by $v^* = q/v$, and multiplying with an (unimportant) overall factor.
This connexion is identical to the well-known duality relation \cite{Potts52,WuWang76} that
relates the partition function of the Potts model on the primal and dual lattices.
For site percolation we do not have such a duality relation. Moreover, the notion of three-terminal
and four-terminal lattices changes slightly for site percolation, since the terminals are now midpoints
of edges; this will be discussed in section~\ref{sec:site}.

For the Potts model we also consider the medial lattices, obtained from the primal lattices
by placing vertices on the midpoints of edges, and placing medial edges cyclically around
the primal faces \cite{BaxterKellandWu76}. Medial lattices are also known as surrounding lattices.
It follows that the medial lattice has faces corresponding to each of the primal faces, and to each
of the primal vertices, with the same degree. Moreover, a pair of mutually dual lattices
have the same medial. We denote medials by the letter ${\cal M}$ so that, for example,
the medial of the ruby lattice is denoted ${\cal M}(3,4,6,4)$.

The medial lattice ${\cal M}(G)$ should not be confused with the covering lattice ${\cal C}(G)$ (also called line graph).
Site percolation on ${\cal C}(G)$ is equivalent to bond percolation on $G$ \cite{EssamFisher61,SykesEssam64}.
When the primal lattice $G$ is planar, ${\cal M}(G)$ is also planar; but ${\cal C}(G)$ will in general be non-planar,
except if $G$ is a cubic lattice.%
\footnote{For example, if $G$ is the hexagonal lattice, ${\cal C}(G) = {\cal M}(G)$ is the kagome lattice.
Site percolation on the kagome lattice is thus equivalent to bond percolation on the hexagonal lattice;
both problems turn out to be exactly solvable.}
We have ${\cal M}(G) = {\cal M}(D(G))$, but the same property does not hold
for covering lattices, unless $G$ is selfdual.
In this paper we only deal with planar lattices, and we shall not consider covering lattices any further.
Also, we shall not consider site percolation problems on medial lattices, although these are independent
of bond percolation problems (except when $G$ is cubic).

Note that the medial of the square lattice is itself a square lattice, ${\cal M}(4^4) = (4^4)$.
The medial of the triangular (and of its dual, hexagonal) lattice is the kagome lattice,
itself an Archimedean lattice. In other words, ${\cal M}(3^6) = {\cal M}(6^3) =  (3,6,3,6)$.
The medial of the kagome lattice is the ruby lattice, ${\cal M}(3,6,3,6) = (3,4,6,4)$.
The remaining seven medial lattices are not Archimedean. Most of them are so-called two-uniform lattices, i.e.,
they contain two different classes of vertices with distinct face environments, while others yet are
three-uniform.

\begin{table}
\begin{center}
 \begin{tabular}{l|ll|ll}
                              & Lattice & Notation & Dual lattice & Notation \\ \hline
 (a) & Triangular & $(3^6)$ & Hexagonal & $(6^3)$ \\
 (b) & Hexagonal & $(6^3)$ & Triangular & $(3^6)$ \\
 (c) & Square & $(4^4)$ & Square & $(4^4)$ \\
 (d) & Kagome & $(3,6,3,6)$ & Dice & $D(3,6,3,6)$ \\
 (e) & Four-eight & $(4,8^2)$ & Union-jack & $D(4,8^2)$ \\
 (f) & Frieze & $(3^3,4^2)$ & Frieze dual & $D(3^3,4^2)$ \\
 (g) & Three-twelve & $(3,12^2)$ & Asanoha & $D(3,12^2)$ \\
 (h) & Cross & $(4,6,12)$ & Bisected hexagonal & $D(4,6,12)$ \\
 (i) & Snub square & $(3^2,4,3,4)$ & Cairo pentagonal & $D(3^2,4,3,4)$ \\
 (j) &  Snub hexagonal & $(3^4,6)$ & Daisy & $D(3^4,6)$ \\
 (k) & Ruby & $(3,4,6,4)$ & Ruby dual & $D(3,4,6,4)$ \\
 \end{tabular}
 \caption{Nomenclature of the Archimedean lattices and their duals (Laves lattices).
  The labels (a)--(k) refer to Figure~\ref{fig:archi}.
 The notation is that of Gr\"unbaum and Shephard \cite{GrunbaumShephard87}.}
 \label{tab:archi}
\end{center}
\end{table}

\section{Graph polynomial and the Temperley-Lieb algebra}
\label{sec:TL}

In this section we shall only be concerned
with the Potts model (\ref{FK_repr}) and the special case of bond percolation, which is obtained
by setting $q=1$ and choosing the probability of an open bond as $p = \frac{v}{1+v}$. Site percolation
requires a few modifications of the general setup and will be discussed in section~\ref{sec:site}.

\subsection{Bases and embeddings}

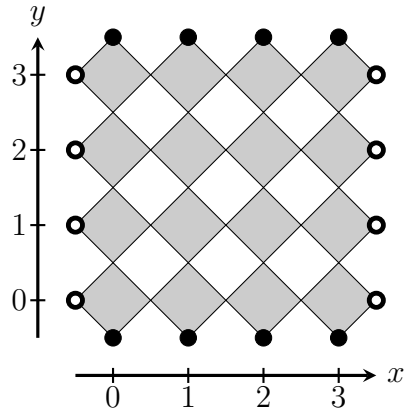
\begin{figure}
\begin{center}

\begin{tikzpicture}[scale=1.0,>=stealth]
\foreach \xpos in {0,1,2,3}
\foreach \ypos in {0,1,2,3}
 \fill[black!20] (\xpos+0.5,\ypos) -- (\xpos+1,\ypos+0.5) -- (\xpos+0.5,\ypos+1) -- (\xpos,\ypos+0.5) -- cycle;
\foreach \xpos in {0,1,2,3}
\foreach \ypos in {0,1,2,3}
 \draw[black] (\xpos+0.5,\ypos) -- (\xpos+1,\ypos+0.5) -- (\xpos+0.5,\ypos+1) -- (\xpos,\ypos+0.5) -- cycle;

\foreach \xpos in {0,1,2,3}
{
 \draw[fill] (\xpos+0.5,0) circle(0.6ex);
 \draw[fill] (\xpos+0.5,4) circle(0.6ex);
}

\foreach \ypos in {0,1,2,3}
{
 \draw[line width=2pt] (0,\ypos+0.5) circle(0.5ex);
 \draw[line width=2pt] (4,\ypos+0.5) circle(0.5ex);
 \draw[fill,white,line width=0pt] (0,\ypos+0.5) circle(0.3ex);
 \draw[fill,white,line width=0pt] (4,\ypos+0.5) circle(0.3ex);
}

\draw[very thick,->] (0,-0.5)--(4,-0.5);
\draw (4,-0.5) node[right] {$x$};
\foreach \xpos in {0,1,2,3}
{
 \draw[thick] (\xpos+0.5,-0.6)--(\xpos+0.5,-0.4);
 \draw (\xpos+0.5,-0.5) node[below] {$\xpos$};
}

\draw[very thick,->] (-0.5,0)--(-0.5,4);
\draw (-0.5,4) node[above] {$y$};
\foreach \ypos in {0,1,2,3}
{
 \draw[thick] (-0.6,\ypos+0.5)--(-0.4,\ypos+0.5);
 \draw (-0.5,\ypos+0.5) node[left] {$\ypos$};
}
 
\end{tikzpicture}
 \caption{Square basis of size $n \times n$ with $n=4$. The horizontal (resp.\ vertical) terminals of the basis are
 shown as white (resp.\ black) circles. The grey squares on the checkerboard are identified by their $(x,y)$ coordinates and
 can contain any arrangement of lattice edges. The white square are either empty, or may contain a diagonal horizontal edge.}
  \label{fig:square-basis}
\end{center}
\end{figure}

The graph polynomial $P_B(q,v)$ for the Potts model depends on a finite part of the lattice,
called the basis $B$, that generates the infinite lattice $G$ by an appropriate set of translations,
called the embedding \cite{Jacobsen13}.
Each regular lattice $G$ admits an infinite number of choices for $B$. We shall be interested
in the simplest possible family of $B$, called {\em square bases} in Ref.~\cite{Jacobsen13}.
They have a checkerboard structure, shown in Figure~\ref{fig:square-basis}, consisting of alternating
grey and white squares.

The grey squares in Figure~\ref{fig:square-basis} can contain any arrangement of lattice edges and
internal vertices. Two adjacent grey squares meet in a common vertex; we shall call such shared
vertices the {\em terminals} of the grey square. The lattices $G$ which can be represented as in Figure~\ref{fig:square-basis}
are called four-terminal lattices. The Potts model on such $G$ is not in general exactly
solvable, as discussed in the Introduction. For the simplest lattices, all the grey squares contain
the same arrangement of edges and internal vertices. For more complicated lattices
it is necessary to let the structure of the grey squares depend on the coordinates $(x,y)$ with
some periodicity.

A rectangular (resp.\ square) basis is an array of $n \times m$ (resp.\ $n \times n$) grey squares.
We shall almost exclusively be interested in square bases. However, for a few lattices we shall need
to decorate the grey squares with a periodicity that is different in the $x$ and $y$ directions, and
in those cases we might need rectangular bases in order to respect that periodicity. The transfer matrix
construction to be described below also allows for decorating (some of) the white squares by a
single horizontal edge. For simplicity, we shall still say that the corresponding lattices are of the
four-terminal type.

The accuracy of the critical manifold
determined by $P_B(q,v) = 0$ increases rapidly with $n$, but unfortunately the same is true for
the computational effort to compute $P_B(q,v)$. To be precise, the contraction-deletion algorithm
described in \cite{Jacobsen12} has time and memory requirements that grow exponentially in $n m$,
whereas for the transfer matrix algorithm of \cite{Jacobsen13} the growth is only exponential in
${\rm min}(n,m)$. The improved transfer matrix algorithm to be described here is again exponential
in ${\rm min}(n,m)$ but with a smaller growth constant, enabling us to access larger sizes.

The algorithm of \cite{Jacobsen12} was capable of computing $P_B(q,v)$ for basis with up to 36 edges.
Since the most interesting lattices have typically at least six edges per grey square, this means that
bases of size $2 \times 2$ or $2 \times 3$ could be handled. The limit of feasibility using the transfer matrix
of \cite{Jacobsen13} was improved to ${\rm min}(n,m) = 4$. In this work we further improve the transfer matrix
algorithm, putting $n \times n$ square bases with $n = 7$ within reach. To be precise, we compute exactly the
two-variable Potts polynomial $P_B(q,v)$ up to $n = 5$; the exact one-variable bond percolation
polynomial $P_B(q=1,v=\frac{p}{1-p})$ up to $n=6$; and roots in $v$ to 50-digit numerical precision of the
equation $P_B(q,v)=0$ for selected values of $q$ up to $n=7$.

The site percolation polynomials $P_B(p)$
can sometimes be computed for even larger $n$, namely up to $n=11$ for the square lattice, and even
$n=16$ for the ruby lattice (see section~\ref{sec:site}). For the site percolation problems we invariably
compute the exact polynomial $P_B(p)$, refraining from any additional gain that might have been obtained
by finding only the relevant root ($0 < p_{\rm c} < 1$) with finite numerical precision.

The vertices situated at a corner of a grey square in Figure~\ref{fig:square-basis}, and not shared between two
distinct grey squares, are called the {\em terminals} of the basis $B$. The embedding of $B$ into $G$ is defined
by gluing distinct copies of $B$ at the terminals.%
\footnote{Note the analogy to how distinct grey squares were glued at
their terminals in order to form the basis $B$.}
This can be done in a variety of ways, but in this work we
are only interested in the simplest possibility, called {\em straight embedding} in Ref.~\cite{Jacobsen13}.
It consists of simply translating the $n \times m$ basis horizontally through multiples of $n {\bf e}_x$,
and vertically through multiples of $m {\bf e}_y$, where ${\bf e}_x$ (resp.\ ${\bf e}_y$) is a unit vector in
the $x$-direction (resp.\ $y$-direction).

\subsection{Graph polynomial}

The graph polynomial $P_B(q,v)$ was initially defined from a deletion-contraction principle \cite{Jacobsen12}.
It was subsequently shown \cite{Jacobsen13} that it can be equivalently written as a linear combination of
conditioned partition functions, similar to (\ref{FK_repr}), defined on a graph which is equal to the basis $B$.
Consider first the partition function $Z$ of the Potts model defined on $B = (V,E)$, where we have imposed toroidal
boundary conditions, i.e., opposite terminals are identified both horizontally and vertically. We can decompose
\begin{equation}
 Z = Z_{\rm 2D} + Z_{\rm 1D} + Z_{\rm 0D} \,,
\end{equation}
where $Z_{\rm 0D}$ is the sum over edge subsets $A \subseteq E$ such that all connected components (clusters)
in the subgraph $G_A = (V,A)$ have trivial homotopy on the torus (i.e., are contractible to a point), and
$Z_{\rm 1D}$ is the sum over terms where there exists both a cluster and a dual cluster with non-trivial homotopy.
In simpler words, $Z_{\rm 2D}$ regroups the terms where there is a cluster that spans both spatial directions;
the terms in $Z_{\rm 1D}$ contain a cluster that spans only one, but not both, of the directions in space; and
in $Z_{\rm 0D}$ there are no spanning clusters.

In this setup the graph polynomial reads \cite{Jacobsen13}
\begin{equation}
 P_B(q,v) = Z_{\rm 2D} - q Z_{\rm 0D} \,.
 \label{PB_cluster}
\end{equation}
The term in $Z$ corresponding to each $A \subseteq E$ can be assigned to either $Z_{\rm 2D}$, $Z_{\rm 1D}$
or $Z_{\rm 0D}$ by using the Euler relation, as explained in \cite{Jacobsen13}. We shall come back to this technical
consideration in section~\ref{sec:gluing}.

\subsection{Loop model formulation}

We shall refer to (\ref{PB_cluster}) as the cluster representation of $P_B(q,v)$. Below we shall use an
equivalent formulation in terms of a loop model \cite{BaxterKellandWu76} defined on the medial
lattice ${\cal M}(B)$. To this end, consider the two possible states of an edge $e = (ij) \in E$ between to
adjacent vertices $i,j$ in $B$:
\begin{equation}
\begin{tikzpicture}[scale=0.7]
 \draw[blue,line width=3pt] (0,0)--(2,0);
 \draw[fill] (0,0) circle(0.6ex) node[left] {$i$};
 \draw[fill] (2,0) circle(0.6ex) node[right] {$j$};
 \draw[red,line width=1.5pt] (0,1)--(0.8,0.2) arc(225:315:2mm) -- (2,1);
 \draw[red,line width=1.5pt] (0,-1)--(0.8,-0.2) arc(135:45:2mm) -- (2,-1);

\begin{scope}[xshift=5cm]
 \draw[blue,line width=3pt,dashed] (0,0)--(2,0);
 \draw[fill] (0,0) circle(0.6ex) node[left] {$i$};
 \draw[fill] (2,0) circle(0.6ex) node[right] {$j$};
 \draw[red,line width=1.5pt] (0,1)--(0.8,0.2) arc(45:-45:2mm) -- (0,-1);
 \draw[red,line width=1.5pt] (2,1)--(1.2,0.2) arc(135:225:2mm) -- (2,-1);
\end{scope}
\end{tikzpicture}
\label{space-like-edge}
\end{equation}
In the left picture we have $e \in A$ and the edge is drawn as a thick blue line. The corresponding
loops (thin red lines) reflect off the edge $e$. In the right picture we have $e \notin A$ and the edge
is shown in dashed line style. The loops then cut through $e$, or equivalently, they reflect off the
corresponding dual edge $e^*$. In both cases ($e \in A$ and $e \notin A$) the loops separate
the clusters in $G_A = (V,A)$ from the dual clusters in $G_{A^*} = (V^*,A^*)$, where by definition
$A^*$ consists of the edges dual to those in $E \setminus A$.

In a transfer matrix formalism the lattice is built up, row by row, starting at the bottom and moving
towards the top. Each row is in turn built up, edge by edge, starting at the left and moving towards
the right. One can think of this as sweeping some imaginary $d-1$ dimensional surface over the
lattice in a number of discrete time steps.  We refer to this surface as the {\em time slice}. Note that
since we are in $d=2$ dimensions it is actually just a curve. More precisely, it is a horizontal line 
each time a row of the lattice has been completed, and a horizontal line with a kink when the row
is only partially completed. At each step, the partially built lattice (the portion below the time slice)
is characterised by the {\em connectivity state} of the clusters or loops intersecting the time slice, in a precise
way that makes it possible to compute the partition function---or, in our case, the conditional partition
functions entering in (\ref{PB_cluster})---in the transfer process. The possible configurations of the
edges ($e \in A$ and $e \notin A$) must be summed over, and each term in the sum induces a
definite operation on the connectivity states.

A detailed description of the transfer matrix formalism in the cluster representation was given
in \cite{Jacobsen13}. We here need the corresponding loop representation, describing the states
and operations on the thin red lines in (\ref{space-like-edge}). We begin with a brief outline,
deferring a more precise description of a number of important points to the following subsections.

For a planar graph one may use \cite{BaxterKellandWu76} the Euler relation to rewrite the partition
function (\ref{FK_repr}) in the loop representation as
\be
 Z = q^{|V|/2} \sum_{A \subseteq E} x^{|A|} n_{\rm loop}^{\ell(A)} \,,
 \label{loop_repr}
\ee
where $x = v / \sqrt{q}$, and $\ell(A)$ denotes the number of closed loops induced by the configuration $A$.
The loop fugacity is $n_{\rm loop} = \sqrt{q}$.
Note that because of (\ref{space-like-edge}) there is a local bijection between configurations of clusters and
loops, so we may use the notation $A \subseteq E$ to specify the configurations of loops as well.

Consider then adding a single edge to the lattice. Imagining for the moment that transfer direction
is upwards (the lattice is built from the bottom to the top), we shall refer to the horizontal edge in (\ref{space-like-edge}) as a
``space-like'' edge. When $e \notin A$ (right picture) the two loop strands just go though and nothing
happens, while for $e \in A$ (left picture) the two strands are being connected and a new partially
completed loop is started out. The sum over both possibilities can be described by an operator
acting on adjacent loop strands at positions $i$ and $i+1$:
\be
 {\sf H}_i = {\sf I} + x {\sf E}_i \,,
 \label{defH}
\ee
where ${\sf E}_i$ are the generators of the Temperley-Lieb algebra \cite{TemperleyLieb71}
defined by
\begin{eqnarray}
 {\sf E}_i^2 = n_{\rm loop} {\sf E}_i \,, \nonumber \\
 {\sf E}_i {\sf E}_{i \pm 1} {\sf E}_i = {\sf E}_i \,, \label{TL} \\
 {\sf E}_i {\sf E}_j = {\sf E}_j {\sf E}_i \quad \mbox{for } |i-j| > 1 \,. \nonumber
\end{eqnarray}
The algebraic relations (\ref{TL}) can be proved graphically by gluing several diagrams of the
type (\ref{space-like-edge}) on top of one another.

For a vertical, or ``time-like'' edge, we would have the graphical correspondence
\begin{equation}
\begin{tikzpicture}[scale=0.7]
 \draw[blue,line width=3pt,dashed] (1,-1)--(1,1);
 \draw[fill] (1,-1) circle(0.6ex) node[below] {$i$};
 \draw[fill] (1,1) circle(0.6ex) node[above] {$j$};
 \draw[red,line width=1.5pt] (0,1)--(0.8,0.2) arc(225:315:2mm) -- (2,1);
 \draw[red,line width=1.5pt] (0,-1)--(0.8,-0.2) arc(135:45:2mm) -- (2,-1);

\begin{scope}[xshift=5cm]
 \draw[blue,line width=3pt] (1,-1)--(1,1);
 \draw[fill] (1,-1) circle(0.6ex) node[below] {$i$};
 \draw[fill] (1,1) circle(0.6ex) node[above] {$j$};
 \draw[red,line width=1.5pt] (0,1)--(0.8,0.2) arc(45:-45:2mm) -- (0,-1);
 \draw[red,line width=1.5pt] (2,1)--(1.2,0.2) arc(135:225:2mm) -- (2,-1);
\end{scope}
\end{tikzpicture}
\label{time-like-edge}
\end{equation}
and since the situation $e \in A$ now corresponds to the right picture, the corresponding operator is
\be
 {\sf V}_i = x {\sf I} + {\sf E}_i \,.
 \label{defV}
\ee

In the following subsections we explain in detail how to adapt the loop representation to the
geometry of Figure~\ref{fig:square-basis} and use it to compute the graph polynomial (\ref{PB_cluster})
rather than the full partition function $Z$. In particular, we explain how to discard configurations
contribution to $Z_{\rm 1D}$, distinguish topologically the contributions to $Z_{\rm 2D}$ and $Z_{\rm 0D}$,
and attribute to each diagram the correct powers of $x = v/\sqrt{q}$ and $n_{\rm loop}=\sqrt{q}$. Based on this,
the graph polynomial (\ref{PB_cluster}) can the be retrieved by changing the $(n_{\rm loop},x)$ variables back
to $(q,v)$ and providing the overall factor $q^{|V|/2}$ in (\ref{loop_repr}).

\subsection{Four-terminal representation}
\label{sec:fourtermrep}

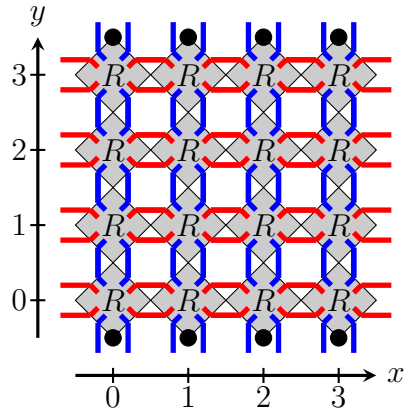
\begin{figure}
\begin{center}

\begin{tikzpicture}[scale=1.0,>=stealth]
\foreach \xpos in {0,1,2,3}
\foreach \ypos in {0,1,2,3}
 \fill[black!20] (\xpos+0.5,\ypos) -- (\xpos+1,\ypos+0.5) -- (\xpos+0.5,\ypos+1) -- (\xpos,\ypos+0.5) -- cycle;
\foreach \xpos in {0,1,2,3}
\foreach \ypos in {0,1,2,3}
 \draw[black] (\xpos+0.5,\ypos) -- (\xpos+1,\ypos+0.5) -- (\xpos+0.5,\ypos+1) -- (\xpos,\ypos+0.5) -- cycle;

\foreach \xpos in {0,1,2,3}
{
 \draw[fill] (\xpos+0.5,0) circle(0.6ex);
 \draw[fill] (\xpos+0.5,4) circle(0.6ex);
}


\foreach \xpos in {0,1,2,3}
\foreach \ypos in {0,1,2,3}
{
 \draw[red,line width=2pt] (\xpos-0.2,\ypos+0.3) -- (\xpos+0.2,\ypos+0.3);
 \draw[red,line width=2pt] (\xpos+0.8,\ypos+0.3) -- (\xpos+1.2,\ypos+0.3);
 \draw[red,line width=2pt] (\xpos-0.2,\ypos+0.7) -- (\xpos+0.2,\ypos+0.7);
 \draw[red,line width=2pt] (\xpos+0.8,\ypos+0.7) -- (\xpos+1.2,\ypos+0.7);
}

\foreach \xpos in {0,1,2,3}
\foreach \ypos in {0,1,2,3}
{
 \draw[blue,line width=2pt] (\xpos+0.3,\ypos-0.2) -- (\xpos+0.3,\ypos+0.2);
 \draw[blue,line width=2pt] (\xpos+0.7,\ypos-0.2) -- (\xpos+0.7,\ypos+0.2);
 \draw[blue,line width=2pt] (\xpos+0.3,\ypos+0.8) -- (\xpos+0.3,\ypos+1.2);
 \draw[blue,line width=2pt] (\xpos+0.7,\ypos+0.8) -- (\xpos+0.7,\ypos+1.2);
}

\foreach \xpos in {0,1,2,3}
\foreach \ypos in {0,1,2,3}
{
 \draw[red,line width=2pt] (\xpos+0.2,\ypos+0.3) -- (\xpos+0.3,\ypos+0.4);
 \draw[red,line width=2pt] (\xpos+0.8,\ypos+0.3) -- (\xpos+0.7,\ypos+0.4);
 \draw[red,line width=2pt] (\xpos+0.2,\ypos+0.7) -- (\xpos+0.3,\ypos+0.6);
 \draw[red,line width=2pt] (\xpos+0.8,\ypos+0.7) -- (\xpos+0.7,\ypos+0.6);
 \draw[blue,line width=2pt] (\xpos+0.3,\ypos+0.2) -- (\xpos+0.4,\ypos+0.3);
 \draw[blue,line width=2pt] (\xpos+0.7,\ypos+0.2) -- (\xpos+0.6,\ypos+0.3);
 \draw[blue,line width=2pt] (\xpos+0.3,\ypos+0.8) -- (\xpos+0.4,\ypos+0.7);
 \draw[blue,line width=2pt] (\xpos+0.7,\ypos+0.8) -- (\xpos+0.6,\ypos+0.7);
 \draw (\xpos+0.5,\ypos+0.5) node{$R$};
}

\draw[very thick,->] (0,-0.5)--(4,-0.5);
\draw (4,-0.5) node[right] {$x$};
\foreach \xpos in {0,1,2,3}
{
 \draw[thick] (\xpos+0.5,-0.6)--(\xpos+0.5,-0.4);
 \draw (\xpos+0.5,-0.5) node[below] {$\xpos$};
}

\draw[very thick,->] (-0.5,0)--(-0.5,4);
\draw (-0.5,4) node[above] {$y$};
\foreach \ypos in {0,1,2,3}
{
 \draw[thick] (-0.6,\ypos+0.5)--(-0.4,\ypos+0.5);
 \draw (-0.5,\ypos+0.5) node[left] {$\ypos$};
}
 
\end{tikzpicture}
 \caption{Square basis of size $n \times n$ with $n=4$ in the loop representation. Terminals of the basis are
 shown as black circles and periodic boundary conditions have been imposed horizontally. The loops live on
 the auxiliary and quantum spaces, shown in red and blue colour respectively. 
 An $\check{R}$-matrix acts inside each grey square.}
 \label{fig:square-basis-loop}
\end{center}
\end{figure}

The graph polynomial $P_B(q,v)$ must be computed in the
square-basis geometry (cf.\ Figure~\ref{fig:square-basis}) which in the loop representation takes
the appearance shown in Figure~\ref{fig:square-basis-loop}. The loops live on the medial lattice 
${\cal M}(B)$, of which a part is shown as red and blue edges in Figure~\ref{fig:square-basis-loop}.
In a nomenclature inspired by the theory of quantum integrable systems, we shall refer to
the (red) horizontal edges as ``auxiliary spaces'' and the (blue) vertical edges as ``quantum spaces''.

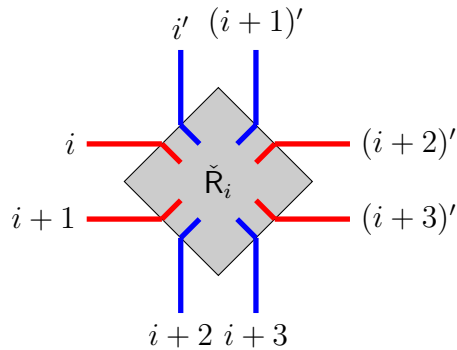
\begin{figure}
\begin{center}

\begin{tikzpicture}[scale=2.5,>=stealth]
\foreach \xpos in {0}
\foreach \ypos in {0}
 \fill[black!20] (\xpos+0.5,\ypos) -- (\xpos+1,\ypos+0.5) -- (\xpos+0.5,\ypos+1) -- (\xpos,\ypos+0.5) -- cycle;
\foreach \xpos in {0}
\foreach \ypos in {0}
 \draw[black] (\xpos+0.5,\ypos) -- (\xpos+1,\ypos+0.5) -- (\xpos+0.5,\ypos+1) -- (\xpos,\ypos+0.5) -- cycle;

\foreach \xpos in {0}
\foreach \ypos in {0}
{
 \draw[red,line width=2pt] (\xpos-0.2,\ypos+0.3) -- (\xpos+0.2,\ypos+0.3);
 \draw[red,line width=2pt] (\xpos+0.8,\ypos+0.3) -- (\xpos+1.2,\ypos+0.3);
 \draw[red,line width=2pt] (\xpos-0.2,\ypos+0.7) -- (\xpos+0.2,\ypos+0.7);
 \draw[red,line width=2pt] (\xpos+0.8,\ypos+0.7) -- (\xpos+1.2,\ypos+0.7);
}

\draw (-0.2,0.7) node[left]{$i$};
\draw (-0.2,0.3) node[left]{$i+1$};
\draw (0.3,-0.2) node[below]{$i+2$};
\draw (0.7,-0.2) node[below]{$i+3$};
\draw (1.2,0.7) node[right]{$(i+2)'$};
\draw (1.2,0.3) node[right]{$(i+3)'$};
\draw (0.3,1.2) node[above]{$i'$};
\draw (0.7,1.2) node[above]{$(i+1)'$};

\foreach \xpos in {0}
\foreach \ypos in {0}
{
 \draw[blue,line width=2pt] (\xpos+0.3,\ypos-0.2) -- (\xpos+0.3,\ypos+0.2);
 \draw[blue,line width=2pt] (\xpos+0.7,\ypos-0.2) -- (\xpos+0.7,\ypos+0.2);
 \draw[blue,line width=2pt] (\xpos+0.3,\ypos+0.8) -- (\xpos+0.3,\ypos+1.2);
 \draw[blue,line width=2pt] (\xpos+0.7,\ypos+0.8) -- (\xpos+0.7,\ypos+1.2);
}

\foreach \xpos in {0}
\foreach \ypos in {0}
{
 \draw[red,line width=2pt] (\xpos+0.2,\ypos+0.3) -- (\xpos+0.3,\ypos+0.4);
 \draw[red,line width=2pt] (\xpos+0.8,\ypos+0.3) -- (\xpos+0.7,\ypos+0.4);
 \draw[red,line width=2pt] (\xpos+0.2,\ypos+0.7) -- (\xpos+0.3,\ypos+0.6);
 \draw[red,line width=2pt] (\xpos+0.8,\ypos+0.7) -- (\xpos+0.7,\ypos+0.6);
 \draw[blue,line width=2pt] (\xpos+0.3,\ypos+0.2) -- (\xpos+0.4,\ypos+0.3);
 \draw[blue,line width=2pt] (\xpos+0.7,\ypos+0.2) -- (\xpos+0.6,\ypos+0.3);
 \draw[blue,line width=2pt] (\xpos+0.3,\ypos+0.8) -- (\xpos+0.4,\ypos+0.7);
 \draw[blue,line width=2pt] (\xpos+0.7,\ypos+0.8) -- (\xpos+0.6,\ypos+0.7);
 \draw (\xpos+0.5,\ypos+0.5) node{$\check{\sf R}_i$};
}
 
\end{tikzpicture}
 \caption{Labeling of the auxiliary and quantum spaces around an $\check{R}$-matrix.}
 \label{fig:labelR}
\end{center}
\end{figure}

The part of ${\cal M}(B)$ inside the grey squares depends of course on the lattice being studied. We denote
it symbolically by the letter $R$. The operator constructing the corresponding part of the lattice is called the
$\check{R}$-matrix and is written $\check{\sf R}_i$. It acts on two auxiliary spaces $(i,i+1)$ coming from the left and two quantum spaces
$(i+2,i+3)$ coming from the bottom of the grey square, and produces outgoing quantum spaces $(i',i'+1)$ on the top
and auxiliary spaces $(i'+2,i'+3)$ on the right of the grey square. This labeling of spaces is shown in Figure~\ref{fig:labelR}.

For each of the lattices to be studied in this paper, $\check{\sf R}_i$ is a definite product of the elementary operators
${\sf H}_j$, ${\sf V}_j$ and ${\sf E}_j$ defined above, with $j=i,i+1,i+2$. This means in practice that the computation
of $P_B(q,v)$ can be adopted to any desired lattice that can be shown to have the four-terminal structure of
Figure~\ref{fig:square-basis-loop} and for which the algebraic expression for $\check{\sf R}_i$ can be provided
(see section~\ref{sec:res-archi}).

To add one row of grey squares to the lattice, one inserts a pair of auxiliary spaces, acts with the product
$\check{\sf R}_{2n-2} \cdots \check{\sf R}_4 \check{\sf R}_2 \check{\sf R}_0$, and removes the pair of auxiliary
spaces. The next subsection contains the precise definitions of the connectivity states, and describes how the elementary operators
act on them, and how auxiliary spaces are inserted and removed.

Note that the labeling of Figure~\ref{fig:labelR} implies that time flows in the North-East direction%
\footnote{This situation will be familiar to readers acquainted with the theory of quantum integrable systems.}
and not simply upwards
as we assumed for pedagogical reasons when discussing (\ref{space-like-edge}) and (\ref{time-like-edge}).
This has an incidence on the interpretation of ``horizontal'' and ``vertical'', since these geometrical notions have now
been rotated $45^{\circ}$ in the clockwise direction. To be precise, the epithet horizontal, or space-like (resp.\ vertical, or time-like)
now means perpendicular (resp.\ parallel) to the direction of time flow.
For instance, the $\check{R}$-matrix that constructs the square lattice is written
\begin{equation}
 \check{\sf R}_i = {\sf H}_{i+1} {\sf V}_i {\sf V}_{i+2} {\sf H}_{i+1} \,.
 \label{R-example-square}
\end{equation}
Many more examples will be given in section~\ref{sec:res-archi}.

As already mentioned it is possible also to perform a restricted set of operations in the white squares. Since after completing
a row of grey squares the auxiliary spaces are no longer at our disposal, this is essentially limited to letting the operator
${\sf H}_i$ act on the two quantum spaces within a white square in Figure~\ref{fig:square-basis-loop}. This has the effect of
adding a horizontal diagonal to the white square.

\subsection{State space}

Since the terminals on the top and bottom of Figure~\ref{fig:square-basis-loop} must eventually be glued together
we actually need a set of two time slices. The first one is a horizontal line at the bottom of the system ($y=-\frac12$),
and the second one is a horizontal line (with a kink when a row is only partially completed) that keeps moving
upwards (and the kink moving to the right, i.e., in the ``North-East direction'', while completing a given row)
until the lattice is completed. After completion the two time slices are glued together
in a precise way (see section~\ref{sec:gluing}) that distinguishes the contributions to $Z_{\rm 2D}$ and $Z_{\rm 0D}$.

\subsubsection{Description of the states}
\label{sec:describe-states}

It is convenient first to describe the state space for a complete row. In this case each of the time slices is a horizontal line
that intersects the $2n$ quantum spaces. No auxiliary spaces are involved in this case.
A connectivity state describes how these $4n$ intersection points are
pairwise connected by the loop strands.

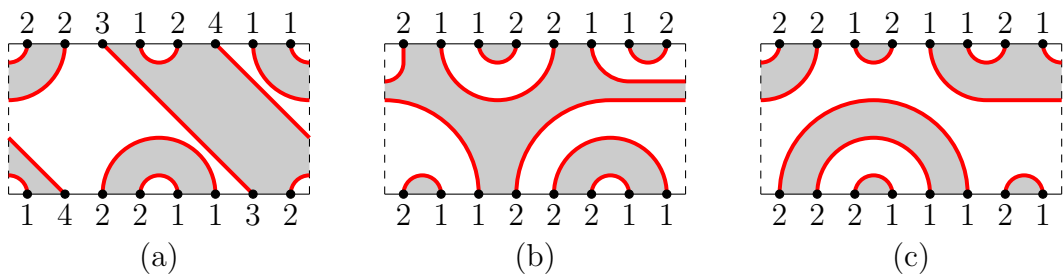
\begin{figure}
\begin{center}

\begin{tikzpicture}[scale=0.5,>=stealth]

\fill[black!20] (1,4) arc(0:-90:15mm) -- ++(0,1) arc(-90:0:5mm) -- cycle;
\fill[black!20] (6,4) arc(180:270:15mm) -- ++(0,1) arc(270:180:5mm) -- cycle;
\fill[black!20] (2,4) -- (6,0) -- (7,0) arc(180:90:5mm) -- ++(0,1) -- (5,4) -- (4,4) arc(0:-180:5mm) -- cycle;
\fill[black!20] (0,0) -- (1,0) -- (-0.5,1.5) -- ++(0,-1) arc(90:0:5mm) -- cycle;
\fill[black!20] (2,0) arc(180:0:15mm) -- (4,0) arc(0:180:5mm) -- cycle;

\draw (-0.5,0)--(7.5,0);
\draw[dashed] (7.5,0)--(7.5,4);
\draw (7.5,4)--(-0.5,4);
\draw[dashed] (-0.5,4)--(-0.5,0);

\draw[red,line width=1.5pt] (1,4) arc(0:-90:15mm);
\draw[red,line width=1.5pt] (0,4) arc(0:-90:5mm);
\draw[red,line width=1.5pt] (6,4) arc(180:270:15mm);
\draw[red,line width=1.5pt] (7,4) arc(180:270:5mm);
\draw[red,line width=1.5pt] (2,4) -- (6,0);
\draw[red,line width=1.5pt] (7,0) arc(180:90:5mm);
\draw[red,line width=1.5pt] (0,0) arc(0:90:5mm);
\draw[red,line width=1.5pt] (1,0) -- (-0.5,1.5);
\draw[red,line width=1.5pt] (7.5,1.5) -- (5,4);
\draw[red,line width=1.5pt] (4,4) arc(0:-180:5mm);
\draw[red,line width=1.5pt] (2,0) arc(180:0:15mm);
\draw[red,line width=1.5pt] (3,0) arc(180:0:5mm);

\foreach \xpos in {0,1,2,3,4,5,6,7}
{
 \draw[fill] (\xpos,0) circle(0.6ex);
 \draw[fill] (\xpos,4) circle(0.6ex);
}

\draw (0,0) node[below]{$1$};
\draw (1,0) node[below]{$4$};
\draw (2,0) node[below]{$2$};
\draw (3,0) node[below]{$2$};
\draw (4,0) node[below]{$1$};
\draw (5,0) node[below]{$1$};
\draw (6,0) node[below]{$3$};
\draw (7,0) node[below]{$2$};

\draw (0,4) node[above]{$2$};
\draw (1,4) node[above]{$2$};
\draw (2,4) node[above]{$3$};
\draw (3,4) node[above]{$1$};
\draw (4,4) node[above]{$2$};
\draw (5,4) node[above]{$4$};
\draw (6,4) node[above]{$1$};
\draw (7,4) node[above]{$1$};

\draw (3.5,-1) node[below]{(a)};

\begin{scope}[xshift=10cm]

\fill[black!20] (0,0) arc(180:0:5mm) -- cycle;
\fill[black!20] (4,0) arc(180:0:15mm) -- ++(-1,0) arc(0:180:5mm) -- cycle;
\fill[black!20] (3,0) arc(180:90:25mm) -- ++(2,0) -- ++(0,0.5) -- ++(-1.5,0) arc(270:180:10mm) -- ++(-1,0) arc(0:-180:15mm) -- ++(-1,0) -- ++(0,-0.5) arc(0:-90:5mm) -- ++(0,-0.5) arc(90:0:25mm) -- cycle;
\fill[black!20] (2,4) arc(180:360:5mm) -- cycle;
\fill[black!20] (6,4) arc(180:360:5mm) -- cycle;

\draw (-0.5,0)--(7.5,0);
\draw[dashed] (7.5,0)--(7.5,4);
\draw (7.5,4)--(-0.5,4);
\draw[dashed] (-0.5,4)--(-0.5,0);

\draw[red,line width=1.5pt] (0,0) arc(180:0:5mm);
\draw[red,line width=1.5pt] (4,0) arc(180:0:15mm);
\draw[red,line width=1.5pt] (5,0) arc(180:0:5mm);
\draw[red,line width=1.5pt] (3,0) arc(180:90:25mm) -- ++(2,0);
\draw[red,line width=1.5pt] (7.5,3) -- ++(-1.5,0) arc(270:180:10mm);
\draw[red,line width=1.5pt] (4,4) arc(0:-180:15mm);
\draw[red,line width=1.5pt] (0,4) -- ++(0,-0.5) arc(0:-90:5mm);
\draw[red,line width=1.5pt] (-0.5,2.5) arc(90:0:25mm);
\draw[red,line width=1.5pt] (2,4) arc(180:360:5mm);
\draw[red,line width=1.5pt] (6,4) arc(180:360:5mm);

\foreach \xpos in {0,1,2,3,4,5,6,7}
{
 \draw[fill] (\xpos,0) circle(0.6ex);
 \draw[fill] (\xpos,4) circle(0.6ex);
}

\draw (0,0) node[below]{$2$};
\draw (1,0) node[below]{$1$};
\draw (2,0) node[below]{$1$};
\draw (3,0) node[below]{$2$};
\draw (4,0) node[below]{$2$};
\draw (5,0) node[below]{$2$};
\draw (6,0) node[below]{$1$};
\draw (7,0) node[below]{$1$};

\draw (0,4) node[above]{$2$};
\draw (1,4) node[above]{$1$};
\draw (2,4) node[above]{$1$};
\draw (3,4) node[above]{$2$};
\draw (4,4) node[above]{$2$};
\draw (5,4) node[above]{$1$};
\draw (6,4) node[above]{$1$};
\draw (7,4) node[above]{$2$};

\draw (3.5,-1) node[below]{(b)};

\end{scope}

\begin{scope}[xshift=20cm]

\fill[black!20] (0,0) arc(180:0:25mm) -- ++(-1,0) arc(0:180:15mm) -- cycle;
\fill[black!20] (2,0) arc(180:0:5mm) -- cycle;
\fill[black!20] (6,0) arc(180:0:5mm) -- cycle;
\fill[black!20] (0,4) arc(0:-90:5mm) -- ++(0,-1) arc(270:360:15mm) -- cycle;
\fill[black!20] (2,4) arc(180:360:5mm) -- cycle;
\fill[black!20] (4,4) arc(180:270:15mm) -- ++(2,0) -- ++(0,1) arc(270:180:5mm) -- ++(-1,0) arc(0:-180:5mm) -- cycle;

\draw (-0.5,0)--(7.5,0);
\draw[dashed] (7.5,0)--(7.5,4);
\draw (7.5,4)--(-0.5,4);
\draw[dashed] (-0.5,4)--(-0.5,0);

\draw[red,line width=1.5pt] (0,0) arc(180:0:25mm);
\draw[red,line width=1.5pt] (1,0) arc(180:0:15mm);
\draw[red,line width=1.5pt] (2,0) arc(180:0:5mm);
\draw[red,line width=1.5pt] (6,0) arc(180:0:5mm);
\draw[red,line width=1.5pt] (0,4) arc(0:-90:5mm);
\draw[red,line width=1.5pt] (1,4) arc(0:-90:15mm);
\draw[red,line width=1.5pt] (2,4) arc(180:360:5mm);
\draw[red,line width=1.5pt] (5,4) arc(180:360:5mm);
\draw[red,line width=1.5pt] (7,4) arc(180:270:5mm);
\draw[red,line width=1.5pt] (4,4) arc(180:270:15mm) -- ++(2,0);

\foreach \xpos in {0,1,2,3,4,5,6,7}
{
 \draw[fill] (\xpos,0) circle(0.6ex);
 \draw[fill] (\xpos,4) circle(0.6ex);
}

\draw (0,0) node[below]{$2$};
\draw (1,0) node[below]{$2$};
\draw (2,0) node[below]{$2$};
\draw (3,0) node[below]{$1$};
\draw (4,0) node[below]{$1$};
\draw (5,0) node[below]{$1$};
\draw (6,0) node[below]{$2$};
\draw (7,0) node[below]{$1$};

\draw (0,4) node[above]{$2$};
\draw (1,4) node[above]{$2$};
\draw (2,4) node[above]{$1$};
\draw (3,4) node[above]{$2$};
\draw (4,4) node[above]{$1$};
\draw (5,4) node[above]{$1$};
\draw (6,4) node[above]{$2$};
\draw (7,4) node[above]{$1$};

\draw (3.5,-1) node[below]{(c)};

\end{scope}

\end{tikzpicture}
 \caption{Three examples of connectivity states for $n=4$. The numbers along the two time slices provide a canonical coding of the
 connectivity state, as explained in the main text. Loops are shown as red solid lines. The corresponding clusters live in the areas
 shaded in grey, whilst the dual clusters live in the white areas.}
 \label{fig:conn-states}
\end{center}
\end{figure}

In Figure~\ref{fig:conn-states} we show a few examples of connectivity states. We call {\em arc} a loop segment that connects two points within the
same time slice, and {\em string} a loop segment that connects points on different time slices. The number of strings is denoted $s$.
Boundary conditions are periodic in the horizontal direction, as shown by the dashed sides of the construction boxes in Figure~\ref{fig:conn-states}.

Suppose that the points are labelled $0,1,2,\ldots,2n-1$ on the bottom time slice and $0',1',2',\ldots,(2n-1)'$ on the upper time slice.
To relate the cluster and loop representations, it is important to observe that points with an even label have a cluster on the right and
a dual cluster on the left (and vice versa for the points with an odd label). The clusters (resp.\ dual clusters) corresponding to the connectivity
states in Figure~\ref{fig:conn-states} are shown in grey (resp.\ white) shading. This relation to clusters imposes important restrictions on the connectivities
of loops. First, arcs connect points of opposite parities. Second, strings connect points of the same parity. In particular one may define a string
to be even or odd, depending on the parity of the points that it connects. Third, $s$ is even and there are as many even as odd strings. 
When $s=0$, consider the configuration of arcs restricted to just one of the two time slices. We call this configuration {\em closed} if the
equivalent clusters are bounded away from the other time slice (in which case one dual cluster will connect the two time slices), and
{\em open} if one cluster connects the two time slices (in which case the dual clusters are bound). Then, fourth, when $s=0$ a connectivity
state consists of two open arc configurations, or of two closed arc configurations.

These concepts are illustrated in Figure~\ref{fig:conn-states}. Figure~\ref{fig:conn-states}a shows a state with $s=2$. The case of $s=0$ is depicted
in Figure~\ref{fig:conn-states}b for two open arc states, and in Figure~\ref{fig:conn-states}c for two closed arc states. The figures also exemplify
a coding of the states which turns out to be convenient for describing the transfer matrix algorithm. On the time slice on the top of the system,
the leftmost (resp.\ rightmost) point of an arc carries the code $1$ (resp.\ $2$).  Arcs on the bottom of the system are coded similarly, but after
a $180^{\circ}$ rotation. Strings are coded as pairs of matching codes ($3,4,\ldots$).

\subsubsection{Action of Temperley-Lieb generators}
\label{sec:action-TL}

The connectivity states provide a representation of the periodic Temperley-Lieb algebra $TL_{2n}(n_{\rm loop})$,
which is faithful for generic values of $n_{\rm loop}$ (see e.g.\ Ref.~\cite{VasseurBraid}). However, to obtain a
finite-dimensional algebra we need to take a quotient (closely related to the so-called Jones-Temperley-Lieb algebra \cite{Jones94}) by
\begin{enumerate}
 \item identifying states in which the strings connect the same points on the top and bottom
time slices, but wind a different number of times around the periodic (horizontal) direction, and
 \item giving a definite weight $n_{\rm wind}$ to each loop that winds the periodic direction.
\end{enumerate}
To take into account the first condition, we choose (for $s > 0$) to draw the string whose anchoring point on the top time slice is the furthest to the left
so that it does not cross the periodic direction. The remaining strings are then drawn in the unique (up to isotopy) way that respects planarity
(i.e., arcs and strings do not cross). This convention at the same time provides a canonical coding of the strings: the anchoring points on the
top time slice are labelled $3,4,5,\ldots,s+2$ from left to right, and the labels on the bottom time slice follow by matching the codes of the points
that are connected by a string. The canonical coding then provides a unique description of the states in the quotient algebra.

Notice that any diagram with a winding loop will contribute to $Z_{\rm 1D}$. We must therefore set $n_{\rm wind} = 0$, and this at the same
time provides a valid choice for the second condition. However, setting just $n_{\rm wind} = 0$ is not sufficient for getting rid of all contributions
to $Z_{\rm 1D}$, so more work will be required (see section~\ref{sec:gluing}).

The Temperley-Lieb generator ${\sf E}_i$, shown graphically in the left picture in (\ref{space-like-edge}) and (\ref{time-like-edge}), now acts on the
connectivity states by contracting the strands at neighbouring positions $i$ and $j = i+1 \mbox{ mod } 2n$ on the top time slice, and subsequently
liking those two points by a new arc (meaning that the points $i$ and $j$ acquire the codes $1$ and $2$ respectively).
The precise meaning of the contraction depends on the codes $(c_i,c_j)$ of the points prior to the action by ${\sf E}_i$:
\begin{enumerate}
 \item If $(c_i,c_j) = (1,2)$ a loop is formed, and the weight $n_{\rm loop}$ must be applied.
 \item If $(c_i,c_j) = (1,1)$ the partner of $c_j$ has its code changed from $2$ to $1$.
 \item If $(c_i,c_j) = (2,2)$ the partner of $c_i$ has its code changed from $1$ to $2$.
 \item If $(c_i,c_j) = (2,1)$ the codes are unchanged, but if $c_i$ is the partner of $c_j$ a winding loop is formed, and the weight
 $n_{\rm wind} = 0$ must be applied.
 \item If $c_i>2$ is a string and $c_j \le 2$ is an arc, then the partner of $c_j$ becomes the new position of the string, and hence has its
 code changed to $c_i$. The same statement holds true with $i$ and $j$ interchanged.
 \item If both of $(c_i,c_j)$ are strings, the partner of $c_i$ has its code changed to $2$ and the partner of $c_j$ gets the code $1$.%
\footnote{Note that this respects the convention that arcs on the bottom time slice have the code 2 (resp.\ 1) assigned to their leftmost (resp.\ rightmost) point,
viz., the convention obtained by rotating through $180^{\circ}$ rotation the one used for arcs on the top time slice.}
\end{enumerate}
Note that in all of these cases, except the last one, the number of strings is conserved by ${\sf E}_i$, so that the codes on the bottom time slice do
not change at all. But in the last case a pair of strings is destroyed and their anchoring points on the bottom time slice are turned into an arc.

\subsubsection{Dimension of the transfer matrix}

To determine the dimension of the transfer matrix we must count the number of states. We still suppose for the time being that
the uppermost row is complete (i.e., there are no auxiliary spaces).

By cutting all the strings (if any) a connectivity state describing the full system of two time slices is transformed into a pair of
{\em reduced states} each associated with one of the time slices. The reduced states consist of arcs and half strings.
When $s=0$ the reduced states can be characterised as open or closed, just like the full states.

It is easy to count the number of reduced states. For $s=0$, out of the $2n$ points there are $n$ with code $1$ and $n$ with
code $2$. A moment's reflection reveals that all the ${2n \choose n}$ possible placements of these codes correspond to
a valid state. For $s=2k>0$ one can similarly convince oneself that specifying the $n-k$ points with code $1$ will uniquely
imply the positions of the arcs and half strings. So there are ${2n \choose n-k}$ reduced states in general.

Conversely, a pair of reduced states with the same number of half strings can be transformed into a full connectivity state by
gluing pairs of half strings. However, for $s=0$ one obtains a valid state only by ``gluing'' (or rather juxtaposing, since nothing is actually
being glued!) two open or two closed half states. Since the set of open and closed half states are bijectively related by performing
a cyclic shift, there are $\frac12 {2n \choose n}$ of each. Moreover, for $s=2k>0$ strings the gluing can be done in $k$ inequivalent ways,
since each half string must be glued to one of the same parity and the cyclic order of strings must be respected.
These observations imply that there are
\be
 {\rm dim}(n) = \frac12 {2n \choose n}^2 + \sum_{k=1}^n k \, {2n \choose n-k}^2
 \label{dimTM}
\ee
connectivity states.

\subsubsection{Bijection between states and integers}
\label{sec:bijection}

To write an efficient transfer matrix algorithm it is desirable to possess a bijection between the integers
$0,1,2,\ldots,{\rm dim}(n)-1$ and the states coded as in Figure~\ref{fig:conn-states}. This is straightforwardly
done provided that one can provide a canonical ordering of the states.

The states can be ordered according to the following criteria:
\begin{enumerate}
 \item The number of strings $s=2k$ with $k=0,1,\ldots,n$
 \item The reduced connectivity state on the bottom time slice.
 \item The cyclic rotation involved in gluing $2k$ half strings on the top time slice to $2k$ half strings on
 the bottom time slice. Note that with the canonical coding of Figure~\ref{fig:conn-states} this amounts to shifting
 cyclically the codes $>2$ on the bottom time slice.
 \item The reduced connectivity state on the top time slice.
\end{enumerate} 
Since the reduced connectivity states have a simple interpretation in terms of binomial coefficients,
they can easily be endowed with a canonical ordering (e.g.\ by ordering them lexicographically).
Alternatively, since the number of reduced states is much less than the total number of states, we
can simply generate the reduced states by hashing techniques and endow them with some {\em ad hoc}
ordering, such as their position in the hash table.

The practical implementation of the bijection in terms of the above criteria, and the ordering of the reduced states,
is most conveniently written in terms of various tables, as outlined in \cite{Blote82,Blote89} for a couple of
related situations.

\subsubsection{Handling auxiliary spaces}
\label{sec:aux-spaces}

To handle a partially completed row of the lattice we need to be able to insert and remove auxiliary spaces.
We also need to count the number of states in the presence of $p$ auxiliary spaces. (The four-terminal
representation requires $p=0,1,2$ but extensions of the formalism to higher values of $p$ may turn out to be of interest for
lattices which cannot be cast in the four-terminal form.)

Suppose now that the $n$ points in the top time slice are initially labelled from left to right:
$$
\begin{tikzpicture}[scale=0.5,>=stealth]
\foreach \xpos in {0,1,2,3}
{
 \draw[blue,line width=2pt] (\xpos,0.0) -- (\xpos,0.4);
 \draw (\xpos,0.0) node[below] {$\xpos$};
}
\draw (4,0.0) node[below] {$\cdots$};
\draw[blue,line width=2pt] (5,0.0) -- (5,0.4);
\draw (6.0,0.0) node[below] {$2n-1$};
\end{tikzpicture}
$$
The bottom time slice simply occupies the $2n$ labels following those of the top time slice, and plays no further role in
the following construction. Accordingly we shall not represent it here.

Inserting the first (lower) auxiliary space amounts to adding two extra points to the left of those in the top time slice.
This can be considered as two copies of the same point which are initially connected. We have now:
$$
\begin{tikzpicture}[scale=0.5,>=stealth]
\draw[red,line width=2pt] (-1.2,1.0) -- (-0.8,1.0);
\draw (-1.2,1.0) node[left] {$2n+1$};
\draw (-0.8,1.0) node[right] {$0$};
\foreach \xpos in {1,2,3,4}
{
 \draw[blue,line width=2pt] (\xpos-1,0.0) -- (\xpos-1,0.4);
 \draw (\xpos-1,0.0) node[below] {$\xpos$};
}
\draw (4,0.0) node[below] {$\cdots$};
\draw[blue,line width=2pt] (5,0.0) -- (5,0.4);
\draw (5.3,0.0) node[below] {$2n$};
\end{tikzpicture}
$$
The $2n$ points of the quantum spaces have had their labels shifted by one, in order to accommodate the labels $0$ and $2n+1$ of
the quantum spaces. In order to connect the latter two points, and respect the conventions for an arc on the top time slice,
we attribute to them the codes $c_{2n+1}=1$ and $c_0 = 2$.

Similarly, the second (upper) auxiliary space is inserted
by adding a pair of points with codes $1$ and $2$ in-between those previously inserted. This looks like:
$$
\begin{tikzpicture}[scale=0.5,>=stealth]
\draw[red,line width=2pt] (-1.2,1.0) -- (-0.8,1.0);
\draw[red,line width=2pt] (-1.2,2.0) -- (-0.8,2.0);
\draw (-1.2,2.0) node[left] {$2n+3$};
\draw (-0.8,2.0) node[right] {$0$};
\draw (-1.2,1.0) node[left] {$2n+2$};
\draw (-0.8,1.0) node[right] {$1$};
\foreach \xpos in {2,3,4,5}
{
 \draw[blue,line width=2pt] (\xpos-2,0.0) -- (\xpos-2,0.4);
 \draw (\xpos-2,0.0) node[below] {$\xpos$};
}
\draw (4,0.0) node[below] {$\cdots$};
\draw[blue,line width=2pt] (5,0.0) -- (5,0.4);
\draw (6.0,0.0) node[below] {$2n+1$};
\end{tikzpicture}
$$
Note that the quantum space labels have again been shifted by one, as have those of the first (lower) auxiliary space.
This is necessary for respecting the cyclic order of the labels upon moving around the top time slice.
So the complete set of four points
inserted to the left of those in the top time slice have the codes $c_{2n+2}=c_{2n+3}=1$ and $c_0=c_1=2$.
To add one complete row of grey squares to the lattice, one then applies the product of operators
$\check{\sf R}_{2n-2} \cdots \check{\sf R}_4 \check{\sf R}_2 \check{\sf R}_0$.
Note that before the action with the factor $\check{\sf R}_i$ the two
rightmost dangling ends of the auxiliary spaces carry the labels $i$ and $i+1$, in agreement with
the conventions of Figure~\ref{fig:labelR}.

After the application of the last factor, $\check{\sf R}_{2n-2}$, the situation is as follows:
$$
\begin{tikzpicture}[scale=1.0,>=stealth]
\foreach \xpos in {0,1,2,3}
\foreach \ypos in {0}
 \fill[black!20] (\xpos+0.5,\ypos) -- (\xpos+1,\ypos+0.5) -- (\xpos+0.5,\ypos+1) -- (\xpos,\ypos+0.5) -- cycle;
\foreach \xpos in {0,1,2,3}
\foreach \ypos in {0}
 \draw[black] (\xpos+0.5,\ypos) -- (\xpos+1,\ypos+0.5) -- (\xpos+0.5,\ypos+1) -- (\xpos,\ypos+0.5) -- cycle;

\foreach \xpos in {0,1,2,3}
\foreach \ypos in {0}
{
 \draw[red,line width=2pt] (\xpos-0.2,\ypos+0.3) -- (\xpos+0.2,\ypos+0.3);
 \draw[red,line width=2pt] (\xpos+0.8,\ypos+0.3) -- (\xpos+1.2,\ypos+0.3);
 \draw[red,line width=2pt] (\xpos-0.2,\ypos+0.7) -- (\xpos+0.2,\ypos+0.7);
 \draw[red,line width=2pt] (\xpos+0.8,\ypos+0.7) -- (\xpos+1.2,\ypos+0.7);
}
\draw (-0.2,0.3) node[left] {$2n+2$};
\draw (-0.2,0.7) node[left] {$2n+3$};
\draw (4.2,0.3) node[right] {$2n+1$};
\draw (4.2,0.7) node[right] {$2n$};

\foreach \xpos in {0,1,2,3}
\foreach \ypos in {0}
{
 \draw[blue,line width=2pt] (\xpos+0.3,\ypos-0.2) -- (\xpos+0.3,\ypos+0.2);
 \draw[blue,line width=2pt] (\xpos+0.7,\ypos-0.2) -- (\xpos+0.7,\ypos+0.2);
 \draw[blue,line width=2pt] (\xpos+0.3,\ypos+0.8) -- (\xpos+0.3,\ypos+1.2);
 \draw[blue,line width=2pt] (\xpos+0.7,\ypos+0.8) -- (\xpos+0.7,\ypos+1.2);
}
\draw (0.3,1.2) node[above] {$0$};
\draw (0.7,1.2) node[above] {$1$};
\draw (1.3,1.2) node[above] {$2$};
\draw (1.7,1.2) node[above] {$3$};
\draw (2.3,1.2) node[above] {$4$};
\draw (2.7,1.2) node[above] {$5$};
\draw (3.5,1.2) node[above] {$\cdots$};

\foreach \xpos in {0,1,2,3}
\foreach \ypos in {0}
{
 \draw[red,line width=2pt] (\xpos+0.2,\ypos+0.3) -- (\xpos+0.3,\ypos+0.4);
 \draw[red,line width=2pt] (\xpos+0.8,\ypos+0.3) -- (\xpos+0.7,\ypos+0.4);
 \draw[red,line width=2pt] (\xpos+0.2,\ypos+0.7) -- (\xpos+0.3,\ypos+0.6);
 \draw[red,line width=2pt] (\xpos+0.8,\ypos+0.7) -- (\xpos+0.7,\ypos+0.6);
 \draw[blue,line width=2pt] (\xpos+0.3,\ypos+0.2) -- (\xpos+0.4,\ypos+0.3);
 \draw[blue,line width=2pt] (\xpos+0.7,\ypos+0.2) -- (\xpos+0.6,\ypos+0.3);
 \draw[blue,line width=2pt] (\xpos+0.3,\ypos+0.8) -- (\xpos+0.4,\ypos+0.7);
 \draw[blue,line width=2pt] (\xpos+0.7,\ypos+0.8) -- (\xpos+0.6,\ypos+0.7);
 \draw (\xpos+0.5,\ypos+0.5) node{$R$};
}
\end{tikzpicture}
$$
The row is basically completed, and the new quantum spaces have come out with the correct labeling $0,1,2,\ldots,2n-1$.
However, the two auxiliary spaces remain open, so the dangling ends on the left and the right will have to be
glued and then removed. To this end, one applies the ``contraction'' operator (discussed above when defining the
Temperley-Lieb generators ${\sf E}_i$) to identify first points $2n+1$ and $2n+2$, and then $2n$ and $2n+3$.
The four auxiliary points are then removed from the state, and we are ready to start all over and add a new row to the lattice.

The number of states ${\rm dim}(n,p)$ in the presence of $p$ auxiliary spaces is an obvious generalisation of (\ref{dimTM}).
We find that
\be
 {\rm dim}(n,p) = \frac12 {2n+2p \choose n+p} {2n \choose n} + \sum_{k=1}^n k \, {2n+2p \choose n+p-k} {2n \choose n-k}
 \label{dim-aux-space}
\ee
We have tabulated these dimensions in Table~\ref{tab:dims} for the values of $n$ and $p$ used in the computations of
section~\ref{sec:res-archi}. These dimensions should be compared with those of the transfer matrix approach
of Ref.~\cite{Jacobsen13}. In order to compute $P_B(q,v)$ for an $n \times n$ square basis, \cite{Jacobsen13} used states which are planar
partitions of the $N=4n$ terminal points in Figure~\ref{fig:square-basis}. The number of such states is given by the Catalan number
\begin{equation}
 {\rm Cat}(N) = \frac{1}{N+1} {2N \choose N} \,.
 \label{Catalan}
\end{equation}
The advantage of the present approach, as announced in the Introduction, is to reduce this to ${\rm dim}(n,2)$, thus
raising the limit of practical feasibility of the computations from $n_{\rm max} = 4$ \cite{Jacobsen13} to $n_{\rm max} = 7$.
This improvement is achieved by the dealing efficiently with the horizontal periodic boundary conditions, leading to the
reduction from $4n$ to $2n$ of the number of terminals implied by Figure~\ref{fig:square-basis-loop}.

\begin{table}
\begin{center}
 \begin{tabular}{l|rrr|r}
 $n$ & ${\rm dim}(n,0)$ & ${\rm dim}(n,1)$ & ${\rm dim}(n,2)$ & Ref.~\cite{Jacobsen13} \\ \hline
 1 & 3 & 10 & 35 & 14 \\
 2 & 36 & 132 & 490 & 1\,430 \\
 3 & 500 & 1\,900 & 7\,245 & 208\,012 \\
 4 & 7\,350 & 28\,420 & 109\,956 & 35\,357\,670 \\
 5 & 111\,132 & 433\,944 & 1\,693\,692 & 6\,564\,120\,420 \\
 6 & 1\,707\,552 & 6\,708\,240 & 26\,332\,020 & 1\,289\,904\,147\,324 \\ 
 7 & 26\,501\,904 & 104\,535\,288 & 411\,945\,105 & 263\,747\,951\,750\,360 \\ \hline
 \end{tabular}
 \caption{Dimension ${\rm dim}(n) = {\rm dim}(n,0)$ of the transfer matrix for a completed row, and dimensions ${\rm dim}(n,p)$
 for a partially completed row with $p=1,2$ auxiliary spaces,
 for the sizes $n$ used in section~\ref{sec:res-archi}. This is compared with the dimension of the transfer matrix of Ref.~\cite{Jacobsen13}.}
 \label{tab:dims}
\end{center}
\end{table}

The bijection between states and integers extends straightforwardly to the case with auxiliary spaces, the only difference
being that the reduced state describing the top time slice contains $2p$ extra points.

\subsection{Implementational details}
\label{sec:implementation}

We now describe some considerations on how to efficiently build the lattice by repeated applications of the transfer matrix.
The general prescription for building a single row has been detailed in section~\ref{sec:aux-spaces}:
\begin{enumerate}
 \item Insert two auxiliary spaces;
 \item Add a row of grey squares, each corresponding to the application of an operator $\check{\sf R}_i$;
 \item Remove the two auxiliary spaces;
 \item (For certain lattices:) Add horizontal edges in the white square by application of the ${\sf H}_i$ operators.
\end{enumerate}
At the beginning of the transfer process, the top and bottom time slices coincide. Therefore the initial state consists of
just a single connectivity state $s_0$ in which, for each $i=0,1,\ldots,2n-1$, the points $i$ and $i+2n$ are connected by a string.
In the conventions of section~\ref{sec:describe-states} this means that the corresponding coding is $c_{i} = c_{i+2n} = 3+i$.
In other words, the initial state is therefore a unit vector where $s_0$ has Boltzmann weight $1$, while all other states
have weight $0$.

For the computation of $P_B(q,v)$, the states are represented as arrays of polynomials in the variables $(n_{\rm loop},x)$, with
coefficients that are non-negative integers. The degrees of the polynomials are $|V|$ in the $n_{\rm loop}$ variable and
$|E|$ in the $x$ variable, where $|V|$ and $|E|$ denote the number of vertices and edges in $B$. For the computation of
the percolation critical polynomial $P_B(1,v)$ it suffices obviously to employ arrays of polynomials in the single $x$ variable. And, finally,
when we only desire to find the roots of $P_B(q,v)$ numerically, the arrays consist of real numbers to the desired numerical
precision. In order to obtain the roots to (at least) 50 digits we use 100-digit real numbers and the Newton-Raphson method, where derivatives
are computed from a first-order finite-difference formula with $\epsilon = 10^{-50}$. In practice we use the CLN library \cite{CLN}
that efficiently handles such high-precision real numbers within our {\sc C++} implementation of the algorithm.

In all cases, the arrays have dimensions ${\rm dim}(n,p)$ given by (\ref{dim-aux-space}), where $p=0,1,2$ depending on the number
of open auxiliary spaces.

The bijection described in section~\ref{sec:bijection} is employed throughout the transfer process to translate back and forth between
connectivity states (on which the action of the fundamental Temperley-Lieb operators ${\sf E}_i$ has been detailed in section~\ref{sec:action-TL})
and integers $0,1,\ldots,{\rm dim}(n,p)-1$ that specify the position in the arrays of the relevant Boltzmann weights.

In the cases where the polynomials $P_B(q,v)$ or $P_B(1,v)$ are computed exactly, the coefficients of the polynomials are very
large integers for all but the smallest values of the size $n$. Although the CLN library \cite{CLN} offers also arbitrary-precision
integers, it is more efficient to compute the result modulo a sufficient number of different primes $p_i$ and reconstitute the
exact result from the Chinese remainder theorem. The standard version of our algorithm (i.e., the one used throughout section~\ref{sec:res-archi})
employs only additions, whereas the ``generic $\check{\sf R}$-matrix'' version described in section~\ref{sec:foureightmedial} uses also
multiplications. Since standard unsigned integers in {\sc C++} lie in the range $0,1,\ldots,2^{32}-1$ we therefore take $p_i < 2^{31}$ in
the former case and $p_i < 2^{16}$ in the latter. The most demanding computations required of the order of 20 different primes within this scheme.

Some consideration on the application of an $\check{\sf R}_i$ operator are also in order. For most lattices---including all those of
section~\ref{sec:res-archi}---we can express $\check{\sf R}_i$ as a product of the elementary operators ${\sf H}_i$, ${\sf V}_i$ and ${\sf E}_i$.
An example has been given in (\ref{R-example-square}). In those cases it is usually numerically most efficient to apply each of the
factors separately, i.e., to build the lattice one edge at a time. (This procedure is known as sparse-matrix factorisation.) Note that the application of each of these elementary operators to a
given input connectivity state produces only two output states. Expanding out the product $\check{\sf R}_i$ by hand would lead instead
to up to fourteen output states (since the number of planar pairings of the eight points appearing in Figure~\ref{fig:labelR} is ${\rm Cat}(4) = 14$),
and with vastly more complicated coefficients. However, in some cases not all $14$ output states are actually generated. This is so
in particular for the kagome lattice, where only $13$ states are generated. Accordingly some of the states stored in the arrays will have
zero weight. For instance, for the kagome lattice we find that only $37 \%$ of the ${\rm dim}(n,2)$ states carry non-zero weight for $n=1$,
but this occupation ratio increases to $74 \%$ for $n=2$, $87 \%$ for $n=3$ and $90 \%$ for $n=4$. It is therefore hardly worth dealing
with this slight waste of memory resources  in this case. 

However, in other situations the occupation ratio may go instead to zero for large $n$. This is notably the case for some of the site
percolation problems of section~\ref{sec:site}. It is then more efficient to avoid storing a lot of zero coefficients in the array, but rather
insert the states that are really generated (with non-zero weight) in a hash table. In particular, the bijection of section~\ref{sec:bijection}
becomes superfluous. We shall describe the hashing version of the algorithm in more details in sections~\ref{sec:foureightmedial} and \ref{sec:site}.

\subsection{Topological considerations}
\label{sec:gluing}

When the basis $B$ has been completely built up, it remains to discuss how to actually compute the graph polynomial,
for which we recall the definition (\ref{PB_cluster}):
$$P_B(q,v) = Z_{\rm 2D} - q Z_{\rm 0D} \,.$$
The end result of the transfer process is a linear combination of connectivity states involving two time slices, such as those shown in
Figure~\ref{fig:conn-states}, with coefficients that are polynomials in $n_{\rm loop}$ and $x$. Each state is described by its number
in the canonical ordering, from which the coding (i.e., the integers $c_i$ shown in Figure~\ref{fig:conn-states}) can be inferred
from the bijection of section~\ref{sec:bijection}. Also, from this coding, one can rather straightforwardly construct a representation of the
pairing of the $4n$ points (namely $2n$ on each time slice) induced by the loops (shown in red in Figure~\ref{fig:conn-states})
which allows, in particular, to ``travel'' along the loops.

The goal is now to identify, for each connectivity state, the top and bottom time slices such that the $i$th point on the top time slice
gets glued to the $i$th point on the bottom time slice. Within this gluing procedure we should, on one hand, be able to distinguish which 
states contribute to $Z_{\rm 2D}$, $Z_{\rm 1D}$ and $Z_{\rm 0D}$ (those of $Z_{\rm 1D}$ are then discarded, since they do not
contribute to $P_B(q,v)$) and, on the other hand, provide some extra powers of $n_{\rm loop} = \sqrt{q}$
that have not been accounted for by the transfer process.

To count the number of loops $P$ and, at the same time, determine whether there is a loop of non-trivial homotopy, we begin by
travelling along each loop. To this end, one starts at some initial reference point and follows its arc or string to the partner point. Then one jumps
to the opposite time slice (since the two are glued) and repeats the process until one comes back to the initial point. During this travel,
a list of the points visited is maintained, as well as the horizontal and vertical winding numbers, $w_x$ and $w_y$, incurred. Note that
$w_x$ changes as the result of an arc or string explicitly crossing the periodic boundary condition, whereas $w_y$ changes when one
jumps from one time slice to the other. If $(w_x,w_y) \neq (0,0)$ the loop has non-trivial homotopy, and we are dealing with a state
that contributes to $Z_{\rm 1D}$ and therefore can be discarded. Note that all non-trivial loops (if any) necessarily have the same homotopy, i.e.,
the same values of $(w_x,w_y)$ up to a global sign change.

Having completed the travel along the first loop, if a non-visited point still exists, this is taken as the new reference point, and we trace
out the next loop. This process terminates when all points have been visited. We now know the number of loops $P$.

For example, the states in Figures~\ref{fig:conn-states}b and \ref{fig:conn-states}c both have a loop with $(w_x,w_y) = (1,0)$. This is easiest
seen by considering the loop in Figure~\ref{fig:conn-states}b (resp.\ Figure~\ref{fig:conn-states}c)
that passes through the third (resp.\ second) point from the left on the bottom time slice.

Suppose now that we have found that all $P$ loops in the state have trivial homotopy. This is the case for the state in Figure~\ref{fig:conn-states}a,
which we shall henceforth use as an example. We then need to find out if the state contributes to $Z_{\rm 2D}$ or to $Z_{\rm 0D}$. To this end
we use a variant of the Euler relation, i.e., we compute the quantity
\begin{equation}
 \chi = E + 2 C - V - P \,,
 \label{euler-relation}
\end{equation}
where the quantities $E$, $C$ and $V$ will be defined below.

Corresponding to a loop configuration (red lines in Figure~\ref{fig:conn-states}) there is a corresponding cluster configuration (grey shading
in Figure~\ref{fig:conn-states}). The cluster configuration can be seen as a hyper graph on the set ${\cal P}_1$ of $2n$ points
(namely $n$ on each time slice) situated to the
right of the (loop) point $i$ and to the left of point $i+1$, for all even $i$. An area with grey shading containing $d+1$ points of the hyper
graph is called a hyper edge of degree $d$. (Note that in the definition of hyper edges we do {\em not} impose the identification of the top and bottom
time slices.) Let now $E$ be the sum of the degrees $d$ of all hyper edges. One may think of $E$ as the equivalent number of usual
(not hyper) edges. For instance, the state in Figure~\ref{fig:conn-states}a can be represented as:
\begin{equation}
\begin{tikzpicture}[scale=0.7,>=stealth]

\fill[black!20] (1,4) arc(0:-90:15mm) -- ++(0,1) arc(-90:0:5mm) -- cycle;
\fill[black!20] (6,4) arc(180:270:15mm) -- ++(0,1) arc(270:180:5mm) -- cycle;
\fill[black!20] (2,4) -- (6,0) -- (7,0) arc(180:90:5mm) -- ++(0,1) -- (5,4) -- (4,4) arc(0:-180:5mm) -- cycle;
\fill[black!20] (0,0) -- (1,0) -- (-0.5,1.5) -- ++(0,-1) arc(90:0:5mm) -- cycle;
\fill[black!20] (2,0) arc(180:0:15mm) -- (4,0) arc(0:180:5mm) -- cycle;

\draw (-0.5,0)--(7.5,0);
\draw[dashed] (7.5,0)--(7.5,4);
\draw (7.5,4)--(-0.5,4);
\draw[dashed] (-0.5,4)--(-0.5,0);

\draw[red,line width=1.5pt] (1,4) arc(0:-90:15mm);
\draw[red,line width=1.5pt] (0,4) arc(0:-90:5mm);
\draw[red,line width=1.5pt] (6,4) arc(180:270:15mm);
\draw[red,line width=1.5pt] (7,4) arc(180:270:5mm);
\draw[red,line width=1.5pt] (2,4) -- (6,0);
\draw[red,line width=1.5pt] (7,0) arc(180:90:5mm);
\draw[red,line width=1.5pt] (0,0) arc(0:90:5mm);
\draw[red,line width=1.5pt] (1,0) -- (-0.5,1.5);
\draw[red,line width=1.5pt] (7.5,1.5) -- (5,4);
\draw[red,line width=1.5pt] (4,4) arc(0:-180:5mm);
\draw[red,line width=1.5pt] (2,0) arc(180:0:15mm);
\draw[red,line width=1.5pt] (3,0) arc(180:0:5mm);

\foreach \xpos in {0.5,2.5,4.5,6.5}
{
 \draw[fill] (\xpos,0) circle(0.6ex);
 \draw[fill] (\xpos,4) circle(0.6ex);
}

\draw (3.5,0.8) node {$e_1$};
\draw (-0.2,3.0) node {$e_2$};
\draw (7.2,3.0) node {$e_2$};
\draw (5.5,2.0) node {$e_3$};
\draw (-0.2,0.8) node {$e_3$};
\end{tikzpicture}
\label{statechi0}
\end{equation}
It has three hyper edges: $e_1$ and $e_2$ each of degree $1$, and $e_3$ of degree $3$.
Therefore $E = 1 + 1 + 3 = 5$.

Let us provide a few details on how to construct the cluster configuration from the loop configuration. The connections between (cluster) points
within a single time slice (top or bottom) can be easily inferred by travelling along the arcs on that time slice. It is more delicate to infer the connections
(if any) from one time slice to the other. It is obvious that if there are $s= 2k > 0$ strings, there will be $k$ such connections, and these can
again be easily inferred by travelling along the strings (this is the case in Figure~\ref{fig:conn-states}a). However, if
$s = 0$ the existence of connections between the clusters on the top and bottom time slices depends
on whether the two reduced states are both closed or both open (see section~\ref{sec:describe-states}).
An example involving open states is provided by the following figure:
\begin{equation}
\begin{tikzpicture}[scale=0.7,>=stealth]

\fill[black!20] (2,5) arc(180:360:15mm) -- ++(-1,0) arc(360:180:5mm) -- cycle;
\fill[black!20] (0,5) arc(0:-90:5mm) -- ++(0,-2) -- ++(2,0) arc(90:0:25mm) -- ++(1,0) arc(180:90:25mm) -- ++(0,2) arc(270:180:5mm) -- ++(-1,0) arc(0:-180:25mm) -- cycle;
\fill[black!20] (0,0) arc(180:0:15mm) -- ++(-1,0) arc(0:180:5mm) -- cycle;
\fill[black!20] (6,0) arc(180:0:5mm) -- cycle;

\draw (-0.5,0)--(7.5,0);
\draw[dashed] (7.5,0)--(7.5,5);
\draw (7.5,5)--(-0.5,5);
\draw[dashed] (-0.5,5)--(-0.5,0);

\draw[red,line width=1.5pt] (1,5) arc(180:360:25mm);
\draw[red,line width=1.5pt] (2,5) arc(180:360:15mm);
\draw[red,line width=1.5pt] (3,5) arc(180:360:5mm);
\draw[red,line width=1.5pt] (0,5) arc(0:-90:5mm);
\draw[red,line width=1.5pt] (7,5) arc(180:270:5mm);

\draw[red,line width=1.5pt] (0,0) arc(180:0:15mm);
\draw[red,line width=1.5pt] (1,0) arc(180:0:5mm);
\draw[red,line width=1.5pt] (6,0) arc(180:0:5mm);
\draw[red,line width=1.5pt] (4,0) arc(0:90:25mm) -- ++(-2,0);
\draw[red,line width=1.5pt] (5,0) arc(180:90:25mm);

\foreach \xpos in {0.5,2.5,4.5,6.5}
{
 \draw[fill] (\xpos,0) circle(0.6ex);
 \draw[fill] (\xpos,5) circle(0.6ex);
}

\foreach \xpos in {-0.5,1.5,3.5,5.5,7.5}
{
 \draw[line width=1.5pt] (\xpos,0) circle(0.6ex);
 \draw[line width=1.5pt] (\xpos,5) circle(0.6ex);
}

\draw (-0.5,5.2) node[above] {$1$};
\draw (0.5,5.2) node[above] {$0$};
\draw (1.5,5.2) node[above] {$1$};
\draw (2.5,5.2) node[above] {$2$};
\draw (3.5,5.2) node[above] {$3$};
\draw (4.5,5.2) node[above] {$2$};
\draw (5.5,5.2) node[above] {$1$};
\draw (6.5,5.2) node[above] {$0$};
\draw (7.5,5.2) node[above] {$1$};

\draw (-0.5,-0.2) node[below] {$1$};
\draw (0.5,-0.2) node[below] {$2$};
\draw (1.5,-0.2) node[below] {$3$};
\draw (2.5,-0.2) node[below] {$2$};
\draw (3.5,-0.2) node[below] {$1$};
\draw (4.5,-0.2) node[below] {$0$};
\draw (5.5,-0.2) node[below] {$1$};
\draw (6.5,-0.2) node[below] {$2$};
\draw (7.5,-0.2) node[below] {$1$};

\end{tikzpicture}
\label{statechi1}
\end{equation}
To determine in general the connections (if any) between the clusters on the top and bottom time slices we proceed as follows.
Consider first the set ${\cal P} = {\cal P}_1 \cup {\cal P}_2$ consisting of $2n$ points on each time slice, which is the union of
the points ${\cal P}_1$ on which the clusters live---i.e., those with an even loop point on the left and an odd loop point on the right,
shown as solid circles in (\ref{statechi1})---and
the points ${\cal P}_2$ on which dual clusters live---i.e., those with an odd loop point on the left and an even loop point on the right,
shown as open circles in (\ref{statechi1}).
We now define a set of integer ``heights'' on ${\cal P}$ as follows. Starting from an arbitrary initial value (chosen as $1$ in (\ref{statechi1})),
and moving along the top (resp.\ bottom) time slice from left to right
(resp.\ from right to left), let the height increase (resp.\ decrease) by one unit each time one crosses a loop opening (resp.\ closing), i.e.,
a loop point with code $c_i = 1$ (resp.\ $c_i = 2$). Since only height differences are defined, this procedure defines the heights only
up to a global translation. The heights corresponding to the example (\ref{statechi1}) are shown next to each point in ${\cal P}$.
It is easy to see that if, for each of the time slices taken separately, the minimum of this height profile (which is $0$ in (\ref{statechi1}) for
both time slices) resides at a point of ${\cal P}_1$ (resp.\ ${\cal P}_2$), the corresponding reduced state is open (resp.\ closed).
In the case of a pair of open reduced states, all the points of ${\cal P}_1$ residing at the minimum height on the top time slice
are incident on the same hyper edge as the corresponding points of minimum height on the bottom time slice. In the example (\ref{statechi1})
there are two such points on the top time slice and one on the bottom time slice, so top and bottom are connected through a hyper
edge of degree $2$. This concludes the construction of the
cluster configuration from the loop configuration.

We finally define $V$ and $C$ as, respectively, the number of vertices and clusters in the hyper graph. Both of these number are defined
{\em after} the identification of the top and bottom time slices. In particular $V = n$. Either of the states (\ref{statechi0}) and (\ref{statechi1})
turn out to have $C=1$. We can then compute $\chi$ from (\ref{euler-relation}), and we find
$\chi = 5 + 2 - 4 - 3 = 0$ for (\ref{statechi0}) and $\chi = 4 + 2 - 4 - 1 = 1$ for (\ref{statechi1}).

In general the possible values are $\chi = 0,1,2$. When $\chi = 0$ the state belongs to the $Z_{\rm 0D}$ class, and when
$\chi = 1$ or $2$ it belongs to the $Z_{\rm 2D}$ class. Moreover, when $\chi = 1$ (resp.\ $\chi = 2$) we can deduce that
the state contains $s = 2k > 0$ strings (resp.\ $s=0$ strings), but we shall not need this fact to compute (\ref{PB_cluster}).

Now that the nature ($Z_{0D}$, $Z_{\rm 1D}$ or $Z_{\rm 2D}$) of each connectivity state has been determined, it remains
only to multiply it by a certain power of $n_{\rm loop}$ that has not been accounted for by the transfer process itself.
First, there is a factor of $\sqrt{q} = n_{\rm loop}$ for each vertex in the basis $B$, coming from the front factor of
(\ref{loop_repr}). The number of vertices should of course be computed up to the identification which is made by imposing
the doubly periodic boundary conditions on $B$ (i.e., gluing the left and right, and the top and bottom).
Second, each state has to be multiplied by $n_{\rm loop}^P$, where we recall that $P$ is the number of loops in the final state.
Finally, the $Z_{\rm 2D}$ configurations should be multiplied by a factor of $q = n_{\rm loop}^2$; this follows from the Euler relation.

\section{Results on Archimedean lattices}
\label{sec:res-archi}

In this section we present our results for the Potts model (and the special case of bond percolation)
on the Archimedean lattices. The hexagonal lattice can be omitted from the discussion since it is
the dual of the triangular lattice, $(6^3) = D(3^6)$, and hence covered by the general remarks on duality
made in section~\ref{sec:res-dual}.

The triangular and square lattices are of the three-terminal type and hence exactly solvable. This means
that the critical curves are exactly known \cite{BaxterTemperleyAshley78,Baxter73,Baxter82}. The exact
solvability will cause $P_B(q,v)$ to shed a small factor \cite{Jacobsen12,Jacobsen13}, corresponding to the exact critical curves.
However, the remaining, large factor in $P_B(q,v)$ will still give important information about additional
critical behaviour in the antiferromagnetic region $v < 0$. This information---and the whole critical
manifold of the other eight Archimedean lattices---is only rendered approximately
by the roots of the critical polynomials, but the accuracy is such that we can use powerful extrapolation
techniques to obtain the ferromagnetic critical points to very high precision, along with a precise global
understanding of the phase diagram in the antiferromagnetic region.

An important feature in the regime $v<0$ is the presence of a so-called Berker-Kadanoff (BK) phase
\cite{Saleur91}. This is a region in the real $(q,v)$ plane throughout which correlation functions
decay as power laws, and where the
temperature variable $v$ is irrelevant in the renormalisation group (RG) sense. The lower and upper
boundaries of the BK phase are a pair of antiferromagnetic transition curves, $v_{-}(q) < v < v_{+}(q)$, that
merge at some value $q_{\rm c}$:
\begin{equation}
 \lim_{q \to q_{\rm c}} v_{-}(q) =  \lim_{q \to q_{\rm c}} v_{+}(q) \,.
\end{equation}
The inequality $q_{\rm c} \le 4$ is guaranteed by quantum group results \cite{Saleur91}. For several lattices---including
the square lattice---one has $q_{\rm c} = 4$ exactly \cite{Saleur91,Jacobsen13}, but there are
indications that on other lattices---including the kagome lattice---one may have $q_{\rm c} < 4$ strictly \cite{Jacobsen13}.
The RG irrelevance of $v$ 
has the consequence that phase transitions inside the BK phase are $v$-independent and will
manifest themselves as vertical rays in the $P_B(q,v) = 0$ manifold. It was found in \cite{Jacobsen12,Jacobsen13}
for several examples that these vertical rays occur when $q$ is equal to a Beraha number 
\begin{equation}
 B_k = \left( 2 \cos(\pi/k) \right)^2
 \label{Beraha}
\end{equation}
with even $k = 4,6,8,\ldots$, but when $q_{\rm c} < 4$ the range of $k$-values is limited by $B_k < q_{\rm c}$.

The results given below provide firm evidence that these characteristics of the BK phase
are generic for the Potts model defined on any two-dimensional lattice. Moreover, we obtain
precise information about the extent of the BK phase and the value of $q_{\rm c}$ for all the lattices
under study.

\begin{table}
\begin{center}
 \begin{tabular}{l|lll}
 Lattice & Vertices & Edges & Parity of $n$ \\ \hline
 Triangular & 2 & 6 & Any \\
 Square &  2 & 4 & Any \\
 Kagome & 3 & 6 & Any \\
 Four-eight & 4 & 6 & Any \\
 Frieze & 2 & 5 & Even \\
 Three-twelve & 6 & 9 & Any \\
 Cross & 3 & $\frac92$ & Even \\
 Snub square & 2 & 5 & Even \\ 
 Snub hexagonal & $\frac{12}{7}$ & $\frac{30}{7}$ & 0 mod 7 \\
                           & 3 & $\frac{15}{2}$ & Even \\
 Ruby & 3 & 6 & Even \\ \hline
  \end{tabular}
 \caption{Number of vertices and edges per grey square (cf.~Figure~\ref{fig:square-basis})
 for each Archimedean lattice, using square bases of size $n \times n$ grey squares. In
 addition we state any parity constraint on $n$. Note that we have two different ways of
 constructing the snub hexagonal lattice.}
  \label{tab:packing}
\end{center}
\end{table}

On a more technical level, we show below how each of the Archimedean lattices can be cast as a
four-terminal lattice, in the precise sense of Figure~\ref{fig:square-basis}. This is done notably by specifying
the corresponding $\check{\sf R}$-matrix. In cases where this construction is not unique, the best
choice for our purposes is the one that allows $n$ to take any value (i.e., with no parity constraints)
and that packs as many vertices and edges as possible into the basis of a given size $n$. To quantify
this latter aspect, we show in Table~\ref{tab:packing} the number of vertices and edges per
$\check{\sf R}$-matrix that were achieved for each lattice (we include any horizontal diagonals on the
white squares in this count). For instance, our largest ($n=7$) computation
on the three-twelve lattice uses a basis of $6 n^2 = 294$ vertices and $9 n^2 = 441$ edges.

For $n \le 5$ (resp.\ $n = 6$) we have computed the exact critical polynomial for the Potts model
(resp.\ for bond percolation only). Our plots of the phase diagrams are based on these polynomials.
As in our preceding work \cite{SJ12,Jacobsen13} the polynomials are available in electronic form as
supplementary material to this paper.%
\footnote{This text file {\tt PB.m} provided can be processed by {\sc Mathematica} or---maybe after minor changes of
formatting---by any symbolic computer algebra program of the reader's liking.}
The degree of $P_B(q,v)$ is $k_q n^2$ in the $q$-variable and $k_v n^2$ in the $v$-variable, where
$k_q$ and $k_v$ can be read from the second and third columns of Table~\ref{tab:packing}. For instance,
our largest ($n=5$) two-variable polynomial for the three-twelve lattice has degree $150$ in $q$, and
degree $225$ in $v$. Moreover, the coefficients are typically 60-digit integers.

\subsection{Triangular lattice $(3^6)$}
\label{sec:triangular}

\begin{figure}
\begin{center}
\begin{tikzpicture}[scale=1.0,>=stealth]
\foreach \xpos in {0,1,2,3}
\foreach \ypos in {0,1,2,3}
 \fill[black!20] (\xpos+0.5,\ypos) -- (\xpos+1,\ypos+0.5) -- (\xpos+0.5,\ypos+1) -- (\xpos,\ypos+0.5) -- cycle;

\foreach \xpos in {0,1,2,3}
\foreach \ypos in {0,1,2,3}
 \draw[blue,ultra thick] (\xpos+0.5,\ypos) -- (\xpos+1,\ypos+0.5) -- (\xpos+0.5,\ypos+1) -- (\xpos,\ypos+0.5) -- cycle;
\foreach \ypos in {0,1,2,3}
{
 \draw[blue,ultra thick] (0,\ypos) -- (4,\ypos);
 \draw[blue,ultra thick] (0,\ypos+0.5) -- (4,\ypos+0.5);
}

\foreach \xpos in {0,1,2,3}
\foreach \ypos in {0,1,2,3}
 \draw[black] (\xpos+0.5,\ypos) -- (\xpos+1,\ypos+0.5) -- (\xpos+0.5,\ypos+1) -- (\xpos,\ypos+0.5) -- cycle;

\draw[very thick,->] (0,-0.5)--(4,-0.5);
\draw (4,-0.5) node[right] {$x$};
\foreach \xpos in {0,1,2,3}
{
 \draw[thick] (\xpos+0.5,-0.6)--(\xpos+0.5,-0.4);
 \draw (\xpos+0.5,-0.5) node[below] {$\xpos$};
}

\draw[very thick,->] (-0.5,0)--(-0.5,4);
\draw (-0.5,4) node[above] {$y$};
\foreach \ypos in {0,1,2,3}
{
 \draw[thick] (-0.6,\ypos+0.5)--(-0.4,\ypos+0.5);
 \draw (-0.5,\ypos+0.5) node[left] {$\ypos$};
}
 
\end{tikzpicture}
 \caption{Four-terminal representation of the triangular lattice.}
 \label{fig:triangular}
\end{center}
\end{figure}
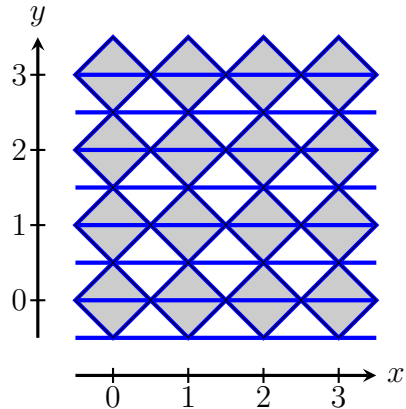

A square basis for the triangular lattice is obtained by placing the following $\check{\sf R}$-matrix
\begin{equation}
 \check{\sf R}_i = {\sf H}_{i+1} {\sf V}_i {\sf H}_{i+1} {\sf V}_{i+2} {\sf H}_{i+1}
 \label{triangular-R}
\end{equation}
inside each grey square in Figure~\ref{fig:square-basis}.
In addition we need horizontal diagonals on all the white squares.
The resulting representation is shown in Figure~\ref{fig:triangular}.
There are 4 vertices and 6 edges per grey square.

The critical polynomials $P_B(q,v)$ invariably factorises for any size $n$ of the
basis, shedding the small factor
\begin{equation}
 P_{\rm tri}(q,v) = v^3 + 3 v^2 - q \,.
 \label{Ptri}
\end{equation}
This is compatible with the fact \cite{BaxterTemperleyAshley78} that the triangular-lattice
Potts model is exactly solvable on the curve $P_{\rm tri}(q,v) = 0$. In particular, for $q=1$
we have the root $v = -1 + 2 \cos(2 \pi / 9)$, meaning that the exact percolation threshold
is
\begin{equation}
 p_{\rm c} = \frac{v}{1+v} = 2 \sin \left( \frac{\pi}{18} \right) \,.
 \label{pc_triangular}
\end{equation}

\begin{figure}
\begin{center}
\includegraphics[width=12cm]{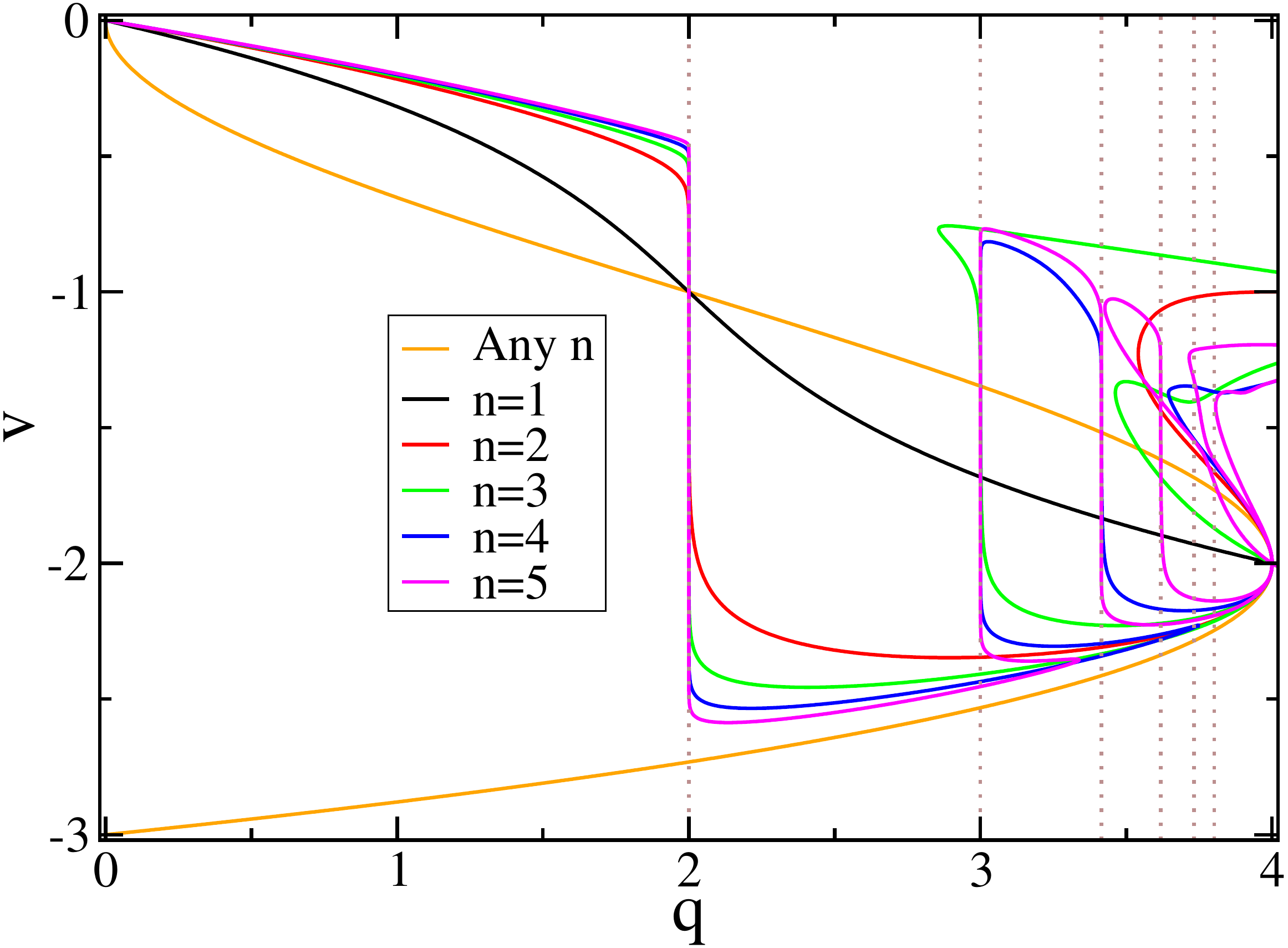}
\caption{Roots of $P_B(q,v)$ for the Potts model on the triangular lattice, using
$n \times n$ square bases. The curve labelled ``any $n$'' corresponds to (\ref{Ptri}).}
\label{fig:triangular-pd}
\end{center}
\end{figure}

The remaining, large factor in $P_B(q,v)$ gives additional information about the critical
manifold in the regime $v < 0$, as we shall now see. Its roots in the real $(q,v)$ plane
are shown in Figure~\ref{fig:triangular-pd}. The lower boundary of the BK phase, denoted
$v_-(q)$, is the lower branch of the cubic (\ref{Ptri}). The corresponding upper boundary
$v_+(q)$ can be seen in Figure~\ref{fig:triangular-pd} as the curve starting from the origin
with near-horizontal slope, passing through the top point of a series of vertical rays, and
extending towards the special point $(q,v)=(4,-2)$. Curiously, the critical polynomials miss
the part of $v_+(q)$ with $2 < q < 3$. Heuristically, this is ``because'' the polynomials have to trace out both
$v_+(q)$, $v_-(q)$ and the vertical rays in a zig-zag fashion that becomes increasingly
complicated upon approaching $q=4$. The vertical rays corresponding to $k=4,6,8,10$
in (\ref{Beraha})---and to a lesser extent $k=12,14$ as well---are clearly visible from the
figure. Here and in the following, we help the visual identification of such vertical
rays by superimposing a number of dotted grey lines on the figures.

In conclusion, Figure~\ref{fig:triangular-pd} provides rather compelling evidence that
the curves $v_\pm(q)$ will merge in $(q,v)=(4,-2)$. In particular $q_{\rm c} = 4$ for
the triangular lattice. It follows almost inevitably that the critical curve to which the
BK phase is RG-attracted must be the middle branch of (\ref{Ptri}). That conclusion
is backed up by a number of other studies \cite{Salas03,Salas06,Salas07,JacSal06}.

We should mention that the triangular-lattice Potts model is exactly solvable on the 
chromatic line $v=-1$ \cite{Nienhuis82,Baxter86,Baxter87}. It follows from
\cite{Baxter86,Baxter87} that $v_+(q) = -1$ for some value $q = 3.819\,671\,731\cdots$
obtained by equating two infinite products. The region near $(q,v)=(4,-1)$ exhibits
some rather complicated physics and would be a suitable subject for a separate study \cite{Salas14}.

\subsection{Square lattice $(4^4)$}

The square lattice can obviously be obtained by removing the diagonal edges from the triangular
lattice. The $\check{\sf R}$-matrix then reads
\begin{equation}
 \check{\sf R}_i = {\sf H}_{i+1} {\sf V}_i {\sf V}_{i+2} {\sf H}_{i+1} \,.
 \label{eq:Rsquare}
\end{equation}
The basis then looks like Figure~\ref{fig:square-basis} with the horizontal edges removed.
There are 2 vertices and 4 edges per grey square.

For any size $n$, the critical polynomial $P_B(q,v)$ factorises, shedding two small factors:
\begin{equation}
 P_{\rm sq}(q,v) = (v^2 - q)(v^2 + 4 v + q) \,.
 \label{Psq}
\end{equation}
The zero set of the first factor describes the selfdual critical point of the square-lattice Potts
model \cite{Baxter73}, while the zero set of the second factor yields two mutually dual
antiferromagnetic critical points \cite{Baxter82}. The model is exactly solvable on these
curves \cite{Baxter73,Baxter82,JacSal06,Ikhlef08,Ikhlef12}. In particular, we find that
$P_{\rm sq}(1,1) = 0$, so the exact percolation threshold is
\begin{equation}
 p_{\rm c} = \frac{1}{2} \,.
\end{equation}

\begin{figure}
\begin{center}
\includegraphics[width=12cm]{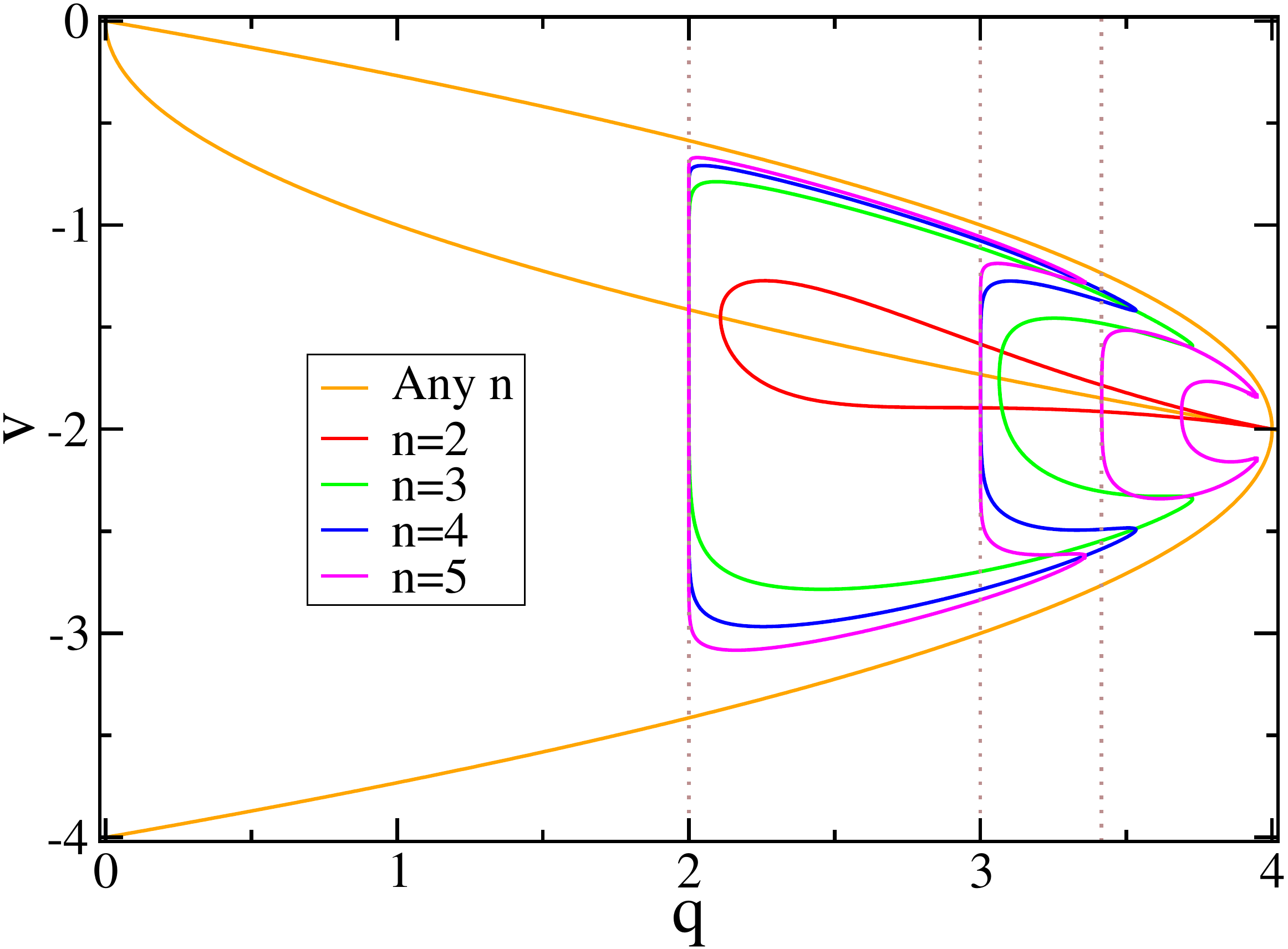}
\caption{Roots of $P_B(q,v)$ for the Potts model on the square lattice, using
$n \times n$ square bases. The curves labelled ``any $n$'' correspond to (\ref{Psq}).}
\label{fig:square-pd}
\end{center}
\end{figure}

The roots of $P_B(q,v)$ in the real $(q,v)$ plane are shown in Figure~\ref{fig:square-pd};
the curves with $n \le 4$ were already reported in \cite{SJ12}.
Of all the Archimedean lattices, the square lattice is the only one where this phase
diagram can be claimed to be completely understood. The boundaries of the BK phase
are given by the second factor in (\ref{Psq}), namely
\begin{equation}
 v_\pm(q) = -2 \pm \sqrt{4-q} \,,
\end{equation}
and in particular $q_{\rm c} = 4$. In the thermodynamical limit, $n \to \infty$,
we expect an infinite set of vertical rays, corresponding to (\ref{Beraha})
with $k=4,6,8,\ldots$. The first few, with $k=4,6,8$, are clearly visible in our
results for $n \le 5$, shown in Figure~\ref{fig:square-pd}, as is the precursor
of the $k=10$ ray (which is not yet in its correct position).

\subsection{Kagome lattice $(3,6,3,6)$}
\label{sec:kagome}

The square and triangular lattices (and the hexagonal lattice, which is the dual
of the triangular) could have been presented in three-terminal form. This fact
actually makes it possible to compute the critical manifolds,
$P_{\rm sq}(q,v) = 0$ and $P_{\rm tri}(q,v) = 0$, exactly \cite{WuLin80}, and it
is closely related to the exact solvability
\cite{Baxter73,Baxter82,JacSal06,Ikhlef08,BaxterTemperleyAshley78}
of the models along these curves.

However, the kagome lattice---and indeed all the remaining Archimedean lattices---is
not of the three-terminal type. Accordingly no exact solution is known to this day.
The graph polynomial method therefore gives only approximate results, which are
however very accurate, in particular in the ferromagnetic region $v > 0$.

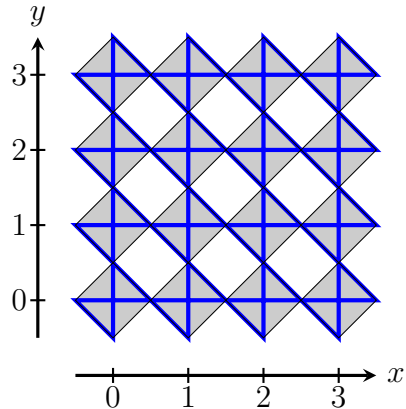
\begin{figure}
\begin{center}
\begin{tikzpicture}[scale=1.0,>=stealth]
\foreach \xpos in {0,1,2,3}
\foreach \ypos in {0,1,2,3}
 \fill[black!20] (\xpos+0.5,\ypos) -- (\xpos+1,\ypos+0.5) -- (\xpos+0.5,\ypos+1) -- (\xpos,\ypos+0.5) -- cycle;
 
\foreach \xpos in {0,1,2,3}
\foreach \ypos in {0,1,2,3}
 \draw[blue,ultra thick] (\xpos+0.5,\ypos) -- (\xpos+0.5,\ypos+1) -- (\xpos+1,\ypos+0.5) -- (\xpos,\ypos+0.5) -- cycle;

\foreach \xpos in {0,1,2,3}
\foreach \ypos in {0,1,2,3}
 \draw[black] (\xpos+0.5,\ypos) -- (\xpos+1,\ypos+0.5) -- (\xpos+0.5,\ypos+1) -- (\xpos,\ypos+0.5) -- cycle;

\draw[very thick,->] (0,-0.5)--(4,-0.5);
\draw (4,-0.5) node[right] {$x$};
\foreach \xpos in {0,1,2,3}
{
 \draw[thick] (\xpos+0.5,-0.6)--(\xpos+0.5,-0.4);
 \draw (\xpos+0.5,-0.5) node[below] {$\xpos$};
}

\draw[very thick,->] (-0.5,0)--(-0.5,4);
\draw (-0.5,4) node[above] {$y$};
\foreach \ypos in {0,1,2,3}
{
 \draw[thick] (-0.6,\ypos+0.5)--(-0.4,\ypos+0.5);
 \draw (-0.5,\ypos+0.5) node[left] {$\ypos$};
}
 
\end{tikzpicture}
 \caption{Four-terminal representation of the kagome lattice.}
 \label{fig:kagome}
\end{center}
\end{figure}

The $\check{\sf R}$-matrix of the kagome lattice can be written as
\begin{equation}
 \check{\sf R}_i = {\sf H}_{i+1} {\sf V}_{i+2} {\sf V}_{i} {\sf E}_{i+1} {\sf V}_{i+2} {\sf V}_i {\sf H}_{i+1} \,.
 \label{eq:Rkagome}
\end{equation}
The resulting representation is shown in Figure~\ref{fig:kagome}.
There are 3 vertices and 6 edges per grey square.

\begin{table}
\begin{center}
 \begin{tabular}{l|l}
 $n$ & $p_{\rm c}$ \\ \hline
 1 & 0.52442971752127479354687968153445507162056741657866 \\
 2 & 0.52440672318823181914323447999258988541033371409674 \\
 3 & 0.52440517271376997270613021015286282832593143635602 \\
 4 & 0.52440502742741472069907568050523723911941122246320 \\
 5 & 0.52440500598061634783869324699127606806572366674050 \\
 6 & 0.52440500130658104881349494422717375226544897430878 \\
 7 & 0.52440499997320890049536436452361870395440274892056 \\ \hline
 $\infty$ & 0.524404999173 (3) \\
 Ref.~\cite{FengDengBlote08} & 0.52440499 (2) \\
 \end{tabular}
 \caption{Bond percolation threshold $p_{\rm c}$ on the kagome lattice.}
  \label{tab:kagome1}
\end{center}
\end{table}

We have computed the (unique) root $P_B(q,v)$ in the ferromagnetic regime, $v > 0$,
for several integer values of $q$. The results for $q=1$ are shown in Table~\ref{tab:kagome1}
to 50-digit numerical precision, in terms of the percolation probability $p = \frac{v}{1+v}$,
for square bases of size $1 \le n \le 7$. A quick glance at the table makes it obvious that
these numbers converge very fast to their expected limit $p_{\rm c}$.

The roots for $1 \le n \le 4$ have already been reported in \cite{SJ12}, leading the authors
to propose a final estimate of $p_{\rm c} = 0.524\, 405\, 00(1)$. 
However, the fact that we now have three more terms in the sequence allows us to employ
powerful extrapolation techniques to obtain a very accurate final value of $p_{\rm c}$.
We have chosen to apply the time-proven Bulirsch-Stoer (BS) extrapolation \cite{BulirschStoer}.
This algorithm requires a parameter $w$ which can be thought of as a correction-to-scaling exponent. 
In an preliminary step we obtain an approximate value for $w$ from a non-linear fit of
the data in Table~\ref{tab:kagome1} to the form
\begin{equation}
 p_{\rm c}(n) = p_{\rm c} + A n^{-w} \,.
\label{extrapol_w}
\end{equation}
The best results are obtained by constraining this fit to the last three available data points.
We do not report the values of $w$ found for each data set to be considered in this paper,
except for the first few examples of each type of problems. We shall however provide a few
general remarks in the Discussion section~\ref{sec:disc}.
Generally speaking we find $w \approx 6$ for the best behaved (bond or site) percolation problems.
Obviously, the higher the value of $w$, the better will be the precision on the final result.

In a second step, we insert the value found for $w$
into the implementation of the BS algorithm described by Monroe \cite{Monroe}. This results
in a series of approximants that are compared among themselves in order to assess a final
value and error bar. We crosscheck our results by repeating the whole procedure with the
last data point being eliminated, in order to ensure that the central value and error bar obtained
from fits on $N-1$ points are compatible (albeit of course less precise) with those obtained on
all $N$ data points.

For some of the lattice for which fewer data points (i.e., sizes $n$) are available, some adaptations
of this general procedure will be necessary. We shall return to this in the following subsections.

In Table~\ref{tab:kagome1} and the following many tables in this paper, we compare our
final result with the most accurate value known from previous numerical work. In the present
case, we find $w \approx 6.36$, and the relative precision on the final value of $p_{\rm c}$ is of the order
$4 \cdot 10^{-11}$, that is, four orders of magnitude
better than the previous result \cite{FengDengBlote08}.

\begin{table}
\begin{center}
 \begin{tabular}{l|l}
 $n$ & $v_{\rm c}$ \\ \hline
 1 & 1.8762692083457608448172661268682642135309588452285 \\
 2 & 1.8764397543028806860142570871053207225112352007846 \\
 3 & 1.8764569161964147459134636690080036024897511662923 \\
 4 & 1.8764589940034619711814113716932244874728362773156 \\
 5 & 1.8764593952716922296679157122640055912513566938545 \\
 6 & 1.8764595053053275343456174924063602310856281037794 \\
 7 & 1.8764595432649855348163610191948592418994287446958 \\ \hline
 $\infty$ & 1.8764595734 (3)\\
 Ref.~\cite{Jensen97} & 1.87646 (5) \\
  \end{tabular}
 \caption{Critical point $v_{\rm c}$ of the $q=3$ state Potts model on the kagome lattice.}
  \label{tab:kagome3}
\end{center}
\end{table}

We next turn to the Ising model ($q=2$). In this case, all the $P_B(q,v)$ are found to
factorise into small factors. The maximum degree of the factors is $d_{\rm max} = 4$ for $n=1,2$;
$d_{\rm max} = 8$ for $n=3,4$; and $d_{\rm max} = 16$ for $n=5$. There is precisely one of 
these factors, namely
\begin{equation}
 -8 - 8 v + 4 v^3 + v^4 \,,
 \label{kagome_factor}
\end{equation}
that possesses a positive root,
\begin{equation}
 v_{\rm c} = \sqrt{3 + 2 \sqrt{3}} - 1 \simeq 1.542\,459\,756\cdots \,.
 \label{kagome_root}
\end{equation}
Moreover, (\ref{kagome_factor}) is a factor in $P_B(q,v)$ for any size $n$. Its physical
root (\ref{kagome_root}) coincides with the exactly known critical point of the kagome-lattice
Ising model \cite{KanoNaya53,Codello10}. Below we shall similarly see that we recover
exact results for the Ising model on any lattice.

The polynomials for the Ising model are often found to simplify under the change of variables
$v = -1 + \sqrt{y}$. Recall that $v = {\rm e}^K - 1$, but rewriting the nearest neighbour interaction
energy as in Ising form, $K \delta_{\sigma_i,\sigma_j} = K_{\rm Ising} (S_i S_j + 1)$ for spins
$S_i = \pm 1$, we find $K = 2 K_{\rm Ising}$ so that $y = {\rm e}^{K_{\rm Ising}}$ is simply the
Boltzmann factor for a pair of aligned Ising spins. In the present
case (\ref{kagome_factor}) simplifies to
\begin{equation}
 -3 - 6t + t^2 \,.
\end{equation}

\begin{table}
\begin{center}
 \begin{tabular}{l|l}
 $n$ & $v_{\rm c}$ \\ \hline
 1 & 2.1558422365136376068815817932185116250325673239278 \\
 2 & 2.1562074529907952231032843370607657724546027337305 \\
 3 & 2.1562475983381240731591377965181755879281862607822 \\
 4 & 2.1562528801542168626963668658051048225566594172221 \\
 5 & 2.1562540028309456631273970233783408172592478843687 \\
 6 & 2.1562543392791356268847391898132756614686640239252 \\
 7 & 2.1562544649475050404839402645792684169318045633213 \\ \hline
 $\infty$ & 2.1562545798 (8) \\
 Ref.~\cite{Jensen97} & 2.1561 (5) \\
  \end{tabular}
 \caption{Critical point $v_{\rm c}$ of the $q=4$ state Potts model on the kagome lattice.}
  \label{tab:kagome4}
\end{center}
\end{table}

The critical points for the $q=3$ and $q=4$ state Potts models are shown
in Tables~\ref{tab:kagome3}--\ref{tab:kagome4}. The exponent appearing
in (\ref{extrapol_w}) is $w \approx 5.36$ for $q=3$, and $w \approx 4.80$ for $q=4$.
In any case, using BS extrapolations as explained above,
we arrive at values for $v_{\rm c}$ which are considerably more accurate than
previous numerical results. Note in particular that numerical simulations
of the Monte Carlo or transfer matrix type are usually particularly difficult for $q=4$
because of the presence of logarithmic corrections to scaling. By contrast, the graph
polynomial method only experiences a slight decrease of $w$, and the precision is almost
as good as for percolation. Thus, for $q=4$ our final value for $v_{\rm c}$ is six orders
of magnitude more precise than the previous result \cite{Jensen97}.

\begin{figure}
\begin{center}
\includegraphics[width=12cm]{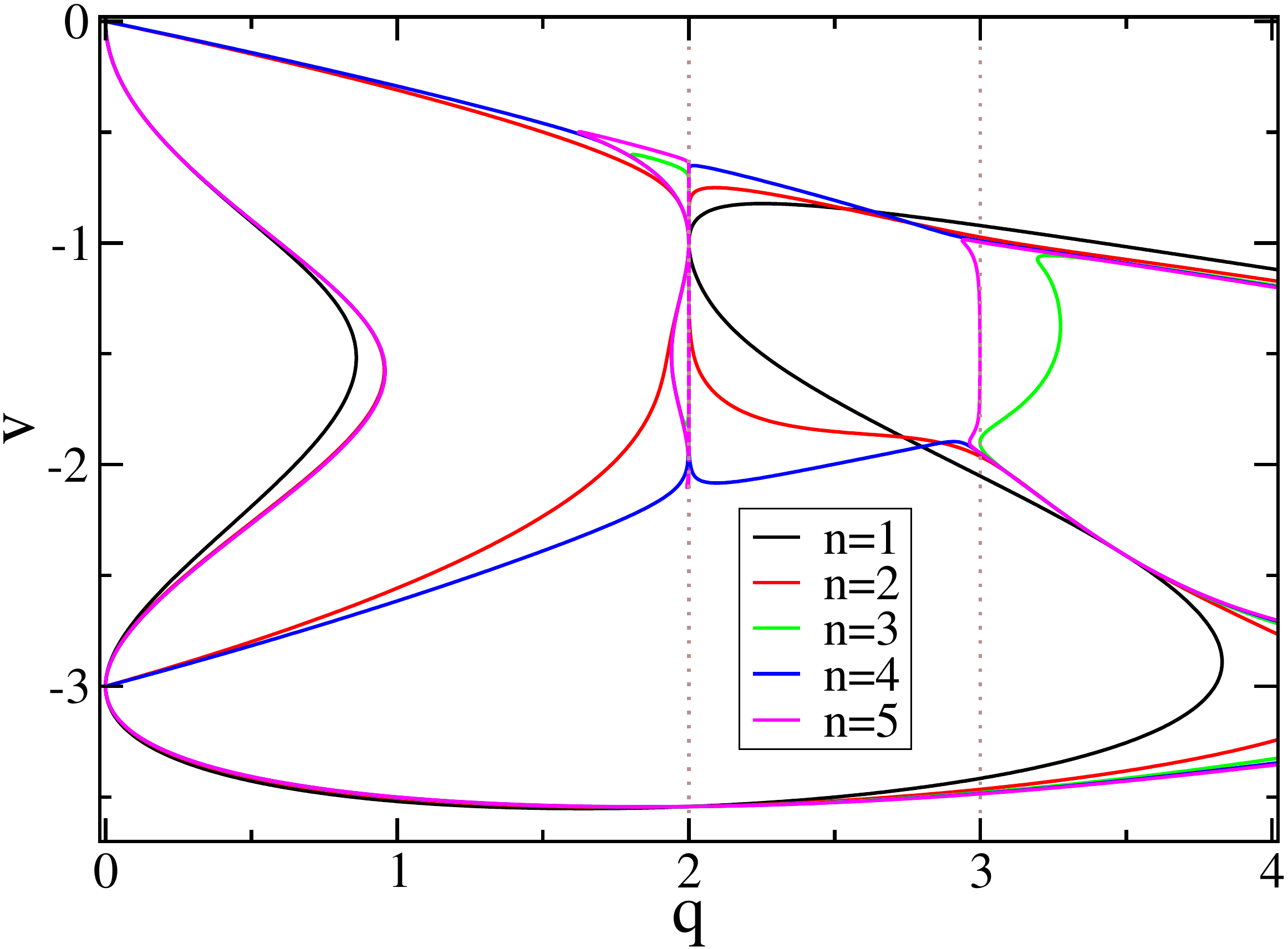}
\caption{Roots of $P_B(q,v)$ for the Potts model on the kagome lattice, using
$n \times n$ square bases.}
\label{fig:kagome-pd}
\end{center}
\end{figure}

The phase diagram on the kagome lattice has been discussed in detail in \cite{Jacobsen12},
and further in \cite{Jacobsen13} based on square-basis $P_B(q,v)$ with $n=1,2,3,4$.
In Figure~\ref{fig:kagome-pd} we show again the roots of $P_B(q,v)$ in the real $(q,v)$ plane, but this time
with the $n=5$ basis included. Because of the extensive treatment of this phase diagram in \cite{Jacobsen12,Jacobsen13}
we shall be rather brief.

The BK phase contains vertical rays at $q=B_4=2$ and $q=B_6=3$. Unlike the triangular and square lattices,
there is no sign of the BK phase widening out towards $q=4$ as $n$ increases. Its rightmost termination might
be close to the $n=3$ arc extending to around $q \approx 3.2$. This arc is confirmed by results from the hexagonal
bases studied in \cite{Jacobsen13}. On the other hand, even allowing for $n \ {\rm mod}\ 2$ parity effects which are
visible elsewhere in the phase diagram, it is curious that this arc is not confirmed by the $n=5$ critical polynomial.
So it might also be that the BK phase in fact terminates right at the $q=3$ vertical ray. In any case, it seems certain
that $q_{\rm c} < 4$ for the kagome lattice.

The upper boundary $v_+(q)$ of the BK phase is the near-straight line emanating from the origin and passing through
the point $(q,v) = (3,-1)$. Indeed, the three-state zero-temperature antiferromagnet on the kagome lattice is equivalent to the corresponding
four-state model on the triangular lattice, which in turn is known to be critical with central charge $c = 2$ \cite{MooreNewman00}.
It is interesting to observe that at finite $n$ the curve $v_+(q)$ is approximated by various pieces from the different $P_B(q,v)$,
and that no single critical polynomial reproduces the curve completely. This is in line with observations already made in
\cite{Jacobsen12,Jacobsen13}. In particular we see clear $n \ {\rm mod}\ 2$ parity effects: the critical polynomials with even $n$
(resp.\ odd $n$) are the only ones to produce the part of $v_+(q)$ with $0 < q \lesssim 1.6$ and $2 < q < 3$ (resp.\ $1.6 \lesssim q < 2$).

Meanwhile, the lower boundary $v_-(q)$ of the BK phase is the near-straight line emanating from $(q,v)=(0,-3)$ and passing
through $\approx (3,-2)$. It is determined by the polynomials with even $n$. Interestingly there is a lower-lying curve, coming out
from $(0,-3)$ with infinite slope, and the space between this curve and $v_-(q)$ is devoid of vertical rays (because it does not
belong to the BK phase).

Finally, the existence of two small enclosed regions, or phases---the first a thin sliver between $(2,-2)$ and $(2,-1)$, and the other a triangular-shaped
region above $(2,-1)$---is confirmed by the new $n=5$ polynomial.

\subsection{Four-eight lattice $(4,8^2)$}
\label{sec:foureight}

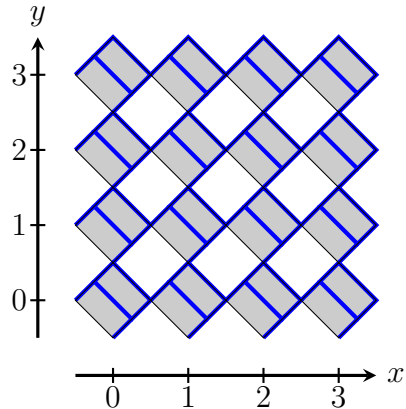
\begin{figure}
\begin{center}
\begin{tikzpicture}[scale=1.0,>=stealth]
\foreach \xpos in {0,1,2,3}
\foreach \ypos in {0,1,2,3}
 \fill[black!20] (\xpos+0.5,\ypos) -- (\xpos+1,\ypos+0.5) -- (\xpos+0.5,\ypos+1) -- (\xpos,\ypos+0.5) -- cycle;
 
\foreach \xpos in {0,1,2,3}
\foreach \ypos in {0,1,2,3}
{
  \draw[blue,ultra thick] (\xpos,\ypos+0.5) -- (\xpos+0.25,\ypos+0.75);
  \draw[blue,ultra thick] (\xpos+0.5,\ypos) -- (\xpos+0.75,\ypos+0.25);
  \draw[blue,ultra thick] (\xpos+0.25,\ypos+0.75) -- (\xpos+0.75,\ypos+0.25) -- (\xpos+1,\ypos+0.5) -- (\xpos+0.5,\ypos+1) -- cycle;
}

\foreach \xpos in {0,1,2,3}
\foreach \ypos in {0,1,2,3}
 \draw[black] (\xpos+0.5,\ypos) -- (\xpos+1,\ypos+0.5) -- (\xpos+0.5,\ypos+1) -- (\xpos,\ypos+0.5) -- cycle;

\draw[very thick,->] (0,-0.5)--(4,-0.5);
\draw (4,-0.5) node[right] {$x$};
\foreach \xpos in {0,1,2,3}
{
 \draw[thick] (\xpos+0.5,-0.6)--(\xpos+0.5,-0.4);
 \draw (\xpos+0.5,-0.5) node[below] {$\xpos$};
}

\draw[very thick,->] (-0.5,0)--(-0.5,4);
\draw (-0.5,4) node[above] {$y$};
\foreach \ypos in {0,1,2,3}
{
 \draw[thick] (-0.6,\ypos+0.5)--(-0.4,\ypos+0.5);
 \draw (-0.5,\ypos+0.5) node[left] {$\ypos$};
}
 
\end{tikzpicture}
 \caption{Four-terminal representation of the four-eight lattice.}
 \label{fig:foureight}
\end{center}
\end{figure}

A four-terminal representation of the four-eight lattice is shown in Figure~\ref{fig:foureight}. The corresponding
$\check{\sf R}$-matrix reads
\begin{equation}
 \check{\sf R}_i = {\sf V}_{i+2} {\sf V}_i {\sf H}_{i+1} {\sf V}_{i+2} {\sf V}_i {\sf H}_{i+1} \,.
\end{equation}
There are 4 vertices and 6 edges per grey square.

\begin{table}
\begin{center}
 \begin{tabular}{l|l}
 $n$ & $p_{\rm c}$ \\ \hline
 1 & 0.6768351988164058635685961953282386597701580217547 \\
 2 & 0.6768110511337950640725367041515342178711568870560 \\
 3 & 0.6768050108863651886629332178738395634982856516173 \\
 4 & 0.6768036936560548693645165737411119820502567917852 \\
 5 & 0.6768033435707186400895193695054260593374332806116 \\
 6 & 0.6768032260648857547522627416623972342414983881124 \\
 7 & 0.6768031780886579080959851245202783704998102947149 \\ \hline
 $\infty$ & 0.6768031269 (6) \\
 Ref.~\cite{Parviainen07} & 0.6768023 (6) \\
 \end{tabular}
 \caption{Bond percolation threshold $p_{\rm c}$ on the four-eight lattice.}
  \label{tab:foureight1}
\end{center}
\end{table}

The approximations to the bond percolation threshold $p_{\rm c}$ are given in Table~\ref{tab:foureight1}.
Since also for this lattice any parity of $n$ is possible, the BS extrapolation scheme produces
very accurate values, improving considerably on the existing numerical results which are
shown for comparison. We note that the exponent appearing in (\ref{extrapol_w}) comes out as $w \approx 4.28$ in
this case, so it is definitely lattice dependent.

In the Ising case ($q=2$) the polynomials $P_B(q,v)$ systematically factorise.
The maximum degree of the factors is $d_{\rm max} = 4$ for $n=1,2$;
$d_{\rm max} = 8$ for $n=3$; $d_{\rm max} = 6$ for $n=4$; and $d_{\rm max} = 16$ for $n=5$. There is precisely one of 
these factors, namely
\begin{equation}
 -4 - 8 v - 6 v^2 + v^4 \,,
 \label{foureight_factor}
\end{equation}
that possesses a positive root,
\begin{equation}
 v_{\rm c} = \frac{1 + \sqrt{5 + 4 \sqrt{2}}}{\sqrt{2}} \simeq 3.015\,445\,388\cdots \,,
 \label{foureight_root}
\end{equation}
and this factor occurs in $P_B(q,v)$ for any size $n$. Its positive
root (\ref{foureight_root}) produces the exactly known critical point of the Ising model
on the four-eight lattice \cite{Utiyama51,Codello10}.

\begin{table}
\begin{center}
 \begin{tabular}{l|l}
 $n$ & $v_{\rm c}$ \\ \hline
 1 & 3.742119707930614518717609546330093725738120066089 \\
 2 & 3.742406812389425084936236849313041587665529581274 \\
 3 & 3.742474558548594455190569943534293743089719291957 \\
 4 & 3.742488803421386923793990079403662434668070157084 \\
 5 & 3.742492503198695522319538480078407840640750408712 \\
 6 & 3.742493724307267658361207275220913244289197872944 \\
 7 & 3.742494216612624056940208879007645446884063978949 \\ \hline
 $\infty$ & 3.742494730 (5) \\
  \end{tabular}
 \caption{Critical point $v_{\rm c}$ of the $q=3$ state Potts model on the four-eight lattice.}
  \label{tab:foureight3}
\end{center}
\end{table}

\begin{table}
\begin{center}
 \begin{tabular}{l|l}
 $n$ & $v_{\rm c}$ \\ \hline
 1 & 4.367630831288118711619621980404618323651758096250 \\
 2 & 4.368211338019043993502048035939247980189947635592 \\
 3 & 4.368344356164380256004179009338136774148698456053 \\
 4 & 4.368371674728465301875011853593201988010643386877 \\
 5 & 4.368378652366770115169979832681074376730845677084 \\
 6 & 4.368380925842698493907585059226287321316844017984 \\
 7 & 4.368381832892126568493945658547177946678474675220 \\ \hline
 $\infty$ & 4.36838276 (2) \\
  \end{tabular}
 \caption{Critical point $v_{\rm c}$ of the $q=4$ state Potts model on the four-eight lattice.}
  \label{tab:foureight4}
\end{center}
\end{table}

The $q=3$ and $q=4$ critical points $v_{\rm c}$ are
shown in Tables~\ref{tab:foureight3}--\ref{tab:foureight4}, and just like in the
percolation case the BS extrapolation produces very accurate final values.

\begin{figure}
\begin{center}
\includegraphics[width=12cm]{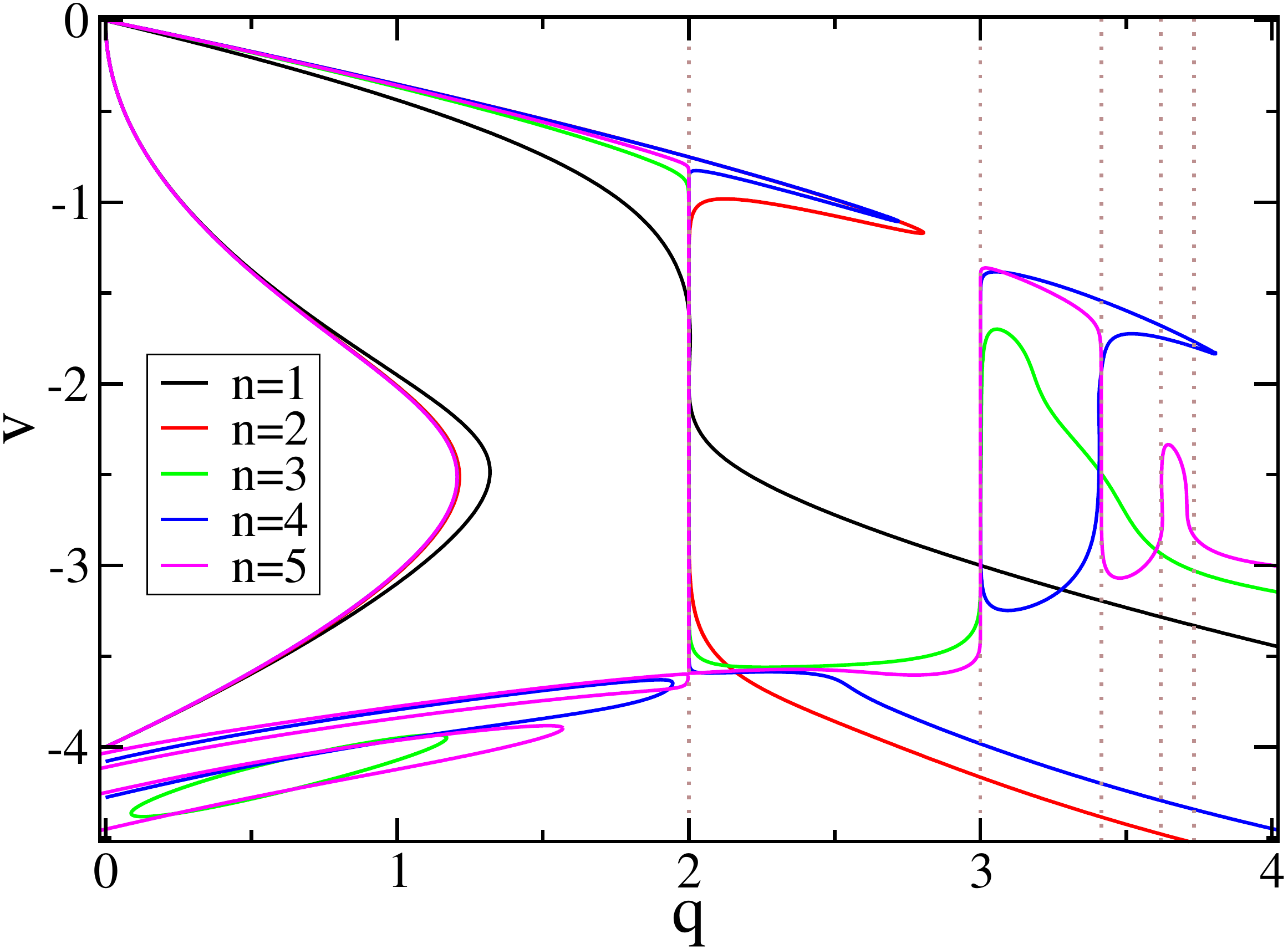}
\caption{Roots of $P_B(q,v)$ for the Potts model on the four-eight lattice, using
$n \times n$ square bases.}
\label{fig:foureight-pd}
\end{center}
\end{figure}

Ref.~\cite{Jacobsen13} already contained a discussion of the phase diagram for the 
four-eight lattice. But to highlight the new $n=5$ results, Figure~\ref{fig:foureight-pd} shows
again the roots of $P_B(q,v)$ in the real $(q,v)$ plane.

The boundaries of the BK phase can be clearly seen. The upper boundary $v_+(q)$
is the almost-straight line emanating from the origin and extending out towards
$\approx (4,-2)$. The lower boundary starts at $\approx (0,-4)$. The zero sets of the
various critical polynomials fill in the curves $v_\pm(q)$ in a zig-zag fashion, while
at the same time providing vertical rays at the Beraha numbers (\ref{Beraha}).
The rays with $k=4,6,8,10,12$ are clearly visible in the figure. It is interesting to
notice the formation of narrow ``fingers'' that tend to close those parts of the BK
boundaries that are not provided by the principal ``zig-zag'' trend. For example,
the lower boundary with $0 < q < 2$ is produced by fingers in the critical polynomials
with $n \ge 3$.

Overall, it seems likely that the BK phase for this lattice will extend all the way out
to $q=4$, and so we can conjecture that $q_{\rm c} = 4$ for the four-eight lattice.

\subsection{Frieze lattice $(3^3,4^2)$}

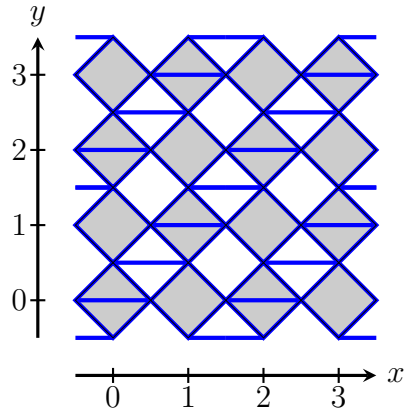
\begin{figure}
\begin{center}
\begin{tikzpicture}[scale=1.0,>=stealth]
\foreach \xpos in {0,1,2,3}
\foreach \ypos in {0,1,2,3}
 \fill[black!20] (\xpos+0.5,\ypos) -- (\xpos+1,\ypos+0.5) -- (\xpos+0.5,\ypos+1) -- (\xpos,\ypos+0.5) -- cycle;
 
\foreach \xpos in {0,1,2,3}
\foreach \ypos in {0,1,2,3}
{
  \draw[blue,ultra thick] (\xpos,\ypos+0.5) -- (\xpos+0.5,\ypos) -- (\xpos+1,\ypos+0.5) -- (\xpos+0.5,\ypos+1) -- cycle;
}
\foreach \xpos in {0,2}
\foreach \ypos in {0,2}
{
  \draw[blue,ultra thick] (\xpos,\ypos) -- (\xpos+0.5,\ypos);
  \draw[blue,ultra thick] (\xpos+1.5,\ypos) -- (\xpos+2,\ypos);
  \draw[blue,ultra thick] (\xpos,\ypos+0.5) -- (\xpos+1,\ypos+0.5);
  \draw[blue,ultra thick] (\xpos+0.5,\ypos+1) -- (\xpos+1.5,\ypos+1);
  \draw[blue,ultra thick] (\xpos+1,\ypos+1.5) -- (\xpos+2,\ypos+1.5);
  \draw[blue,ultra thick] (\xpos,\ypos+2) -- (\xpos+0.5,\ypos+2);
  \draw[blue,ultra thick] (\xpos+1.5,\ypos+2) -- (\xpos+2,\ypos+2);
}

\foreach \xpos in {0,1,2,3}
\foreach \ypos in {0,1,2,3}
 \draw[black] (\xpos+0.5,\ypos) -- (\xpos+1,\ypos+0.5) -- (\xpos+0.5,\ypos+1) -- (\xpos,\ypos+0.5) -- cycle;

\draw[very thick,->] (0,-0.5)--(4,-0.5);
\draw (4,-0.5) node[right] {$x$};
\foreach \xpos in {0,1,2,3}
{
 \draw[thick] (\xpos+0.5,-0.6)--(\xpos+0.5,-0.4);
 \draw (\xpos+0.5,-0.5) node[below] {$\xpos$};
}

\draw[very thick,->] (-0.5,0)--(-0.5,4);
\draw (-0.5,4) node[above] {$y$};
\foreach \ypos in {0,1,2,3}
{
 \draw[thick] (-0.6,\ypos+0.5)--(-0.4,\ypos+0.5);
 \draw (-0.5,\ypos+0.5) node[left] {$\ypos$};
}
 
\end{tikzpicture}
 \caption{Four-terminal representation of the frieze lattice.}
 \label{fig:frieze}
\end{center}
\end{figure}

The frieze lattice is the first example of a lattice which cannot be represented in four-terminal
form by using the same $\check{\sf R}$-matrix in all grey squares $(x,y)$. Instead we have:
\begin{equation}
 \check{\sf R}_i = \left \lbrace
 \begin{array}{ll}
  {\sf H}_{i+1} {\sf V}_i {\sf H}_{i+1} {\sf V}_{i+2} {\sf H}_{i+1} & \mbox{for $x+y$ even} \\
  {\sf H}_{i+1} {\sf V}_{i+2} {\sf V}_i {\sf H}_{i+1} & \mbox{for $x+y$ odd} \\
 \end{array} \right.
\end{equation}
In addition, there are horizontal diagonals on the white squares with coordinates $(x+\frac12,y+\frac12)$ for
$x+y$ even. The resulting basis, shown in Figure~\ref{fig:frieze}, needs $n$ to be even in
order for it to produce the frieze lattice upon tiling. It has 2 vertices and 5 edges per grey square.

The restriction that $n$ need to be even is somewhat problematic for our
approach. We have now only three data points ($n=2,4,6$) for the
extrapolations, instead of the usual seven ($n=1,2,\ldots,7$) when no
parity constraint is operative. Fitting first the three data points to
the form (\ref{extrapol_w}) we find that $w \approx 3.03$. This three-point
fit also provides the central value shown in Table~\ref{tab:frieze1}. Next, based on other
lattices for which more data points are available, we can estimate that
this value of $w$ is likely to shift by around $1/10$ upon increasing the size.
Performing next a two-point fit to the two largest sizes, with a fixed value of
$w$ within the range $3.03 \pm 0.10$, we get some alternative extrapolated values
from which we can judge the size of the error bar on the central value.

Obviously this procedure leads to final results which
are less precise than those of the preceding sections; but they still have
a precision superior to that of existing numerical results.

\begin{table}
\begin{center}
 \begin{tabular}{l|l}
 $n$ & $p_{\rm c}$ \\ \hline
 2 & 0.41963138921485282522020802393581141495177970672968  \\
 4 & 0.41963933385520484749314071759551986794774805977647 \\
 6 & 0.41964011556577399022999609393742526619676365808443 \\ \hline
 $\infty$ & 0.41964044 (1) \\
 Ref.~\cite{Parviainen07} & 0.4196419 (4) \\
 \end{tabular}
 \caption{Bond percolation threshold $p_{\rm c}$ on the frieze lattice.}
  \label{tab:frieze1}
\end{center}
\end{table}

The approximations for the percolation threshold $p_{\rm c}$ are shown in
Table~\ref{tab:frieze1}, and those for the critical points $v_{\rm c}$ for the
$q=3$ and $q=4$ Potts models are given in Tables~\ref{tab:frieze3}--\ref{tab:frieze4}.

\begin{table}
\begin{center}
 \begin{tabular}{l|l}
 $n$ & $v_{\rm c}$ \\ \hline
 2 & 1.2060925735117353155747857952640056241243199926123 \\
 4 & 1.2060633270192625886214470713076706696973022455589 \\
 6 & 1.2060607188311827112162217288312021603769835821224 \\ \hline
 $\infty$ & 1.20605973 (5) \\
  \end{tabular}
 \caption{Critical point $v_{\rm c}$ of the $q=3$ state Potts model on the frieze lattice.}
  \label{tab:frieze3}
\end{center}
\end{table}

\begin{table}
\begin{center}
 \begin{tabular}{l|l}
 $n$ & $v_{\rm c}$ \\ \hline
 2 & 1.3761421373604738385559000230091418072845484528956 \\
 4 & 1.3760828367736083803069388239875587480765505652209 \\
 6 & 1.3760777196319081822843041399511989912220832991296 \\ \hline
 $\infty$ & 1.37607584 (7) \\
  \end{tabular}
 \caption{Critical point $v_{\rm c}$ of the $q=4$ state Potts model on the frieze lattice.}
  \label{tab:frieze4}
\end{center}
\end{table}

For the Ising model ($q=2$) we find the now-familiar factorisation of the polynomials $P_B(q,v)$.
The maximum degree of the factors is $d_{\rm max} = 2$ for $n=2$ and $d_{\rm max} = 12$ for n=4.
But the recurrent factor that determines the ferromagnetic critical point is simply
\begin{equation}
 -1 + v \,
 \label{frieze_factor}
\end{equation}
so that
\begin{equation}
 v_{\rm c} = 1 \,.
 \label{frieze_root}
\end{equation}
in agreement with \cite{Codello10}.

\begin{figure}
\begin{center}
\includegraphics[width=12cm]{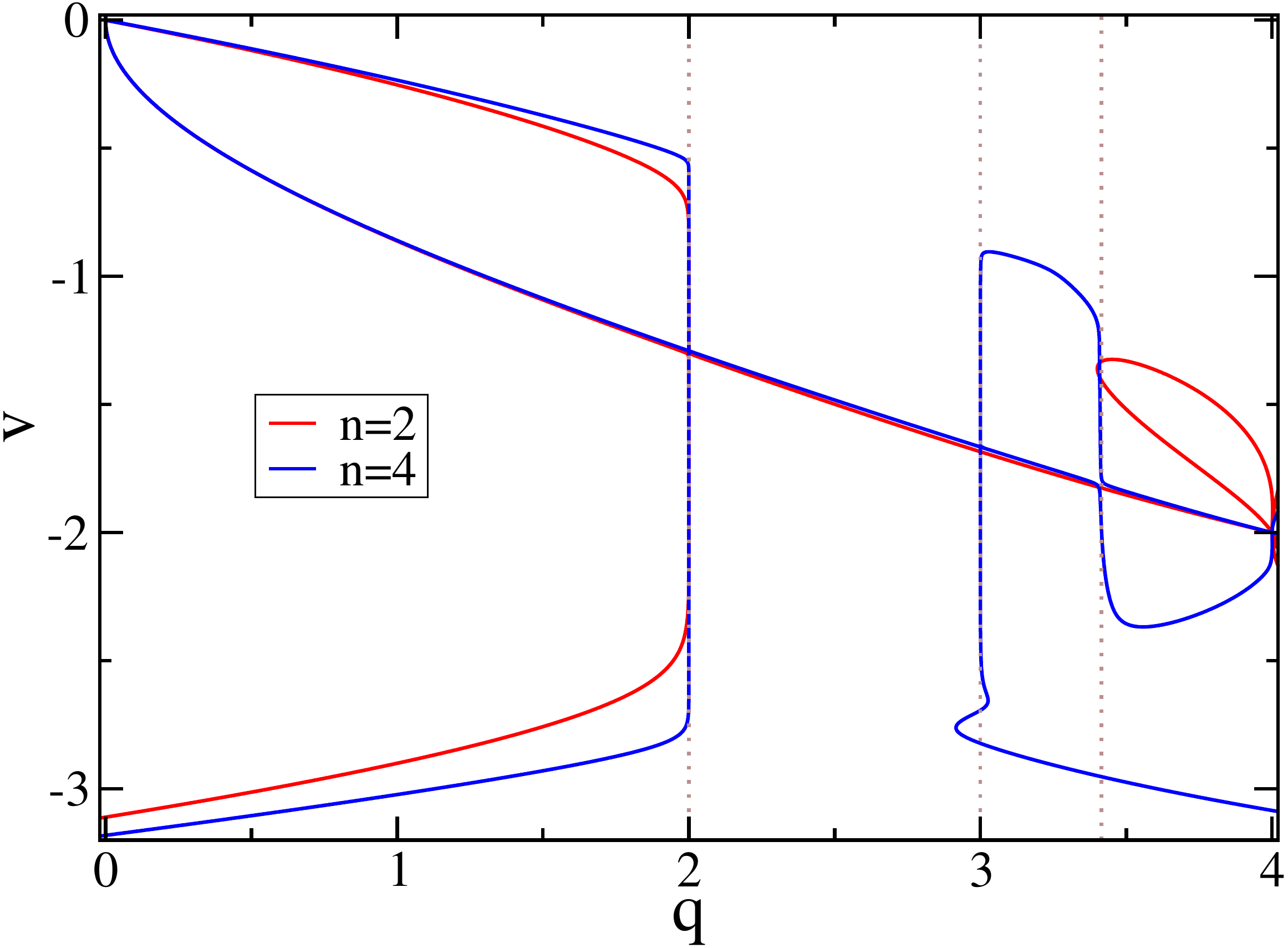}
\caption{Roots of $P_B(q,v)$ for the Potts model on the frieze lattice, using
$n \times n$ square bases.}
\label{fig:frieze-pd}
\end{center}
\end{figure}

The obtainable information on the phase diagram also suffers from the parity restriction,
since the polynomials $P_B(q,v)$ are now at our disposal only for $n=2,4$. It is
nevertheless clear from Figure~\ref{fig:frieze-pd} that the usual features are present. The extent of the BK phase can
be judged from the vertical rays at the Beraha numbers with $k=4,6,8$. Inside the BK phase
we have a critical curve coming out of the origin with a vertical tangent and extending to the
point $(q,v)=(4,-2)$. Since this is the analytical continuation of the critical curve in the
ferromagnetic regime $v > 0$, it must be the RG attractor governing the BK phase. We
should therefore have $q_{\rm c} = 4$ for the frieze lattice.

The curves $v_\pm(q)$ bounding the BK phase are only partially represented by the roots
of the polynomials $P_B(q,v)$. Indeed the parts with $2 < q < 3$ are missing altogether.
We notice however that the $n=4$ curve develops a small bulge near the bottom of the
$q=3$ vertical ray, and it is conceivable that for higher $n$ this will develop into narrow
``fingers'' that will close the curves $v_\pm(q)$---a situation that was seen clearly in Figure~\ref{fig:foureight-pd}
for the four-eight lattice.

\subsection{Three-twelve lattice $(3,12^2)$}
\label{sec:threetwelve}

\begin{figure}
\begin{center}
\begin{tikzpicture}[scale=1.0,>=stealth]
\foreach \xpos in {0,1,2,3}
\foreach \ypos in {0,1,2,3}
 \fill[black!20] (\xpos+0.5,\ypos) -- (\xpos+1,\ypos+0.5) -- (\xpos+0.5,\ypos+1) -- (\xpos,\ypos+0.5) -- cycle;
 
\foreach \xpos in {0,1,2,3}
\foreach \ypos in {0,1,2,3}
{
  \draw[blue,ultra thick,rounded corners=0.5mm] (\xpos+0.125,\ypos+0.625) -- (\xpos+0.5,\ypos+0.5) -- (\xpos+0.625,\ypos+0.125) -- cycle;
  \draw[blue,ultra thick,rounded corners=0.5mm] (\xpos+1,\ypos+0.5) -- (\xpos+0.5,\ypos+1) -- (\xpos+0.625,\ypos+0.625) -- cycle;
  \draw[blue,ultra thick] (\xpos+0.5,\ypos+0.5) -- (\xpos+0.625,\ypos+0.625);
  \draw[blue,ultra thick] (\xpos,\ypos+0.5) -- (\xpos+0.125,\ypos+0.625);
  \draw[blue,ultra thick] (\xpos+0.5,\ypos) -- (\xpos+0.625,\ypos+0.125);
}

\foreach \xpos in {0,1,2,3}
\foreach \ypos in {0,1,2,3}
 \draw[black] (\xpos+0.5,\ypos) -- (\xpos+1,\ypos+0.5) -- (\xpos+0.5,\ypos+1) -- (\xpos,\ypos+0.5) -- cycle;

\draw[very thick,->] (0,-0.5)--(4,-0.5);
\draw (4,-0.5) node[right] {$x$};
\foreach \xpos in {0,1,2,3}
{
 \draw[thick] (\xpos+0.5,-0.6)--(\xpos+0.5,-0.4);
 \draw (\xpos+0.5,-0.5) node[below] {$\xpos$};
}

\draw[very thick,->] (-0.5,0)--(-0.5,4);
\draw (-0.5,4) node[above] {$y$};
\foreach \ypos in {0,1,2,3}
{
 \draw[thick] (-0.6,\ypos+0.5)--(-0.4,\ypos+0.5);
 \draw (-0.5,\ypos+0.5) node[left] {$\ypos$};
}
 
\end{tikzpicture}
 \caption{Four-terminal representation of the three-twelve lattice.}
 \label{fig:threetwelve}
\end{center}
\end{figure}
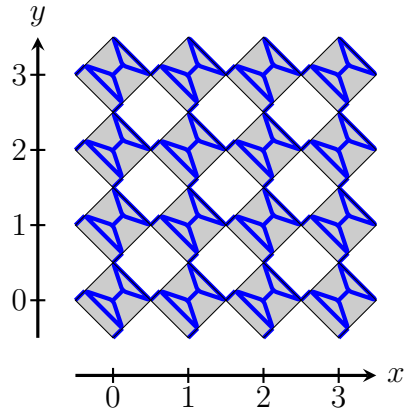

With the three-twelve lattice we are back in the category of lattices that can be represented
in four-terminal form for any parity of $n$. The appropriate $\check{\sf R}$-matrix reads
\begin{equation}
 \check{\sf R} = {\sf H}_{i+1} {\sf V}_i {\sf H}_{i+1} {\sf V}_i {\sf E}_{i+2}
  {\sf H}_{i+1} {\sf V}_i {\sf H}_{i+1} {\sf V}_{i+2} {\sf V}_i \,.
\end{equation}
The corresponding basis is depicted in Figure~\ref{fig:threetwelve}. It possesses
6 vertices and 9 edges per grey square, the highest numbers that we have attained
among the Archimedean lattices. Accordingly we can expect the most accurate
results for the critical points of this lattice (except, obviously, for the exactly solvable
three-terminal cases).

Results for the percolation threshold $p_{\rm c}$ are given in Table~\ref{tab:threetwelve1},
using square bases of size $n \le 7$. Those with $n \le 4$ have already been reported in
\cite{SJ12} where the final estimate $p_{\rm c} = 0.740\,420\,800(1)$ was proposed.
The three extra data points allow us to improve considerably on this, adding four more digits
to the final result reported in Table~\ref{tab:threetwelve1}. We have again benefited from
a very high value $w \approx 6.39$ of the parameter appearing in (\ref{extrapol_w}).
This brings the relative
precision to $1 \cdot 10^{-12}$, that is, more than four orders of magnitude better than
the best available numerical result \cite{Ding10}.
Note that the value of $w$ is almost coincident with that of the kagome lattice; this is
presumably linked to the fact that this lattice has the same symmetry group as the three-twelve lattice.

\begin{table}
\begin{center}
 \begin{tabular}{l|l}
 $n$ & $p_{\rm c}$ \\ \hline
 1 & 0.74042331791989696781390025991096022103832189014608 \\
 2 & 0.74042099242999609235990682258592153549751047668010 \\
 3 & 0.74042081882197909432814081145369695940339872717580 \\
 4 & 0.74042080213011204414779539731474889355503786568625 \\
 5 & 0.74042079963976341893709572470832241553387184495527 \\
 6 & 0.74042079909690334068260706338305717610656002517013 \\
 7 & 0.74042079894276532366605540512642419991201536276834 \\ \hline
 $\infty$ & 0.7404207988509 (8) \\
 Ref.~\cite{Ding10} & 0.74042077 (2) \\
 \end{tabular}
 \caption{Bond percolation threshold $p_{\rm c}$ on the three-twelve lattice.}
  \label{tab:threetwelve1}
\end{center}
\end{table}

The graph polynomials $P_B(q,v)$ factorise as usually in the $q=2$ Ising case.
The maximum degree of the factors is $d_{\rm max} = 4$ for $n=1,2$;
$d_{\rm max} = 8$ for $n=3,4$; and $d_{\rm max} = 16$ for $n=5$. There is precisely one of 
these factors, namely
\begin{equation}
 -8 - 8 v - 6 v^2 - 2 v^3 + v^4 \,,
 \label{threetwelve_factor}
\end{equation}
that possesses a positive root,
\begin{equation}
 v_{\rm c} = \frac12 \left(1 + \sqrt{3} + \sqrt{2 (6 + 5 \sqrt{3})} \right)
 \simeq 4.073\,446\,135\cdots \,.
 \label{threetwelve_root}
\end{equation}
Moreover, (\ref{threetwelve_factor}) is a factor in $P_B(q,v)$ for any size $n$. Its positive
root (\ref{threetwelve_root}) furnishes the exactly known critical point
\cite{Syozi55,Syozi72,Codello10erratum}.

The results for the critical point $v_{\rm c}$ of the Potts models
with $q=3$ and $q=4$ are shown in Tables~\ref{tab:threetwelve3}--\ref{tab:threetwelve4}.
Again we obtain very high accuracies on the final estimates.

\begin{table}
\begin{center}
 \begin{tabular}{l|l}
 $n$ & $v_{\rm c}$ \\ \hline
 1 & 5.0330225148727450936191152069088523071556984046386 \\
 2 & 5.0330723130708872393918569395936083805639594860270 \\
 3 & 5.0330776369208258020214559125337418019531815564544 \\
 4 & 5.0330782997119322552115261369184585114757976531754 \\
 5 & 5.0330784300991523816473701505184903317147518415418 \\
 6 & 5.0330784662543648528166987041595243471021156868329 \\
 7 & 5.0330784788156319031365051678892523632559589908672 \\ \hline
 $\infty$ & 5.03307848898 (7)  \\
  \end{tabular}
 \caption{Critical point $v_{\rm c}$ of the $q=3$ state Potts model on the three-twelve lattice.}
  \label{tab:threetwelve3}
\end{center}
\end{table}

\begin{table}
\begin{center}
 \begin{tabular}{l|l}
 $n$ & $v_{\rm c}$ \\ \hline
 1 & 5.8573948279836477826193148233190408839405294733071 \\
 2 & 5.8574980277679771830251052175741883454287604168277 \\
 3 & 5.8575099292060853584152486686816804379692602098265 \\
 4 & 5.8575115251380370204992586915287203141213189319694 \\
 5 & 5.8575118674921102618186189385720483562901709368620 \\
 6 & 5.8575119704409279852620284760092589856034565012177 \\
 7 & 5.8575120089261340455809871849023068234150587067062 \\ \hline
 $\infty$ & 5.8575120444 (3) \\
  \end{tabular}
 \caption{Critical point $v_{\rm c}$ of the $q=4$ state Potts model on the three-twelve lattice.}
  \label{tab:threetwelve4}
\end{center}
\end{table}

The phase diagram shown in Figure~\ref{fig:threetwelve} has a very intricate structure,
containing more features than for any of the other lattices considered so far.
It was discussed in detail in \cite{Jacobsen13} for $n \le 4$, but we gain extra
information---and confirmation of some of the salient features---from the $n=5$
critical polynomial $P_B(q,v)$ now available.

We see of course the usual vertical rays characterising the BK phase, visible at
the Beraha numbers (\ref{Beraha}) with $k=4,6,8$ and building up at $k=10$ as well.
The extent of the BK phase can be estimated from those rays. It seems clear that
it extends out to the point $(q,v)=(4,-2)$, and thus we have $q_{\rm c}=4$ for the
three-twelve lattice.

To discuss the phase diagram in some more detail, we first focus on the region $0 \le q \lesssim 2$.
We first note the formation of fingers in the $n=4,5$ curves in the range $0 \le q \lesssim 1.2$.
To the left of the lower edge of the $q=2$ vertical ray one sees the formation of a triangular-shaped
enclosed region with $1.3 \lesssim q \lesssim 2$; this is visible in particular in the small wrinkle
that develops in the $n=5$ curve. Moreover, there is a tiny enclosed sliver with $q \le 2 \lesssim 2.01$
that is consistently visible for any parity of $n$. These two regions (of triangular and sliver shape)
near $q=2$ are reminiscent of similar features on the kagome lattice, and we believe that they
subsist in the thermodynamical limit.

\begin{figure}
\begin{center}
\includegraphics[width=12cm]{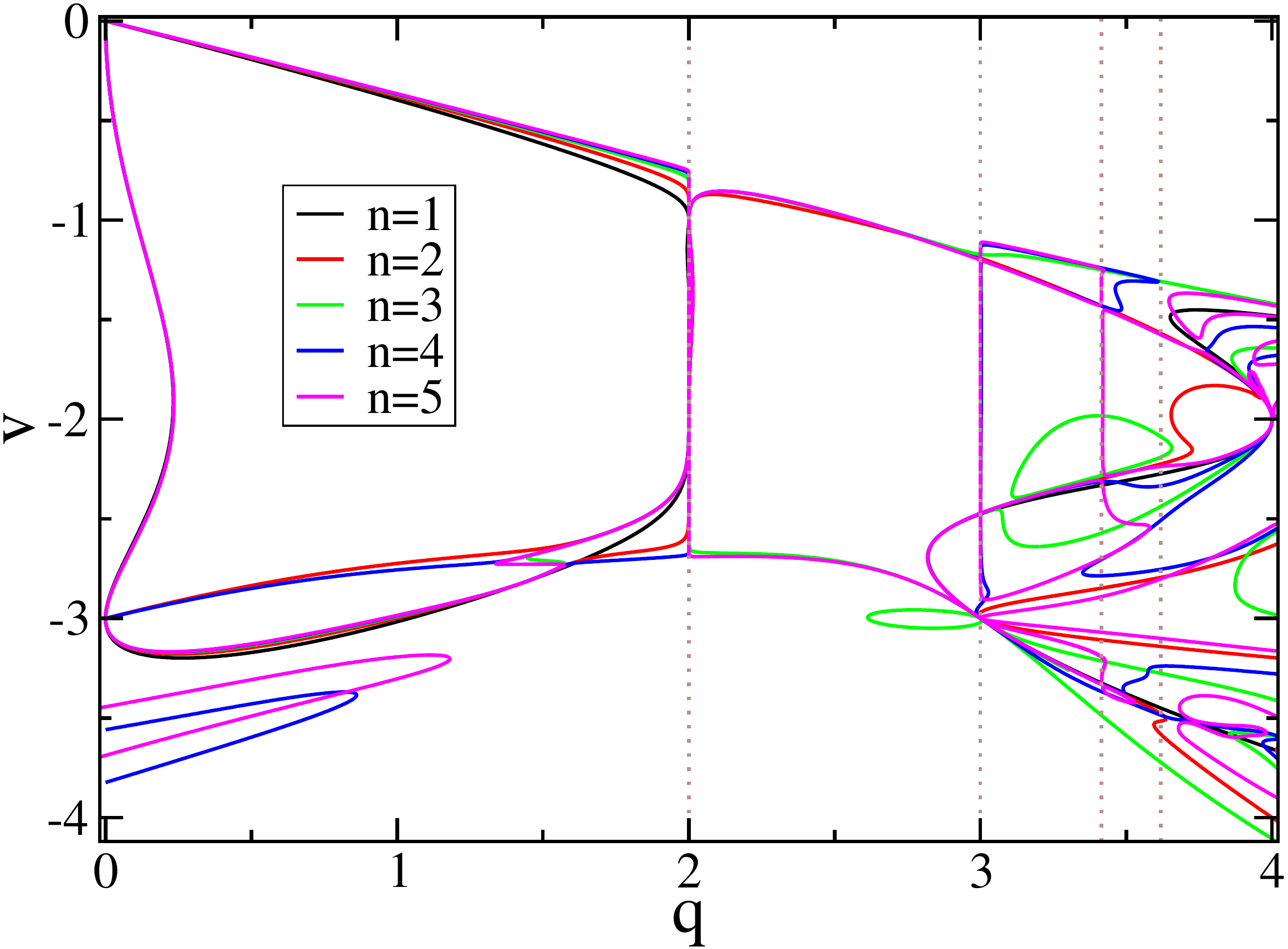}
\caption{Roots of $P_B(q,v)$ for the Potts model on the three-twelve lattice, using
$n \times n$ square bases.}
\label{fig:threetwelve-pd}
\end{center}
\end{figure}

For $2.8 \lesssim q \le 4$ the situation is quite complicated. There are notably a couple of curves
inside the BK phase, such that the vertical rays with $k=6,8$ are cut into three pieces. Just above
$(q,v)=(4,-2)$ one observes several extra curves developing for large $n$ which clearly tend to
extend this structure to higher values of $k$. A similar situation happens near the lower boundary
of the BK phase. Moreover, in the region $3 \lesssim q \lesssim 4$ and $-4 \lesssim v \lesssim -3$
many curves are building up and accumulating at $(q,v)=(3,-3)$. Clearly this latter point must have
a very distinguished role in the phase diagram.

\subsection{Cross lattice $(4,6,12)$}
\label{sec:cross}

\begin{figure}
\begin{center}
\begin{tikzpicture}[scale=1.0,>=stealth]
\foreach \xpos in {0,1,2,3}
\foreach \ypos in {0,1,2,3}
 \fill[black!20] (\xpos+0.5,\ypos) -- (\xpos+1,\ypos+0.5) -- (\xpos+0.5,\ypos+1) -- (\xpos,\ypos+0.5) -- cycle;
 
\foreach \xpos in {0,2}
\foreach \ypos in {0,2}
{
  \draw[blue,ultra thick] (\xpos,\ypos+0.5) -- (\xpos+0.5,\ypos+1) -- (\xpos+1,\ypos+0.5) -- (\xpos+0.75,\ypos+0.25) -- cycle;
  \draw[blue,ultra thick] (\xpos+0.5,\ypos) -- (\xpos+0.75,\ypos+0.25);
  \draw[blue,ultra thick] (\xpos+1,\ypos+0.5) -- (\xpos+1.5,\ypos);
  \draw[blue,ultra thick] (\xpos+1.5,\ypos+1) -- (\xpos+2,\ypos+0.5);
  \draw[blue,ultra thick] (\xpos+0.25,\ypos+1.75) -- (\xpos+0.5,\ypos+2) -- (\xpos+1,\ypos+1.5) -- (\xpos+0.75,\ypos+1.25) -- cycle;
  \draw[blue,ultra thick] (\xpos,\ypos+1.5) -- (\xpos+0.25,\ypos+1.75);
  \draw[blue,ultra thick] (\xpos+0.5,\ypos+1) -- (\xpos+0.75,\ypos+1.25);
  \draw[blue,ultra thick] (\xpos+1.25,\ypos+1.75) -- (\xpos+1.5,\ypos+2) -- (\xpos+2,\ypos+1.5) -- (\xpos+1.5,\ypos+1) -- cycle;
  \draw[blue,ultra thick] (\xpos+1,\ypos+1.5) -- (\xpos+1.25,\ypos+1.75);
  
}

\foreach \xpos in {0,1,2,3}
\foreach \ypos in {0,1,2,3}
 \draw[black] (\xpos+0.5,\ypos) -- (\xpos+1,\ypos+0.5) -- (\xpos+0.5,\ypos+1) -- (\xpos,\ypos+0.5) -- cycle;

\draw[very thick,->] (0,-0.5)--(4,-0.5);
\draw (4,-0.5) node[right] {$x$};
\foreach \xpos in {0,1,2,3}
{
 \draw[thick] (\xpos+0.5,-0.6)--(\xpos+0.5,-0.4);
 \draw (\xpos+0.5,-0.5) node[below] {$\xpos$};
}

\draw[very thick,->] (-0.5,0)--(-0.5,4);
\draw (-0.5,4) node[above] {$y$};
\foreach \ypos in {0,1,2,3}
{
 \draw[thick] (-0.6,\ypos+0.5)--(-0.4,\ypos+0.5);
 \draw (-0.5,\ypos+0.5) node[left] {$\ypos$};
}
 
\end{tikzpicture}
 \caption{Four-terminal representation of the cross lattice.}
 \label{fig:cross}
\end{center}
\end{figure}
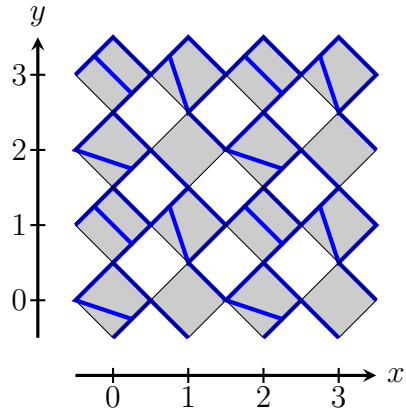

To cast the cross lattice in four-terminal form, four different types of grey squares are needed.
The $\check{\sf R}$-matrix takes the form
\begin{equation}
 \check{\sf R}_i = \left \lbrace
 \begin{array}{ll}
   {\sf H}_{i+1} {\sf V}_{i+2} {\sf V}_i {\sf H}_{i+1} {\sf V}_{i+2} & \mbox{for $x$ even and $y$ even} \\
   {\sf H}_{i+1} {\sf E}_{i+2} {\sf E}_i {\sf H}_{i+1} & \mbox{for $x$ odd and $y$ even} \\ 
   {\sf H}_{i+1} {\sf V}_{i+2} {\sf V}_i {\sf H}_{i+1} {\sf V}_{i+2} {\sf V}_i & \mbox{for $x$ even and $y$ odd} \\
   {\sf H}_{i+1} {\sf V}_{i+2} {\sf V}_i {\sf H}_{i+1} {\sf V}_i & \mbox{for $x$ odd and $y$ odd} \\
 \end{array} \right.
\end{equation}
and the resulting basis is illustrated in Figure~\ref{fig:cross}. The reader is invited to check
carefully from the figure that each vertex is indeed surrounded by a square, a hexagon and a dodecagon.
Obviously we must require that $n$ be even. There is on average 3 vertices and $\frac{9}{2}$ edges
per grey square.

The approximations to the percolation threshold $p_{\rm c}$ are shown in Table~\ref{tab:cross1}.
Despite the parity constraint on $n$, we are still able to provide a final estimate that is more
precise than those obtainable from numerical simulations \cite{Parviainen07}.

\begin{table}
\begin{center}
 \begin{tabular}{l|l}
 $n$ & $p_{\rm c}$ \\ \hline
 2 & 0.69377849010809934126953435770375111640557701680580 \\
 4 & 0.69373853603434634159328968721278581288678588137638 \\
 6 & 0.69373378891455375106650642653023991489166486978424 \\ \hline
 $\infty$ & 0.6937314 (1) \\
 Ref.~\cite{Parviainen07} & 0.6937338 (7) \\
 \end{tabular}
 \caption{Bond percolation threshold $p_{\rm c}$ on the cross lattice.}
  \label{tab:cross1}
\end{center}
\end{table}

For $q=2$, the graph polynomials $P_B(q,v)$ factorise, and
the maximum degree of the factors is $d_{\rm max} = 8$ for both $n=2$ and $n=4$.
The unique common factor possessing a positive root is
\begin{equation}
 -32 - 128 v - 208 v^2 - 176 v^3 - 84 v^4 - 24 v^5 + 4 v^7 + v^8 \,.
 \label{cross_factor}
\end{equation}
After some algebra this leads to the critical coupling
\begin{eqnarray}
 v_{\rm c} &=& \exp \left(2 \mbox{ arctanh } \sqrt{\frac{2}{1 + \sqrt{3} + \sqrt{-4 + 6 \sqrt{3}}}} \right) - 1 \\
 &\simeq& 3.216\,563\,123\cdots \,. \nonumber
 \label{cross_root}
\end{eqnarray}
in agreement with \cite{Codello10}.

The results for the critical points $v_{\rm c}$ for the three and four-state Potts
models appear in Tables~\ref{tab:cross3}--\ref{tab:cross4}.

\begin{table}
\begin{center}
 \begin{tabular}{l|l}
 $n$ & $v_{\rm c}$ \\ \hline
 2 & 3.9586332997059740720181837928768500290543859070058 \\
 4 & 3.9591763776237344298683368980854435922679135046889 \\
 6 & 3.9592412531275763639725384140180205407837422105870 \\ \hline
 $\infty$ & 3.959273 (1) \\
  \end{tabular}
 \caption{Critical point $v_{\rm c}$ of the $q=3$ state Potts model on the cross lattice.}
  \label{tab:cross3}
\end{center}
\end{table}

\begin{table}
\begin{center}
 \begin{tabular}{l|l}
 $n$ & $v_{\rm c}$ \\ \hline
 2 & 4.5935119229510043784589758419016606560974164157721 \\
 4 & 4.5946031686522078767110561721497673809936709287964 \\
 6 & 4.5947347336947799761277325102560255207753903194732 \\ \hline
 $\infty$ & 4.594801 (2) \\
  \end{tabular}
 \caption{Critical point $v_{\rm c}$ of the $q=4$ state Potts model on the cross lattice.}
  \label{tab:cross4}
\end{center}
\end{table}

The phase diagram---as witnessed by the roots of the critical polynomials $P_B(q,v)$---combines
a number of features familiar from the four-eight lattice. There is a finger in the $n=4$ curve with $0 \le q \lesssim 1.7$
that tends to bridge the gap between the point $(q,v)=(0,-4)$ and the lower edge of the $q=2$
vertical ray. Similarly, wrinkles in both curves ($n=2,4$) tends to delimit the BK phase from above
on the interval $2 < q < 3$. As a result, we observe vertical rays at the Beraha numbers (\ref{Beraha})
with $k=4,6,8$. Presumably the BK phase extends out to $q_{\rm c} = 4$ for this lattice.

\begin{figure}
\begin{center}
\includegraphics[width=12cm]{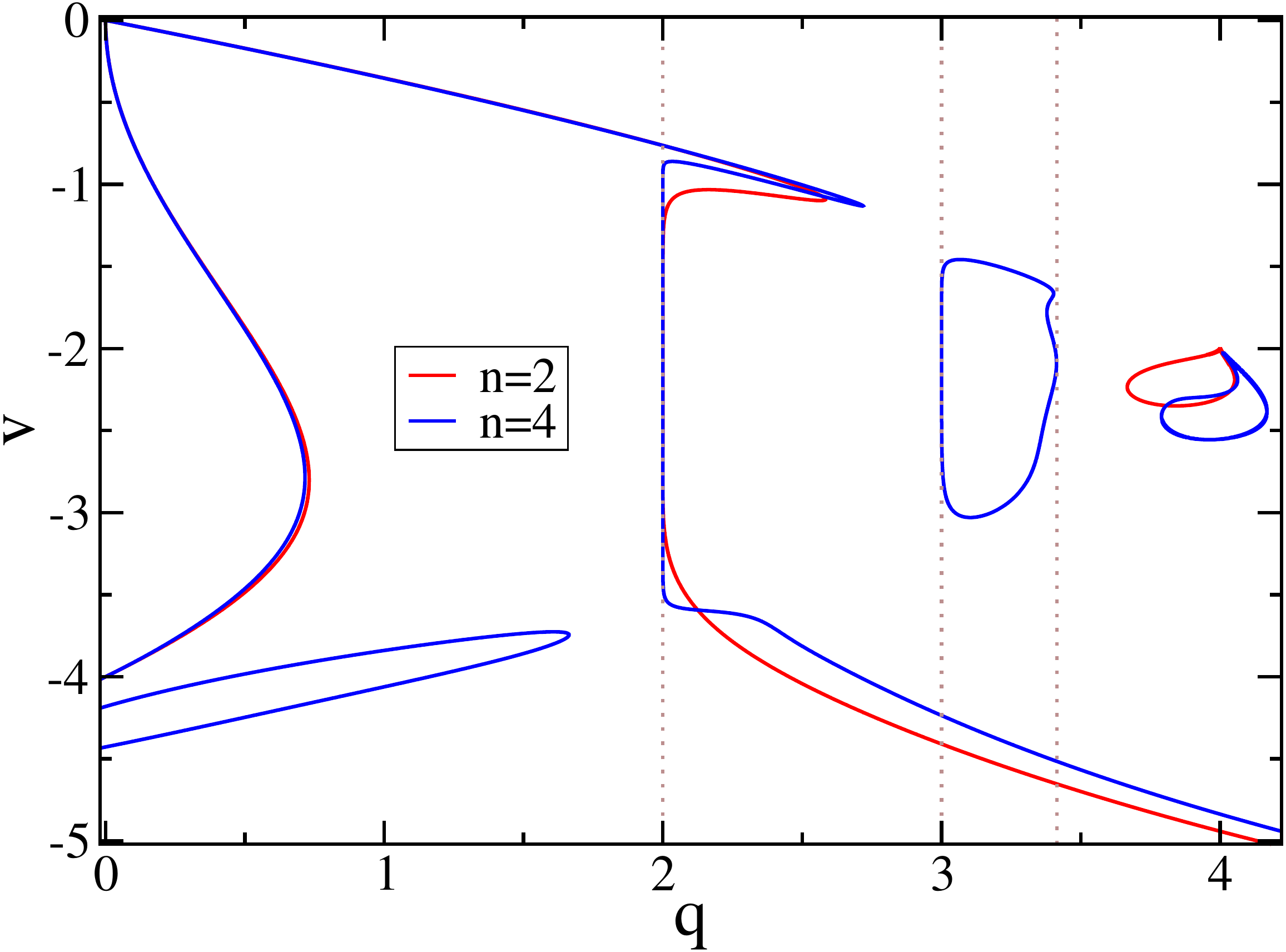}
\caption{Roots of $P_B(q,v)$ for the Potts model on the cross lattice, using
$n \times n$ square bases.}
\label{fig:cross-pd}
\end{center}
\end{figure}

\subsection{Snub square lattice $(3^2,4,3,4)$}

\begin{figure}
\begin{center}
\begin{tikzpicture}[scale=1.0,>=stealth]
\foreach \xpos in {0,1,2,3}
\foreach \ypos in {0,1,2,3}
 \fill[black!20] (\xpos+0.5,\ypos) -- (\xpos+1,\ypos+0.5) -- (\xpos+0.5,\ypos+1) -- (\xpos,\ypos+0.5) -- cycle;
 
\foreach \xpos in {0,1,2,3}
\foreach \ypos in {0,1,2,3}
{
  \draw[blue,ultra thick] (\xpos,\ypos+0.5) -- (\xpos+0.5,\ypos) -- (\xpos+1,\ypos+0.5) -- (\xpos+0.5,\ypos+1) -- cycle;
}
\foreach \xpos in {0,2}
\foreach \ypos in {0,2}
{
  \draw[blue,ultra thick] (\xpos+0.5,\ypos) -- (\xpos+0.5,\ypos+1);
  \draw[blue,ultra thick] (\xpos+1,\ypos+0.5) -- (\xpos+2,\ypos+0.5);
  \draw[blue,ultra thick] (\xpos,\ypos+1.5) -- (\xpos+1,\ypos+1.5);
  \draw[blue,ultra thick] (\xpos+1.5,\ypos+1) -- (\xpos+1.5,\ypos+2);
}

\foreach \xpos in {0,1,2,3}
\foreach \ypos in {0,1,2,3}
 \draw[black] (\xpos+0.5,\ypos) -- (\xpos+1,\ypos+0.5) -- (\xpos+0.5,\ypos+1) -- (\xpos,\ypos+0.5) -- cycle;

\draw[very thick,->] (0,-0.5)--(4,-0.5);
\draw (4,-0.5) node[right] {$x$};
\foreach \xpos in {0,1,2,3}
{
 \draw[thick] (\xpos+0.5,-0.6)--(\xpos+0.5,-0.4);
 \draw (\xpos+0.5,-0.5) node[below] {$\xpos$};
}

\draw[very thick,->] (-0.5,0)--(-0.5,4);
\draw (-0.5,4) node[above] {$y$};
\foreach \ypos in {0,1,2,3}
{
 \draw[thick] (-0.6,\ypos+0.5)--(-0.4,\ypos+0.5);
 \draw (-0.5,\ypos+0.5) node[left] {$\ypos$};
}
 
\end{tikzpicture}
 \caption{Four-terminal representation of the snub square lattice.}
 \label{fig:snubsquare}
\end{center}
\end{figure}
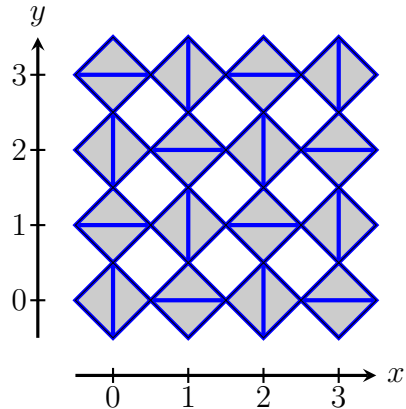

The four-terminal representation of the snub square lattice is quite straightforward to obtain.
The $\check{\sf R}$-matrix reads
\begin{equation}
 \check{\sf R} = \left \lbrace
 \begin{array}{ll}
 {\sf H}_{i+1} {\sf V}_{i+2} {\sf H}_{i+1} {\sf V}_i {\sf H}_{i+1} & \mbox{for $x+y$ even} \\
 {\sf H}_{i+1} {\sf V}_i {\sf H}_{i+1} {\sf V}_{i+2} {\sf H}_{i+1} & \mbox{for $x+y$ odd} \\
 \end{array} \right.
 \label{R-snubsquare}
\end{equation}
and Figure~\ref{fig:snubsquare} contains a rendering of the corresponding basis.
There are 2 vertices and 5 edges per grey square.

\begin{table}
\begin{center}
 \begin{tabular}{l|l}
 $n$ & $p_{\rm c}$ \\ \hline
 2 & 0.41414477068891106678269962067496175696956628442687 \\
 4 & 0.41413831583758918250111117204245967836572650308666 \\
 6 & 0.41413794448601941204552140015728270649261591711977 \\ \hline
 $\infty$ & 0.4141378476 (7) \\
 Ref.~\cite{Parviainen07} & 0.4141374 (5) \\
 \end{tabular}
 \caption{Bond percolation threshold $p_{\rm c}$ on the snub square lattice.}
  \label{tab:snubsquare1}
\end{center}
\end{table}

The bond percolation thresholds $p_{\rm c}$ obtained from the critical polynomials
are reported in Table~\ref{tab:snubsquare1}, and the corresponding results for the
critical point $v_{\rm c}$ of the Potts model with $q=3$ and $q=4$ are shown in
Tables~\ref{tab:snubsquare3}--\ref{tab:snubsquare4}.

The Ising model polynomials $P_B(q,v)$ with $q=2$ factorise,
the maximum degree of the factors being $d_{\rm max} = 6$ for $n=2$, and
$d_{\rm max} = 12$ for $n=4$.
There is a unique common factor possessing a positive root
\begin{equation}
 -4 - 8 v - 4 v^2 + 4 v^3 + 8 v^4 + 4 v^5 + v^6 \,.
 \label{snubsquare_factor}
\end{equation}
If we define $\omega = 37 + 27 \sqrt{2} + 3 \sqrt{315 + 222 \sqrt{2}}$, the critical
coupling can be written
\begin{equation}
 v_{\rm c} = \frac13 \left( -2 -2 w^{-1/3} + w^{1/3} \right)
 \simeq 0.980\,730\,864\cdots
 \label{snubsquare_root}
\end{equation}
and this coincides with the result of \cite{Codello10}.

\begin{table}
\begin{center}
 \begin{tabular}{l|l}
 $n$ & $v_{\rm c}$ \\ \hline
 2 & 1.18529390865581786407849700528021223954317049051598 \\
 4 & 1.18531433669635871167720996722549305108279483104371 \\
 6 & 1.18531541517566214804035962759914618124246371903443 \\ \hline
 $\infty$ & 1.185315678 (3) \\
  \end{tabular}
 \caption{Critical point $v_{\rm c}$ of the $q=3$ state Potts model on the snub square lattice.}
  \label{tab:snubsquare3}
\end{center}
\end{table}

\begin{table}
\begin{center}
 \begin{tabular}{l|l}
 $n$ & $v_{\rm c}$ \\ \hline
 2 & 1.35449677473518648766976021945792542217215508930616 \\
 4 & 1.35453610524182819267343590204899630391908189104331 \\
 6 & 1.35453810944807886217841873099783060879140651150399 \\ \hline
 $\infty$ & 1.354538584 (6) \\
  \end{tabular}
 \caption{Critical point $v_{\rm c}$ of the $q=4$ state Potts model on the snub square lattice.}
  \label{tab:snubsquare4}
\end{center}
\end{table}

\begin{figure}
\begin{center}
\includegraphics[width=12cm]{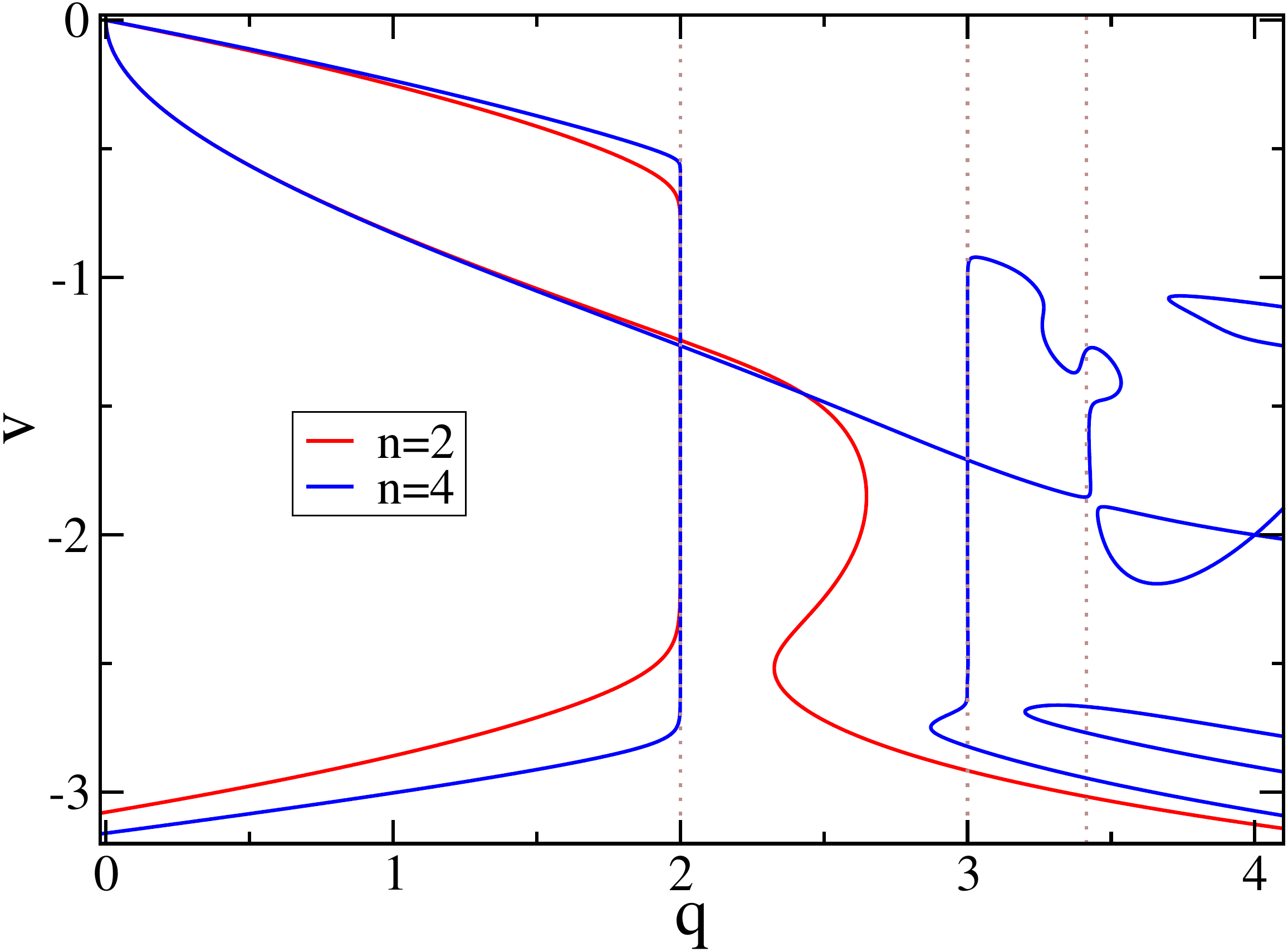}
\caption{Roots of $P_B(q,v)$ for the Potts model on the snub square lattice, using
$n \times n$ square bases.}
\label{fig:snubsquare-pd}
\end{center}
\end{figure}

It remains to discuss the phase diagram, shown in Figure~\ref{fig:snubsquare-pd}. As usual,
the extent of the BK phase can be seen from the vertical rays, here appearing at Beraha
numbers (\ref{Beraha}) with $k=4,6,8$. An interesting feature for this lattice is that
there is a transition curve in the middle of the BK phase, reminiscent of what was found for the
frieze lattice (see Figure~\ref{fig:frieze-pd}). This curve emanates from the origin
with vertical slope and extends out to $(q,v) = (4,-2)$. For finite $n$ there are some gaps
in this curve, here visible for $n=4$ near $k=8$, but these should disappear in the thermodynamical
limit. The closure of the BK phase is ensured by a wrinkle developing near the lower edge of the 
$q=3$ vertical ray in the $n=4$ curve, and by fingers developing at $q \approx 4$.

\subsection{Snub hexagonal lattice $(3^4,6)$}
\label{sec:snubhex}

The snub hexagonal lattice is a depleted version of the triangular lattice, where $1/7$ of
the vertices and their adjacent edges have been erased. We have constructed this lattice
in two different ways.

The first construction closely parallels the one that we have used in section~\ref{sec:triangular}
for the triangular lattice. The only difference is that the ${\sf H}$ and ${\sf V}$
operators corresponding to erased edges have been replaced by just one of their two terms
(the identity ${\sf Id}$ or the Temperley-Lieb generator ${\sf E}$, as the case may be).
There are then $\frac{12}{7}$ vertices and $\frac{30}{7}$ edges per grey square; note that
these numbers are compatible with the coordination number being $5$.

This depletion representation is a bit easier to state than to actually write down. But in formal
terms we arrive, after drawing things carefully, at the following $\check{\sf R}$-matrix:
\begin{equation}
 \check{\sf R} = \left \lbrace
 \begin{array}{ll}
  {\sf H}_{i+1} {\sf V}_i {\sf H}_{i+1} {\sf V}_{i+2} {\sf H}_{i+1} & \mbox{for $5 y + x = 0,2,4$ mod 7} \\
  {\sf H}_{i+1} {\sf V}_i {\sf H}_{i+1} {\sf E}_{i+2} {\sf Id}_{i+1} & \mbox{for $5 y + x = 1$ mod 7} \\
  {\sf Id}_{i+1} {\sf E}_i {\sf H}_{i+1} {\sf V}_{i+2} {\sf H}_{i+1} & \mbox{for $5 y + x = 3$ mod 7} \\
  {\sf Id}_{i+1} {\sf V}_i {\sf Id}_{i+1} {\sf E}_{i+2} {\sf H}_{i+1} & \mbox{for $5 y + x = 5$ mod 7} \\
  {\sf H}_{i+1} {\sf E}_i {\sf Id}_{i+1} {\sf V}_{i+2} {\sf Id}_{i+1} & \mbox{for $5 y + x = 6$ mod 7} \\
 \end{array} \right.
 \label{snubhex-R}
\end{equation}
In addition there are horizontal diagonals in the write squares at coordinates $(x+\frac12,y+\frac12)$
when $5 y + x = 0,1,4,5,6$ mod 7. Note that the first line in (\ref{snubhex-R}) corresponds to undepleted
grey squares, i.e., it coincides with (\ref{triangular-R}). Subsequent lines are obtained from the first one
by depletion, i.e., replacing some of the ${\sf H}$ operators by the identity ${\sf Id}$, and some of the
${\sf V}$ operators by the Temperley-Lieb generator ${\sf E}$.

Note that this construction will just disconnect the erased spins from the remainder of the lattice.
This means that when computing the critical polynomial from (\ref{snubhex-R}), the true $P_B(q,v)$
will be multiplied by a spurious factor of $q$ per erased spin. This will obviously not change the
set of roots $P_B(q,v) = 0$. We take the convention (here and elsewhere) of dividing such spurious
factors out of the polynomials that are provided in electronic form in the supplementary material.

Note that (\ref{snubhex-R}) only makes sense when $n$ is a multiple of 7. This means that, using this construction,
the only computations that we can perform in practice is to find numerically the roots in $v$ of $P_B(q,v)=0$
with $n=7$. While this agrees nicely with the upper bound $n_{\rm max} = 7$ for the feasibility of the computations,
it entails two inconveniences. First, our inability to compute the full polynomial $P_B(q,v)$ with $n=7$ implies
that we have no access to the phase diagram in the real $(q,v)$ plane. Second, the fact that we get just one single
estimate for $v_{\rm c}$ means that the only sensible proposal for the $n \to \infty$ result is the $n=7$ value itself.
In particular, an error bar on this result can only be obtained by making the ({\em a priori} not unlikely) assumption that the distance
between the $n=7$ value and the would-be extrapolation is comparable to that of other lattices for which we have been
able to perform the extrapolation carefully.

\begin{table}
\begin{center}
 \begin{tabular}{l|l}
 $n$ & $p_{\rm c}$ \\ \hline
 2 & 0.43435240230711099756452570147287096581374932914479 \\
 4 & 0.43433086696991174675056512033952861575538482310455 \\
 6 & 0.43432861831549797141228558205641632097611421130662 \\ \hline
 7 & 0.43432809783895257624033939574442324419941437511711 \\ \hline
 $\infty$ & 0.43432764 (3) \\
 Ref.~\cite{Parviainen07} & 0.4343062 (5) \\
 \end{tabular}
 \caption{Bond percolation threshold $p_{\rm c}$ on the snub hexagonal lattice.}
  \label{tab:snubhexagonal1}
\end{center}
\end{table}

To elaborate on this last remark, consider for instance the $n=7$ estimate for the percolation threshold shown in Table~\ref{tab:snubhexagonal1}.
Let us recall that the relative deviation between the $n=7$ estimate and the $n \to \infty$ extrapolated value
is $2 \cdot 10^{-9}$ for the kagome lattice (see Table~\ref{tab:kagome1}) and $1 \cdot 10^{-10}$ for the
three-twelve lattice (see Table~\ref{tab:threetwelve1}). However, these two lattices benefited from very high values
($w \approx 6.36$ and $w \approx 6.39$ respectively) of the correction-to-scaling exponent in (\ref{extrapol_w}).
For the four-eight lattice (see Table~\ref{tab:foureight1}) we found a smaller value $w \approx 4.28$ and accordingly
the relative precision of the $n=7$ estimate comes out as $8 \cdot 10^{-8}$.
Recall also that the basis for the kagome and four-eight (resp.\ three-twelve)
lattices has 6 (resp.\ 9) edges per grey square. From the number of edges in the basis, we thus have no {\em a priori}
reason to believe that the convergence
properties of the snub hexagonal lattice (whose basis has $\frac{30}{7} \simeq 4.3$ edges per grey square)
would be significantly worse than any of those lattices. However, since the value of $w$ for the snub hexagonal lattice
is presently unknown, we could conservatively assume that
the $n=7$ data point has a relative precision of only $8 \cdot 10^{-7}$, and use this assumption to provide a
tentative error bar:
\begin{equation}
 p_{\rm c} = 0.4343281 (4) \,. \qquad \mbox{(tentative result)}
 \label{pcsnubhextentative}
\end{equation}
Noting that even this conservative estimate is in significant disagreement with the
numerical result of Parviainen \cite{Parviainen07} (see Table~\ref{tab:snubhexagonal1}) we therefore turn to
a different way on constructing the snub hexagonal lattice.

\begin{figure}
\begin{center}
\begin{tikzpicture}[scale=1.0,>=stealth]
\foreach \xpos in {0,1,2,3}
\foreach \ypos in {0,2,4,6}
 \fill[black!20] (\xpos+0.5,\ypos) -- (\xpos+1,\ypos+0.5) -- (\xpos+0.5,\ypos+1) -- (\xpos,\ypos+0.5) -- cycle;

\foreach \xpos in {0,2}
\foreach \ypos in {0,4}
{
 \fill[red!20] (\xpos+0.5,\ypos+1) -- (\xpos+1,\ypos+0.5) -- (\xpos+1.5,\ypos+1) -- (\xpos+1.5,\ypos+2) -- (\xpos+1,\ypos+2.5) -- (\xpos+0.5,\ypos+2) -- cycle;
 \fill[red!20] (\xpos,\ypos+2.5) -- (\xpos+0.5,\ypos+3) -- (\xpos+0.5,\ypos+4) -- (\xpos,\ypos+4.5) -- cycle;
 \fill[red!20] (\xpos+2,\ypos+2.5) -- (\xpos+1.5,\ypos+3) -- (\xpos+1.5,\ypos+4) -- (\xpos+2,\ypos+4.5) -- cycle;
}

\foreach \xpos in {0,1,2,3}
\foreach \ypos in {0,2,4,6}
{
  \draw[blue,ultra thick] (\xpos,\ypos+0.5) -- (\xpos+0.5,\ypos) -- (\xpos+1,\ypos+0.5) -- (\xpos+0.5,\ypos+1) -- cycle;
}
\foreach \xpos in {0,2}
\foreach \ypos in {0,4}
{
  \draw[blue,ultra thick] (\xpos+0.5,\ypos) -- (\xpos+0.5,\ypos+1);
  \draw[blue,ultra thick] (\xpos+1,\ypos+0.5) -- (\xpos+2,\ypos+0.5);
  \draw[blue,ultra thick] (\xpos,\ypos+2.5) -- (\xpos+1,\ypos+2.5);
  \draw[blue,ultra thick] (\xpos+1.5,\ypos+2) -- (\xpos+1.5,\ypos+3);
}

\foreach \xpos in {0,2}
\foreach \ypos in {0,4}
{
 \draw[blue,ultra thick] (\xpos+0.5,\ypos+1) -- (\xpos+1.5,\ypos+1) -- (\xpos+1.5,\ypos+2) -- (\xpos+0.5,\ypos+2) -- cycle;
 \draw[blue,ultra thick] (\xpos+0.5,\ypos+2) -- (\xpos+1.5,\ypos+1);
 \draw[blue,ultra thick] (\xpos,\ypos+4) -- (\xpos+0.5,\ypos+4) -- (\xpos+0.5,\ypos+3) -- (\xpos,\ypos+3);
 \draw[blue,ultra thick] (\xpos,\ypos+3.5) -- (\xpos+0.5,\ypos+3);
 \draw[blue,ultra thick] (\xpos+2,\ypos+4) -- (\xpos+1.5,\ypos+4) -- (\xpos+1.5,\ypos+3) -- (\xpos+2,\ypos+3);
 \draw[blue,ultra thick] (\xpos+1.5,\ypos+4) -- (\xpos+2,\ypos+3.5);
}

\foreach \xpos in {0,1,2,3}
\foreach \ypos in {0,2,4,6}
 \draw[black] (\xpos+0.5,\ypos) -- (\xpos+1,\ypos+0.5) -- (\xpos+0.5,\ypos+1) -- (\xpos,\ypos+0.5) -- cycle;

\draw[very thick,->] (0,-0.5)--(4,-0.5);
\draw (4,-0.5) node[right] {$x$};
\foreach \xpos in {0,1,2,3}
{
 \draw[thick] (\xpos+0.5,-0.6)--(\xpos+0.5,-0.4);
 \draw (\xpos+0.5,-0.5) node[below] {$\xpos$};
}

\draw[very thick,->] (-0.5,0)--(-0.5,8);
\draw (-0.5,8) node[above] {$y$};
\foreach \ypos in {0,1,2,3}
{
 \draw[thick] (-0.6,2*\ypos+0.5)--(-0.4,2*\ypos+0.5);
 \draw (-0.5,2*\ypos+0.5) node[left] {$\ypos$};
}
 
\end{tikzpicture}
 \caption{Four-terminal representation of the snub hexagonal lattice.}
 \label{fig:snubhexagonal}
\end{center}
\end{figure}
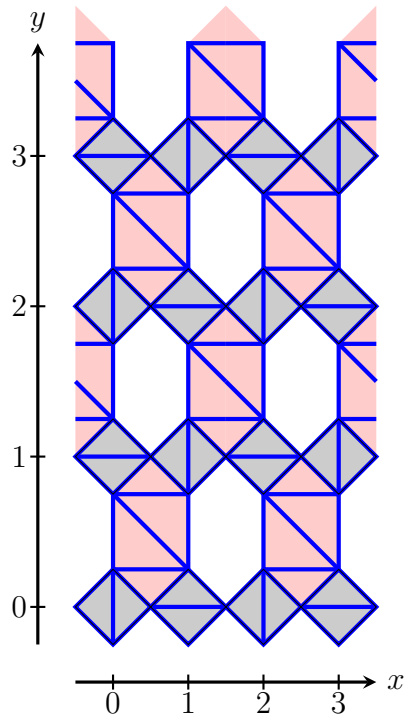

The second construction is based on the four-terminal representation shown in Figure~\ref{fig:snubhexagonal}.%
\footnote{The author is grateful to Chris Scullard for pointing out this representation.}
It generalises the general four-terminal setup of Figure~\ref{fig:square-basis} by stretching the white squares
vertically, so that they are now hexagons. To make these hexagons more visible in the figure we have
shaded them alternately using white and pink colours, but we shall still refer to them as ``white hexagons''.
The $\check{\sf R}$-matrix describing the grey squares
is exactly the same as that of the snub square lattice, see Eq.~(\ref{R-snubsquare}). 
The essential characteristics of the white hexagons is that---just like the white squares used previously---the lattice structure inside them
can be constructed without the use of auxiliary spaces. The corresponding operator (the analogue of $\check{\sf R}_i$) reads
\begin{equation}
 {\sf O}_i \equiv {\sf H}_{i+1} {\sf V}_{i+2}  {\sf H}_{i+1} {\sf V}_i {\sf H}_{i+1} \,.
\end{equation}
After each even row (i.e., $y = \frac12 \mbox{ mod } 2$) the operator ${\sf O}_i$ acts at positions $i = 0 \mbox{ mod } 4$, and
after each odd row (i.e., $y = \frac32 \mbox{ mod } 2$) it acts at $i = 2 \mbox{ mod } 4$. Note that ${\cal O}_i$ and ${\cal O}_{i+4}$
commute, since each operator acts on precisely four adjacent strands, and so we do not need to specify the order of factors in
these products. Moreover, since ${\cal O}_i$ now contains operators of the type ${\sf V}$ that propagate the system upwards, it is
crucial for this representation that every other white hexagon be empty. This is indeed the case, as shown by the alternating
white and pink shadings of the ``white'' hexagons.

The representation of Figure~\ref{fig:snubhexagonal} makes sense for even $n$. It contains $3$ vertices and $\frac{15}{2}$ edges
per grey square. This is a significant improvement over the first construction described in this subsection. We note that the basis
with $n=2$ coincides with that used in \cite[Figure21a]{Scullard12} in which only percolation (not the general $q$-state Potts model) was
studied. We have checked that our result for $P_B(1,v)$ with $n=2$ is proportional to \cite[Eq.~(27)]{Scullard12} after the usual
change of variables, $p = v/(1+v)$.

\begin{table}
\begin{center}
 \begin{tabular}{l|l}
 $n$ & $v_{\rm c}$ \\ \hline
 2 & 1.25822914824213089627783168669624673560080841355215 \\
 4 & 1.25831385989581118739654002272081781643725758811430 \\
 6 & 1.25832226936396475004073969036992612800230543696904 \\ \hline
 7 & 1.25832368118331690296415028885813907747723413111257 \\ \hline
 $\infty$ & 1.25832577 (4) \\
  \end{tabular}
 \caption{Critical point $v_{\rm c}$ of the $q=3$ state Potts model on the snub hexagonal lattice.}
  \label{tab:snubhexagonal3}
\end{center}
\end{table}

\begin{table}
\begin{center}
 \begin{tabular}{l|l}
 $n$ & $v_{\rm c}$ \\ \hline
 2 & 1.42904479893946258801335842598986168793665913956403 \\
 4 & 1.42921237939109852302475951646616375686887686926371 \\
 6 & 1.42922900017262995076127341190422753009029194456598 \\ \hline
 7 & 1.42923133461518622283008319875598743145981594637169 \\ \hline
 $\infty$ & 1.42923591 (9) \\
  \end{tabular}
 \caption{Critical point $v_{\rm c}$ of the $q=4$ state Potts model on the snub hexagonal lattice.}
  \label{tab:snubhexagonal4}
\end{center}
\end{table}

The percolation thresholds $p_{\rm c}$ using both the first and the second constructions are shown in Table~\ref{tab:snubhexagonal1}.
We have separated the results using the two different constructions by a horizontal line in this and the following two tables.
Extrapolating the results with $n=2,4,6$ leads to the estimate for the $n \to \infty$ limit shown in Table~\ref{tab:snubhexagonal1}. This is in agreement with the tentative result (\ref{pcsnubhextentative}), but is obviously more precise
since we now take advantage of a well-controlled extrapolation procedure. Note that we find $w \approx 2.94$,
significantly lower than for the four-eight lattice, so our precautions in arriving at (\ref{pcsnubhextentative}) were
justified. The final result of Table~\ref{tab:snubhexagonal1} enhances our disagreement with \cite{Parviainen07}
for this lattice.

For the Ising model ($q=2$) the polynomials $P_B(q,v)$ factorise as usual. The degree of the largest factor is
$d_{\rm max} = 8$ for $n=2$ and $d_{\rm max} = 12$ for $n=4$. The recurrent factor that leads to a positive
real root is
\begin{equation}
 -8 - 8 v - 4 v^2 + 4 v^3 + 8 v^4 + 4 v^5 + v^6 \,.
 \label{snubhexagonal_factor}
\end{equation}
If we define $\omega = 37 + 27 \sqrt{3} + 3 \sqrt{6 (66 + 37 \sqrt{3})}$,
the critical coupling can be written in the same way as (\ref{snubsquare_root}), viz.
\begin{equation}
 v_{\rm c} = \frac13 \left( -2 -2 w^{-1/3} + w^{1/3} \right)
 \simeq 1.050\,155\,297\cdots \,,
 \label{snubhexagonal_root}
\end{equation}
and this coincides with the result of \cite{Codello10}. Using the first construction, we have
performed a $50$-digit numerical computation to verify that this number is also a root of the
$n=7$ polynomial. This agreement provides compelling evidence that both constructions
of $P_B(q,v)$ are correct.

Our results for the critical point $v_{\rm c}$ in the $q=3$ and $q=4$ state Potts models
are shown in Tables~\ref{tab:snubhexagonal3}--\ref{tab:snubhexagonal4}.
Also in those cases have we based the $n \to \infty$ extrapolations on the results
coming from the second construction (with $n=2,4,6$). Using the same kind of reasoning as
discussed above for percolation, we observe that the $n=7$ results are compatible with
these extrapolated values.

\begin{figure}
\begin{center}
\includegraphics[width=12cm]{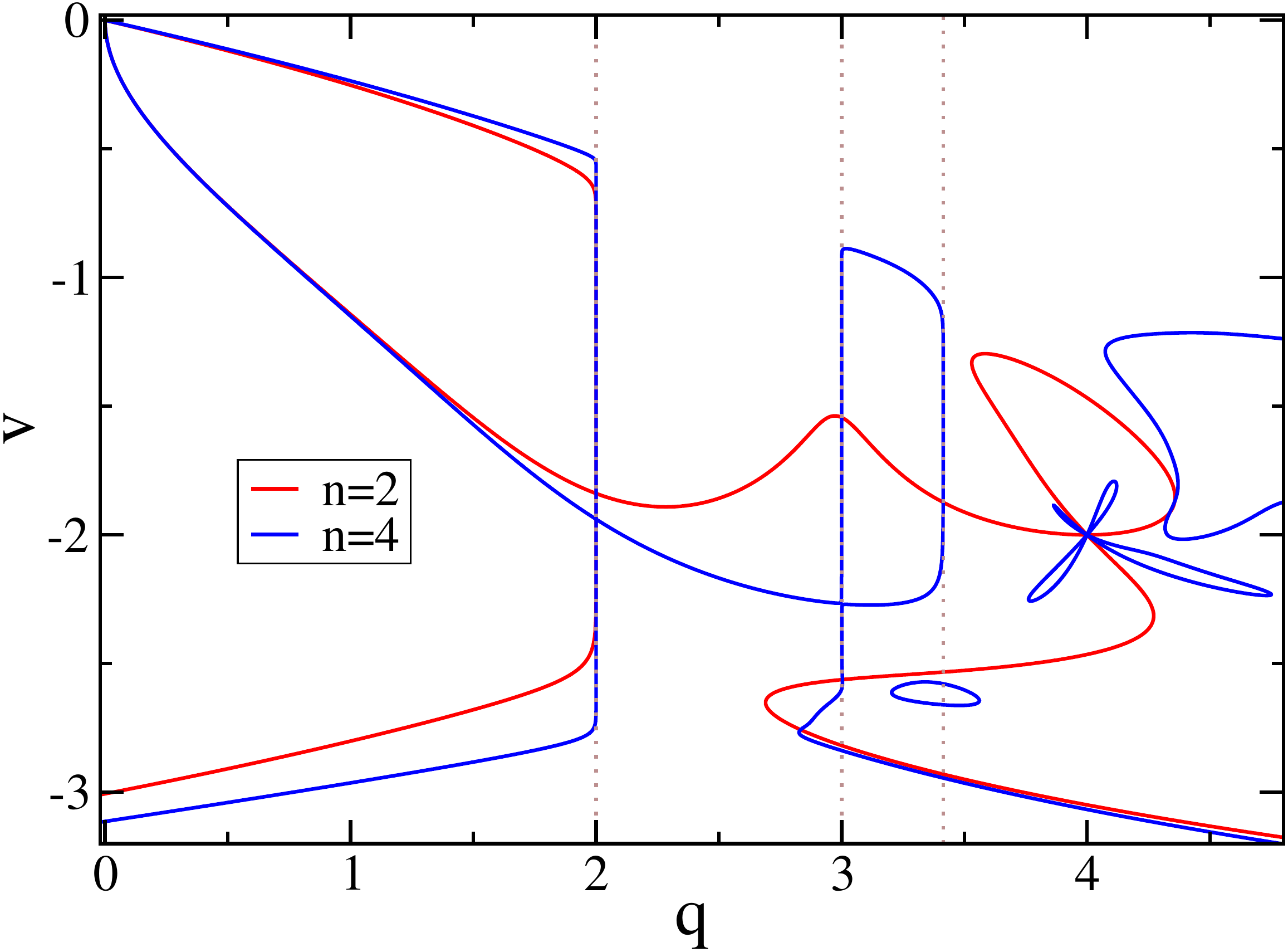}
\caption{Roots of $P_B(q,v)$ for the Potts model on the snub hexagonal lattice, using
$n \times n$ square bases.}
\label{fig:snubhex-pd}
\end{center}
\end{figure}

The phase diagram of the snub hexagonal lattice is shown in Figure~\ref{fig:snubhex-pd}.
The extent of the BK phase can be estimated from the vertical rays at $q=B_k$ with $k=4,6,8$.
Its upper and lower boundaries $v_\pm(q)$ are interrupted by a hiatus for $2 < q < 3$ with these
bases, but the wrinkle developing near the bottom of the $q=3$ ray in the $n=4$ curve indicates
that this gap may be partly filled in at larger sizes. The curve coming out of $(q,v)=(0,0)$ with
infinite slope cuts through the BK phase. We also note a rather rich structure near $q=4$
where a flower-like structure with four petals grows out of the point $(q,v)=(4,-2)$. There are thus
$4$ (resp.\ $8$) branches of the curve coming out that point for $n=2$ (resp.\ $n=4$), and this
number might well continue to grow for larger $n$.

\subsection{Ruby lattice $(3,4,6,4)$}
\label{sec:ruby}

\begin{figure}
\begin{center}
\begin{tikzpicture}[scale=1.0,>=stealth]
\foreach \xpos in {0,1,2,3}
\foreach \ypos in {0,1,2,3}
 \fill[black!20] (\xpos+0.5,\ypos) -- (\xpos+1,\ypos+0.5) -- (\xpos+0.5,\ypos+1) -- (\xpos,\ypos+0.5) -- cycle;
 
\foreach \xpos in {0,2}
\foreach \ypos in {0,2}
{
  \draw[blue,ultra thick] (\xpos+0.5,\ypos+1) -- (\xpos,\ypos+0.5) -- (\xpos+0.5,\ypos) -- (\xpos+1,\ypos+0.5) -- (\xpos+0.25,\ypos+0.75);
  \draw[blue,ultra thick] (\xpos+1.5,\ypos) -- (\xpos+2,\ypos+0.5) -- (\xpos+1.5,\ypos+1) -- (\xpos+1,\ypos+0.5) -- (\xpos+1.75,\ypos+0.25);
  \draw[blue,ultra thick] (\xpos+0.5,\ypos+1) -- (\xpos+1,\ypos+1.5) -- (\xpos+0.5,\ypos+2) -- (\xpos,\ypos+1.5) -- (\xpos+0.75,\ypos+1.25);
  \draw[blue,ultra thick] (\xpos+1.5,\ypos+2) -- (\xpos+1,\ypos+1.5) -- (\xpos+1.5,\ypos+1) -- (\xpos+2,\ypos+1.5) -- (\xpos+1.25,\ypos+1.75);
}
\foreach \ypos in {0,1,2,3}
 \draw[blue,ultra thick] (0,\ypos) -- (4,\ypos);

\foreach \xpos in {0,1,2,3}
\foreach \ypos in {0,1,2,3}
 \draw[black] (\xpos+0.5,\ypos) -- (\xpos+1,\ypos+0.5) -- (\xpos+0.5,\ypos+1) -- (\xpos,\ypos+0.5) -- cycle;

\draw[very thick,->] (0,-0.5)--(4,-0.5);
\draw (4,-0.5) node[right] {$x$};
\foreach \xpos in {0,1,2,3}
{
 \draw[thick] (\xpos+0.5,-0.6)--(\xpos+0.5,-0.4);
 \draw (\xpos+0.5,-0.5) node[below] {$\xpos$};
}

\draw[very thick,->] (-0.5,0)--(-0.5,4);
\draw (-0.5,4) node[above] {$y$};
\foreach \ypos in {0,1,2,3}
{
 \draw[thick] (-0.6,\ypos+0.5)--(-0.4,\ypos+0.5);
 \draw (-0.5,\ypos+0.5) node[left] {$\ypos$};
}
 
\end{tikzpicture}
 \caption{Four-terminal representation of the ruby dual lattice.}
 \label{fig:rubydual}
\end{center}
\end{figure}
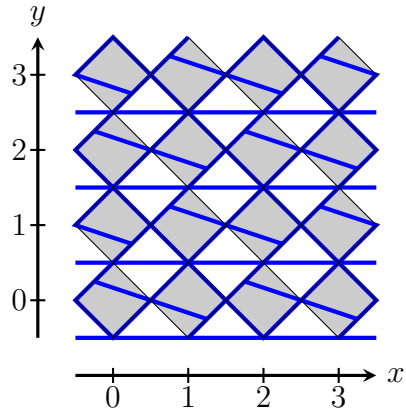

The ruby lattice is most simply treated by considering the corresponding dual lattice.
A four-terminal representation of the basis of the ruby dual lattice is shown in
Figure~\ref{fig:rubydual}. The reader can check that it is indeed a quadrangulation,
with each of the quadrangles being surrounded by vertices of degrees 3, 4, 6 and 4.
The $\check{\sf R}$-matrix reads
\begin{equation}
 \check{\sf R} = \left \lbrace
 \begin{array}{ll}
 {\sf V}_{i} {\sf H}_{i+1} {\sf V}_{i+2} {\sf V}_i {\sf H}_{i+1} & \mbox{for $x+y$ even} \\
 {\sf H}_{i+1} {\sf V}_{i+2} {\sf V}_{i} {\sf H}_{i+1} {\sf V}_{i+2} & \mbox{for $x+y$ odd} \\
 \end{array} \right.
\end{equation}
and there are horizontal diagonals on all the white squares. Clearly this representation
of the ruby dual lattice requires $n$ to be even. 
There are 3 faces (corresponding to vertices of the ruby lattice itself) and 6 edges per grey square.

The graph polynomials $P_B(q,v)$ for the ruby lattice (and their specialisation
$P_B(p)$ to the percolation case)
are then found from those of the dual by
a simple duality transformation. Details on duality will be deferred to section~\ref{sec:res-dual}.

\begin{table}
\begin{center}
 \begin{tabular}{l|l}
 $n$ & $p_{\rm c}$ \\ \hline
 2 & 0.52483166874192588456166387740108217826717357914594 \\
 4 & 0.52483306819526131752968049973689841123051783280705 \\
 6 & 0.52483166919384159966725735129503659944406282355902 \\ \hline
 $\infty$ & 0.5248311 (1) \\
 Ref.~\cite{Parviainen07} & 0.5248326 (5) \\
 \end{tabular}
 \caption{Bond percolation threshold $p_{\rm c}$ on the ruby lattice.}
  \label{tab:ruby1}
\end{center}
\end{table}

Bond percolation thresholds obtained from $P_B(p)$ are reported in Table~\ref{tab:ruby1}.
Uncharacteristically for the graph polynomial approach the convergence to the thermodynamical
limit is here seen to be non-monotonic. This is presumably due to the smallest result ($n=2$)
straying away from the general trend. Combined with the parity constraint on $n$ this negatively
impacts the precision of the extrapolated threshold $p_{\rm c}$ for this lattice.

For the Ising case ($q=2$), the graph polynomials $P_B(q,v)$ factorise once again, and
the maximum degree of the factors is $d_{\rm max} = 6$ for $n=2$,
and $d_{\rm max} = 10$ for $n=4$.
There is a unique common factor possessing a positive root:
\begin{equation}
 -8 - 8 v + 4 v^3 + v^4 \,.
 \label{ruby_factor}
\end{equation}
The critical coupling reads
\begin{eqnarray}
 v_{\rm c} = \sqrt{3 + 2 \sqrt{3}} -1 \simeq 1.542\,459\,756\cdots
 \label{ruby_root}
\end{eqnarray}
in agreement with \cite{Codello10}.
Rather curiously, this is identical to the result (\ref{kagome_root}) found for the Ising model
on the kagome lattice.

\begin{table}
\begin{center}
 \begin{tabular}{l|l}
 $n$ & $v_{\rm c}$ \\ \hline
 2 & 1.87394547154498212745580696195918007388348035664800 \\
 4 & 1.87391585478227655442007815269131135001811822203444 \\
 6 & 1.87392197066479367079676783453126504408425961116844 \\ \hline
 $\infty$ & 1.8739245 (6) \\
  \end{tabular}
 \caption{Critical point $v_{\rm c}$ of the $q=3$ state Potts model on the ruby lattice.}
  \label{tab:ruby3}
\end{center}
\end{table}

\begin{table}
\begin{center}
 \begin{tabular}{l|l}
 $n$ & $v_{\rm c}$ \\ \hline
 2 & 2.15108293916710713480143126975686391827111703551182 \\
 4 & 2.15100732273079876867673763885401274828091470382521 \\
 6 & 2.15101806147233059726929249029635229313711666783623 \\ \hline
 $\infty$ & 2.1510225 (9) \\
  \end{tabular}
 \caption{Critical point $v_{\rm c}$ of the $q=4$ state Potts model on the ruby lattice.}
  \label{tab:ruby4}
\end{center}
\end{table}

The critical points for the $q=3$ and $q=4$ models are shown in Tables~\ref{tab:ruby3}--\ref{tab:ruby4}.
As for percolation the convergence is non-monotonic, preventing us from attaining the usual
precision in the final results.

\begin{figure}
\begin{center}
\includegraphics[width=12cm]{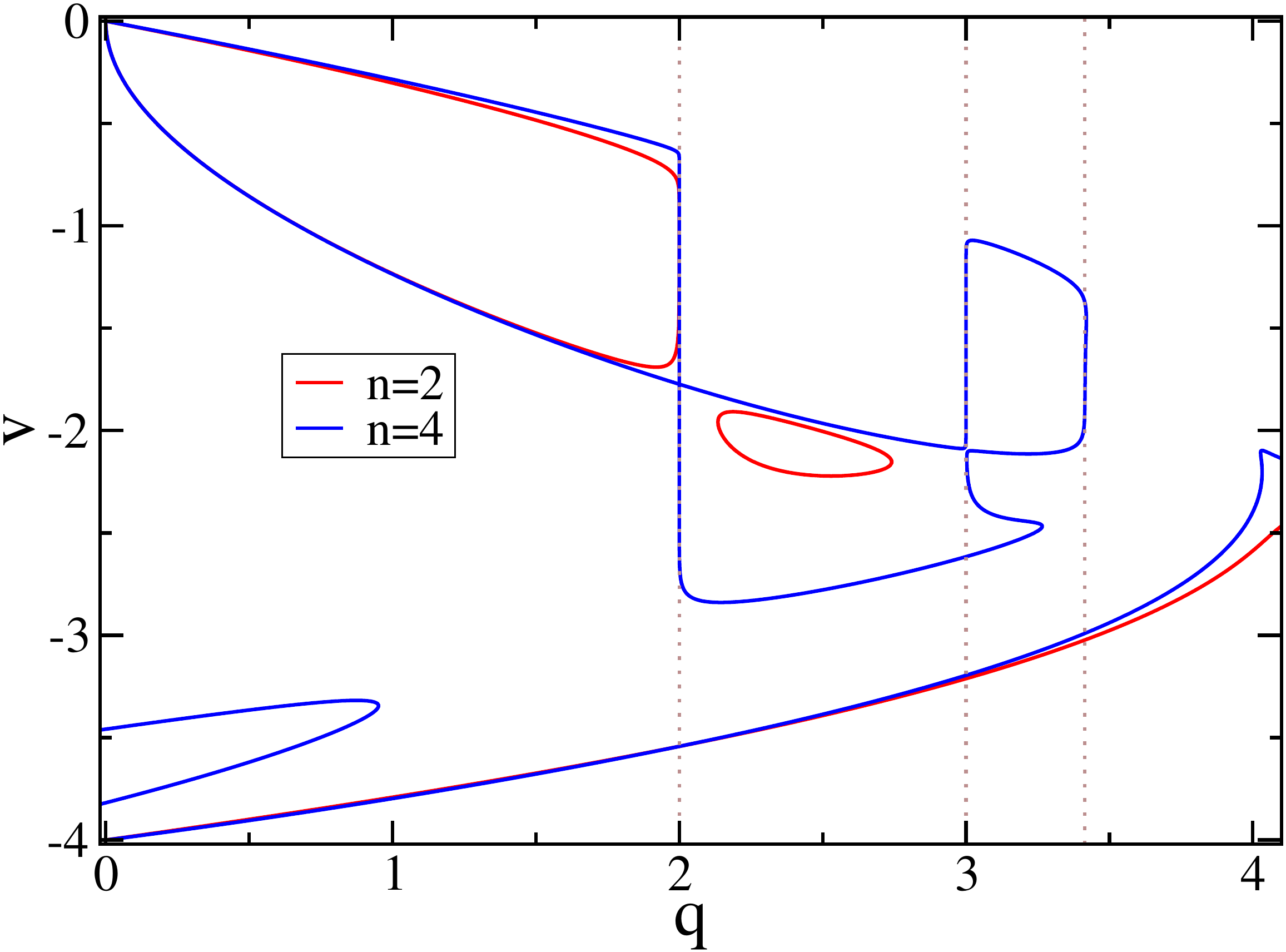}
\caption{Roots of $P_B(q,v)$ for the Potts model on the ruby lattice, using
$n \times n$ square bases.}
\label{fig:ruby-pd}
\end{center}
\end{figure}

The phase diagram for the ruby lattice is shown in Figure~\ref{fig:ruby-pd}. The extent of the BK phase
can be judged from the vertical rays at the Beraha numbers (\ref{Beraha}) with $k=4,6,8$. The upper and
lower boundaries of the BK phase are only partially brought out by the largest size $n=4$. In particular,
the upper boundary between $q=2$ and $q=3$ is still missing. The lower boundary between $q=0$ and $q=2$
is partially provided by the finger protruding from $q=0$, and one notes the formation of a wrinkle to the right
of $q=3$. Interestingly there is another curve, emanating from $(q,v)= (0,-4)$ that lies below the lower boundary
of the BK phase. This curve extends towards the point $(q,v) = (4,-2)$, and presumably the boundaries of the
BK phase will also join at that point. We therefore conjecture that $q_{\rm c} = 4$ for this lattice.

\subsection{Factorisable cases}
\label{sec:factorisation}

A remarkable property of the critical polynomial $P_B(q,v)$ is that it usually factorises in exactly solvable cases \cite{Jacobsen12,Jacobsen13}.
This is true in particular for the exactly solvable lattices (square, triangular and hexagonal). This is also true for the Ising
model ($q=2$) on any of the lattices considered. Note however that some cases also exist where $P_B(q,v)$ fails
to factorise, even though the model is known to be exactly solvable. This is so in particular for the zero-temperature
antiferromagnetic three-state model, $(q,v)=(3,-1)$, on the kagome lattice and for the entire chromatic line, $v=-1$,
on the triangular lattice. Indeed, Baxter has shown that the latter model is integrable \cite{Baxter86,Baxter87}, and the
former model (three-state kagome) is equivalent to a special case of the latter (four-state triangular) by means of an
exact mapping \cite{MooreNewman00}.

Conversely, it is compelling to consider any systematic factorisation of $P_B(q,v)$
as evidence that the model may be exactly solvable (by ``systematic factorisation'' we mean a factorisation that occurs for any
value of the size $n$).

In this section we examine exhaustively the issue of factorisation for all the Archimedian lattices. We consider
the following cases: Integer $q=0,1,2,3,4$, the chromatic polynomial $v=-1$, the flow polynomial $v=-q$ \cite{Wu88,Salas13}, and
the limit $(q,v) \to (0,0)$ with fixed $w=v/q$ that correspond to spanning forests \cite{JSS05} with weight $1/w$ per component tree.

For $q=0$, $P_B(q,v)$ factorises for all the lattices, producing a root $v=0$. This is consistent with the observation that for all
the lattices there is a branch of the critical curve going through the point $(q,v)=(0,0)$. This describes the problem of spanning
trees, which can in turn be related to free (symplectic) fermions with central charge $c=-2$ \cite{CJSSS04,JSS05,JacSal05}.
Moreover, the triangular, kagome and three-twelve lattices have a root $(q,v)=(0,-3)$.
And the square, four-eight, cross and ruby lattice have a root $(q,v)=(0,-4)$. 
By duality, the models with $(q,v)=(0,v_{\rm c}$) are equivalent \cite{JSS05} to models of spanning forests on the
corresponding dual lattice with weight $v_{\rm c}$ per component tree. These models can in turn be formulated
as interacting fermionic theories \cite{CJSSS04,JacSal05}, and we conjecture that they are in fact exactly solvable.
It follows, still by duality, that the spanning tree problem factorises on the square lattice with $w=-1/4$ and on the
hexagonal lattice with $w=-1/3$. We find moreover that spanning trees on the cross lattice factorise with $w=-1/3$.

The case of the snub square and snub hexagonal lattices is interesting. For $q=0$, the critical polynomials of both these
lattices shed the small factor $8+5v+v^2$, and we conjecture that the corresponding roots $v=(-5 \pm i \sqrt{7})/2$ are
loci of exact solvability.

The Ising case ($q=2$) has been extensively discussed in the preceding sections. Cases where $P_B(2,v)$
has a negative integer root in $v$ occur only for $v=-1$ (the chromatic polynomial) and $v=-2$ (the flow polynomial).
More precisely, $v=-1$ factorises for the triangular, kagome, frieze, three-twelve, snub square, snub hexagonal and ruby lattices.
And $v=-2$ factorises for the hexagonal, four-eight, frieze, three-twelve, cross, snub square and snub hexagonal lattices.

For $q=3$, $P_B(q,v)$ has a root at $v=-3$ for the four-eight and three-twelve lattices. Note that these are three-flow
problems or, equivalently, three-colouring problems of the corresponding dual lattices.

Finally, for $q=4$ there is a root at $v=-2$ for all lattices except the kagome lattice. Motivated by this, and by the phase
diagrams reported in the preceding sections, we conjecture that the BK phase extends to $(q,v)=(4,-2)$ for all the Archimedian
lattices, except the kagome lattice.

\section{Results on dual lattices}
\label{sec:res-dual}

The Potts model partition function admits the duality transformation \cite{Potts52,Wu88}
\begin{equation}
 v \to v^* := q/v \,.
\end{equation}
It is a consequence of (\ref{PB_cluster}) that the same is true for the graph polynomial $P_B(q,v)$.
To see this, note that the configurations contributing to $Z_{\rm 2D}$ are in bijection with those
contributing to $Z^*_{\rm 0D}$ on the dual lattice, and vice versa.
It follows that
\begin{equation}
 P_{B^*}(q,v^*) = \frac{(v^*)^{|E|}}{q^{|V|}} P_B(q,q/v^*)
 \label{PB-dual}
\end{equation}
is the graph polynomial on the dual lattice, corresponding to the dual basis $B^*$. Here
$|V|$ and $|E|$ denote respectively the number of vertices and edges in the basis $B$.

All the results given in section~\ref{sec:res-archi} can therefore be applied to the dual
Archimedian lattices (Laves lattices) as well, simply by making the change of variables
(\ref{PB-dual}). Note that in the case of the ruby lattice (section~\ref{sec:ruby}) we have
already anticipated on this relation, because it is easier to represent the ruby dual lattice
in the required four-terminal form than the ruby lattice itself. A few duality arguments were also
used in section~\ref{sec:factorisation}.

\section{Results on medial lattices}
\label{sec:res-medial}

We have also computed the graph polynomial $P_B(q,v)$ for all the medials of the Archimedian lattices.
Medial lattices were defined and discussed in section~\ref{sec:lattices}. We recall that the
square lattice is its own medial, ${\cal M}(4^4) = (4^4)$, the triangular and hexagonal lattices
have the same medial which is the kagome lattice, ${\cal M}(3^6) = {\cal M}(6^3) =  (3,6,3,6)$,
and the medial of the kagome lattice is the ruby lattice, ${\cal M}(3,6,3,6) = (3,4,6,4)$.

\begin{table}
\begin{center}
 \begin{tabular}{l|lll}
 Lattice & Vertices & Edges & Parity of $n$ \\ \hline
 Four-eight medial & 6 & 12 & Any \\
 Frieze medial & $\frac{5}{2}$ & 5 & Any${}^\dagger$ \\
 Three-twelve medial & 9 & 18 & Any \\
 Cross medial & $\frac{9}{2}$ & 9 & Even \\
 Snub square medial & $\frac{5}{2}$ & 5 & Even \\
 Snub hexagonal medial & $\frac{15}{7}$ & $\frac{30}{7}$ & 0 mod 7 \\
                                        & $\frac{5}{2}$ & 5 & 0 mod 3${}^\dagger$ \\
 Ruby medial & $\frac{3}{2}$ & 3 & 0 mod 4 \\
  \hline
  \end{tabular}
 \caption{Number of vertices and edges per grey square (cf.~Figure~\ref{fig:square-basis})
 for the medials of Archimedean lattices studied here, using square bases of size $n \times n$ grey squares
 (${}^\dagger$or rectangular bases of size $n \times 2n$). In addition we state any parity constraint on $n$.
 For the snub hexagonal medial lattices two different constructions are provided.}
  \label{tab:packingmedial}
\end{center}
\end{table}

So we shall consider in the following subsections the remaining seven medial lattices.
Some of these require specific tricks---which might be of independent interest---such as
avoiding the introduction of intermediate points by acting in each grey square with a
generic Temperley-Lieb operator, and deleting and contracting some of the edges by
formally setting the coupling constants to $x=0$ or $x=\infty$.

The degree of the critical polynomials $P_B(q,v)$ is $k_q n^2$ in the $q$-variable and $k_v n^2$ in the $v$-variable, where
$k_q$ and $k_v$ are tabulated in the second and third columns of Table~\ref{tab:packingmedial}.
We have $k_v = 2 k_q$ throughout, since all the medial lattices are four-regular (i.e., all their vertices are of degree $4$).

The polynomials that we have obtained explicitly are available in electronic form as
supplementary material to this paper.%
\footnote[1]{In the form of a text file {\tt PB.m} that can be processed by {\sc Mathematica} or any other
symbolic computer algebra software.}

\subsection{Four-eight medial lattice ${\cal M}(4,8^2)$}
\label{sec:foureightmedial}

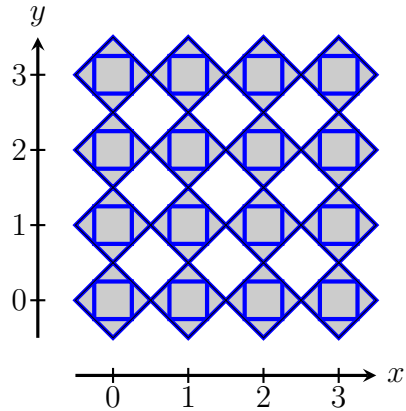
\begin{figure}
\begin{center}
\begin{tikzpicture}[scale=1.0,>=stealth]
\foreach \xpos in {0,1,2,3}
\foreach \ypos in {0,1,2,3}
 \fill[black!20] (\xpos+0.5,\ypos) -- (\xpos+1,\ypos+0.5) -- (\xpos+0.5,\ypos+1) -- (\xpos,\ypos+0.5) -- cycle;

\foreach \xpos in {0,1,2,3}
\foreach \ypos in {0,1,2,3}
{
 \draw[blue,ultra thick] (\xpos+0.5,\ypos) -- (\xpos+1,\ypos+0.5) -- (\xpos+0.5,\ypos+1) -- (\xpos,\ypos+0.5) -- cycle;
 \draw[blue,ultra thick] (\xpos+0.25,\ypos+0.25) -- (\xpos+0.75,\ypos+0.25) -- (\xpos+0.75,\ypos+0.75) -- (\xpos+0.25,\ypos+0.75) -- cycle;
}

\foreach \xpos in {0,1,2,3}
\foreach \ypos in {0,1,2,3}
 \draw[black] (\xpos+0.5,\ypos) -- (\xpos+1,\ypos+0.5) -- (\xpos+0.5,\ypos+1) -- (\xpos,\ypos+0.5) -- cycle;

\draw[very thick,->] (0,-0.5)--(4,-0.5);
\draw (4,-0.5) node[right] {$x$};
\foreach \xpos in {0,1,2,3}
{
 \draw[thick] (\xpos+0.5,-0.6)--(\xpos+0.5,-0.4);
 \draw (\xpos+0.5,-0.5) node[below] {$\xpos$};
}

\draw[very thick,->] (-0.5,0)--(-0.5,4);
\draw (-0.5,4) node[above] {$y$};
\foreach \ypos in {0,1,2,3}
{
 \draw[thick] (-0.6,\ypos+0.5)--(-0.4,\ypos+0.5);
 \draw (-0.5,\ypos+0.5) node[left] {$\ypos$};
}
 
\end{tikzpicture}
 \caption{Four-terminal representation of the four-eight medial lattice.}
 \label{fig:foureightmedial}
\end{center}
\end{figure}

A four-terminal representation of the four-eight medial lattice is shown in Figure~\ref{fig:foureightmedial}.
At first sight it does not appear feasible to write the corresponding $\check{\sf R}$-matrix in the usual
form, viz., as a product of the single-edge operators ${\sf V}_i$, ${\sf H}_{i+1}$ and ${\sf V}_{i+2}$
(or, more generally, the Temperley-Lieb generators ${\sf E}_i$, ${\sf E}_{i+1}$ and ${\sf E}_{i+2}$)
acting within the unit cell shown in Figure~\ref{fig:labelR}. The problem is that we need an intermediate
point in the spin representation, or two intermediate strands in the loop representation that would be
situated between those labeled $i+1$ and $i+2$ in Figure~\ref{fig:labelR}. These intermediate strands
can however be eliminated once the grey square is completed, and they are not needed for connecting
among themselves the grey squares of which the lattice consists.

To avoid dealing with intermediate points, the most efficient solution is to write down directly the entire
$\check{\sf R}$-matrix that propagates strands $i,i+1,i+2,i+3$ into $i',(i+1)',(i+2)',(i+3)'$. Recall that a
single-edge operator, such as ${\sf V}_i$, consists of two terms (${\sf Id}_i$ and ${\sf E}_i$), since there
are ${\rm Cat}(2) = 2$ possible planarity-respecting pairings of the four points $i,i+1,i',(i+1)'$. Similarly, the entire
$\check{\sf R}$-matrix contains in general fourteen terms, corresponding to the ${\rm Cat}(4) = 14$ pairings of eight points
that respect planarity. We have therefore written a version of the algorithm in which a lattice is specified
by supplying the fourteen terms of a generic $\check{\sf R}$-matrix, each of which are polynomials in
$n_{\rm loop}$ and $x$ with integer coefficients. Further remarks on this version can be found in section~\ref{sec:implementation}.

In the case at hand we remark that $\check{\sf R}_i = ({\sf B}_i)^2$, where ${\sf B}_i$ denotes the bow tie
operator, first discussed for the special case of percolation in \cite[section 3.2]{SJ12} and subsequently
generalised to the Potts model in \cite[section 3.2]{Jacobsen13}. Squaring the explicit expression
\cite[Eq.~(27)]{Jacobsen13} we therefore obtain the fourteen polynomials defining $\check{\sf R}_i$, 
each of which contains up to a maximum of 10 monomials $x^a n_{\rm loop}^b$.

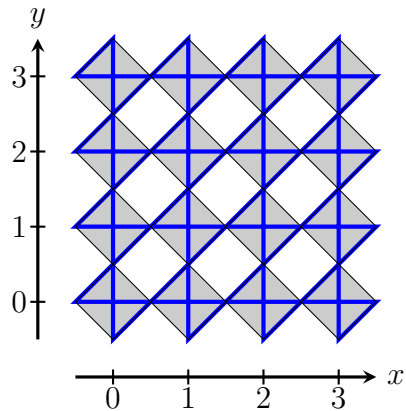
\begin{figure}
\begin{center}
\begin{tikzpicture}[scale=1.0,>=stealth]
\foreach \xpos in {0,1,2,3}
\foreach \ypos in {0,1,2,3}
 \fill[black!20] (\xpos+0.5,\ypos) -- (\xpos+1,\ypos+0.5) -- (\xpos+0.5,\ypos+1) -- (\xpos,\ypos+0.5) -- cycle;

\foreach \xpos in {0,1,2,3}
\foreach \ypos in {0,1,2,3}
 \draw[blue,ultra thick] (\xpos,\ypos+0.5) -- (\xpos+1,\ypos+0.5) -- (\xpos+0.5,\ypos) -- (\xpos+0.5,\ypos+1) -- cycle;

\foreach \xpos in {0,1,2,3}
\foreach \ypos in {0,1,2,3}
 \draw[black] (\xpos+0.5,\ypos) -- (\xpos+1,\ypos+0.5) -- (\xpos+0.5,\ypos+1) -- (\xpos,\ypos+0.5) -- cycle;

\draw[very thick,->] (0,-0.5)--(4,-0.5);
\draw (4,-0.5) node[right] {$x$};
\foreach \xpos in {0,1,2,3}
{
 \draw[thick] (\xpos+0.5,-0.6)--(\xpos+0.5,-0.4);
 \draw (\xpos+0.5,-0.5) node[below] {$\xpos$};
}

\draw[very thick,->] (-0.5,0)--(-0.5,4);
\draw (-0.5,4) node[above] {$y$};
\foreach \ypos in {0,1,2,3}
{
 \draw[thick] (-0.6,\ypos+0.5)--(-0.4,\ypos+0.5);
 \draw (-0.5,\ypos+0.5) node[left] {$\ypos$};
}
 
\end{tikzpicture}
 \caption{Alternative four-terminal representation of the kagome lattice.}
 \label{fig:kagome_alt}
\end{center}
\end{figure}

To test the general algorithm we have also investigated the case where each $\check{\sf R}$-matrix is 
a single bow tie operator, $\check{\sf R}_i = {\sf B}_i$. This can be represented as in Figure~\ref{fig:kagome_alt}.
Obviously this is just a rotated version of the four-terminal representation of the kagome lattice shown in
Figure~\ref{fig:kagome}. We have validated the ``generic $\check{\sf R}$-matrix'' algorithm by verifying that
in this case it gives the very same critical polynomials $P_B(q,v)$ as those obtained in section~\ref{sec:kagome}.
The choice of transfer direction made in section~\ref{sec:kagome} is slightly more efficient for dealing with the
kagome lattice, since $\check{\sf R}_i$ involves the application of six operators with each two terms ($6 \times 2 = 12$)
rather than a single operator with fourteen terms. Moreover, the approach with two terms per operator involves
only very simple coefficients ($1$ or $x$).

\begin{table}
\begin{center}
 \begin{tabular}{l|l}
 $n$ & $p_{\rm c}$ \\ \hline
 1 & 0.54490357617280539732583696992078943358822055517888 \\
 2 & 0.54482333432473326673827916421900410932649253360415 \\
 3 & 0.54480395308638647849435301431084308963838444393987 \\
 4 & 0.54479979248458363100469869489549952372478071895348 \\
 5 & 0.54479869412491430661835928269715088171915677271830 \\
 6 & 0.54479832681576092745539465910054157272244608840374 \\
 7 & 0.54479817718177358100731028878250145618621980837755 \\ \hline
 $\infty$ & 0.544798017 (4) \\
 Ref.~\cite{Ziff-unpub,WikiPercThres} & 0.5447979 (3) \\
 \end{tabular}
 \caption{Bond percolation threshold $p_{\rm c}$ on the four-eight medial lattice.}
  \label{tab:foureightmedial1}
\end{center}
\end{table}

Returning to the four-eight medial lattice, we remark that in Figure~\ref{fig:foureightmedial} there are $6$ vertices and
$12$ edges per grey square (see Table~\ref{tab:packingmedial}). The approximations to the bond percolation
threshold $p_c$ obtained from the unique positive root of $P_B(p)$ are shown in Table~\ref{tab:foureightmedial1}.

For the Ising model ($q=2$) the polynomials $P_B(q,v)$ always factorise. The maximum degree of the factors is
$d_{\rm max} = 8$ for $n=1,2$, $d_{\rm max} = 16$ for $n=3$, and $d_{\rm max} = 12$ for $n=4$. One
of these factors, namely
\begin{equation}
 -64 - 128 v - 160 v^2 - 96 v^3 - 8 v^4 + 32 v^5 + 24 v^6 + 8 v^7 + v^8 \,,
\end{equation}
occurs systematically for any $n$. By the change of variables $v = -1 + \sqrt{y}$ this simplifies to
\begin{equation}
 -23 - 20 y - 18 y^2 - 4 y^3 + y^4 \,.
\end{equation}
The unique positive root is
\begin{equation}
 v_{\rm c} = -1 + \left( 1 + \sqrt{2} + \sqrt{10 + 8 \sqrt{2}} \right)^{1/2} \simeq 1.651\,582\,692\cdots \,.
\end{equation}
We expect this to be the exact critical point, although we are not aware of any exact solution of the Ising model
on the four-eight medial lattice.

\begin{table}
\begin{center}
 \begin{tabular}{l|l}
 $n$ & $v_{\rm c}$ \\ \hline
 1 & 1.9922041076260751644586139440972263436652635871358 \\
 2 & 1.9926404728708502640118089145818512148977796047886 \\
 3 & 1.9927384787472335781145657627229386691955129596069 \\
 4 & 1.9927586076549435383693884954495153641240804012262 \\
 5 & 1.9927637904129256320901803185635371343359435001717 \\
 6 & 1.9927654949220578625901153681198029149913267082580 \\
 7 & 1.9927661807712696317912790872926681763051395367055 \\ \hline
 $\infty$ & 1.99276689 (2) \\
  \end{tabular}
 \caption{Critical point $v_{\rm c}$ of the $q=3$ state Potts model on the four-eight medial lattice.}
  \label{tab:foureightmedial3}
\end{center}
\end{table}

\begin{table}
\begin{center}
 \begin{tabular}{l|l}
 $n$ & $v_{\rm c}$ \\ \hline
 1 & 2.2755212087361303992802668673758851902898410610266 \\
 2 & 2.2763771760824237004043542724461772715320090614142 \\
 3 & 2.2765630817464703256746144539003180310447940465275 \\
 4 & 2.2766003034722388602617749274377947499719604673753 \\
 5 & 2.2766097352625615580128595130617901964243752604729 \\
 6 & 2.2766128023187296094751641612708914426382050828433 \\
 7 & 2.2766140251425216149873240698999853443282508981560 \\ \hline
 $\infty$ & 2.27661527 (5) \\
  \end{tabular}
 \caption{Critical point $v_{\rm c}$ of the $q=4$ state Potts model on the four-eight medial lattice.}
  \label{tab:foureightmedial4}
\end{center}
\end{table}

Moreover, the critical points for the $q=3$ and $q=4$ state Potts models are given in Tables~\ref{tab:foureightmedial3}--\ref{tab:foureightmedial4}.

\begin{figure}
\begin{center}
\includegraphics[width=12cm]{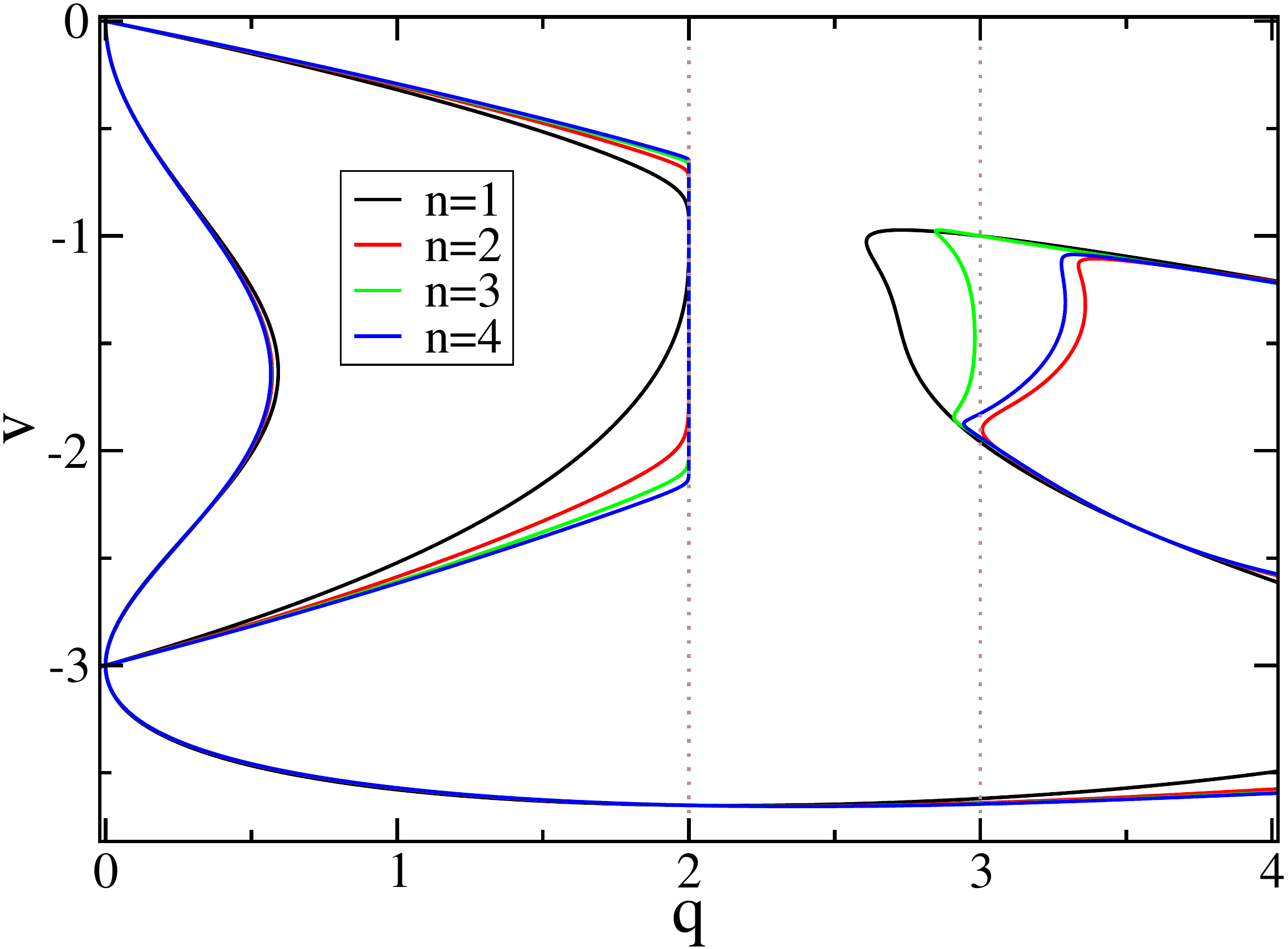}
\caption{Roots of $P_B(q,v)$ for the Potts model on the four-eight medial lattice, using
$n \times n$ square bases.}
\label{fig:foureightmedial-pd}
\end{center}
\end{figure}

The phase diagram for the four-eight medial lattice is rather simple; see Figure~\ref{fig:foureightmedial-pd}.
There is a clear vertical ray at the Beraha number (\ref{Beraha}) with $k=4$; and possibly the $n=3$ polynomial also indicates
that a ray will emerge at $k=6$, although the $n=4$ result fails to confirm this. It seems likely that the BK phase will extend
to the arc near $q \approx 3.3$, although its upper and lower boundaries are not visible on the interval $2 < q < 3$ with these
critical polynomials. In any case, the absence of vertical rays for $k>6$ is a clear sign that $q_{\rm c} < 4$ for this lattice.
We also note that all curves go through the point $(q,v)=(0,-3)$ exactly.

\subsection{Frieze medial lattice ${\cal M}(3^3,4^2)$}

\begin{figure}
\begin{center}
\begin{tikzpicture}[scale=1.0,>=stealth]
\foreach \xpos in {0,1,2,3}
\foreach \ypos in {0,1,2,3}
 \fill[black!20] (\xpos+0.5,\ypos) -- (\xpos+1,\ypos+0.5) -- (\xpos+0.5,\ypos+1) -- (\xpos,\ypos+0.5) -- cycle;

\foreach \xpos in {0,1,2,3}
\foreach \ypos in {0,2}
 \draw[blue,ultra thick] (\xpos,\ypos+0.5) -- (\xpos+1,\ypos+0.5) -- (\xpos+0.5,\ypos+1) -- (\xpos+0.5,\ypos) -- cycle;

\foreach \xpos in {0,1,2,3}
\foreach \ypos in {1,3}
 \draw[blue,ultra thick] (\xpos,\ypos+0.5) -- (\xpos+0.5,\ypos) -- (\xpos+1,\ypos+0.5) -- (\xpos+0.5,\ypos+1) -- cycle;
 
\foreach \xpos in {0,1,2,3}
\foreach \ypos in {0,1,2,3}
 \draw[black] (\xpos,\ypos+0.5) -- (\xpos+0.5,\ypos) -- (\xpos+1,\ypos+0.5) -- (\xpos+0.5,\ypos+1) -- cycle;
 
\draw[very thick,->] (0,-0.5)--(4,-0.5);
\draw (4,-0.5) node[right] {$x$};
\foreach \xpos in {0,1,2,3}
{
 \draw[thick] (\xpos+0.5,-0.6)--(\xpos+0.5,-0.4);
 \draw (\xpos+0.5,-0.5) node[below] {$\xpos$};
}

\draw[very thick,->] (-0.5,0)--(-0.5,4);
\draw (-0.5,4) node[above] {$y$};
\foreach \ypos in {0,1,2,3}
{
 \draw[thick] (-0.6,\ypos+0.5)--(-0.4,\ypos+0.5);
 \draw (-0.5,\ypos+0.5) node[left] {$\ypos$};
}
 
\end{tikzpicture}
 \caption{Four-terminal representation of the frieze medial lattice.}
 \label{fig:friezemedial}
\end{center}
\end{figure}
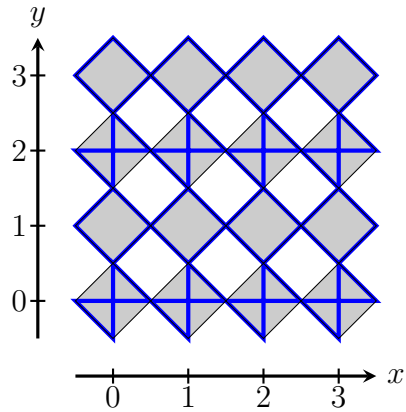

For the frieze medial lattice we can use the four-terminal representation depicted in Figure~\ref{fig:friezemedial}. It consists of alternating rows
of grey squares of the types used in the kagome and square lattices. Therefore, the $\check{\sf R}$-matrix is given by
(\ref{eq:Rkagome}) on even rows, and by (\ref{eq:Rsquare}) on odd rows. It is thus convenient to use rectangular bases of
size $n \times 2n$ grey squares, for any parity of $n$. There are $\frac{5}{2}$ vertices and $5$ edges per grey square.

The corresponding bond percolation thresholds $p_c$ are shown in Table~\ref{tab:friezemedial1}.

\begin{table}
\begin{center}
 \begin{tabular}{l|l}
 $n$ & $p_{\rm c}$ \\ \hline
 1 & 0.51252671239052872734284920277093949452459399814557 \\
 2 & 0.51252555030832881329338864757758450264150947803966 \\
 3 & 0.51252505197721914539499792825297076237457413214236 \\
 4 & 0.51252476509531218613179917728236141170706073572125 \\
 5 & 0.51252466184456998405019250652673909947600021592946 \\
 6 & 0.51252462541470314484904781495993557924808658518665 \\
 7 & 0.51252461131296365065118129973998555670342068841944 \\ \hline
 $\infty$ & 0.5125245984 (9) \\
 \end{tabular}
 \caption{Bond percolation threshold $p_{\rm c}$ on the frieze medial lattice.}
  \label{tab:friezemedial1}
\end{center}
\end{table}

When $q=2$ we obtain as usual a factorisation of $P_B(q,v)$. The maximum degree of the factors
is $d_{\rm max} = 8$ for $n=1,2$, $d_{\rm max} = 16$ for $n=3,4$, and $d_{\rm max} = 32$ for $n=5$.
The factor relevant for determining the critical point simplifies upon setting $v = -1 + \sqrt{y}$ and becomes
\begin{equation}
 1 - 20 y - 10 y^2 - 4 y^3 + y^4 \,.
\end{equation}
Its physically relevant solution has an expression in terms of
cube roots, which is however to lengthy to be reported here.
It corresponds to $v_{\rm c} \simeq 1.479\,990\,057\cdots$ in the original variable.

\begin{table}
\begin{center}
 \begin{tabular}{l|l}
 $n$ & $v_{\rm c}$ \\ \hline
 1 & 1.80713435658088600169174675586705957329793973933928 \\
 2 & 1.80714787637307386116716755890184038031063366743447 \\
 3 & 1.80715275038905311055716653443917611922430053521537 \\
 4 & 1.80715499141679873814200897705704030717174184979977 \\
 5 & 1.80715574557129979201879822669803898126598246764060 \\
 6 & 1.80715600281645714745051778969055570176041887478573 \\
 7 & 1.80715610046331273129256920306307290435340513061371 \\ \hline
 $\infty$ & 1.807156187 (2) \\
 \end{tabular}
 \caption{Critical point $v_{\rm c}$ of the $q=3$ state Potts model on the frieze medial lattice.}
  \label{tab:friezemedial3}
\end{center}
\end{table}

\begin{table}
\begin{center}
 \begin{tabular}{l|l}
 $n$ & $v_{\rm c}$ \\ \hline
 1 & 2.08199719749764750866564040310148994909037162782967 \\
 2 & 2.08202742833616593271859950614838779880367025625026 \\
 3 & 2.08203865409278753166570381108592447912012172754718 \\
 4 & 2.08204360154127867025506347751737245230091088735469 \\
 5 & 2.08204524006200808869976649798749897664563660237247 \\
 6 & 2.08204579575822659373529481034339920792570530958012 \\
 7 & 2.08204600668989993066296166093996927319153647466394 \\ \hline
 $\infty$ & 2.08204619 (3) \\
 \end{tabular}
 \caption{Critical point $v_{\rm c}$ of the $q=4$ state Potts model on the frieze medial lattice.}
  \label{tab:friezemedial4}
\end{center}
\end{table}

The critical points $v_{\rm c}$ for the $q=3$ and $q=4$ state Potts models appear in Tables~\ref{tab:friezemedial3}--\ref{tab:friezemedial4}.

\begin{figure}
\begin{center}
\includegraphics[width=12cm]{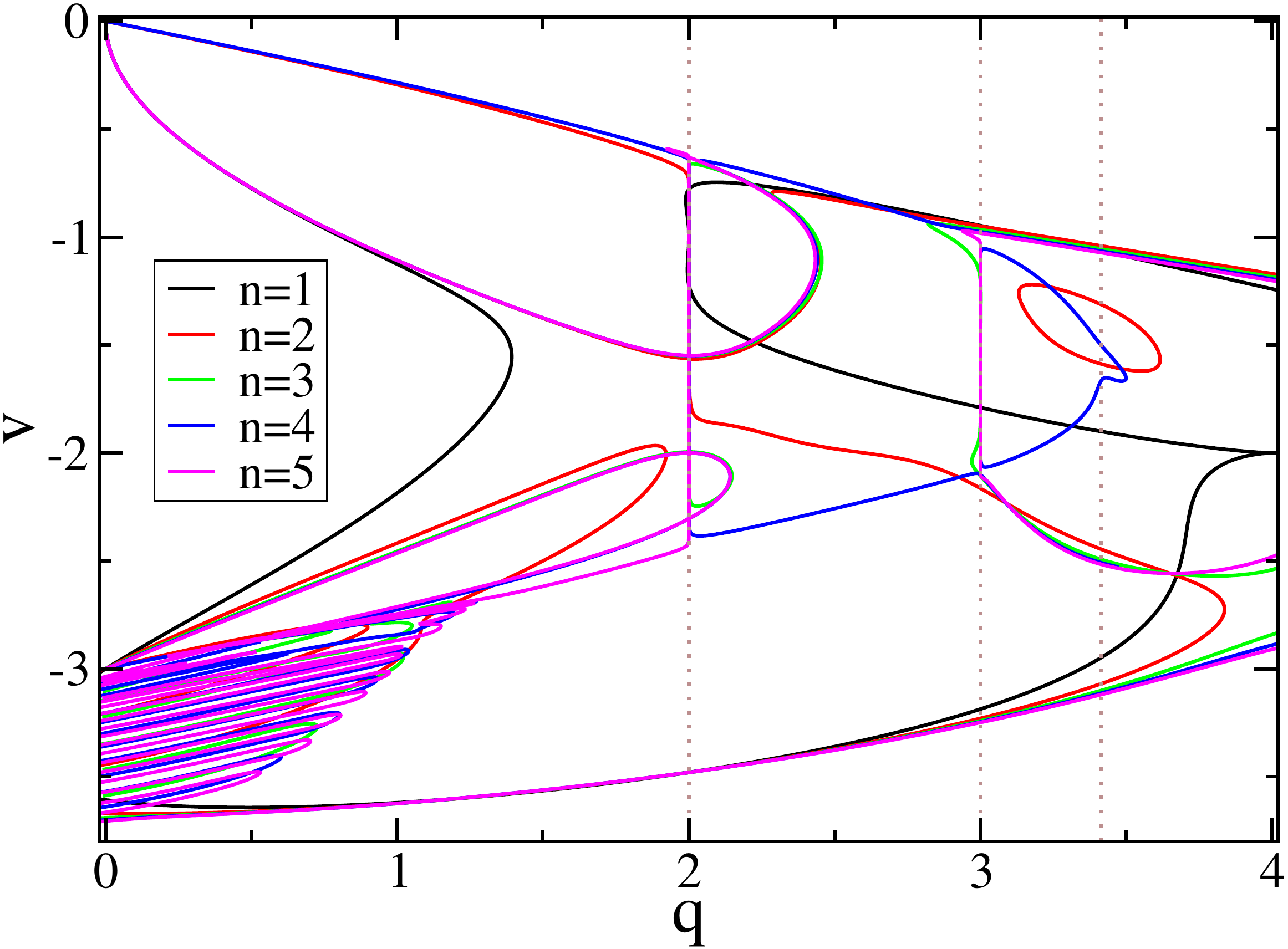}
\caption{Roots of $P_B(q,v)$ for the Potts model on the frieze medial lattice, using
$n \times 2n$ rectangular bases.}
\label{fig:friezemedial-pd}
\end{center}
\end{figure}

The phase diagram of the frieze medial lattice, shown in Figure~\ref{fig:friezemedial-pd}, is exceedingly complicated, and arguably even more
complicated than that of the three-twelve lattice (see Figure~\ref{fig:threetwelve-pd}). The upper limit $v_+(q)$ of the BK phase contains an almost straight
part from the origin to the neighbourhood of $(q,v)=(3,-1)$.  The almost straight continuation to higher $q$ cannot be the upper limit of the BK
phase, though, since it is not adjacent to the characteristic vertical rays. Rather, the continuation of $v_+(q)$ must be provided by the $n=4$
arc that bends around at $(q,v) \approx (3.5,-1.6)$. For $n=2$ this arc is prefigured by a bubble that is not connected to the rest of the curves.
It seems likely that all even $n$ will participate to this part of the phase diagram. Note in particular that the $n=4$ arc has a small wrinkle
at $q \simeq B_8$ which would most likely turn into a vertical ray for higher (even) $n$. One can therefore believe that $q_{\rm c} > B_8$
for this lattice, but it is yet unclear whether $q_{\rm c}$ might be as large as $4$.
After the $n=4$ arc bends around, it traces out the lower limit $v_-(q)$ of the BK phase that continues to the point $(q,v) = (0,-3)$ through which
all curves pass exactly. The extent of the BK phase can be judged from the vertical rays at $q=B_k$ with $k=4,6$ (and maybe $8$ as just mentioned).

Two further branches of the curve come out of $(q,v) = (0,-3)$ and trace out a finger that extends a little further than the vertical ray at $q=2$.
There is a similar, broader finger coming out of $(q,v) = (0,0)$, whose upper side coincides with $v_+(q)$.

Apart from these features, there is almost horizontal branch coming out of $(q,v) \approx (0,-3.7)$ and extending towards large $q$. Similarly,
the bottom and top of the vertical ray at $q=3$ connect to branches that extend towards large $q$.

\begin{figure}
\begin{center}
\includegraphics[width=12cm]{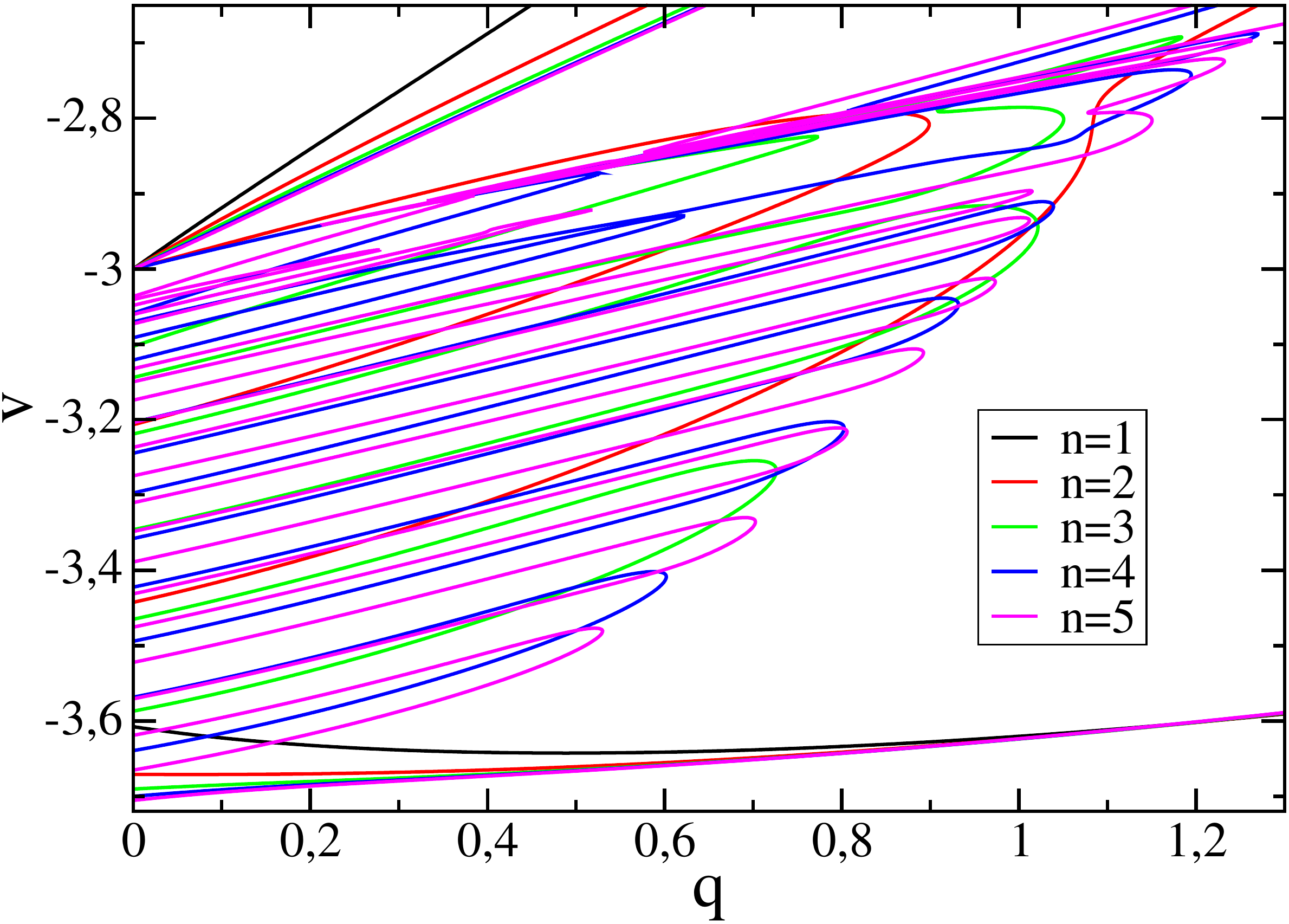}
\caption{Close-up on a region of Figure~\ref{fig:friezemedial-pd}.}
\label{fig:friezemedial-pd-zoom}
\end{center}
\end{figure}

A close-up of the region near $(q,v) = (0,-3)$ is shown in Figure~\ref{fig:friezemedial-pd-zoom}. Throughout this region the curves become
increasingly dense as $n$ increases. Counting the number of fingers emanating from the $v$-axis in the range $-3.7 < v < 3$ gives compelling
evidence for the conjecture that the curves will, in fact, become space-filling in this region when $n \to \infty$. This is a novel feature, not seen
in the phase diagrams for any of the other lattices. It is very reminiscent of a recent study of the phase diagram of the Potts model on a family
of non-planar graphs, called the generalised Petersen graphs \cite{Salas13-1}, where a number of ``critical regions'' (marked by a $\star$
in \cite[Figure~2]{Salas13-1}) were identified throughout which the two dominant eigenvalues of the transfer matrix are exactly
degenerate in norm. Such critical regions also exist in two dimensions, and in particular for the $q$-state Potts model on the triangular lattice
close to the point $(q,v) = (0,-3)$, as well as for $q > 4$ \cite{Salas14}.

This capability of the $P_B(q,v) = 0$ curves to be space-filling might also offer a new interpretation of the thin fingers emanating from
the $v$-axis in many of the preceding figures (see, e.g., Figure~\ref{fig:threetwelve-pd}). Might it be that these fingers will also become
more numerous and tend to fill out space for larger (i.e., not accessible in this paper) values of $n$? We leave this question for
future investigations.

\subsection{Three-twelve medial lattice ${\cal M}(3,12^2)$}
\label{sec:threetwelvemedial}

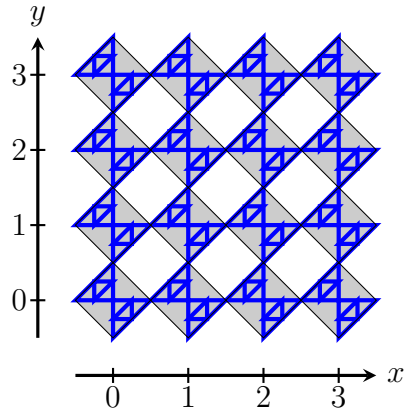
\begin{figure}
\begin{center}
\begin{tikzpicture}[scale=1.0,>=stealth]
\foreach \xpos in {0,1,2,3}
\foreach \ypos in {0,1,2,3}
 \fill[black!20] (\xpos+0.5,\ypos) -- (\xpos+1,\ypos+0.5) -- (\xpos+0.5,\ypos+1) -- (\xpos,\ypos+0.5) -- cycle;

\foreach \xpos in {0,1,2,3}
\foreach \ypos in {0,1,2,3}
{
 \draw[blue,ultra thick] (\xpos,\ypos+0.5) -- (\xpos+1,\ypos+0.5) -- (\xpos+0.5,\ypos) -- (\xpos+0.5,\ypos+1) -- cycle;
 \draw[blue,ultra thick] (\xpos+0.25,\ypos+0.5) -- (\xpos+0.5,\ypos+0.75) -- (\xpos+0.25,\ypos+0.75) -- cycle;
 \draw[blue,ultra thick] (\xpos+0.5,\ypos+0.25) -- (\xpos+0.75,\ypos+0.5) -- (\xpos+0.75,\ypos+0.25) -- cycle;
}

\foreach \xpos in {0,1,2,3}
\foreach \ypos in {0,1,2,3}
 \draw[black] (\xpos+0.5,\ypos) -- (\xpos+1,\ypos+0.5) -- (\xpos+0.5,\ypos+1) -- (\xpos,\ypos+0.5) -- cycle;

\draw[very thick,->] (0,-0.5)--(4,-0.5);
\draw (4,-0.5) node[right] {$x$};
\foreach \xpos in {0,1,2,3}
{
 \draw[thick] (\xpos+0.5,-0.6)--(\xpos+0.5,-0.4);
 \draw (\xpos+0.5,-0.5) node[below] {$\xpos$};
}

\draw[very thick,->] (-0.5,0)--(-0.5,4);
\draw (-0.5,4) node[above] {$y$};
\foreach \ypos in {0,1,2,3}
{
 \draw[thick] (-0.6,\ypos+0.5)--(-0.4,\ypos+0.5);
 \draw (-0.5,\ypos+0.5) node[left] {$\ypos$};
}
 
\end{tikzpicture}
 \caption{Four-terminal representation of the three-twelve medial lattice.}
 \label{fig:threetwelvemedial}
\end{center}
\end{figure}

A four-terminal representation of the three-twelve medial lattice is shown in Figure~\ref{fig:threetwelvemedial}.
This lattice is also known as the $2 \times 2$ kagome subnet \cite{WikiPercThres}, since it can be obtained
by replacing each of the triangles of the kagome lattice by an equilateral made of $2 \times 2 = 4$ triangles. In our representation there are
$9$ vertices and $18$ edges per grey square---the highest numbers for any lattice considered in this paper.
This means that the basis for the $n=7$ numerical computation encompasses $18 \times 7^2 = 882$ edges,
as mentioned in the abstract.

The $\check{\sf R}$-matrix can only be computed in the sparse-matrix
factorisation scheme if one inserts intermediate points. In this case, three such points (or six loop strands) are needed.
Just like in section~\ref{sec:foureightmedial}, it is therefore advantageous to use the 
``generic $\check{\sf R}$-matrix'' version of the algorithm. The fourteen weights can be readily computed separately,
from a transfer matrix that builds a single grey square, using time slices of width five points (or ten loop strands).
Alternatively, this may be done by hand defining a transfer process where time flows in the North-West (rather than the usual
North-East) direction, and using the fact that $\check{\sf R}_i = \tilde{\sf S}_i {\sf S}_i$, where ${\sf S}_i$ denotes the $2 \times 2$ subnet
operator, and $\tilde{\sf S}_i$ is its time-reflected counterpart.
Either way, we obtain the fourteen polynomials defining $\check{\sf R}_i$, 
each of which contains up to a maximum of 30 monomials $x^a n_{\rm loop}^b$.
These are obviously too lengthy to be reproduced here, but are available upon request from the author.

The bond percolation thresholds $p_c$ are shown in Table~\ref{tab:threetwelvemedial1}.
As for the three-twelve lattice itself, we obtain a very substantial improvement on the
precision of the existing results, here by more than four orders of magnitude.

\begin{table}
\begin{center}
 \begin{tabular}{l|l}
 $n$ & $p_{\rm c}$ \\ \hline
 1 & 0.60087024823863130165397946806873749105812018535152 \\
 2 & 0.60086257369370252495888862646970191568249607022723 \\
 3 & 0.60086202873714154724433811970387661781266525679198 \\
 4 & 0.60086197705173039926972062364498505003371758838282 \\
 5 & 0.60086196938791403077982454158989347857185134707268 \\
 6 & 0.60086196771854575664675151979737081972939591265171 \\
 7 & 0.60086196724363425411431529583585404028066985304771 \\ \hline
 $\infty$ & 0.600861966960 (2) \\
 Ref.~\cite{Ding10} & 0.60086193 (3) \\
 \end{tabular}
 \caption{Bond percolation threshold $p_{\rm c}$ on the three-twelve medial lattice.}
  \label{tab:threetwelvemedial1}
\end{center}
\end{table}

For the Ising model ($q=2$) we have the usual factorisation of $P_B(q,v)$. The maximum degree of the factors is
$d_{\rm max} = 8$ for $n=1,2$, $d_{\rm max} = 16$ for $n=3,4$. The relevant factor for determining the critical point
has degree $8$ in the $v$-variable, but changing variables through $v = -1 + \sqrt{y}$ we find the simper polynomial
\begin{equation}
 -83 - 32 y - 6 y^2 - 8 y^3 + y^4 \,.
\end{equation}
The unique root that corresponds to a positive value of $v$ reads
\begin{equation}
 v_{\rm c} = -1 + \left(2 + \sqrt{3} + \sqrt{2 (6 + 5 \sqrt{3})} \right)^{1/2} \simeq 2.024\,382\,957\cdots \,.
\end{equation}
Once again, we expect this to be the exact critical point, although we are not aware of any exact solution of the Ising model
on the three-twelve medial lattice.

\begin{table}
\begin{center}
 \begin{tabular}{l|l}
 $n$ & $v_{\rm c}$ \\ \hline
 1 & 2.4051388771937835004651461060627027673916305236110 \\
 2 & 2.4052102689837864921984641492409181211413650400579 \\
 3 & 2.4052176562543970511035491279473526360116379644946 \\
 4 & 2.4052185619190028113565896890585979366394650085930 \\
 5 & 2.4052187382745471274900934254447806792951661256278 \\
 6 & 2.4052187868761485331493140625659769323839252659120 \\
 7 & 2.4052188036960929537799936461532303199029211244332 \\ \hline
 $\infty$ & 2.40521881719 (7) \\
  \end{tabular}
 \caption{Critical point $v_{\rm c}$ of the $q=3$ state Potts model on the three-twelve medial lattice.}
  \label{tab:threetwelvemedial3}
\end{center}
\end{table}

\begin{table}
\begin{center}
 \begin{tabular}{l|l}
 $n$ & $v_{\rm c}$ \\ \hline
 1 & 2.7176916926829055691824338364721435254533774542481 \\
 2 & 2.7178413379520479881792003331093087495269320730958 \\
 3 & 2.7178581401378053862927765036267716737456525642391 \\
 4 & 2.7178603686953558608313898940710121620288427512401 \\
 5 & 2.7178608441674816937339597619859108606883960648273 \\
 6 & 2.7178609868443710417441879453116940146104031301600 \\
 7 & 2.7178610401487845292293333134497014710797779301550 \\ \hline
 $\infty$ & 2.7178610889 (3) \\
  \end{tabular}
 \caption{Critical point $v_{\rm c}$ of the $q=4$ state Potts model on the three-twelve medial lattice.}
  \label{tab:threetwelvemedial4}
\end{center}
\end{table}

The critical points $v_{\rm c}$ for the $q=3$ and $q=4$ state Potts models are given in Tables~\ref{tab:threetwelvemedial3}--\ref{tab:threetwelvemedial4}.

\begin{figure}
\begin{center}
\includegraphics[width=12cm]{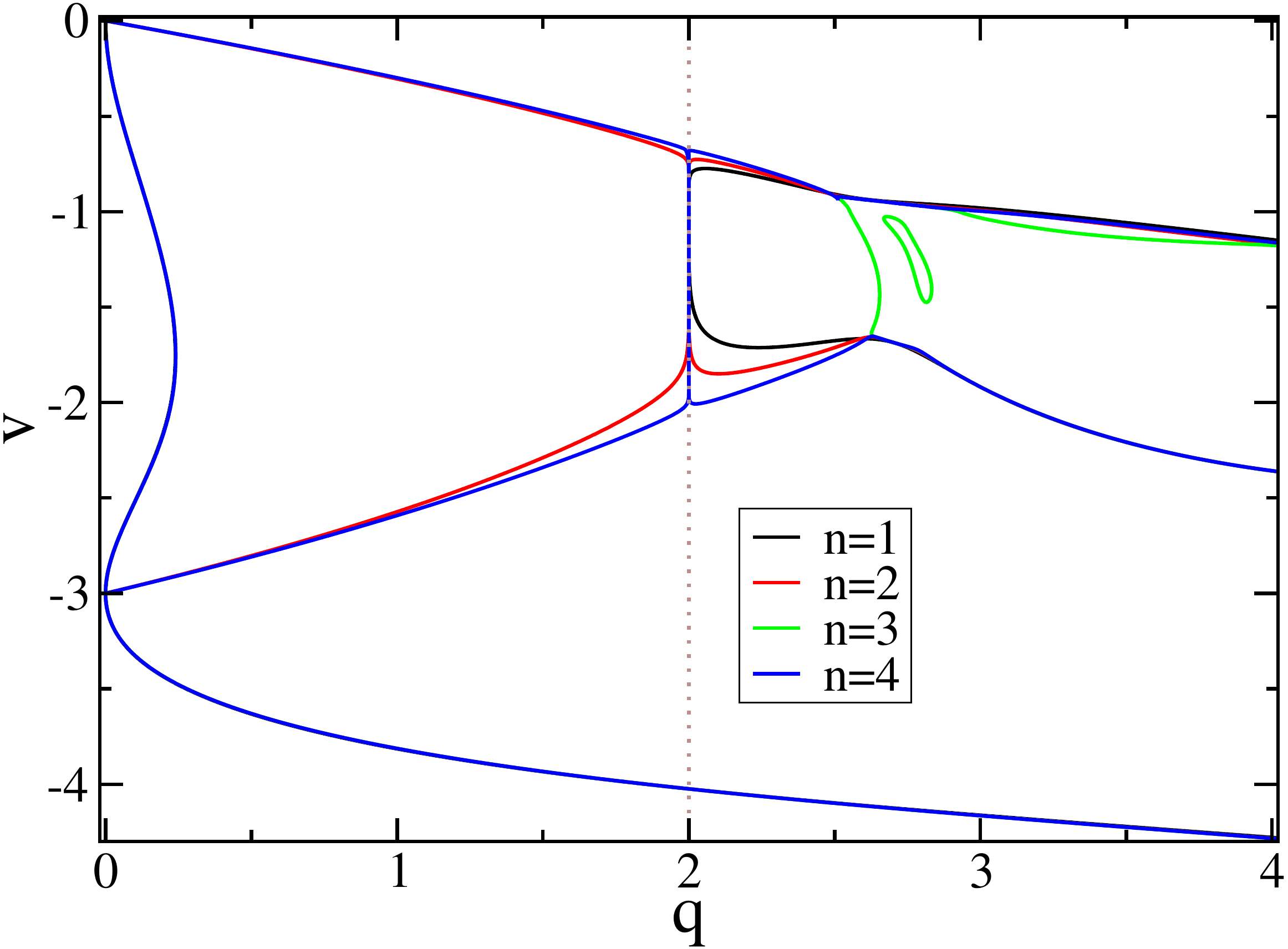}
\caption{Roots of $P_B(q,v)$ for the Potts model on the three-twelve medial lattice, using
$n \times n$ square bases.}
\label{fig:threetwelvemedial-pd}
\end{center}
\end{figure}

Despite the very large size of the bases, the phase diagram of the three-twelve medial lattice (see Figure~\ref{fig:threetwelvemedial-pd})
is not very complicated. The upper and lower boundaries $v_\pm(q)$ of the BK phase are a couple of curves going out of the points
$(q,v) = (0,0)$ and $(0,-3)$, respectively, with finite slopes. They join via an arc at $q \approx 2.6$ which is however only visible in the
$n=3$ result.  Accordingly the only vertical ray is at $q = B_4 = 2$. The role of the elongated bubble at $q \approx 2.7$ in the $n=3$ curve
is not clear and would have to be confirmed at larger sizes. An additional, lowest lying curve goes out of $(q,v) = (0,-3)$ vertically and continues
to large $q$; this curve is outside the BK phase since it does not touch the vertical ray at $q=2$.

Note that the large degree of the polynomials $P_B(q,v)$ makes the computation very memory demanding, so that, at variance with
the general rule, we have not computed the case $n=5$ for this lattice.

\subsection{Cross medial lattice ${\cal M}(4,6,12)$}
\label{sec:crossmedial}

\begin{figure}
\begin{center}
\begin{tikzpicture}[scale=1.0,>=stealth]
\foreach \xpos in {0,1,2,3}
\foreach \ypos in {0,1,2,3}
 \fill[black!20] (\xpos+0.5,\ypos) -- (\xpos+1,\ypos+0.5) -- (\xpos+0.5,\ypos+1) -- (\xpos,\ypos+0.5) -- cycle;

\foreach \xpos in {0,1,2,3}
\foreach \ypos in {0,2}
{
 \draw[blue,ultra thick] (\xpos+0.5,\ypos) -- (\xpos+1,\ypos+0.5) -- (\xpos+0.5,\ypos+1) -- (\xpos,\ypos+0.5) -- cycle;
 \draw[blue,ultra thick] (\xpos+0.25,\ypos+0.25) -- (\xpos+0.75,\ypos+0.25) -- (\xpos+0.75,\ypos+0.75) -- (\xpos+0.25,\ypos+0.75) -- cycle;
}

\foreach \xpos in {0,2}
\foreach \ypos in {1,3}
{
 \draw[blue,ultra thick] (\xpos+0.5,\ypos) -- (\xpos+1,\ypos+0.5) -- (\xpos+0.5,\ypos+1) -- (\xpos,\ypos+0.5) -- cycle;
 \draw[blue,ultra thick] (\xpos+0.25,\ypos+0.25) -- (\xpos+0.75,\ypos+0.25) -- (\xpos+0.75,\ypos+0.75) -- (\xpos+0.25,\ypos+0.75) -- cycle;
 \draw[blue,ultra thick] (\xpos+1,\ypos+0.5) -- (\xpos+1.05,\ypos+0.55);
 \draw[blue,ultra thick] (\xpos+1.45,\ypos+0.95) -- (\xpos+1.5,\ypos+1);
 \draw[blue,ultra thick,decorate,decoration={coil,segment length=1.5mm,amplitude=1.5mm}] (\xpos+1.05,\ypos+0.55) -- (\xpos+1.45,\ypos+0.95);
 \draw[blue,ultra thick] (\xpos+1.5,\ypos) -- (\xpos+1.55,\ypos+0.05);
 \draw[blue,ultra thick] (\xpos+1.95,\ypos+0.45) -- (\xpos+2,\ypos+0.5);
 \draw[blue,ultra thick,decorate,decoration={coil,segment length=1.5mm,amplitude=1.5mm}] (\xpos+1.55,\ypos+0.05) -- (\xpos+1.95,\ypos+0.45);
}

\foreach \xpos in {0,1,2,3}
\foreach \ypos in {0,1,2,3}
 \draw[black] (\xpos+0.5,\ypos) -- (\xpos+1,\ypos+0.5) -- (\xpos+0.5,\ypos+1) -- (\xpos,\ypos+0.5) -- cycle;

\draw[very thick,->] (0,-0.5)--(4,-0.5);
\draw (4,-0.5) node[right] {$x$};
\foreach \xpos in {0,1,2,3}
{
 \draw[thick] (\xpos+0.5,-0.6)--(\xpos+0.5,-0.4);
 \draw (\xpos+0.5,-0.5) node[below] {$\xpos$};
}

\draw[very thick,->] (-0.5,0)--(-0.5,4);
\draw (-0.5,4) node[above] {$y$};
\foreach \ypos in {0,1,2,3}
{
 \draw[thick] (-0.6,\ypos+0.5)--(-0.4,\ypos+0.5);
 \draw (-0.5,\ypos+0.5) node[left] {$\ypos$};
}
 
\end{tikzpicture}
 \caption{Four-terminal representation of the cross medial lattice. The coil-like symbols 
 \begin{tikzpicture}
 \draw[blue,ultra thick] (0,0) -- (0.2,0);
 \draw[blue,ultra thick] (0.8,0) -- (1,0);
 \draw[blue,ultra thick,decorate,decoration={coil,segment length=1.5mm,amplitude=1.5mm}] (0.2,0) -- (0.8,0);
 \end{tikzpicture}
 indicate couplings of infinite strength ($x = \infty$),
 which amount to identifying the corresponding end points.}
 \label{fig:crossmedial}
\end{center}
\end{figure}
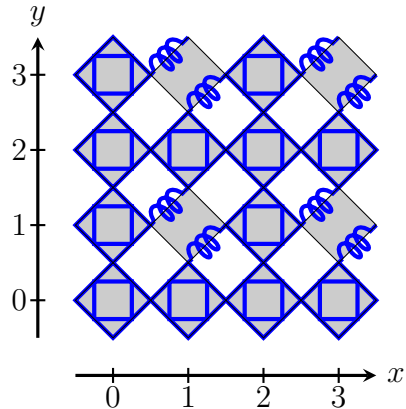

Figure~\ref{fig:crossmedial} shows a four-terminal representation of the cross medial lattice. The $\check{\sf R}$-matrix
is identical to that of the four-eight medial lattice (see section~\ref{sec:foureightmedial}) for grey squares where
at least one of $x$ and $y$ is even. When both $x$ and $y$ are odd the $\check{R}$-matrix is simply the identity.
This corresponds formally to setting the coupling constant to infinity on two of the edges, and is represented
by a coil-like symbol in the figure. We have therefore $\frac{9}{2}$ vertices and $9$ edges per grey square.
Note that this representation is defined only for even $n$.

The reader might want to verify the presence of hexagons and dodecagons in Figure~\ref{fig:crossmedial},
apart from the obvious squares. All of these polygons share each of their edges with a triangle, as they should,
since the underlying $(4,6,12)$-lattice is a cubic graph.

\begin{table}
\begin{center}
 \begin{tabular}{l|l}
 $n$ & $p_{\rm c}$ \\ \hline
 2 & 0.55937724024723794256962108611047506283318610938956 \\
 4 & 0.55932369573589496957440303330493818296148431924351 \\
 6 & 0.55931723978375762066932675944175473744208276838550 \\ \hline
 $\infty$ & 0.5593140 (2) \\
 Ref.~\cite{Ziff-unpub,WikiPercThres} & 0.559315 (1) \\
 \end{tabular}
 \caption{Bond percolation threshold $p_{\rm c}$ on the cross medial lattice.}
  \label{tab:crossmedial1}
\end{center}
\end{table}

The percolation thresholds $p_{\rm c}$ are given in Table~\ref{tab:crossmedial1}.

For the Ising model, the largest degree of the factors is $d_{\rm max} = 16$ for both $n=2$ and $n=4$.
The factor relevant for determining $v_{\rm c}$ simplifies upon setting $v = -1 + \sqrt{y}$ and becomes
\begin{equation}
 -647 - 2192 y - 2700 y^2 - 1952 y^3 - 594 y^4 - 80 y^5 - 28 y^6 + y^8 \,.
\end{equation}
The relevant root cannot be written simply, but reads numerically $v_{\rm c} \simeq 1.726\,376\,028\cdots$.

\begin{table}
\begin{center}
 \begin{tabular}{l|l}
 $n$ & $v_{\rm c}$ \\ \hline
 2 & 2.0661192719652071527256369125640867737169955530184 \\
 4 & 2.0664267060174541103892930833168592835252221310632 \\
 6 & 2.0664638219492513119758793783459064888440440205083 \\ \hline
 $\infty$ & 2.0664824 (5) \\
  \end{tabular}
 \caption{Critical point $v_{\rm c}$ of the $q=3$ state Potts model on the cross medial lattice.}
  \label{tab:crossmedial3}
\end{center}
\end{table}

\begin{table}
\begin{center}
 \begin{tabular}{l|l}
 $n$ & $v_{\rm c}$ \\ \hline
 2 & 2.3476220668229052461072103127486172282397282948400 \\
 4 & 2.3482125007252188243553453531545206790126600180109 \\
 6 & 2.3482844812797068752986152370611930911819486107845 \\ \hline
 $\infty$ & 2.3483209 (7) \\
  \end{tabular}
 \caption{Critical point $v_{\rm c}$ of the $q=4$ state Potts model on the cross medial lattice.}
  \label{tab:crossmedial4}
\end{center}
\end{table}

The critical points for the $q=3$ and $q=4$ state Potts models are displayed in Tables~\ref{tab:crossmedial3}--\ref{tab:crossmedial4}.

\begin{figure}
\begin{center}
\includegraphics[width=12cm]{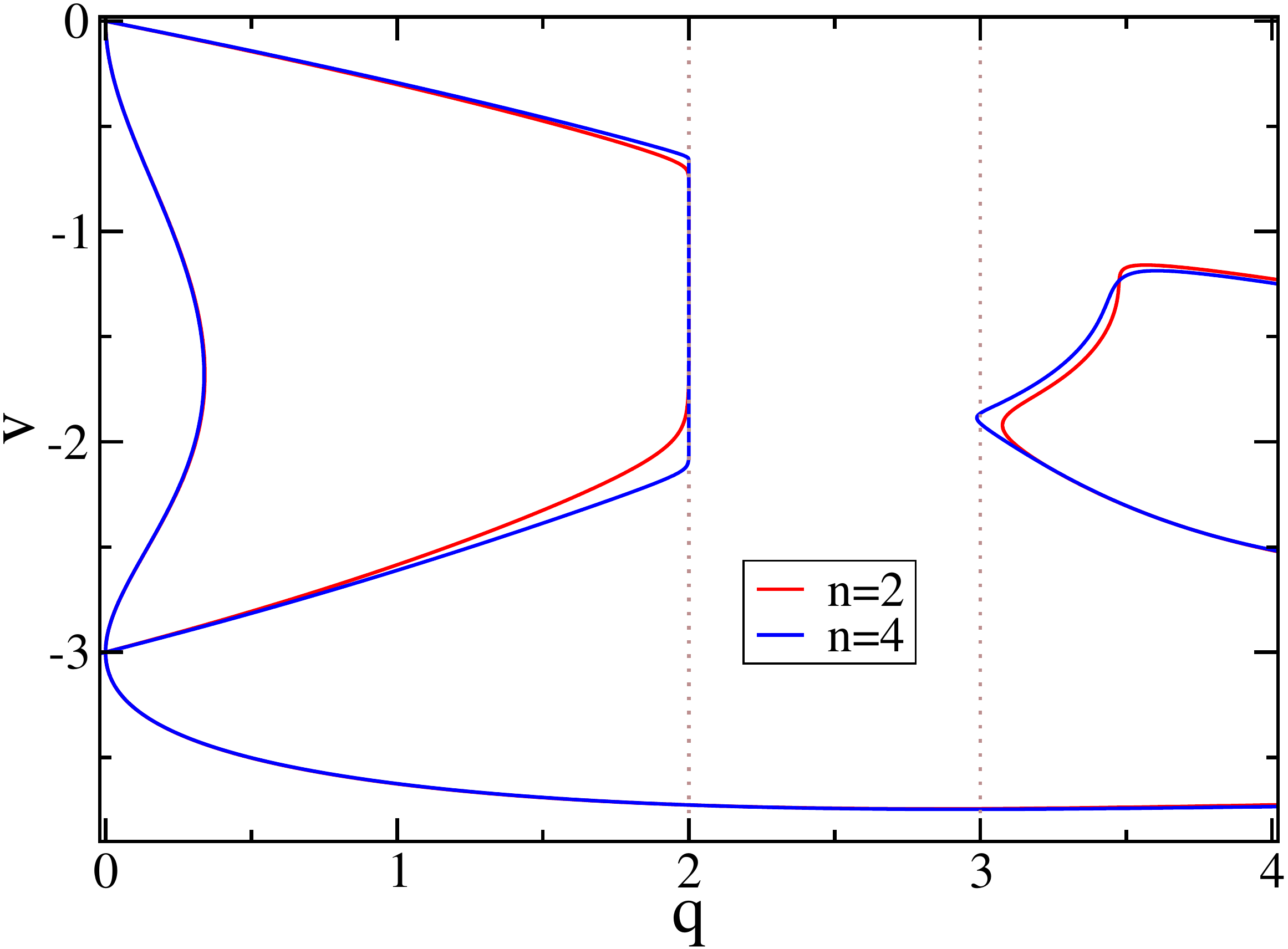}
\caption{Roots of $P_B(q,v)$ for the Potts model on the cross medial lattice, using
$n \times n$ square bases.}
\label{fig:crossmedial-pd}
\end{center}
\end{figure}

The phase diagram of the cross medial lattice, shown in Figure~\ref{fig:crossmedial-pd}, is rather simple.
The upper and lower boundaries $v_\pm(q)$ of the BK phase go out of the points $(q,v) = (0,0)$ and
$(0,-3)$ with finite slope. Those curves do not continue beyond the vertical ray
at $q = B_4 = 2$, but from the experience with other lattices we can safely assume that this is an
artefact of our choice of bases. In particular, since the $q=2$ ray has finite length, the BK phase must
extend further to the right. Presumably it ends at the arc extending from $q \approx 3.0$ to $q \approx 3.4$.
If so, we would expect a further vertical ray at $q = B_6 = 3$ to build up for larger bases (note that the
lower end point of the arc in the $n=4$ curve is conspicuously close to $q=3$).
Finally, there is another, lowest lying curve going out of $(q,v) = (0,-3)$ vertically which is outside the BK phase.

\subsection{Snub square medial lattice ${\cal M}(3^2,4,3,4)$}
\label{sec:snubsquaremedial}

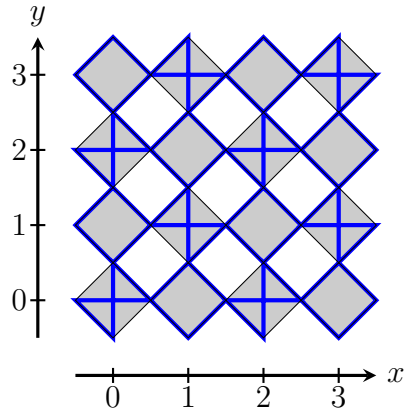
\begin{figure}
\begin{center}
\begin{tikzpicture}[scale=1.0,>=stealth]
\foreach \xpos in {0,1,2,3}
\foreach \ypos in {0,1,2,3}
 \fill[black!20] (\xpos+0.5,\ypos) -- (\xpos+1,\ypos+0.5) -- (\xpos+0.5,\ypos+1) -- (\xpos,\ypos+0.5) -- cycle;

\foreach \xpos in {0,2}
\foreach \ypos in {0,2}
{
 \draw[blue,ultra thick] (\xpos,\ypos+0.5) -- (\xpos+1,\ypos+0.5) -- (\xpos+0.5,\ypos+1) -- (\xpos+0.5,\ypos) -- cycle;
 \draw[blue,ultra thick] (\xpos+1,\ypos+1.5) -- (\xpos+2,\ypos+1.5) -- (\xpos+1.5,\ypos+1) -- (\xpos+1.5,\ypos+2) -- cycle;
 \draw[blue,ultra thick] (\xpos+1,\ypos+0.5) -- (\xpos+1.5,\ypos) -- (\xpos+2,\ypos+0.5) -- (\xpos+1.5,\ypos+1) -- cycle;
 \draw[blue,ultra thick] (\xpos,\ypos+1.5) -- (\xpos+0.5,\ypos+1) -- (\xpos+1,\ypos+1.5) -- (\xpos+0.5,\ypos+2) -- cycle;
}

\foreach \xpos in {0,1,2,3}
\foreach \ypos in {0,1,2,3}
 \draw[black] (\xpos+0.5,\ypos) -- (\xpos+1,\ypos+0.5) -- (\xpos+0.5,\ypos+1) -- (\xpos,\ypos+0.5) -- cycle;

\draw[very thick,->] (0,-0.5)--(4,-0.5);
\draw (4,-0.5) node[right] {$x$};
\foreach \xpos in {0,1,2,3}
{
 \draw[thick] (\xpos+0.5,-0.6)--(\xpos+0.5,-0.4);
 \draw (\xpos+0.5,-0.5) node[below] {$\xpos$};
}

\draw[very thick,->] (-0.5,0)--(-0.5,4);
\draw (-0.5,4) node[above] {$y$};
\foreach \ypos in {0,1,2,3}
{
 \draw[thick] (-0.6,\ypos+0.5)--(-0.4,\ypos+0.5);
 \draw (-0.5,\ypos+0.5) node[left] {$\ypos$};
}
 
\end{tikzpicture}
 \caption{Four-terminal representation of the snub square medial lattice.}
 \label{fig:snubsquaremedial}
\end{center}
\end{figure}

A four-terminal representation of the snub square medial lattice is shown in Figure~\ref{fig:snubsquaremedial}.
Its $\check{\sf R}$-matrix is a mixture of known ingredients: It is given by $\check{\sf R}_i$ of the square lattice,
eq.~(\ref{eq:Rsquare}), when $x+y$ is odd; by that of the kagome lattice, eq.~(\ref{eq:Rkagome}), when $x$ and $y$
are both even; and by the alternative kagome representation discussed in section~\ref{sec:foureightmedial} and
depicted in Figure~\ref{fig:kagome_alt} when $x$ and $y$ are both odd. This representation contains $\frac52$ vertices
and $5$ edges per grey square. Once again we must require $n$ to be even.

The approximate percolation thresholds $p_{\rm c}$ obtained from the positive root of $P_B(p)$ are
displayed in Table~\ref{tab:snubsquaremedial1}.

\begin{table}
\begin{center}
 \begin{tabular}{l|l}
 $n$ & $p_{\rm c}$ \\ \hline
 2 & 0.51268177007416705104691620130181855898975396855381 \\
 4 & 0.51268265882717082159880177253322350153155873133318 \\
 6 & 0.51268281383012303050037916209722508801359877315417 \\ \hline
 $\infty$ & 0.512682929 (8) \\
 \end{tabular}
 \caption{Bond percolation threshold $p_{\rm c}$ on the snub square medial lattice.}
  \label{tab:snubsquaremedial1}
\end{center}
\end{table}

The polynomials $P_B(q,v)$ for the Ising model ($q=2$) factorise, and the maximum degree of the factors
is $d_{\rm min} = 16$ for $n=2,4$. However, if we perform the change of variables $v = -1 + \sqrt{y}$,
the polynomial determining $y$ is only of degree $8$:
\begin{equation}
 1 - 72 y - 304 y^2 - 320 y^3 - 226 y^4 - 88 y^5 - 16 y^6 + y^8 \,.
\end{equation}
The largest positive root in $y$ corresponds to the unique positive root in $v$, which is
$v_{\rm c} \simeq 1.480\,593\,024\cdots$.

\begin{table}
\begin{center}
 \begin{tabular}{l|l}
 $n$ & $v_{\rm c}$ \\ \hline
 2 & 1.8076912391671183952391787449901823591647820014398 \\
 4 & 1.8076887583089533312504288196364050004751764498759 \\
 6 & 1.8076880495316679240666496789871554448835168665187 \\ \hline
 $\infty$ & 1.80768699 (4) \\
  \end{tabular}
 \caption{Critical point $v_{\rm c}$ of the $q=3$ state Potts model on the snub square medial lattice.}
  \label{tab:snubsquaremedial3}
\end{center}
\end{table}

\begin{table}
\begin{center}
 \begin{tabular}{l|l}
 $n$ & $v_{\rm c}$ \\ \hline
 2 & 2.0825184394311798898958915154426893632024045546203 \\
 4 & 2.0825149857011588692125250315589456916982608757655 \\
 6 & 2.0825137172947986829507608673283764173966544463916 \\ \hline
 $\infty$ & 2.0825105 (3) \\
  \end{tabular}
 \caption{Critical point $v_{\rm c}$ of the $q=4$ state Potts model on the snub square medial lattice.}
  \label{tab:snubsquaremedial4}
\end{center}
\end{table}

Critical points of the $q=3$ and $q=4$ state Potts models are given in Tables~\ref{tab:snubsquaremedial3}--\ref{tab:snubsquaremedial4}.

\begin{figure}
\begin{center}
\includegraphics[width=12cm]{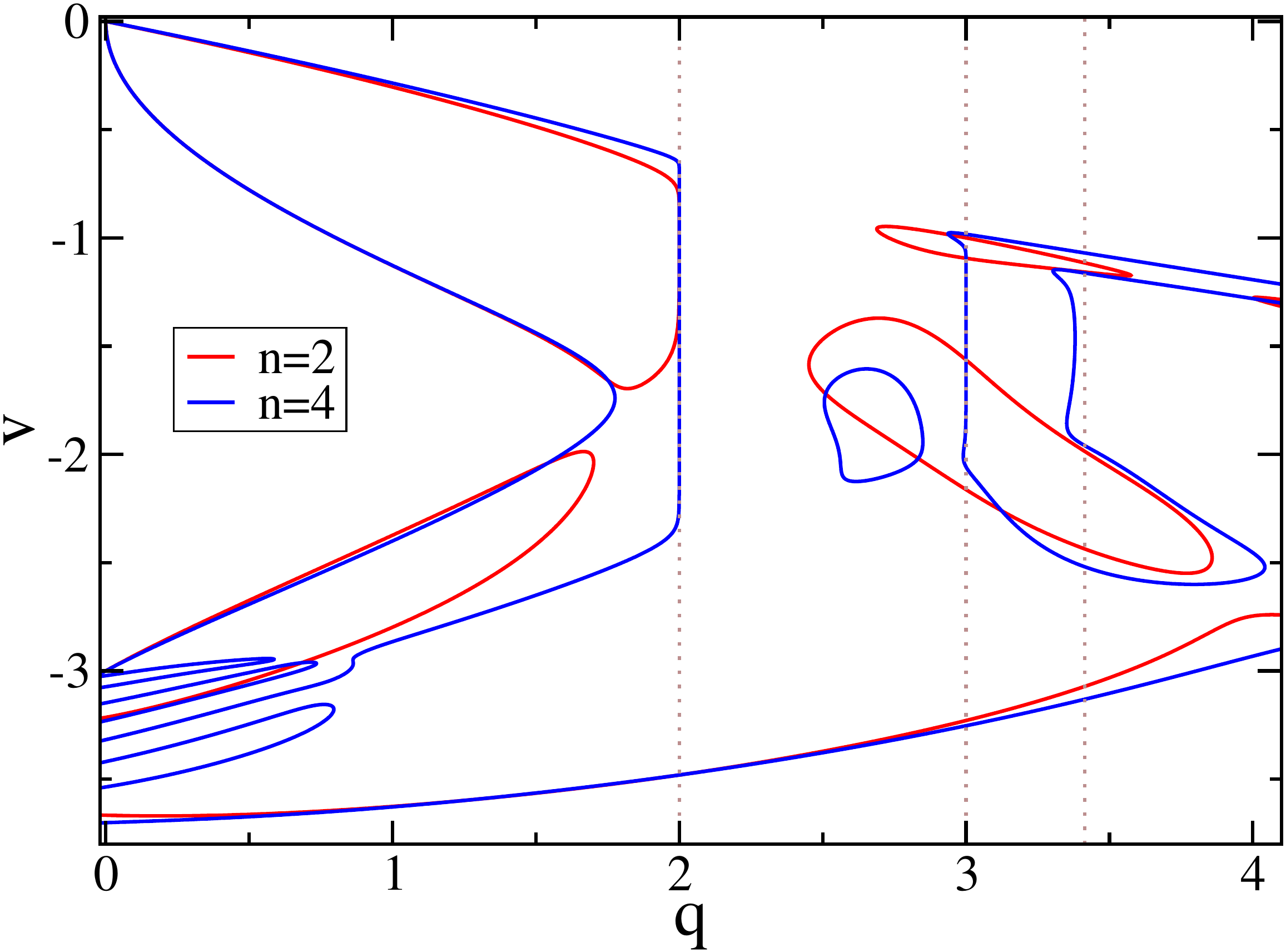}
\caption{Roots of $P_B(q,v)$ for the Potts model on the snub square medial lattice, using
$n \times n$ square bases.}
\label{fig:snubsquaremedial-pd}
\end{center}
\end{figure}

The phase diagram of the snub square medial lattice is shown in Figure~\ref{fig:snubsquaremedial-pd}.
There are vertical rays at $q=B_k$ with $k=4,6,8$. The gap in the rendering of the boundaries $v_\pm(q)$
of the BK phase for $2 < q < 3$ is partly filled out by the bubble appearing in the $n=4$ curve.
The curve coming out of $(q,v)=(0,0)$ with infinite slope bends around and goes through the point
$(q,v)=(-3,0)$. Just below the latter point we notice the formation of several narrow fingers which might
well tend to be space filling at larger $n$, in analogy with what was observed for the frieze medial
lattice (see Figure~\ref{fig:friezemedial-pd-zoom}). Finally, in the bottom of the phase diagram there
is an almost-horizontal curve that lies below the BK phase and extends out to large $q$.

\subsection{Snub hexagonal medial lattice ${\cal M}(3^4,6)$}
\label{sec:snubhexmedial}

Recall from section~\ref{sec:snubhex} that for the snub hexagonal lattice we have computed $P_B(q,v)$ using two different methods.
Also for the corresponding medial lattice shall we present two distinct constructions.

The first construction proceeds in analogy with that of the snub hexagonal lattice itself. Just like the latter
was obtained in Section~\ref{sec:snubhex} as a depleted version of the triangular lattice, we can construct its medial by depleting the kagome (medial of the
triangular) lattice. The ${\sf H}$ and ${\sf V}$ operators corresponding to erased edges are
replaced by just one of their two terms (the identity ${\sf Id}$ or the Temperley-Lieb generator ${\sf E}$, as the case may be).
We thus obtain $\frac{15}{7}$ vertices and $\frac{30}{7}$ edges per grey square.

In formal terms, the $\check{\sf R}$-matrix reads:
\begin{equation}
 \check{\sf R} = \left \lbrace
 \begin{array}{ll}
  {\sf H}_{i+1} {\sf V}_{i+2} {\sf V}_{i} {\sf E}_{i+1} {\sf V}_{i+2} {\sf V}_i {\sf H}_{i+1} & \mbox{for $3 y + x = 0,3,6$ mod 7} \\
  {\sf Id}_{i+1} {\sf V}_{i+2} {\sf V}_{i} {\sf E}_{i+1} {\sf V}_{i+2} {\sf V}_i {\sf H}_{i+1} & \mbox{for $3 y + x = 1$ mod 7} \\
  {\sf E}_{i+1} {\sf V}_{i+2} {\sf E}_i {\sf E}_{i+1} & \mbox{for $3y + x = 2$ mod 7} \\
  {\sf E}_{i+1} {\sf E}_{i+2} {\sf V}_i {\sf E}_{i+1} & \mbox{for $3y + x = 4$ mod 7} \\
  {\sf H}_{i+1} {\sf V}_{i+2} {\sf V}_{i} {\sf E}_{i+1} {\sf V}_{i+2} {\sf V}_i {\sf Id}_{i+1} & \mbox{for $3 y + x = 5$ mod 7} \\
 \end{array} \right.
 \label{snubhexmedial-R}
\end{equation}

The first line in (\ref{snubhexmedial-R}) corresponds to undepleted bow tie patterns
in the grey squares, i.e., it coincides with (\ref{eq:Rkagome}). Subsequent lines are obtained from the first one
by depletion, i.e., replacing some of the ${\sf H}$ operators by the identity ${\sf Id}$, and some of the
${\sf V}$ operators by the Temperley-Lieb generator ${\sf E}$.

As with the snub hexagonal lattice itself (see section~\ref{sec:snubhex}), the representation
(\ref{snubhex-R}) only makes sense when $n$ is a multiple of 7.
So the only computation that we can perform in practice is to find numerically the roots in $v$ of $P_B(q,v)=0$
with $n=7$. We therefore turn now to an alternative construction.

\begin{figure}
\begin{center}
\begin{tikzpicture}[scale=1.0,>=stealth]
\foreach \xpos in {0,1,2,3,4,5}
\foreach \ypos in {0,1,2,3,4,5}
 \fill[black!20] (\xpos+0.5,\ypos) -- (\xpos+1,\ypos+0.5) -- (\xpos+0.5,\ypos+1) -- (\xpos,\ypos+0.5) -- cycle;

\foreach \xpos in {0,3}
{
 \draw[blue,ultra thick] (\xpos,0.5) -- (\xpos+1,0.5) -- (\xpos+0.5,1) -- (\xpos+0.5,0) -- cycle;
 \draw[blue,ultra thick] (\xpos+0.5,1) -- (\xpos+1,1.5) -- (\xpos+0.5,2);
 \draw[blue,ultra thick] (\xpos,2.5) -- (\xpos+1,2.5) -- (\xpos+0.5,2) -- (\xpos+0.5,3) -- cycle;
 \draw[blue,ultra thick] (\xpos+0.5,3) -- (\xpos,3.5) -- (\xpos+0.5,4);
 \draw[blue,ultra thick] (\xpos,4.5) -- (\xpos+1,4.5) -- (\xpos+0.5,5) -- (\xpos+0.5,4) -- cycle;
 \draw[blue,ultra thick] (\xpos,5.5) -- (\xpos+1,5.5) -- (\xpos+0.5,6) -- (\xpos+0.5,5) -- cycle;

 \draw[blue,ultra thick] (\xpos+1,0.5) -- (\xpos+2,0.5) -- (\xpos+1.5,1) -- (\xpos+1.5,0) -- cycle;
 \draw[blue,ultra thick] (\xpos+1,1.5) -- (\xpos+2,1.5) -- (\xpos+1.5,2) -- (\xpos+1.5,1) -- cycle;
 \draw[blue,ultra thick] (\xpos+1,2.5) -- (\xpos+2,2.5) -- (\xpos+1.5,3) -- (\xpos+1.5,2) -- cycle;
 \draw[blue,ultra thick] (\xpos+1.5,3) -- (\xpos+2,3.5) -- (\xpos+1.5,4);
 \draw[blue,ultra thick] (\xpos+1,4.5) -- (\xpos+2,4.5) -- (\xpos+1.5,4) -- (\xpos+1.5,5) -- cycle;
 \draw[blue,ultra thick] (\xpos+1.5,5) -- (\xpos+1,5.5) -- (\xpos+1.5,6);

 \draw[blue,ultra thick] (\xpos+2,0.5) -- (\xpos+3,0.5) -- (\xpos+2.5,0) -- (\xpos+2.5,1) -- cycle;
 \draw[blue,ultra thick] (\xpos+2.5,1) -- (\xpos+2,1.5) -- (\xpos+2.5,2);
 \draw[blue,ultra thick] (\xpos+2,2.5) -- (\xpos+3,2.5) -- (\xpos+2.5,3) -- (\xpos+2.5,2) -- cycle;
 \draw[blue,ultra thick] (\xpos+2,3.5) -- (\xpos+3,3.5) -- (\xpos+2.5,4) -- (\xpos+2.5,3) -- cycle;
 \draw[blue,ultra thick] (\xpos+2,4.5) -- (\xpos+3,4.5) -- (\xpos+2.5,5) -- (\xpos+2.5,4) -- cycle;
 \draw[blue,ultra thick] (\xpos+2.5,5) -- (\xpos+3,5.5) -- (\xpos+2.5,6);
 
 \draw[blue,ultra thick] (\xpos,1) -- (\xpos+0.5,1);
 \draw[blue,ultra thick] (\xpos+2.5,1) -- (\xpos+3,1);
 \draw[blue,ultra thick] (\xpos,2) -- (\xpos+0.5,2);
 \draw[blue,ultra thick] (\xpos+2.5,2) -- (\xpos+3,2);
 \draw[blue,ultra thick] (\xpos+0.5,3) -- (\xpos+1.5,3);
 \draw[blue,ultra thick] (\xpos+0.5,4) -- (\xpos+1.5,4);
 \draw[blue,ultra thick] (\xpos+1.5,5) -- (\xpos+2.5,5);
 \draw[blue,ultra thick] (\xpos+1.5,6) -- (\xpos+2.5,6);
}

\foreach \xpos in {0,1,2,3,4,5}
\foreach \ypos in {0,1,2,3,4,5}
 \draw[black] (\xpos+0.5,\ypos) -- (\xpos+1,\ypos+0.5) -- (\xpos+0.5,\ypos+1) -- (\xpos,\ypos+0.5) -- cycle;

\draw[very thick,->] (0,-0.5)--(6,-0.5);
\draw (6,-0.5) node[right] {$x$};
\foreach \xpos in {0,1,2,3,4,5}
{
 \draw[thick] (\xpos+0.5,-0.6)--(\xpos+0.5,-0.4);
 \draw (\xpos+0.5,-0.5) node[below] {$\xpos$};
}

\draw[very thick,->] (-0.5,0)--(-0.5,6);
\draw (-0.5,6) node[above] {$y$};
\foreach \ypos in {0,1,2,3,4,5}
{
 \draw[thick] (-0.6,\ypos+0.5)--(-0.4,\ypos+0.5);
 \draw (-0.5,\ypos+0.5) node[left] {$\ypos$};
}

\end{tikzpicture}
 \caption{Four-terminal representation of the snub hexagonal medial lattice.}
 \label{fig:snubhexmedial}
\end{center}
\end{figure}
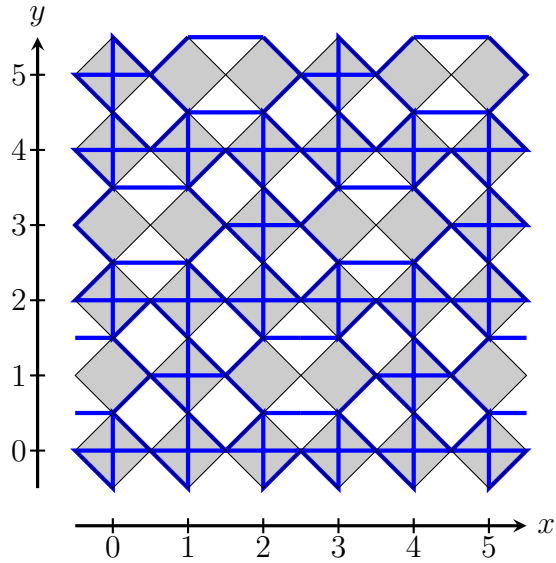

This second construction is based on the four-terminal representation shown in Figure~\ref{fig:snubhexmedial}.
The periodicity of the tiling is $3$ horizontally and $6$ vertically, so we can use it with rectangular bases
of size $n \times 2n$ provided that $n = 0 \mbox{ mod } 3$. Like the snub square medial lattice (see
Figure~\ref{fig:snubsquaremedial}) it uses both the kagome bow tie and its rotated counterpart.
More precisely, for even $y$ the $\check{\sf R}$-matrix is given by (\ref{eq:Rkagome}) when $x+y=0,1 \mbox{ mod } 3$,
and is that of the alternative kagome representation (see Figure~\ref{fig:kagome_alt}) when $x+y=2 \mbox{ mod } 3$.
For odd $y$ we have
\begin{equation}
 \check{\sf R}_i = \left \lbrace
  \begin{array}{ll}
  {\sf E}_{i+2} {\sf V}_i {\sf H}_{i+1} & \mbox{for $x+y = 0$ mod 3} \\
  {\sf H}_{i+1} {\sf V}_{i+2} {\sf E}_i & \mbox{for $x+y = 1$ mod 3} \\
  \mbox{Eq.~(\ref{eq:Rkagome})}   & \mbox{for $x+y = 2$ mod 3} \\
 \end{array} \right.
\end{equation}
Moreover, one must place a horizontal diagonal on the white squares having coordinates $(x+\frac12,y+\frac12)$ whenever
$x + 2 \lfloor y/2 \rfloor = 2 \mbox{ mod } 3$. The reader should check that each hexagon (which contains two of the horizontal diagonals)
shares each of its edges with a pentagon and each of its vertices with a triangle.

This construction provides $\frac52$ vertices and $5$ edges per grey square. The packing density has therefore been improved
with respect to the first construction. However, due to the parity constraint on $n$ we can now only attain $n=6$, instead of $n=7$.

\begin{table}
\begin{center}
 \begin{tabular}{l|l}
 $n$ & $p_{\rm c}$ \\ \hline
 3 & 0.52475737762284412234924602775712228115650254824673 \\
 6 & 0.52475180898647697016696829398686817893243747112660 \\ \hline
 7 & 0.52475071084639438962284356792715186350322362744063 \\ \hline
 $\infty$ & 0.5247495 (5) \\
 \end{tabular}
 \caption{Bond percolation threshold $p_{\rm c}$ on the snub hexagonal medial lattice.}
 \label{tab:snubhexmedial1}
\end{center}
\end{table}

In Table~\ref{tab:snubhexmedial1} we show the percolation thresholds $p_{\rm c}$ using both the first and the second constructions.
We have separated the results using the two different constructions by a horizontal line in this and the following two tables.
Extrapolating the results with $n=3,6$ leads to the estimate for the $n \to \infty$ limit shown in the last line of Table~\ref{tab:snubhexmedial1}. We have here supposed that the parameter $w$ in (\ref{extrapol_w}) is in the same range, up to
a confidence interval of $\pm 0.2$, as that found for the closely related snub {\em square} medial lattice (see section~\ref{sec:snubsquaremedial}). The distance of the $n=7$ result to the final value is consistent with our scaling analysis.

In the case of the Ising model ($q=2$) the polynomials $P_B(q,v)$ with $n=3$ factorises. The degrees of the largest factors
are $1, 4, 6, 12$ and $d_{\rm max} = 44$. We note that the value of $d_{\rm max}$ is unusually large.
The factor leading to a positive real root is the one of degree $4$. It simplifies upon setting $v = -1 + \sqrt{y}$, becoming
\begin{equation}
 -3 - 6y + y^2 \,.
 \label{snubhexmedialfactor}
\end{equation}
Our conjecture for the critical point is thus
\begin{equation}
 v = -1 + \sqrt{3 + 2 \sqrt{3}} \simeq 1.542\,459\,756\cdots
 \label{snubhexmedialroot}
\end{equation}
and this is seen to coincide with the result (\ref{ruby_root}) for the ruby lattice.
By means of a $50$-digit numerical computation we have verified that the
number (\ref{snubhexmedialroot}) is also a root of the $n=6$ polynomial,
and of the $n=7$ polynomial arising from the first construction.
This provides compelling evidence that, on one hand, the factor (\ref{snubhexmedialfactor})
is recurrent for all of the critical polynomials, and, on the other hand, that our two different
constructions lead to consistent results.

\begin{table}
\begin{center}
 \begin{tabular}{l|l}
 $n$ & $v_{\rm c}$ \\ \hline
 3 & 1.87441224209231315831992290889569424963206638843069 \\
 6 & 1.87444613270433948971046627661542407584315443408254 \\ \hline
 7 & 1.87445210254248307019647275740812225826172420927313 \\ \hline
 $\infty$ & 1.874472 (5) \\
  \end{tabular}
 \caption{Critical point $v_{\rm c}$ of the $q=3$ state Potts model on the snub hexagonal medial lattice.}
  \label{tab:snubhexmedial3}
\end{center}
\end{table}

\begin{table}
\begin{center}
 \begin{tabular}{l|l}
 $n$ & $v_{\rm c}$ \\ \hline
 3 & 2.15207135493272709295185607677925040508589248866351 \\
 6 & 2.15213853284896945377792212011213016797276961979614 \\ \hline
 7 & 2.15214995613273843394699263476921693525626146414824 \\ \hline
 $\infty$ & 2.15219 (1) \\
  \end{tabular}
 \caption{Critical point $v_{\rm c}$ of the $q=4$ state Potts model on the snub hexagonal medial lattice.}
  \label{tab:snubhexmedial4}
\end{center}
\end{table}

The results for the critical point $v_{\rm c}$ in the $q=3$ and $q=4$ state Potts models
are shown in Tables~\ref{tab:snubhexmedial3}--\ref{tab:snubhexmedial4}.
Also in those cases have we based the $n \to \infty$ extrapolations on the results
coming from the second construction (with $n=3,6$), and the approach is similar to that
described above for $q=1$. Again, the $n=7$ results confirm our scaling analysis.

\begin{figure}
\begin{center}
\includegraphics[width=12cm]{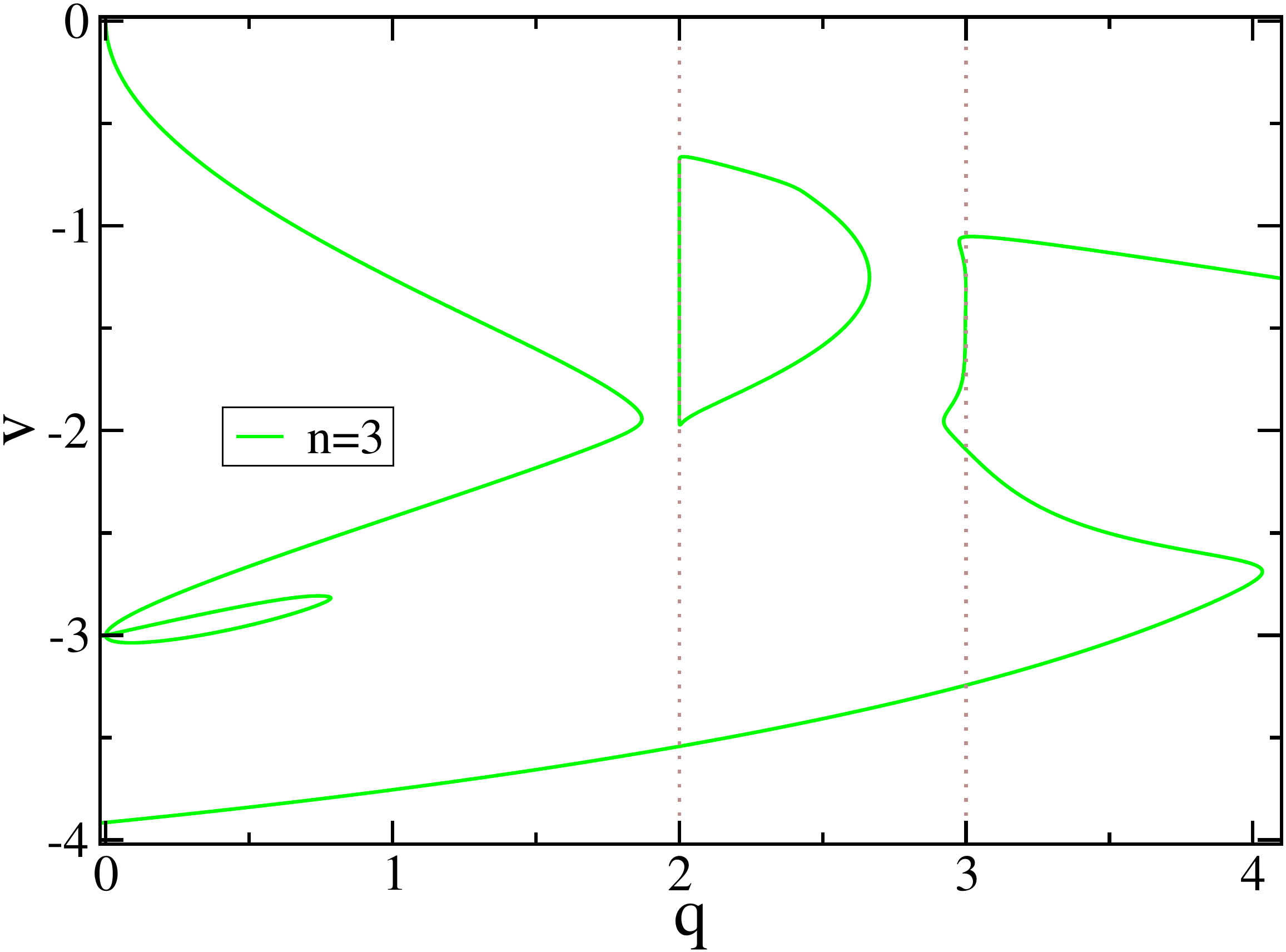}
\caption{Roots of $P_B(q,v)$ for the Potts model on the snub hexagonal medial lattice, using
$n \times 2n$ rectangular bases.}
\label{fig:snubhexmedial-pd}
\end{center}
\end{figure}

The phase diagram of the snub hexagonal medial lattice is shown in Figure~\ref{fig:snubhexmedial-pd}.
Unfortunately this is based on the single size $n=3$, so we cannot make very firm statements about the
convergence properties. The usual vertical rays are visible at $q=B_k$ with $k=4,6$. The rightmost arc
of the bubble containing the $q=2$ vertical ray may well be a precursor of the $v_\pm(q)$ curves on the
interval $2 < q < 3$. On the other hand, the part of $v_+(q)$ with $0 < q < 2$ is clearly missing with this
choice of basis. There is a narrow finger close to the point $(q,v)=(0,-3)$ through which the curve
passes twice.

\subsection{Ruby medial lattice ${\cal M}(3,4,6,4)$}

\begin{figure}
\begin{center}
\begin{tikzpicture}[scale=1.0,>=stealth]
\foreach \xpos in {0,1,2,3}
\foreach \ypos in {0,1,2,3}
 \fill[black!20] (\xpos+0.5,\ypos) -- (\xpos+1,\ypos+0.5) -- (\xpos+0.5,\ypos+1) -- (\xpos,\ypos+0.5) -- cycle;

\draw[blue,ultra thick] (0,2.5) -- (1.5,4);
\draw[blue,ultra thick] (0,1.5) -- (2.5,4);
\draw[blue,ultra thick] (0,0.5) -- (0.5,1);
\draw[blue,ultra thick] (0,0.5) -- (0.5,1);
\draw[blue,ultra thick] (2,2.5) -- (2.5,3);
\draw[blue,ultra thick] (0.5,0) -- (1,0.5);
\draw[blue,ultra thick] (2.5,2) -- (3,2.5);
\draw[blue,ultra thick] (1.5,0) -- (4,2.5);
\draw[blue,ultra thick] (2.5,0) -- (4,1.5);

\draw[blue,ultra thick] (0,0.5) -- (0.5,0);
\draw[blue,ultra thick] (0,1.5) -- (1.5,0);
\draw[blue,ultra thick] (0,2.5) -- (0.5,2);
\draw[blue,ultra thick] (2,0.5) -- (2.5,0);
\draw[blue,ultra thick] (0.5,3) -- (1,2.5);
\draw[blue,ultra thick] (2.5,1) -- (3,0.5);
\draw[blue,ultra thick] (0.5,4) -- (4,0.5);
\draw[blue,ultra thick] (1.5,4) -- (4,1.5);

\draw[blue,ultra thick] (1,0.5) -- (2,0.5);
\draw[blue,ultra thick] (3,0.5) -- (4,0.5);
\draw[blue,ultra thick] (1.5,1) -- (2.5,1);
\draw[blue,ultra thick] (1.5,2) -- (2.5,2);
\draw[blue,ultra thick] (1,2.5) -- (2,2.5);
\draw[blue,ultra thick] (3,2.5) -- (4,2.5);
\draw[blue,ultra thick] (0,3) -- (0.5,3);
\draw[blue,ultra thick] (3.5,3) -- (4,3);
\draw[blue,ultra thick] (0,4) -- (0.5,4);
\draw[blue,ultra thick] (3.5,4) -- (4,4);

\draw[blue,ultra thick] (0.5,1) -- (0.7,1);
\draw[blue,ultra thick] (1.3,1) -- (1.5,1);
\draw[blue,ultra thick,decorate,decoration={coil,segment length=1.5mm,amplitude=1.5mm}] (0.7,1) -- (1.3,1);

\draw[blue,ultra thick] (0.5,2) -- (0.7,2);
\draw[blue,ultra thick] (1.3,2) -- (1.5,2);
\draw[blue,ultra thick,decorate,decoration={coil,segment length=1.5mm,amplitude=1.5mm}] (0.7,2) -- (1.3,2);

\draw[blue,ultra thick] (2.5,3) -- (2.7,3);
\draw[blue,ultra thick] (3.3,3) -- (3.5,3);
\draw[blue,ultra thick,decorate,decoration={coil,segment length=1.5mm,amplitude=1.5mm}] (2.7,3) -- (3.3,3);

\draw[blue,ultra thick] (2.5,4) -- (2.7,4);
\draw[blue,ultra thick] (3.3,4) -- (3.5,4);
\draw[blue,ultra thick,decorate,decoration={coil,segment length=1.5mm,amplitude=1.5mm}] (2.7,4) -- (3.3,4);

\foreach \xpos in {0,1,2,3}
\foreach \ypos in {0,1,2,3}
 \draw[black] (\xpos+0.5,\ypos) -- (\xpos+1,\ypos+0.5) -- (\xpos+0.5,\ypos+1) -- (\xpos,\ypos+0.5) -- cycle;

\draw[very thick,->] (0,-0.5)--(4,-0.5);
\draw (4,-0.5) node[right] {$x$};
\foreach \xpos in {0,1,2,3}
{
 \draw[thick] (\xpos+0.5,-0.6)--(\xpos+0.5,-0.4);
 \draw (\xpos+0.5,-0.5) node[below] {$\xpos$};
}

\draw[very thick,->] (-0.5,0)--(-0.5,4);
\draw (-0.5,4) node[above] {$y$};
\foreach \ypos in {0,1,2,3}
{
 \draw[thick] (-0.6,\ypos+0.5)--(-0.4,\ypos+0.5);
 \draw (-0.5,\ypos+0.5) node[left] {$\ypos$};
}
 
\end{tikzpicture}
 \caption{Four-terminal representation of the ruby medial lattice.}
 \label{fig:rubymedial}
\end{center}
\end{figure}
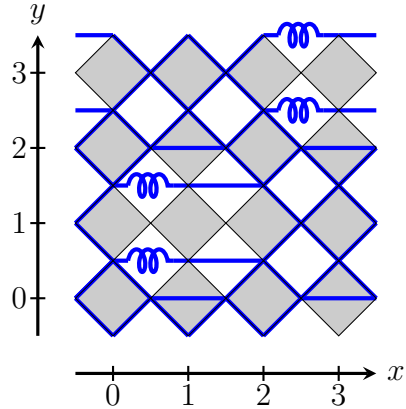

The ruby medial lattice can be represented in four-terminal form as shown in Figure~\ref{fig:rubymedial}. This representation is
valid only when $n$ is a multiply of 4, and so we shall be limited to studying the case $n=4$ in the following.

As discussed in section~\ref{sec:fourtermrep} one can act on the two strands in a white square by the operator
${\sf H}_i$ in order to produce a horizontal diagonal. This has been done here on the white squares with coordinates
$(x,y) = (\frac32,\frac12), (\frac32,\frac32), (\frac75,\frac52), (\frac72,\frac72)$ mod 4. But for the present lattice we shall
also need a slight variation of this trick, namely to act with the operator ${\sf E}_i$ instead. This corresponds to formally
setting the coupling strength $x = \infty$, and renormalising by a factor of $x$, meaning that the two sites sitting across
the horizontal diagonal of the white square will be effectively identified, or contracted. This operation is represented by
a coil-like symbol in Figure~\ref{fig:rubymedial}, and we use it in the white squares with coordinates
$(x,y) = (\frac12,\frac12), (\frac12,\frac32), (\frac52,\frac52), (\frac52,\frac72)$ mod 4. The reader may want to verify
from the figure that each triangle is surrounded by six squares (three of which share an edge with the triangle, and the
other three share only a vertex), and similarly that each hexagon is surrounded by twelve squares (that again alternate between
edge-sharing and vertex-sharing). Each hexagon has been represented as an octagon with two $x=\infty$ edges.

Similarly, some of the edges in the grey squares have coupling strength $x=0$, i.e., they reduce to operators ${\sf Id}$ or ${\sf E}$.
Thus, each hexagon comprises two spins and one dual spin that are ``free'', in the sense that they do not interact with the remainder
of the lattice. Since each $4 \times 4$ pattern contains two hexagons, this leads to a extraneous factor $q^6$ by which we must divide
in order to form $P_B(q,v)$. Obviously the occurrences of $x=0$ or $x=\infty$ do not count as edges of the lattice that we are
investigating, and therefore diminish the efficiency of the representation. In the present case we have therefore a rather modest
number of $\frac32$ vertices and $3$ edges per grey square.

The $\check{\sf R}$-matrix acting on the grey squares can be formally described as
\begin{equation}
 \check{\sf R}_i = \left \lbrace
 \begin{array}{ll}
 {\sf H}_{i+1} {\sf V}_i {\sf V}_{i+2} {\sf H}_{i+1} & \mbox{if } (x,y)=(0,0) \mbox{ mod } 2 \\
 {\sf H}_{i+1} {\sf V}_i {\sf V}_{i+2} {\sf H}_{i+1} & \mbox{if } (x,y)=(1,3) \mbox{ or } (3,1) \mbox{ mod } 4 \\
 {\sf E}_i {\sf H}_{i+1} {\sf V}_{i+2} {\sf H}_{i+1} & \mbox{if } (x,y)=(1,0) \mbox{ or } (3,2) \mbox{ mod } 4 \\
 {\sf H}_{i+1} {\sf V}_i {\sf H}_{i+1} {\sf E}_{i+2} & \mbox{if } (x,y)=(3,0) \mbox{ or } (1,2) \mbox{ mod } 4 \\
 {\sf V}_i {\sf E}_{i+2} {\sf H}_{i+1} & \mbox{if } (x,y)=(0,1) \mbox{ or } (2,3) \mbox{ mod } 4 \\
 {\sf H}_{i+1} {\sf E}_i {\sf V}_{i+2} & \mbox{if } (x,y)=(2,1) \mbox{ or } (0,3) \mbox{ mod } 4 \\
 {\sf E}_i {\sf E}_{i+2} & \mbox{if } (x,y)=(1,1) \mbox{ or } (3,3) \mbox{ mod } 4 \\ 
 \end{array} \right.
\end{equation}

\begin{table}
\begin{center}
 \begin{tabular}{l|l}
 $n$ & $p_{\rm c}$ \\ \hline
 4 & 0.51276405773159089981168194057612883705509718962342 \\
 \end{tabular}
 \caption{Bond percolation threshold $p_{\rm c}$ on the ruby medial lattice.}
 \label{tab:rubymedial1}
\end{center}
\end{table}

\begin{table}
\begin{center}
 \begin{tabular}{l|l}
 $n$ & $v_{\rm c}$ \\ \hline
 4 & 1.80131354317808391632479082679158473077866941417858 \\
 \end{tabular}
 \caption{Critical point $v_{\rm c}$ of the $q=3$ state Potts model on the ruby medial lattice.}
  \label{tab:rubymedial3}
\end{center}
\end{table}

\begin{table}
\begin{center}
 \begin{tabular}{l|l}
 $n$ & $v_{\rm c}$ \\ \hline
 4 & 2.07313974259180806296373118024496389902474086148828 \\
  \end{tabular}
 \caption{Critical point $v_{\rm c}$ of the $q=4$ state Potts model on the ruby medial lattice.}
  \label{tab:rubymedial4}
\end{center}
\end{table}

The $n=4$ result for the percolation threshold $p_{\rm c}$ is shown in Table~\ref{tab:rubymedial1}, and those for
the critical points $v_{\rm c}$ of the $q=3$ and $q=4$ state Potts models are reported in Tables~\ref{tab:rubymedial3}--\ref{tab:rubymedial4}.
From these single data points, which moreover do not correspond to a very large size of the basis compared to the other lattices
treated in this work, it does not seem reasonable to provide a final result with error bars for the $n \to \infty$ limit.

For the Ising model ($q=2$) the factorisation of $P_B(q,v)$ with $n=4$ contains factors of maximum degree $d_{\rm max} = 20$.
The factor determining $v_{\rm c}$ is of degree $16$, and by setting $y = -1 + \sqrt{y}$ it becomes simpler:
\begin{equation}
-3 - 6 y - 66 y^2 - 174 y^3 - 194 y^4 - 58 y^5 - 10 y^6 - 2 y^7 + y^8 \,.
\end{equation}
The largest real root in $y$ determines the critical coupling as $v_{\rm c} \simeq1.477\,488\,025\cdots$.
Based on our experience from the Archimedean lattices we can assume this value to be the exact result.

\begin{figure}
\begin{center}
\includegraphics[width=12cm]{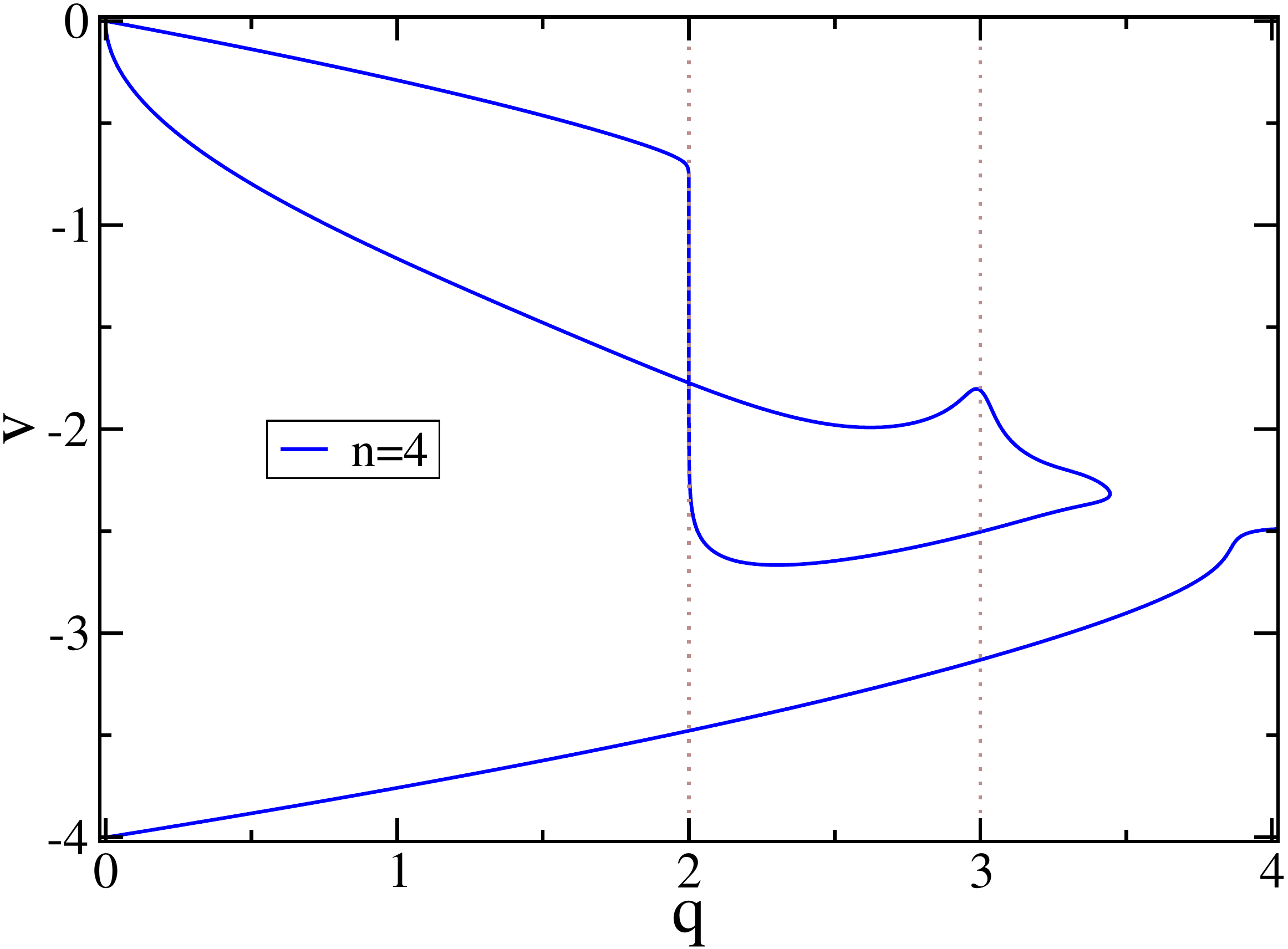}
\caption{Roots of $P_B(q,v)$ for the Potts model on the ruby medial lattice, using
$n \times n$ square bases.}
\label{fig:rubymedial-pd}
\end{center}
\end{figure}

The phase diagram in shown in Figure~\ref{fig:rubymedial-pd}.
There is a clear vertical ray at the Beraha number (\ref{Beraha}) with $k=4$, and another incipient ray with $k=6$.
Note also that the $n=4$ curve passes through the point $(q,v)=(0,-4)$ exactly. The curve emanating from that
point is presumably a good approximation to the lower boundary of the BK phase. The value of $q_{\rm c}$ is
however difficult to estimate from just one curve.

\subsection{Factorisable cases}
\label{sec:factorisation_medial}

In analogy with Section~\ref{sec:factorisation} we now discuss the cases of exact factorisation for the seven
medial lattices which are not themselves Archimedean (i.e., those discussed in section~\ref{sec:res-medial}).
Statements including the words ``all'' or ``none'' thus refer to those seven lattices only.

Note that for none of the medial lattices does $P_B(q,v)$ factorise for generic values of $q$ and $v$.
This presumably means that the Potts model is not solvable on these lattices along any {\em curve}.
Conversely, factorisation does occur for isolated (integer) values of $q$ and $v$, as we now discuss.

For $q=0$, $P_B(q,v)$ factorises for all the lattices, producing a root $v=0$. The resulting free fermion
theories describe spanning trees \cite{CJSSS04,JSS05}. Moreover, the four-eight medial, frieze medial, three-twelve medial,
cross medial, snub square medial and snub hexagonal medial lattices have a root $(q,v)=(0,-3)$.
And the ruby medial lattice has a root $(q,v)=(0,-4)$. These are models of spanning forests \cite{JSS05}
on the corresponding dual lattices, with weight $v$ per component tree.

Unlike the case of Archimedean lattices, we have found no medial lattices with a size-independent (finite) slope with which the
curves pass through the origin $(q,v) = (0,0)$.

There are a couple of unusual cases for $q=0$. The four-eight medial and cross medial lattices both shed the small factor
$8 + 3 v + v^2$ with roots $v = (-3 \pm i \sqrt{23})/2$. Similarly, the three-twelve medial lattice sheds the factor
$6 + 3 v + v^2$ with roots $(-3 \pm i \sqrt{15})/2$. All of these cases presumably provide loci of exact solvability.

In the Ising case ($q=2$) there is a root in $v=-1$ (chromatic polynomial) for all the medial lattices. Unlike the case of Archimedean lattices,
there are no medial lattices with a root in $v=-2$.

For $q=3$ there is a root in $v=-1$ (chromatic polynomial) for the four-eight medial lattice provided that $n$ is odd
(we have checked this for $n=1,3$), while for $n$ even (i.e., for $n=2,4$) the curves in Figure~\ref{fig:foureightmedial-pd} do not even come close
to the point $(q,v)=(3,-1)$. However, we have seen in this study that parity effects in $n$ are extremely common, so we feel
confident in conjecturing that the three-colouring problem on the four-eight medial lattice is exactly solvable.%
\footnote{On the other hand, $v=-1$ factorises for the snub square medial lattice with $n=2$, but not with $n=4$.
In this case the factorisation appears to be a fortuity for $n=2$ rather than a systematic phenomenon.}

For $q=3$ the frieze medial and snub hexagonal medial lattices both shed the small factor
$3 + 3 v + v^2$ whose roots, $v = (-3 \pm i \sqrt{3})/2$, are presumably loci of exact solvability.

Finally, for $q=4$ there is a root at $v=-2$ for the frieze medial lattice.

\section{Site percolation}
\label{sec:site}

The site percolation problem can be seen as the $q \to 1$ limit of a Potts model only when the
latter is generalised to include multi-spin interactions \cite{KunzWu78}. It follows that site percolation is
not dual to a site percolation on the dual lattice. Therefore one might in principle want to study
site percolation on all the lattices on which we have treated the Potts model above, as well as
on their corresponding medial lattices. This should be possible using the techniques
exposed this far, combined with a few extra tricks that we expose below. For practical reasons we
shall however limit the study to a subclass of Archimedean lattices and their duals, namely those
having only cubic and quartic vertices, discarding also those cases \cite{SJ12} for 
which the site percolation problem is exactly solvable. This amounts to treating the seven
lattices listed in Table~\ref{tab:packingsite}.

We have already recalled in Section~\ref{sec:lattices} that bond percolation on a cubic lattice $G$
is equivalent to site percolation on the corresponding medial lattice ${\cal M}(G)$ \cite{EssamFisher61,SykesEssam64}.
By duality this extends to cases where $G$ is a triangulation. It follows in particular from
Section~\ref{sec:triangular} that site percolation is exactly solvable on the kagome lattice, with $p_{\rm c} = 1 - 2 \sin(\pi/18)$,
cf.\ Eq.~(\ref{pc_triangular}).

By the same token, site percolation on the four-eight medial, the three-twelve medial and the cross medial lattices are equivalent to
bond percolation on the corresponding original (i.e., non-medial) lattices, discussed in Sections~\ref{sec:foureight},
\ref{sec:threetwelve} and \ref{sec:cross} respectively. The site percolation thresholds for
these three lattices can therefore be read from Tables~\ref{tab:foureight1}, \ref{tab:threetwelve1}
and \ref{tab:cross1}.

Apart from the kagome lattice, there are a few more lattices where site percolation is exactly solvable by relatively elementary tricks.
For instance, the problem on the three-twelve lattice follows from that on the kagome lattice upon replacing $p$ by $p^2$.
And site percolation on the triangular lattice is dealt with by noticing that the percolation hulls (that live on the dual lattice) describe
the well-known \cite{Nienhuis82} O($n$) loop model on the hexagonal lattice with $n=1$. Its critical point in the dense phase has
monomer fugacity $K=1$. It follows that $p_{\rm c} = 1/2$.

Site percolation on these three exactly solvable lattices (i.e., kagome, three-twelve and triangular) was discussed in \cite{SJ12}
from the point of view of graph polynomials $P_B(p)$. It was found that indeed the smallest possible bases provide the exact
threshold $p_{\rm c}$ and that larger bases lead to a factorisation of the exact result.

In the remainder of this section we discuss how to compute $P_B(p)$ for site percolation on other lattices
by using a four-terminal representation and the periodic TL transfer matrix approach.
This treatment of the site percolation problem has some advantages over bond percolation, but also some inconveniences.
These aspects are most vividly illustrated in the case of the square lattice. Recall that in Section~\ref{sec:foureightmedial} we
have introduced a method in which a generic Temperley-Lieb operator acts within a grey square (see Figure~\ref{fig:labelR})
by giving specific weights to the fourteen possible planar pairings of the strands $i,j,k,l := i,i+1,i+2,i+3$ and $i',j',k',l'$.
Consider now lodging one site of the percolation problem inside each grey square of the four-terminal representation
(see Figure~\ref{fig:square-basis-loop}). When that site is occupied it must connect the grey square onto the four surrounding
grey squares that it touches at its corners. Treating the loops as hulls of the percolation clusters, this is accomplished by
choosing the pairing $(ii')(jk)(ll')(j'k')$. On the other hand, when the site is empty it must disconnect the grey square from
its surroundings. This is done by taking the pairing $(ij)(kl)(i'j')(k'l')$. The Boltzmann weights corresponding to an occupied
(resp.\ an empty) site is taken as $p$ (resp.\ $1-p$), or equivalently as $v$ (resp.\ $1$), where we have set $v = p/(1-p)$ as usual.

The advantage of this approach is that the TL operator is, in fact, not generic at all: it only gives a non-zero weight to two
out of the fourteen possible pairings. Acting repeatedly on all grey squares of the basis therefore only produces a relatively
small subset of all possible elements of the (periodic) TL algebra. We can exploit this by abandoning the approach of storing
all Boltzmann weights in tables (using in particular the bijection of section~\ref{sec:bijection} between connectivity states and integers),
since many of those weights would be zero anyway. Instead, we simply insert the states that are produced by the transfer matrix
(with non-zero weight) in a hash table. Still for the square lattice, this enables us to treat $n \times n$ bases as large as $n=11$.

The main inconvenience is that each grey square can accommodate only a relatively small number of sites (e.g., just one for the 
square lattice). Moreover, we have found no meaningful way of using the white squares. We shall also see below that the rate
of convergence of the estimates for $p_{\rm c}$ is noticeably slower for site percolation than for bond percolation. Still, our
method leads to final results that are generally at least as precise as those of the best available Monte Carlo simulations.

Throughout this section we have computed the exact percolation polynomials $P_B(p)$ for all sizes $n$ discussed. These
polynomials are available in electronic form as supplementary material to this article.%
\footnote{This supplementary material takes the form of a text file {\tt PB.m} that can be processed by {\sc Mathematica} or similar
programs for symbolic algebra manipulations.}
In the following subsections we
tabulate as usual the positive root $p_{\rm c} \in (0,1)$ of $P_B(p)$ to 50 decimal digits, but it should be kept in mind that
in all cases these numbers are in fact known to arbitrary precision. In contrast with the bond percolation thresholds and Potts 
model critical points tabulated in the preceding sections, we have here not pushed the computations to larger sizes by
seeking a purely numerical evaluation of the root $p_{\rm c}$.

\begin{table}
\begin{center}
 \begin{tabular}{l|ll|rrr}
 Lattice & Vertices & Parity of $n$ & $n_{\rm max}$ & $|V|_{\rm max}$ & $d_{\rm max}$ \\ \hline
 Hexagonal & 2 & Any & 8 & 128 & 311\,467\,520 \\
 Square & 1 & Any & 11 & 121 & 1\,770\,114\,092 \\
 Four-eight & 4 & Any & 7 & 196 & 199\,753\,311 \\
 Cross & 3 & Any & 8 & 192 & 605\,394\,138 \\
 Ruby & $\frac{3}{4}$ & 0 mod 4 & 16 & 192 & 843\,378\,845 \\
 Cairo pentagonal & $\frac{3}{2}$ & Even & 8 & 96 & 32\,215\,001 \\
 Frieze dual & $\frac{3}{2}$ &  Any${}^\dagger$ & 9 & 243 & 339\,644\,725 \\
  \hline
  \end{tabular}
 \caption{Number of vertices per grey square (cf.~Figure~\ref{fig:square-basis})
 for site percolation on various lattices, using square bases of size $n \times n$ grey squares
 (${}^\dagger$or rectangular bases of size $n \times 2n$). In addition we state any parity constraint on $n$.
 The right part of the table shows the largest size $n_{\rm max}$ for which we have been able to compute
 the polynomial $P_B(p)$, the corresponding number of vertices $|V|_{\rm max}$ in $B$, and the dimension
 $d_{\rm max}$ of the transfer matrix.}
 \label{tab:packingsite}
\end{center}
\end{table}

The degree of the critical polynomials $P_B(p)$ for site percolation on the lattices studied below
is shown in Table~\ref{tab:packingsite}. We also provide the largest size $n_{\rm max}$ for which we have been able to compute
the polynomial $P_B(p)$ for each lattice, as well as the corresponding number of vertices $|V|_{\rm max}$ in $B$ (which is also
the degree of $P_B(p)$). The table also gives the dimension $d_{\rm max}$ of the largest transfer matrix used in the computation,
i.e., the maximum number of states required at any intermediate stage. These dimensions should be compared with the number
${\rm dim}(n,2)$ in the generic case (see Table~\ref{tab:dims}), from which the advantage of the hashing approach can be judged.

\subsection{Hexagonal lattice}

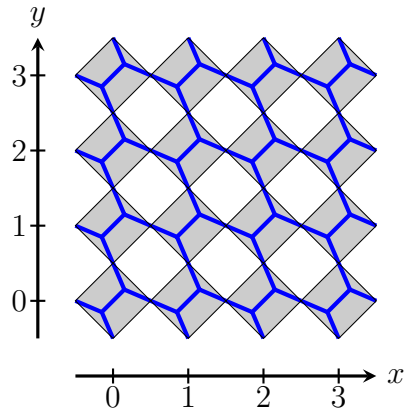
\begin{figure}
\begin{center}
\begin{tikzpicture}[scale=1.0,>=stealth]
\foreach \xpos in {0,1,2,3}
\foreach \ypos in {0,1,2,3}
 \fill[black!20] (\xpos+0.5,\ypos) -- (\xpos+1,\ypos+0.5) -- (\xpos+0.5,\ypos+1) -- (\xpos,\ypos+0.5) -- cycle;

\foreach \xpos in {0,1,2,3}
\foreach \ypos in {0,1,2,3}
{
 \draw[blue,ultra thick] (\xpos,\ypos+0.5) -- (\xpos+0.35,\ypos+0.35) -- (\xpos+0.5,\ypos);
 \draw[blue,ultra thick] (\xpos+0.5,\ypos+1) -- (\xpos+0.65,\ypos+0.65) -- (\xpos+1,\ypos+0.5);
 \draw[blue,ultra thick] (\xpos+0.35,\ypos+0.35) -- (\xpos+0.65,\ypos+0.65);
}

\foreach \xpos in {0,1,2,3}
\foreach \ypos in {0,1,2,3}
 \draw[black] (\xpos+0.5,\ypos) -- (\xpos+1,\ypos+0.5) -- (\xpos+0.5,\ypos+1) -- (\xpos,\ypos+0.5) -- cycle;

\draw[very thick,->] (0,-0.5)--(4,-0.5);
\draw (4,-0.5) node[right] {$x$};
\foreach \xpos in {0,1,2,3}
{
 \draw[thick] (\xpos+0.5,-0.6)--(\xpos+0.5,-0.4);
 \draw (\xpos+0.5,-0.5) node[below] {$\xpos$};
}

\draw[very thick,->] (-0.5,0)--(-0.5,4);
\draw (-0.5,4) node[above] {$y$};
\foreach \ypos in {0,1,2,3}
{
 \draw[thick] (-0.6,\ypos+0.5)--(-0.4,\ypos+0.5);
 \draw (-0.5,\ypos+0.5) node[left] {$\ypos$};
}
 
\end{tikzpicture}
 \caption{Four-terminal representation for site percolation on the hexagonal lattice.}
 \label{fig:hexagonalsite}
\end{center}
\end{figure}

Figure~\ref{fig:hexagonalsite} shows a four-terminal representation for site percolation on the hexagonal lattice. There
are $2$ vertices per grey square.

It is useful at this stage to point out the key differences with bond percolation. In bond percolation, the ``conducting units''
are the edges. Since edges meet at vertices, each of the four terminals of the grey squares must be situated at the position of
a vertex. On the other hand, in site percolation the ``conducting units'' are the vertices. It is convenient to represent an
occupied site instead as a colouring of its adjacent half-edges, so that site percolation clusters become connected components
(clusters) of coloured half-edges. It follows from this picture that each of the four terminals of the grey squares must be situated at
the mid point of an edge.

The loop strands of Figure~\ref{fig:labelR} must turn around the clusters of coloured half-edges. It follows that
each choice of occupancy of sites within a grey square induces a
planar pairings of the strands $i,j,k,l := i,i+1,i+2,i+3$ and $i',j',k',l'$.
In the case of Figure~\ref{fig:hexagonalsite} the $\check{\sf R}$-matrix becomes
\begin{eqnarray}
 \check{\sf R}_i &=& (ij)(kl)(i'j')(k'l') + v \, (il)(jk)(i'j')(k'l') \nn \\
 &+& v \, (ij)(kl)(i'l')(j'k') + v^2 \, (ii')(jk)(j'k')(ll') \,,
 \label{Rsitehex}
\end{eqnarray}
where the first term corresponds to both sites within the grey square being empty, the
second and third term correspond to one site being occupied and the other empty, and
the fourth term comes from the case where both sites are occupied. This can be expressed
more elegantly in terms of TL generators:
\begin{equation}
 \check{\sf R}_i = {\sf E}_{i+2} {\sf E}_i + v \, ({\sf E}_{i+2} {\sf E}_i {\sf E}_{i+1} + {\sf E}_{i+1} {\sf E}_{i+2} {\sf E}_i) + v^2 \, {\sf E}_{i+1} \,.
\end{equation}

\begin{table}
\begin{center}
 \begin{tabular}{l|l}
 $n$ & $p_{\rm c}$ \\ \hline
 1 & 0.7071067811865475244008443621048490392848359376885 \\
 2 & 0.6971069014219768583833477251437189263456173513781 \\
 3 & 0.6971928819498649590842656967290585989087604117228 \\
 4 & 0.6970613429377088940378961155082870171465188648927 \\
 5 & 0.6970449503839377097642274158922485690092452929594 \\
 6 & 0.6970416734307503900742086237111348168897232567095 \\
 7 & 0.6970407718280774909951027270176125447620620155082 \\
 8 & 0.6970404617236920725796530320181513631207407839926 \\ \hline
 $\infty$ & 0.697040230 (5) \\
 Ref.~\cite{FengDengBlote08} & 0.6970402 (1) \\
 \end{tabular}
 \caption{Site percolation threshold $p_{\rm c}$ on the hexagonal lattice.}
  \label{tab:site-hexagonal}
\end{center}
\end{table}

The fact that (\ref{Rsitehex}) contains only four terms out of fourteen possible means that
the computation of $P_B(p)$ can be accomplished for square bases of size up to $n=8$.
The corresponding thresholds $p_{\rm c}$ are shown in Table~\ref{tab:site-hexagonal}.

The exponent appearing in (\ref{extrapol_w}) here comes out as $w \approx 6.35$, which is
the same value as found for bond percolation on the kagome and three-twelve lattices.
This seems reasonable, since all those lattices have the same three-fold rotational symmetry.%
\footnote{In particular $w$ does not seem to depend on the nature of the percolation problem (bond or site),
provided that the lattice has the same symmetry. Obviously, this remark does not hold true when comparing
problems that are exactly solvable with those which are not.}
The high value of $w$ again entails a high precision of the final value, more than two orders
of magnitude better than the best numerical result \cite{FengDengBlote08}.

Note that graph polynomials for this lattice were previously studied in \cite{SJ12} for 
hexagonal bases with up to $54$ sites (compared to the square bases with up to $128$ sites used here).

\subsection{Square lattice}
\label{sec:square_site}

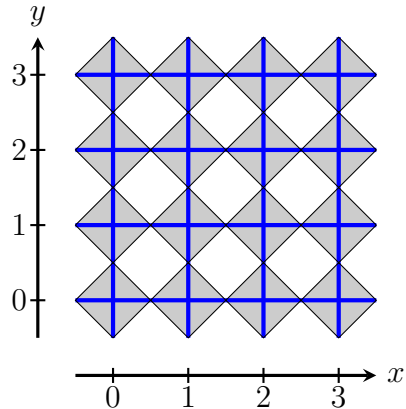
\begin{figure}
\begin{center}
\begin{tikzpicture}[scale=1.0,>=stealth]
\foreach \xpos in {0,1,2,3}
\foreach \ypos in {0,1,2,3}
 \fill[black!20] (\xpos+0.5,\ypos) -- (\xpos+1,\ypos+0.5) -- (\xpos+0.5,\ypos+1) -- (\xpos,\ypos+0.5) -- cycle;

\foreach \xpos in {0,1,2,3}
\foreach \ypos in {0,1,2,3}
{
 \draw[blue,ultra thick] (\xpos,\ypos+0.5) -- (\xpos+1,\ypos+0.5);
 \draw[blue,ultra thick] (\xpos+0.5,\ypos) -- (\xpos+0.5,\ypos+1);
}

\foreach \xpos in {0,1,2,3}
\foreach \ypos in {0,1,2,3}
 \draw[black] (\xpos+0.5,\ypos) -- (\xpos+1,\ypos+0.5) -- (\xpos+0.5,\ypos+1) -- (\xpos,\ypos+0.5) -- cycle;

\draw[very thick,->] (0,-0.5)--(4,-0.5);
\draw (4,-0.5) node[right] {$x$};
\foreach \xpos in {0,1,2,3}
{
 \draw[thick] (\xpos+0.5,-0.6)--(\xpos+0.5,-0.4);
 \draw (\xpos+0.5,-0.5) node[below] {$\xpos$};
}

\draw[very thick,->] (-0.5,0)--(-0.5,4);
\draw (-0.5,4) node[above] {$y$};
\foreach \ypos in {0,1,2,3}
{
 \draw[thick] (-0.6,\ypos+0.5)--(-0.4,\ypos+0.5);
 \draw (-0.5,\ypos+0.5) node[left] {$\ypos$};
}
 
\end{tikzpicture}
 \caption{Four-terminal representation for site percolation on the square lattice.}
 \label{fig:squaresite}
\end{center}
\end{figure}

Site percolation on the square lattice was already discussed as an example in the introduction to 
section~\ref{sec:site}. The four-terminal representation is shown in Figure~\ref{fig:squaresite}.
It has $1$ vertex per grey square.

The $\check{\sf R}$-matrix reads
\begin{equation}
 \check{\sf R}_i = (ij)(kl)(i'j')(k'l') + v \, (ii')(jk)(ll')(j'k') \,,
 \label{Rsitesquare}
\end{equation}
as discussed previously. In terms of TL generators this reads
\begin{equation}
 \check{\sf R}_i = {\sf E}_{i+2} {\sf E}_i + v \, {\sf E}_{i+1} \,.
 \label{RsitesquareTL}
\end{equation}
Since there are only two terms out of fourteen possible,
we have been able to obtain $P_B(p)$ for bases of size up to $n=11$. The
corresponding thresholds $p_{\rm c}$ are given in Table~\ref{tab:site-square}.

\begin{table}
\begin{center}
 \begin{tabular}{l|l}
 $n$ & $p_{\rm c}$ \\ \hline
  1 & 0.50000000000000000000000000000000000000000000000000 \\
  2 & 0.54119610014619698439972320536638942006107206337802 \\
  3 & 0.58651145511267563565455897660690173482430062489384 \\
  4 & 0.59067211233102829689590201143951286962111713272216 \\
  5 & 0.59198825651833384461096868021192887904787477719722 \\
  6 & 0.59239507081770423769385580764250543411218819923508 \\
  7 & 0.59256103742766484893896496885851283129444654611347 \\
  8 & 0.59263900074535204810167646273223073775874995672901 \\
  9 & 0.59267976548917331887514046884728174207445322469521 \\
10 & 0.59270280369294408906688315667089943939640050817027 \\
11 & 0.59271663223297437516334096268406621556972040226157 \\ \hline
 $\infty$ & 0.59274601 (2) \\
 Ref.~\cite{FengDengBlote08} & 0.59274605 (3) \\
 \end{tabular}
 \caption{Site percolation threshold $p_{\rm c}$ on the square lattice.}
  \label{tab:site-square}
\end{center}
\end{table}

Since we have here eleven data points---the highest number for any of the problems
treated in this paper---we have taken particular care to get the best possible
extrapolation out of them. Applying the usual procedure we first found
$w \approx 4.07$ from a non-linear fit (\ref{extrapol_w}) to the last three
data points. However, it was easily detected that this choice of $w$ led to
some unnecessary spread on the BS approximants. 
In fact, fitting successively $w$ from the last three data points
among the first 9, 10 or all 11 points, we found $w \approx 4.139$, $w
\approx 4.098$ and $w \approx 4.074$, indicating that the true $w$
might be slightly lower. Repeating then the BS procedure while moving
down $w$ in steps of $0.01$ we have checked that the choice $w = 4.03 \pm 0.01$
minimises the spread of the approximants, and hence the final value
given in Table~\ref{tab:site-square} is based on this choice. This
is consistent with (and slightly more accurate than)
the best numerical result \cite{FengDengBlote08}.

\subsection{Four-eight lattice}

The four-terminal representation for site percolation on the four-eight lattice is shown in
Figure~\ref{fig:foureightsite}. It can accommodate $4$ vertices per grey square.

\begin{figure}
\begin{center}
\begin{tikzpicture}[scale=1.0,>=stealth]
\foreach \xpos in {0,1,2,3}
\foreach \ypos in {0,1,2,3}
 \fill[black!20] (\xpos+0.5,\ypos) -- (\xpos+1,\ypos+0.5) -- (\xpos+0.5,\ypos+1) -- (\xpos,\ypos+0.5) -- cycle;

\foreach \xpos in {0,1,2,3}
\foreach \ypos in {0,1,2,3}
{
 \draw[blue,ultra thick] (\xpos+0.2,\ypos+0.5) -- (\xpos+0.5,\ypos+0.2) -- (\xpos+0.8,\ypos+0.5) -- (\xpos+0.5,\ypos+0.8) -- cycle;
 \draw[blue,ultra thick] (\xpos,\ypos+0.5) -- (\xpos+0.2,\ypos+0.5);
 \draw[blue,ultra thick] (\xpos+0.8,\ypos+0.5) -- (\xpos+1,\ypos+0.5);
 \draw[blue,ultra thick] (\xpos+0.5,\ypos) -- (\xpos+0.5,\ypos+0.2);
 \draw[blue,ultra thick] (\xpos+0.5,\ypos+0.8) -- (\xpos+0.5,\ypos+1);
}

\foreach \xpos in {0,1,2,3}
\foreach \ypos in {0,1,2,3}
 \draw[black] (\xpos+0.5,\ypos) -- (\xpos+1,\ypos+0.5) -- (\xpos+0.5,\ypos+1) -- (\xpos,\ypos+0.5) -- cycle;

\draw[very thick,->] (0,-0.5)--(4,-0.5);
\draw (4,-0.5) node[right] {$x$};
\foreach \xpos in {0,1,2,3}
{
 \draw[thick] (\xpos+0.5,-0.6)--(\xpos+0.5,-0.4);
 \draw (\xpos+0.5,-0.5) node[below] {$\xpos$};
}

\draw[very thick,->] (-0.5,0)--(-0.5,4);
\draw (-0.5,4) node[above] {$y$};
\foreach \ypos in {0,1,2,3}
{
 \draw[thick] (-0.6,\ypos+0.5)--(-0.4,\ypos+0.5);
 \draw (-0.5,\ypos+0.5) node[left] {$\ypos$};
}
 
\end{tikzpicture}
 \caption{Four-terminal representation for site percolation on the four-eight lattice.}
 \label{fig:foureightsite}
\end{center}
\end{figure}

The $\check{\sf R}$-matrix is now
\begin{eqnarray}
 \check{\sf R} &=& (1+4v+2v^2) \,  (ij)(kl)(i'j')(k'l') + v^4 \, (ii')(jk)(j'k')(ll') \nonumber \\
 &+& v^2 \left[ (ii')(jj')(kl)(k'l') + (ij)(i'j')(kk')(ll') + \right. \nonumber \\
 & & \quad \left. (il)(jk)(i'j')(k'l') + (ij)(kl)(i'l')(j'k') \right] \nonumber \\
 &+& v^3 \left[ (ii')(j'k')(jl')(kl) + (ii')(jk)(j'l)(k'l') + \right. \nonumber \\
 & & \quad \left. (ik')(jk)(i'j')(ll') + (i'k)(j'k')(ij)(ll') \right] \,.
\end{eqnarray}
With ten terms out of fourteen possible, the advantage of the hashing scheme over
the complete tabulation of states is less pronounced than for the lattices treated
previously. Accordingly we can treat square bases of size up to $n=7$. The
results for the threshold $p_{\rm c}$ are given in Table~\ref{tab:site-foureight}.

\begin{table}
\begin{center}
 \begin{tabular}{l|l}
 $n$ & $p_{\rm c}$ \\ \hline
 1 & 0.7336147478371355308558999643084317148034022526034 \\
 2 & 0.7312492379002034814136736960461227248789777597490 \\
 3 & 0.7301157282458717016250281495894718983091568806878 \\
 4 & 0.7298412248145118289566094607464823352336030277592 \\
 5 & 0.7297673439318198743116752465745127856997428287266 \\
 6 & 0.7297429799645414198759930896484250713568213365103 \\
 7 & 0.7297332402222111151444172392979287587234589502354 \\ \hline
 $\infty$ & 0.7297232 (5) \\
 Ref.~\cite{SudingZiff99} & 0.729724 (3) \\
 \end{tabular}
 \caption{Site percolation threshold $p_{\rm c}$ on the four-eight lattice.}
  \label{tab:site-foureight}
\end{center}
\end{table}

\subsection{Cross lattice}

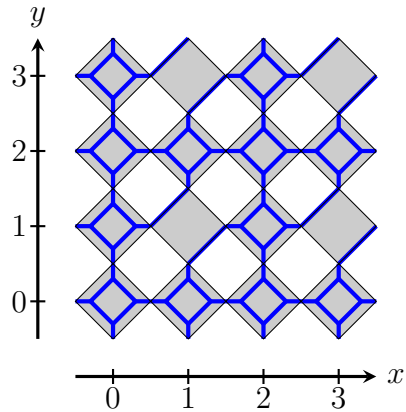
\begin{figure}
\begin{center}
\begin{tikzpicture}[scale=1.0,>=stealth]
\foreach \xpos in {0,1,2,3}
\foreach \ypos in {0,1,2,3}
 \fill[black!20] (\xpos+0.5,\ypos) -- (\xpos+1,\ypos+0.5) -- (\xpos+0.5,\ypos+1) -- (\xpos,\ypos+0.5) -- cycle;

\foreach \xpos in {0,2}
\foreach \ypos in {0,2}
{
 \draw[blue,ultra thick] (\xpos+0.2,\ypos+0.5) -- (\xpos+0.5,\ypos+0.2) -- (\xpos+0.8,\ypos+0.5) -- (\xpos+0.5,\ypos+0.8) -- cycle;
 \draw[blue,ultra thick] (\xpos,\ypos+0.5) -- (\xpos+0.2,\ypos+0.5);
 \draw[blue,ultra thick] (\xpos+0.8,\ypos+0.5) -- (\xpos+1,\ypos+0.5);
 \draw[blue,ultra thick] (\xpos+0.5,\ypos) -- (\xpos+0.5,\ypos+0.2);
 \draw[blue,ultra thick] (\xpos+0.5,\ypos+0.8) -- (\xpos+0.5,\ypos+1);

 \draw[blue,ultra thick] (\xpos+1.2,\ypos+0.5) -- (\xpos+1.5,\ypos+0.2) -- (\xpos+1.8,\ypos+0.5) -- (\xpos+1.5,\ypos+0.8) -- cycle;
 \draw[blue,ultra thick] (\xpos+1,\ypos+0.5) -- (\xpos+1.2,\ypos+0.5);
 \draw[blue,ultra thick] (\xpos+1.8,\ypos+0.5) -- (\xpos+2,\ypos+0.5);
 \draw[blue,ultra thick] (\xpos+1.5,\ypos) -- (\xpos+1.5,\ypos+0.2);
 \draw[blue,ultra thick] (\xpos+1.5,\ypos+0.8) -- (\xpos+1.5,\ypos+1);

 \draw[blue,ultra thick] (\xpos+0.2,\ypos+1.5) -- (\xpos+0.5,\ypos+1.2) -- (\xpos+0.8,\ypos+1.5) -- (\xpos+0.5,\ypos+1.8) -- cycle;
 \draw[blue,ultra thick] (\xpos,\ypos+1.5) -- (\xpos+0.2,\ypos+1.5);
 \draw[blue,ultra thick] (\xpos+0.8,\ypos+1.5) -- (\xpos+1,\ypos+1.5);
 \draw[blue,ultra thick] (\xpos+0.5,\ypos+1) -- (\xpos+0.5,\ypos+1.2);
 \draw[blue,ultra thick] (\xpos+0.5,\ypos+1.8) -- (\xpos+0.5,\ypos+2);

 \draw[blue,ultra thick] (\xpos+1,\ypos+1.5) -- (\xpos+1.5,\ypos+2);
 \draw[blue,ultra thick] (\xpos+1.5,\ypos+1) -- (\xpos+2,\ypos+1.5);
}

\foreach \xpos in {0,1,2,3}
\foreach \ypos in {0,1,2,3}
 \draw[black] (\xpos+0.5,\ypos) -- (\xpos+1,\ypos+0.5) -- (\xpos+0.5,\ypos+1) -- (\xpos,\ypos+0.5) -- cycle;

\draw[very thick,->] (0,-0.5)--(4,-0.5);
\draw (4,-0.5) node[right] {$x$};
\foreach \xpos in {0,1,2,3}
{
 \draw[thick] (\xpos+0.5,-0.6)--(\xpos+0.5,-0.4);
 \draw (\xpos+0.5,-0.5) node[below] {$\xpos$};
}

\draw[very thick,->] (-0.5,0)--(-0.5,4);
\draw (-0.5,4) node[above] {$y$};
\foreach \ypos in {0,1,2,3}
{
 \draw[thick] (-0.6,\ypos+0.5)--(-0.4,\ypos+0.5);
 \draw (-0.5,\ypos+0.5) node[left] {$\ypos$};
}
 
\end{tikzpicture}
 \caption{Four-terminal representation for site percolation on the cross lattice.}
 \label{fig:crosssite}
\end{center}
\end{figure}

We give a four-terminal representation for site percolation on the cross lattice in
Figure~\ref{fig:crosssite}. It requires $n$ to be even and hosts $3$ vertices per grey square.
The $\check{R}$-matrix is the same as for the four-eight lattice, except when $x$ and $y$ are
both odd in which case it is replaced by the identity operator. One may check the presence
of faces of degree 6 and 12.

Table~\ref{tab:site-cross} provides the results for the site percolation threshold $p_{\rm c}$.

\begin{table}
\begin{center}
 \begin{tabular}{l|l}
 $n$ & $p_{\rm c}$ \\ \hline
 2 & 0.7486176795231741957833806597419001398221099848029 \\
 4 & 0.7478954923957336800827109221019572017788873033627 \\
 6 & 0.7478142908979452473592066469359302051298398370653 \\
 8 & 0.7478045322937317100218711372257136728840658082387 \\ \hline
 $\infty$ & 0.7478008 (2) \\
 Ref.~\cite{SudingZiff99} & 0.747806 (4) \\
 \end{tabular}
 \caption{Site percolation threshold $p_{\rm c}$ on the cross lattice.}
  \label{tab:site-cross}
\end{center}
\end{table}

\subsection{Ruby lattice}
\label{sec:rubysite}

\begin{figure}
\begin{center}
\begin{tikzpicture}[scale=1.0,>=stealth]
\foreach \xpos in {0,1,2,3}
\foreach \ypos in {0,1,2,3}
 \fill[black!20] (\xpos+0.5,\ypos) -- (\xpos+1,\ypos+0.5) -- (\xpos+0.5,\ypos+1) -- (\xpos,\ypos+0.5) -- cycle;

\foreach \xpos in {2,3}
\foreach \ypos in {0,3}
{
 \draw[blue,ultra thick] (\xpos,\ypos+0.5) -- (\xpos+1,\ypos+0.5);
 \draw[blue,ultra thick] (\xpos+0.5,\ypos) -- (\xpos+0.5,\ypos+1);
}

\foreach \xpos in {0,1}
\foreach \ypos in {1,2}
{
 \draw[blue,ultra thick] (\xpos,\ypos+0.5) -- (\xpos+1,\ypos+0.5);
 \draw[blue,ultra thick] (\xpos+0.5,\ypos) -- (\xpos+0.5,\ypos+1);
}

\draw[blue,ultra thick] (0,0.5) -- (0.5,0.5) -- (0.5,1);
\draw[blue,ultra thick] (2,0.5) -- (1.5,0.5) -- (1.5,1);
\draw[blue,ultra thick] (2,1.5) -- (2.5,1.5) -- (2.5,1);
\draw[blue,ultra thick] (4,1.5) -- (3.5,1.5) -- (3.5,1);

\draw[blue,ultra thick] (2,2.5) -- (2.5,2.5) -- (2.5,3);
\draw[blue,ultra thick] (4,2.5) -- (3.5,2.5) -- (3.5,3);
\draw[blue,ultra thick] (0,3.5) -- (0.5,3.5) -- (0.5,3);
\draw[blue,ultra thick] (2,3.5) -- (1.5,3.5) -- (1.5,3);

\foreach \ypos in {0,3}
{
 \draw[blue,ultra thick] (0.5,\ypos+0.5) -- (0.7,\ypos+0.5);
 \draw[blue,ultra thick] (1.3,\ypos+0.5) -- (1.5,\ypos+0.5);
 \draw[blue,ultra thick,decorate,decoration={coil,segment length=1.5mm,amplitude=1.5mm}] (0.7,\ypos+0.5) -- (1.3,\ypos+0.5);
}

\foreach \ypos in {1,2}
{
 \draw[blue,ultra thick] (2.5,\ypos+0.5) -- (2.7,\ypos+0.5);
 \draw[blue,ultra thick] (3.3,\ypos+0.5) -- (3.5,\ypos+0.5);
 \draw[blue,ultra thick,decorate,decoration={coil,segment length=1.5mm,amplitude=1.5mm}] (2.7,\ypos+0.5) -- (3.3,\ypos+0.5);
}

\foreach \xpos in {0,1,2,3}
\foreach \ypos in {0,1,2,3}
 \draw[black] (\xpos+0.5,\ypos) -- (\xpos+1,\ypos+0.5) -- (\xpos+0.5,\ypos+1) -- (\xpos,\ypos+0.5) -- cycle;

\draw[very thick,->] (0,-0.5)--(4,-0.5);
\draw (4,-0.5) node[right] {$x$};
\foreach \xpos in {0,1,2,3}
{
 \draw[thick] (\xpos+0.5,-0.6)--(\xpos+0.5,-0.4);
 \draw (\xpos+0.5,-0.5) node[below] {$\xpos$};
}

\draw[very thick,->] (-0.5,0)--(-0.5,4);
\draw (-0.5,4) node[above] {$y$};
\foreach \ypos in {0,1,2,3}
{
 \draw[thick] (-0.6,\ypos+0.5)--(-0.4,\ypos+0.5);
 \draw (-0.5,\ypos+0.5) node[left] {$\ypos$};
}
 
\end{tikzpicture}
 \caption{Four-terminal representation for site percolation on the ruby lattice.}
 \label{fig:rubysite}
\end{center}
\end{figure}
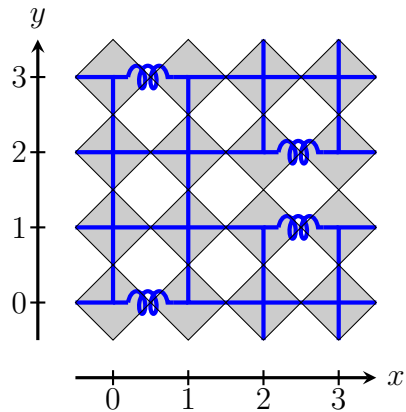

The four-terminal representation for site percolation on the ruby lattice requires some new tricks. It is shown
in Figure~\ref{fig:rubysite}, where we have supposed that $n = 0$ mod $4$. As usual the coil-like symbol indicates an edge of infinite strength, meaning that the
two vertices at its end points have been identified. The states of the two grey squares linked by a coil are now
correlated: if the site in one of the squares is empty (resp.\ occupied) the same is true in the adjacent square,
since the two sites have been identified. Identifying such conglomerates of two grey squares by the coordinates
$(x,y)$ of the leftmost one, we have the following $\check{\sf R}$-matrix:
\begin{equation}
 \check{\sf R}_i = \left \lbrace
 \begin{array}{ll}
 {\sf E}_{i+4} {\sf E}_{i+2} {\sf E}_i + v \, {\sf E}_{i+3} {\sf E}_{i+4} {\sf E}_{i+1} {\sf E}_{i+2} & \mbox{if } x+y=0 \mbox{ mod } 4 \\
 {\sf E}_{i+4} {\sf E}_{i+2} {\sf E}_i + v \, {\sf E}_{i+2} {\sf E}_{i+3} {\sf E}_{i} {\sf E}_{i+1} & \mbox{if } x+y=3 \mbox{ mod } 4 \\
 \end{array} \right.
 \label{Rrubyconglomerate}
\end{equation}
whereas the remaining grey squares (those without coils in Figure~\ref{fig:rubysite}) are described by (\ref{RsitesquareTL}), the usual $\check{\sf R}$-matrix
of the square lattice.

As a result we have an average of $\frac34$ sites per grey square. While this might seem mediocre, it should be remembered
that the $\check{\sf R}$-matrix (\ref{Rrubyconglomerate}) for a conglomerate of two grey squares acts on six pairs of points,
labelled $i,i+1,\ldots,i+5$ and $i',(i+1)',\ldots,(i+5)'$, which may in general accommodate ${\rm Cat}(6) = 132$ different connectivities. However,
(\ref{Rrubyconglomerate}) contains only two terms, so that the number of connectivity states actually generated by the transfer process
is only a very small subset of the total number of states respecting planarity. Accordingly we can attain a size of $n=16$, the largest
for any of the problems studied in this article.
Table~\ref{tab:site-ruby} gives the corresponding results for the threshold $p_{\rm c}$.

\begin{table}
\begin{center}
 \begin{tabular}{r|l}
 $n$ & $p_{\rm c}$ \\ \hline
    4 & 0.6217033170149886495775240445635207543697869760484 \\
    8 & 0.6218440760093613233443689965452296246219054789997 \\
  12 & 0.6218163965094603058947746474408604618101513181220 \\
  16 & 0.6218132249218005882705833731451247391116085233506 \\ \hline
 $\infty$ & 0.62181207 (7) \\
 Ref.~\cite{SudingZiff99} & 0.621819 (3) \\
 \end{tabular}
 \caption{Site percolation threshold $p_{\rm c}$ on the ruby lattice.}
  \label{tab:site-ruby}
\end{center}
\end{table}

\subsection{Cairo pentagonal lattice $D(3^2,4,3,4)$}

\begin{figure}
\begin{center}
\begin{tikzpicture}[scale=1.0,>=stealth]
\foreach \xpos in {0,1,2,3}
\foreach \ypos in {0,1,2,3}
 \fill[black!20] (\xpos+0.5,\ypos) -- (\xpos+1,\ypos+0.5) -- (\xpos+0.5,\ypos+1) -- (\xpos,\ypos+0.5) -- cycle;

\foreach \xpos in {0,2}
\foreach \ypos in {0,2}
{
 \draw[blue,ultra thick] (\xpos+1,\ypos+0.5) -- (\xpos+2,\ypos+0.5);
 \draw[blue,ultra thick] (\xpos+1.5,\ypos) -- (\xpos+1.5,\ypos+1);
 \draw[blue,ultra thick] (\xpos,\ypos+1.5) -- (\xpos+1,\ypos+1.5);
 \draw[blue,ultra thick] (\xpos+0.5,\ypos+1) -- (\xpos+0.5,\ypos+2);
 \draw[blue,ultra thick] (\xpos,\ypos+0.5) -- (\xpos+0.5,\ypos);
 \draw[blue,ultra thick] (\xpos+0.5,\ypos+1) -- (\xpos+1,\ypos+0.5);
 \draw[blue,ultra thick] (\xpos+0.25,\ypos+0.25) -- (\xpos+0.75,\ypos+0.75);
 \draw[blue,ultra thick] (\xpos+1,\ypos+1.5) -- (\xpos+1.5,\ypos+2);
 \draw[blue,ultra thick] (\xpos+1.5,\ypos+1) -- (\xpos+2,\ypos+1.5);
 \draw[blue,ultra thick] (\xpos+1.25,\ypos+1.75) -- (\xpos+1.75,\ypos+1.25);
}

\foreach \xpos in {0,1,2,3}
\foreach \ypos in {0,1,2,3}
 \draw[black] (\xpos+0.5,\ypos) -- (\xpos+1,\ypos+0.5) -- (\xpos+0.5,\ypos+1) -- (\xpos,\ypos+0.5) -- cycle;

\draw[very thick,->] (0,-0.5)--(4,-0.5);
\draw (4,-0.5) node[right] {$x$};
\foreach \xpos in {0,1,2,3}
{
 \draw[thick] (\xpos+0.5,-0.6)--(\xpos+0.5,-0.4);
 \draw (\xpos+0.5,-0.5) node[below] {$\xpos$};
}

\draw[very thick,->] (-0.5,0)--(-0.5,4);
\draw (-0.5,4) node[above] {$y$};
\foreach \ypos in {0,1,2,3}
{
 \draw[thick] (-0.6,\ypos+0.5)--(-0.4,\ypos+0.5);
 \draw (-0.5,\ypos+0.5) node[left] {$\ypos$};
}
 
\end{tikzpicture}
 \caption{Four-terminal representation for site percolation on the Cairo pentagonal lattice.}
 \label{fig:cairosite}
\end{center}
\end{figure}
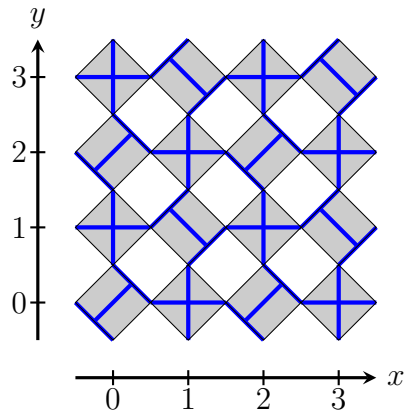

Figure~\ref{fig:cairosite} provides a four-terminal representation of the Cairo pentagonal lattice.
For the convenience of the drawing certain pairs of half-edges make an angle at the junction between
neighboring grey squares, but such a pair should of course just be considered a single edge.
The reader may verify that the lattice indeed consists of pentagons, and that going around each pentagon
the degrees of the vertices are 3, 3, 4, 3 and 4 as they should be. There is on average $\frac32$
vertices per grey square, and we must take $n$ even to respect the alternation of patterns.

The $\check{\sf R}$-matrix is given by that of the square lattice, Eq.~(\ref{Rsitesquare}), when $x+y$ is odd;
by that of the hexagonal lattice, Eq.~(\ref{Rsitehex}), when $x$ and $y$ are both even; and by a rotated
version of the latter when $x$ and $y$ are both odd.

\begin{table}
\begin{center}
 \begin{tabular}{l|l}
 $n$ & $p_{\rm c}$ \\ \hline
 2 & 0.6405124488065491504828379012985494181683890014987 \\
 4 & 0.6500236759295320258852272011860502886365174620953 \\
 6 & 0.6501636204301126095600239860376239822997457709415 \\
 8 & 0.6501786803344847213746555063732068861224526222353 \\ \hline
 $\infty$ & 0.6501834 (2) \\
 Ref.~\cite{Parviainen04} & 0.650184 (5) \\
 \end{tabular}
 \caption{Site percolation threshold $p_{\rm c}$ on the Cairo pentagonal lattice.}
  \label{tab:site-cairo}
\end{center}
\end{table}

Results for the thresholds $p_{\rm c}$ are given in Table~\ref{tab:site-cairo}.

\subsection{Frieze dual lattice $D(3^3,4^2)$}

\begin{figure}
\begin{center}
\begin{tikzpicture}[scale=1.0,>=stealth]
\foreach \xpos in {0,1,2,3}
\foreach \ypos in {0,1,2,3}
 \fill[black!20] (\xpos+0.5,\ypos) -- (\xpos+1,\ypos+0.5) -- (\xpos+0.5,\ypos+1) -- (\xpos,\ypos+0.5) -- cycle;

\foreach \xpos in {0,1,2,3}
\foreach \ypos in {0,2}
{
 \draw[blue,ultra thick] (\xpos,\ypos+0.5) -- (\xpos+1,\ypos+0.5);
 \draw[blue,ultra thick] (\xpos+0.5,\ypos) -- (\xpos+0.5,\ypos+1);
}

\foreach \xpos in {0,1,2,3}
\foreach \ypos in {1,3}
{
 \draw[blue,ultra thick] (\xpos,\ypos+0.5) -- (\xpos+0.5,\ypos+1);
 \draw[blue,ultra thick] (\xpos+0.5,\ypos) -- (\xpos+1,\ypos+0.5);
 \draw[blue,ultra thick] (\xpos+0.25,\ypos+0.75) -- (\xpos+0.75,\ypos+0.25);
}

\foreach \xpos in {0,1,2,3}
\foreach \ypos in {0,1,2,3}
 \draw[black] (\xpos+0.5,\ypos) -- (\xpos+1,\ypos+0.5) -- (\xpos+0.5,\ypos+1) -- (\xpos,\ypos+0.5) -- cycle;

\draw[very thick,->] (0,-0.5)--(4,-0.5);
\draw (4,-0.5) node[right] {$x$};
\foreach \xpos in {0,1,2,3}
{
 \draw[thick] (\xpos+0.5,-0.6)--(\xpos+0.5,-0.4);
 \draw (\xpos+0.5,-0.5) node[below] {$\xpos$};
}

\draw[very thick,->] (-0.5,0)--(-0.5,4);
\draw (-0.5,4) node[above] {$y$};
\foreach \ypos in {0,1,2,3}
{
 \draw[thick] (-0.6,\ypos+0.5)--(-0.4,\ypos+0.5);
 \draw (-0.5,\ypos+0.5) node[left] {$\ypos$};
}
 
\end{tikzpicture}
 \caption{Four-terminal representation for site percolation on the frieze dual lattice.}
 \label{fig:friezedualsite}
\end{center}
\end{figure}
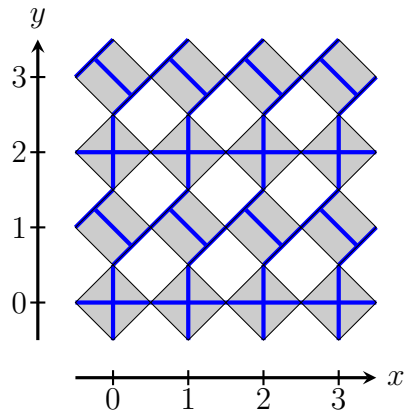

A four-terminal representation of the frieze dual lattice is shown in Figure~\ref{fig:friezedualsite}.
Its $\check{\sf R}$-matrix is that of the square lattice, Eq.~(\ref{Rsitesquare}), when $y$ is even;
and that of the rotated hexagonal lattice, cf.\ Eq.~(\ref{Rsitehex}), when $y$ is odd. All faces are
pentagons, and the reader may verify from the figure that the degrees of the surrounding vertices
are $(3^3,4^2)$ as they should be. There is on average $\frac32$ vertices per grey square.
Any parity of $n$ defines a valid basis, provided that we use $n \times 2n$ rectangular bases in
order to respect the alternation between rows.

The percolation thresholds $p_{\rm c}$ are displayed in Table~\ref{tab:site-friezedual}.

\begin{table}
\begin{center}
 \begin{tabular}{l|l}
 $n$ & $p_{\rm c}$ \\ \hline
 1 & 0.6180339887498948482045868343656381177203091798058 \\
 2 & 0.6378128305160144243967540410580253259072131959399 \\
 3 & 0.6450649046998213833516533917200464266392184415175 \\
 4 & 0.6463326652770031214369536045900433370913902143502 \\
 5 & 0.6467492557573062279714940630727565301149079566625 \\
 6 & 0.6469070727071462993287212489961496026254166173237 \\
 7 & 0.6469742627385098566946369955909988391276996580357 \\
 8 & 0.6470059040321607623572113208752443456723332267969 \\
 9 & 0.6470222011016419470419039106312292679463616691412 \\ \hline
 $\infty$ & 0.6470471 (2) \\
 Ref.~\cite{Parviainen04} & 0.647084 (5) \\
 \end{tabular}
 \caption{Site percolation threshold $p_{\rm c}$ on the frieze dual lattice.}
  \label{tab:site-friezedual}
\end{center}
\end{table}

\section{Discussion}
\label{sec:disc}

In this paper we have presented a new algorithm for the computation of the graph polynomial $P_B(q,v)$
associated with the $q$-state Potts model. This polynomial was
introduced in \cite{Jacobsen12} as a generalization of the bond percolation polynomial $P_B(1,v)$ studied in
\cite{ScullardZiff06,ScullardZiff08,ScullardZiff10}. Just like properties of $P_B(1,v)$ were investigated for
increasingly larger bases in \cite{Scullard11,Scullard11-2,Scullard12}, we have here pursued the effort to
compute $P_B(q,v)$ for the largest possible bases. This was achieved by means of a reformulation of the
defining Eq.~(\ref{PB_cluster}) within
the periodic Temperley-Lieb algebra, and the use of powerful transfer matrix techniques. Thus, the computations
in \cite{Jacobsen12} for bases of size up to $|E|=36$ edges using the original deletion-contraction definition of
$P_B(q,v)$---and that were improved to $|E|=144$ in \cite{Jacobsen13} by means of a first, non-periodic transfer matrix
approach (and even to $|E|=243$ using supercomputer facilities \cite{Jacobsen13})---were here carried to $|E|=882$ through
the application of the novel transfer matrix formulation. Presumably these sizes can be pushed still farther by working
out a parallelised version of the present algorithm and using supercomputer facilities \cite{Scullard14}.

These technical improvements have enabled us to refine the precision of the bond percolation thresholds $p_{\rm c}$
(for $q=1$) and critical temperatures $v_{\rm c}$ (for general $q$) much beyond what was previously possible.
For instance, from numerical simulations the values of $p_{\rm c}$ are known to a precision of the order $10^{-8}$ for the
best-studied lattices. While the previous work on graph polynomials \cite{Jacobsen13} yielded a comparable precision,
this has now been carried to the order $10^{-13}$, which seems beyond the possibilities of Monte Carlo simulations
for many years to come. This progress was made possible because of the extremely rapid convergence of the
estimates coming from bases of finite size $n$, a fact that enabled us to extrapolate to the $n \to \infty$ limit using
standard acceleration of convergence techniques.

%
It may appear surprising at first sight that the use of extrapolation
techniques may add, in the most favourable cases, 2 or even 3 extra
digits of precision to those that appear to have converged from a
mere visual inspection of the last two finite-$n$ results. We stress that
this kind of accuracy relies both on the rapid and well-behaved
convergence---and in particular on the high value of $w$ in
(\ref{extrapol_w})---and on the fact that our finite-$n$ data are
exact results, i.e., without error bars. The tables in this paper
report these results with 50-digit numerical precision, enabling
sceptical readers to check our final results with their own
extrapolation procedures.

%
The comments about the precision of the results also apply to the values of $v_{\rm c}$ for general $q$, in particular for the
notably tricky case of $q=4$ where numerical simulations of the Monte Carlo or transfer matrix type are hampered
by strong logarithmic corrections to scaling, whilst the graph polynomial method seems to encounter no noticeable
loss of precision.

To illustrate the versatility of the graph polynomial method we have also extended the study of the most common
lattices \cite{Jacobsen13} to a considerably larger set consisting of all Archimedean lattices, their duals, and their medials. For some of these
lattices parity constraints on $n$ applied, implying fewer data points and hence less precise extrapolations. It nevertheless remains true
that we have been able to improve---often significantly---on the precision of the bond percolation thresholds for
essentially all of the lattices studied here, and that have previously been addressed using other methods (see \cite{WikiPercThres} for
a review). Other lattices, notably some of the medial lattices, have to our knowledge never been investigated before.

The $q$-state Potts model appears to have been studied only on a more restricted class of lattices than percolation.
Therefore, many of our results for the Potts model are new and cannot be compared to existing work.

An important feature of the graph polynomial $P_B(q,v)$ is that its factorisation appears to signal cases of exact
solvability \cite{Jacobsen12,Jacobsen13}. This is nowhere more conspicuous than for the Ising model ($q=2$),
for which our factorised results successfully reproduce the known critical temperatures $v_{\rm c}$ for all of the
Archimedean lattices \cite{Codello10}. We have presented this type of results also for all the medial lattices,
which do not appear to have been much studied before.

More generally, in sections~\ref{sec:factorisation} and \ref{sec:factorisation_medial} we have systematically searched
for cases where $P_B(q,v)$ factorises. We have thus identified many cases of presumed exact solvability, including
models of spanning forests, chromatic and flow polynomials, and Potts models with integer values of $q$. Each of
these cases deserves a specific study, opening possibilities for much future work. We should also point out that
the three-state antiferromagnet, $(q,v)=(3,-1)$, appears to play a special role for many of the lattices studied: even
when the curves $P_B(q,v)=0$ do not pass though this point exactly, they often come increasingly close upon increasing
the size $n$.

Apart from producing extremely precise numerical values for $p_{\rm c}$ and $v_{\rm c}$, the graph polynomials have
also emerged as a powerful tool for studying the full phase diagram of the Potts model in the real $(q,v)$ plane.
While this was realised already in the first study of the kagome lattice \cite{Jacobsen12}, and increasingly so with the
extension to some of the other most well-known lattices \cite{Jacobsen13}, we have here seen many new features
emerge as results of the improved precision and the greater variety of lattices being studied. In particular, the presence
of phase transitions at the Beraha numbers, $q = B_k$ with even index $k=4,6,8,\ldots$ in (\ref{Beraha}), has been
firmly established for all the lattices. These transitions take place inside the Berker-Kadanoff phase \cite{Saleur91}
whose extent can thus be inferred from the size and positions of the vertical rays at the Beraha numbers.

Beyond this, the phase diagram inside the antiferromagnetic region $v < 0$ turns out to be dauntingly complicated and highly
lattice dependent. Often extra curves exist outside the BK phase, or cutting through it, and the question whether these
curves are critical and, if so, what is the precise nature of the criticality, remains in most cases open. In particular, numerical
transfer matrix studies would be a precious help to assess whether the parts of these curves with $q \in [0,4]$ enjoy conformal
invariance. A first such study was made for the kagome lattice in \cite[Figure~12]{Jacobsen12}, but clearly much more work
is needed. The cases where extra curves are present inside the BK phase are particularly interesting, since then presumably
the relevant critical theory consists of conformal excitations on top of a highly excited level inside the BK phase that could
only be accessed by changing from the loop (or FK cluster) representation to another (e.g., RSOS height) representation
of the Temperley-Lieb algebra. Another intriguing feature is the possibility of the critical curves to be space filling, the strongest
evidence for which is provided by the frieze medial lattice (see in particular Figure~\ref{fig:friezemedial-pd-zoom}).
This is reminiscent of the ``critical regions'' found in recent transfer matrix studies of the $q$-state Potts model \cite{Salas13-1},
including on planar lattices \cite{Salas14}. Finally we note that most of the phase diagrams tend to become increasingly
complicated as $q \to 4$ (see Figure~\ref{fig:threetwelve-pd} for a vivid illustration), and that for $q > 4$---a regime which
we have chosen not to discuss in the present paper---the curves again appear to develop a space filling behaviour for several lattices.
This should again be compared with the outcome of other studies of the critical behaviour of antiferromagnetic Potts models
close to \cite{Baxter86,Baxter87}, at \cite{Deng11} and beyond \cite{Salas13-1,Deng13} $q=4$. Obviously,
a detailed study of all these aspects for at least one of the lattices presented here would add much substance to
these observations \cite{Salas14}.

On a more technical level, we have shown how to write all the lattices under investigation in the four-terminal representation
of Figure~\ref{fig:square-basis}. In many cases, these representations are not at all obvious to come by (see section~\ref{sec:cross}
and Figure~\ref{fig:cross} for an example). The importance of such representations transcends the applications made in the present paper. In particular,
these representations provide, for each lattice, an efficient sparse-matrix factorised construction of the transfer matrix that
can be immediately applied in numerical studies. It forms part of the interest of studying so many different lattices, that we
have discovered several tricks by which the basic four-terminal representation can be adapted to new situations. These tricks
include the addition of horizontal diagonals on the white squares (section~\ref{sec:triangular}), their extension to ``white hexagonals''
with even more structure (section~\ref{sec:snubhex}), the ``generic $\check{\sf R}$-matrix algorithm'' that avoids having to add
extra auxiliary points inside the grey squares (section~\ref{sec:foureightmedial}), the use of contracted edges (section~\ref{sec:crossmedial}),
and the possibility of having correlated vertices (section~\ref{sec:rubysite}). Some of these improvements of the general method
have turned out particularly useful in order to be able to access also site percolation on a variety of lattices.

We finally comment on the values of the parameter $w$ appearing in the finite-size scaling form (\ref{extrapol_w}). For simplicity
we focus here on the case of (bond or site) percolation on the lattices for which we have a sufficiently (at least seven) large number
of sizes $n$ to allow for a precise determination of the effective values of $w$ (the numbers quoted are based on the three largest sizes).
We have seen that $w$ takes essentially the same values for bond percolation on the kagome ($w \approx 6.36$), the three-twelve
($w \approx 6.39$), and the three-twelve medial lattice ($w \approx 6.38$), as well as for site percolation on the hexagonal lattice
($w \approx 6.35$). It is remarkable that all those lattices have a three-fold rotational symmetry. Another value of $w$ is taken
for bond percolation on the four-eight ($w \approx 4.28$), the four-eight medial ($w \approx 4.29$), and the frieze medial lattice
($w \approx 4.59$), as well as for site percolation on the square ($w \approx 4.07$), the four-eight ($w \approx 4.40$), and the
frieze dual lattice ($w \approx 4.25$). Although there is a larger spread in those values, it is tempting to speculate that they might
in fact be finite-size approximations to a common value applying to the lattices with a four-fold rotational symmetry. We stress that
it is the value for site percolation on the square lattice, $w \approx 4.07$, that is determined with the largest reliability, since it is
based on eleven sizes $n$ (see section~\ref{sec:square_site}). It is conceivable that the exact value might be $4 + \frac{5}{48} \simeq 4.104\cdots$,
where $2h = \frac{5}{48}$ is the critical exponent giving the co-dimension of a bulk percolation cluster. This corresponds in the transfer
matrix formalism \cite{JacobsenReview} to setting the weight of winding loops $n_{\rm wind} = 0$, which, as we have seen in
section~\ref{sec:action-TL}, is the proper condition to discard contributions from $Z_{\rm 1D}$ in (\ref{PB_cluster}).
If the value of $w$, and more generally of further correction-to-scaling exponents, could be determined exactly one should be able
to produce even more precise extrapolations $p_{\rm c}$ from the existing data.%
\footnote{See \cite{GuoBlote02} for an example of what gain in precision in the determination of the critical point
can be obtained from a thorough understanding of the finite-size scaling exponents.}
We leave this important question for future work.

\section*{Acknowledgments}

The author's work was supported by the Agence Nationale de la Recherche
(grant ANR-10-BLAN-0414:~DIME) and the Institut Universitaire de
France.  He warmly thanks C.~R.~Scullard for collaboration
on related projects, and for many helpful comments and encouragement throughout
the process of writing up this long paper. The author would like to express his gratitude for the hospitality of
the Galileo Galilei Institute for Theoretical Physics, Florence,
where a crucial part of this work was carried out (in April 2012). He also thanks A.J.~Guttmann
for discussions about extrapolation methods.
The numerical computations reported here were performed at the High Performance
Computing center at New York University and made possible by a
generous donation from the Dell Corporation and the kind permission of A.D.~Sokal.

\section*{References}
\bibliographystyle{iopart-num}
\bibliography{SJ}

\end{document}